\documentclass{birkmono_lieb}
\usepackage{amssymb,amsthm,amsmath,graphicx}
\usepackage{amsbsy}
\usepackage{makeidx}
\makeindex

 \newtheorem{thm}{Theorem}[chapter]
 \newtheorem{corollary}[thm]{Corollary}
 \newtheorem{lem}[thm]{Lemma}
 
 \theoremstyle{definition}
 
 \theoremstyle{remark}

\newcommand{\beqa}{\begin{eqnarray}}
\newcommand{\eeqa}{\end{eqnarray}}

\def\uprho{\raise1pt\hbox{$\rho$}}

\def\R{{\mathbb R}}

\def\mfr#1/#2{\hbox{${\frac{#1}{#2}}$}}

\newcommand{\brtf}{\bar\rho}
\newcommand{\Egp}{E^{\rm GP}}

\def\R{{\mathbb R}}

\def\sc{{\mathcal C}}
\def\E{{\mathcal E}}

\def\x{{\bf x}}
\def\p{{\bf p}}
\def\z{{\bf z}}
\def\q{{\bf q}}
\def\y{{\bf y}}
\def\k{{\bf k}}
\def\0{{\bf 0}}

\def\const{{\rm const.\,}}

\def\mfr#1/#2{\hbox{$\frac{{#1}}{{#2}}$}}
\def\uprho{\raise1pt\hbox{$\rho$}}
\def\upchi{\raise1pt\hbox{$\chi$}}
\def\dlambda{\lower1pt\hbox{$\lambda$}}
\newcommand{\xij}{|\x_i-\x_j|}

\newcommand{\rtf}{\rho^{\rm TF}}

\newcommand{\mtf}{\mu^{\rm TF}}

\newcommand{\hn}{{\mathord{\widehat{n}}}}
\newcommand{\an}{{\mathord{a}}^{\phantom{*}}}
\newcommand{\bn}{{\mathord{b}}^{\phantom{*}}}
\newcommand{\cA}{{\mathord{\mathcal A}}}
\newcommand{\cB}{{\mathord{\mathcal B}}}
\newcommand{\vp}{\mathord{\hbox{\boldmath $\varphi$}}}
\newcommand{\Z}{\mathbb{Z}}

\newcommand{\beq}{\begin{equation}}
\newcommand{\eeq}{\end{equation}}

\newcommand{\eps}{\varepsilon}
\newcommand{\K}{{\mathcal K}}
\newcommand{\X}{{\bf X}}

\newcommand{\Tr}{{\rm Tr}\, }
\newcommand{\half}{\mbox{$\frac{1}{2}$}}

\newcommand{\pgp}{\phi^{\rm GP}}
\newcommand{\al}{\alpha}
\newcommand{\rmax}{\rho_{\al,{\rm max}}}
\newcommand{\rmin}{\rho_{\al,{\rm min}}}
\newcommand{\g}{g}
\newcommand{\dn}{{\mathord{d}}^{\phantom{*}}}
\newcommand{\cn}{{\mathord{c}}^{\phantom{*}}}
\newcommand{\cP}{{\mathord{\mathcal P}}}

\newcommand{\cK}{{\mathord{\mathcal K}}}
\renewcommand{\u}{{\bf u}}

\newcommand{\vecp}{{\p}}

\newcommand{\vecx}{{\x}}
\newcommand{\vecy}{{\y}}

\newcommand{\ulx}{\x}
\newcommand{\ulk}{\k}
\newcommand{\ulp}{{\bf p}}
\newcommand{\ubp}{{\bf p}}
\newcommand{\ubr}{\x}
\newcommand{\ulr}{\x}
\newcommand{\ulq}{{\bf q}}

\newcommand{\txk}{k}
\newcommand{\Ap}{\alpha}

\newcommand\bfnab\nabla
\newcommand\niv{{\bf n}}
\newcommand\D{d}

\newcommand{\aoo}{{a_{\bf 0}^{\phantom *}}}
\newcommand{\aos}{a_{\bf 0}^*}
\newcommand{\ak}{{a_{\bf k}^{\phantom *}}}
\newcommand{\aks}{a_{\bf k}^*}
\newcommand{\aq}{a_{\bf q}}

\newcommand{\no}{n_{\bf 0}}
\newcommand{\kk}{{\bf k}}
\newcommand{\pp}{{\bf p}}
\newcommand{\qq}{{\bf q}}
\newcommand{\hi}{\mathcal{H}}
\newcommand{\hip}{{\mathcal{H}'}}

\newcommand{\be}{\begin{equation}}
\newcommand{\ee}{\end{equation}}
\newcommand{\bea}{\begin{eqnarray}}
\newcommand{\eea}{\end{eqnarray}}
\newcommand{\atd}{a_{2\rm D}}
\newcommand{\xp}{x}

\def\Tr{\operatorname{tr}}

\def\o1{{\mathrm{o}(1)}}

\begin{document}

\pagenumbering{roman}
\title{ \vspace{-6cm} $\phantom{x}$ \\ 
{\bf {\underline{\underline{$ {\cal THE\quad MATHEMATICS\quad OF\quad THE\quad
        BOSE}$}}}} \\ \vspace{1truecm} {\bf {\underline{\underline{${\cal
    GAS\quad AND \quad ITS\quad CONDENSATION}$}}}} \\ \vspace{5cm} } 
\author{{\Large Elliott H. Lieb, Robert Seiringer,} \\ \\
{\Large Jan Philip Solovej and Jakob Yngvason}}
\date{}
\maketitle

\noindent
\begin{tabular}{ll}
Elliott H. Lieb  & Robert Seiringer \\
Departments of Mathematics and Physics & Department of Physics\\
Princeton University & Princeton University \\ 
Jadwin Hall, P.O. Box 708 & Jadwin Hall, P.O. Box 708\\
Princeton, NJ 08544 & Princeton, NJ 08544\\
USA & USA\\
\texttt{lieb@math.princeton.edu} & \texttt{rseiring@math.princeton.edu}
\\
\\
\\
Jan Philip Solovej & Jakob Yngvason \\
Department of Mathematics & Institut f\"ur Theoretische Physik\\
University of Copenhagen & Universit\"at Wien\\
Universitetsparken 5 & Boltzmanngasse 5\\
2100 Copenhagen & 1090 Wien\\
Denmark & Austria\\
\texttt{solovej@math.ku.dk} & \texttt{yngvason@thor.thp.univie.ac.at}
\end{tabular}

\vspace{3truecm}

\noindent
This book was first published by Birkh\"auser Verlag (Basel-Boston-Berlin)
in 2005 under the title \lq\lq The Mathematics of the Bose Gas and its
Condensation\rq\rq\ as number 34 of its \lq\lq Oberwolfach Seminar\rq\rq\ series.

\medskip
\noindent The present version differs from the originally published version in
several respects. Various minor errors have been corrected, Chapter 9
has been slightly expanded, references have been updated, and Chapter
12 has been added in which results obtained after the publication are
noted.

\bigskip

\noindent ISBN 3-7643-7336-9

\bigskip

\noindent $\copyright$ 2005 Elliott H. Lieb, Robert Seiringer, Jan Philip Solovej
and Jakob Yngvason. This material may be copied for non-commercial
purposes, provided the source is properly acknowledged.

\newpage
\baselineskip=14pt

\setcounter{tocdepth}{1}
\tableofcontents
\newpage
\section*{Preface}
\addcontentsline{toc}{section}{Preface}
\markboth{Preface}{Preface}\thispagestyle{empty}

\renewcommand{\baselinestretch}{1.2}

The mathematical study of the Bose gas goes back to the first quarter
of the twentieth century, with the invention of quantum mechanics. The
name refers to the Indian physicist S.N.  Bose\index{Bose} who
realized in 1924 that the statistics governing photons (essentially
invented by Max Planck in 1900) is determined (using modern
terminology) by restricting the physical Hilbert space to be the
symmetric tensor product of single photon states. Shortly afterwards,
Einstein\index{Einstein} applied this idea to massive particles, such
as a gas of atoms, and discovered the phenomenon that we now call
Bose-Einstein condensation\index{Bose-Einstein condensation}. At that
time this was viewed as a mathematical curiosity with little
experimental interest, however.

The peculiar properties of liquid Helium
\index{Helium}
(first liquefied by
Kammerlingh Onnes
\index{Kammerlingh Onnes}
in 1908) were eventually viewed as an experimental
realization of Bose-Einstein statistics
\index{Bose-Einstein statistics}
applied to Helium atoms. The
unresolved mathematical problem was that the atoms in liquid Helium
are far from the kind of non-interacting particles envisaged in
Einstein's theory, and the question that needed to be resolved was
whether Bose-Einstein condensation really takes place in a strongly
interacting system --- or even in a weakly interacting system.

That question is still with us, three quarters of a century later!

The first systematic and semi-rigorous mathematical treatment of the
problem was due to Bogoliubov in 1947, but that theory, while
intuitively appealing and undoubtedly correct in many aspects, has
major gaps and some flaws.  The 1950's and 1960's brought a renewed
flurry of interest in the question, but while theoretical intuition
benefited hugely from this activity the mathematical structure did not
significantly improve.

The subject was largely quiescent until the 1990's when experiments on
low density (and, therefore, weakly interacting instead of strongly
interacting, as in the case of liquid Helium) gases showed for the
first time an unambiguous manifestation of Bose-Einstein condensation.
This created an explosion of activity in the physics community as can
be seen from the web site \\
\qquad \texttt{http://bec01.phy.georgiasouthern.edu/bec.html/bibliography.html},
\\
which
contains a bibliography of several thousand papers related to BEC
written in the last 10 years.

At more or less the same time some progress was made in obtaining
rigorous mathematical proofs of some of the properties proposed in the
50's and 60's.  A general proof of Bose-Einstein condensation for
interacting gases still eludes us, but we are now in a much stronger
position to attack this problem rigorously. These notes, which are an
extension of our 2004 Oberwolfach course, summarize some rigorous
results that have been obtained by us in the past decade.  Most of
them are about the ground state energy in various models and
dimensions, but we do have a few results about the occurrence (and
non-occurrence) of condensation.

This pedagogical summary has several antecedents. It has grown
organically as new results emerged. The first one was \cite{LYbham},
followed by \cite{LSYdoeb}, \cite{S4}, \cite{L3}, \cite{LSYnn},
\cite{LSSY}, and
\cite{LSSY2}.  Apart from this stream, there was another pedagogical
survey going back to the 60's \cite{EL2} that dealt with Bogoliubov
theory and other things.
Some of that material is reproduced in Appendices~\ref{chap2} and~\ref{chap3}.

 There is, of course, a large body of rigorous work by other people on
 various aspects of BEC that was not covered in the Oberwolfach course
 and is not mentioned in these notes. The subject can be approached
 from many angles and our aim was not to give a complete overview of
 the subject but to focus on themes where we have been able to make
 some contributions. The recent Physics Reports article \cite{zagrebnov} on
 the Bogoliubov model is a good source of references to some other
 approaches and results. There exist also several reviews, e.g.,
 \cite{DGPS,BEC,Legg, C, Yu} and even monographs \cite{PS,PiSt2003} on
 the fascinating physics of the Bose gas and its condensation.

\vfill

\addcontentsline{toc}{section}{Acknowledgments}
\subsection*{Acknowledgments}
Our thanks go to Letizzia Wastavino for producing the two figures and
to Michael Aizenman and Kai Schnee for permission to include the
material in Chapters~\ref{xychap} and~\ref{3dto2d}, respectively. We
are also grateful to Daniel Ueltschi, Bruno Nachtergaele and Valentin
Zagrebnov for very useful comments on various parts of these notes.
Finally, we thank the Oberwolfach Institute for the opportunity to give
this course and we thank the participants for pointing out numerous
typos in a previous version of the manuscript. The work was supported
in part by US NSF grants PHY 0139984-A01 (EHL), PHY 0353181 (RS) and
DMS-0111298 (JPS); by an A.P.~Sloan Fellowship (RS); by EU grant
HPRN-CT-2002-00277 (JPS and JY); by the Alexander von Humboldt
Foundation (EHL); by FWF grant P17176-N02 (JY); by MaPhySto -- A
Network in Mathematical Physics and Stochastics funded by The Danish
National Research Foundation (JPS), and by grants from the Danish
research council (JPS).


\chapter{Introduction}\label{intro}
\thispagestyle{empty}
\pagenumbering{arabic}\setcounter{page}{1}

\section{The Ideal Bose Gas}\label{ideal}

Schr\"odinger's equation\index{Schr\"odinger equation} of 1926 defined
a new mechanics whose Hamiltonian\index{Hamiltonian} is based on
classical mechanics. The ``ideal" gas\index{ideal gas} of particles
consists of the following ingredients: A collection of $N \gg 1$
non-interacting particles in a large box $\Lambda \subset \R^3$ and
volume $V= L^3$. We are interested in the ``thermodynamic
limit''\index{thermodynamic!limit}, which means that we will take
$N\to \infty$ and $L\to \infty$ in such a way that the density $\rho
=N/V$ is held fixed.

The fact that the particles are non-interacting\index{non-interacting particles}
means that the classical energy, or Hamiltonian
$H$\index{Hamiltonian}, is entirely kinetic energy \index{kinetic
  energy} and this, in turn, is
\begin{equation}
H =  \frac{1}{2m} \sum_{i=1}^N \p_i^2
\end{equation}
where $m$ is the particle mass \index{mass} and $\p_i$ is the momentum
\index{momentum} of particle $i$.  The lowest (or {\it ground
  state}\index{ground state!energy}) energy of this classical system
is, of course, $0$. The {\it thermodynamic
  properties}\index{thermodynamic!properties} are determined from the
{\it partition function}\index{partition function}
\begin{equation}
Z_N= \frac {1}{h^{3N}N!} \int_{\Lambda^N} \prod_{i=1}^N d\x_i  \, \int_{\R^{3N}}
\prod_{i=1}^N d\p_i \,
e^{-\beta H} = \frac {V^N}{h^{3N} N!} \left(\frac {2m}\beta\right)^{3N/2}
\end{equation}
(with $\beta = 1/k_{\rm B } \ T$, $k_{\rm B}$ is Boltzmann's constant
and $T$ is the temperature,\index{temperature}
and with $h$ an arbitrary constant with
the dimension of momentum times length) in terms of which the {\it
  free energy}\index{free energy}
 is given as
\begin{equation}
F = - \frac 1\beta \ln Z_N .
\end{equation}
The pressure\index{pressure} $p$ is
\beq p=-\partial F/\partial V=\rho k_{\rm B}T
\eeq
where Stirling's approximation,  $\ln N!\approx N\ln N$, has been used.
The average energy is
\beq E=-\frac{\partial}{\partial\beta}\ln Z=\frac 32 Nk_{\rm B}T.
\eeq

In quantum mechanics, the Hamiltonian is an operator obtained by
replacing each $\p_j$ by $-{\rm i} \hbar \nabla_j$, acting on the Hilbert
space\index{Hilbert space} $L^2(\Lambda)$, with appropriate boundary conditions. The
eigenvalues of $\p^2 = -\hbar^2 \Delta$ (with $\Delta$ =Laplacian
\index{Laplacian}
=$\nabla^2$), for a box with periodic boundary conditions\index{boundary condition},
are
$(2\pi\hbar)^2{\bf n}^2/L^2$, where ${\bf n}$ is a vector with
integer components. If the statistics of the particles is disregarded,
the partition function\index{partition function}, which in the quantum case is given by \beq
\label{partf} Z_N= \frac 1{N!} \Tr\, e^{-\beta
  H}, \eeq factorizes as $Z_N = Z_1^N /N!$. This equals the classical
expression in the thermodynamic limit, if one takes $h=2\pi \hbar$.

Taking the statistics\index{particle statistics} of the particles into account, we have to
restrict the trace in (\ref{partf}) to the symmetric or
anti-symmetric subspace of the total Hilbert space\index{Hilbert space} $\bigotimes^N
L^2(\Lambda)$, depending on whether we intend to describe bosons or
fermions. This makes the prefactor $1/N!$ superfluous, but one has to
face the problem that $Z_N$ is no longer determined by $Z_1$. For this
reason, it is more convenient to pass to the grand-canonical partition
function\index{partition function!grand-canonical} (or generating function)
\begin{equation}
 \Xi = \sum_{N\geq 0} Z_N z^N,
\end{equation}
where $z=e^{\beta \mu}$ is
the fugacity\index{fugacity} for chemical potential\index{chemical potential} $\mu$. The chemical potential
\index{chemical potential} is
then determined by the average particle number, \beq \langle N\rangle
= z \frac {\partial }{\partial z} \ln \Xi.  \eeq The grand-canonical
\index{partition function!grand-canonical}
partition function $\Xi$ can be calculated because it factorizes into
the contributions from the single particle energy levels. For bosons the
result is \beq \Xi =
\prod_{i\geq 0} \frac{1}{1- \exp\left(-\beta(\eps_i - \mu)\right)},
\eeq
where $\eps_0\leq \eps_1\leq \dots$ denote the single-particle
energy levels (given, in this case, by $(2\pi\hbar)^2{\bf n}^2/(2m L^2)$, with
${\bf n} \in {\mathbb Z}^3$). Note that in this ``free particle"
bosonic case it is necessary that $\mu< \eps_0$. In the thermodynamic limit $\varepsilon_0\to0$.

{F}or fixed $\mu<0$,
the average particle number, is given by
\beq\label{sumn} \langle N
\rangle = \sum_{i\geq 0} \frac { 1}{\exp\left(\beta(\eps_i -
    \mu)\right)-1}.
\eeq
In the thermodynamic limit, the sum becomes
an integral (more precisely, $L^{-3} \sum_{\bf  p} \to (2\pi \hbar)^{-3}
\int_{{\mathbb  R}^3}$), and we have
\beq \lim_{L\to \infty} \frac {\langle N
  \rangle}{L^3} \equiv \rho= h^{-3} \int d\p\, \frac {
  1}{\exp\left(\beta(\p^2/(2m) - \mu)\right)-1}.
\eeq
This is a
monotonously increasing function of $\mu$, which is bounded as $\mu\to
0$, however, by the {\it critical density}\index{critical density}
\beq \rho_c(\beta) =
g_{3/2}(1) (2\pi \hbar^2 \beta/m)^{-3/2}.  \eeq Here
$g_{3/2}(1)=\sum_{\ell=1}^\infty \ell^{-3/2} \approx 2.612$. That is,
the density seems to be bounded by this value. This is absurd, of
course. This phenomenon was discovered by Einstein
\cite{Einstein},
and the resolution
is that the particles exceeding the critical number all go into the lowest
energy state. In mathematical terms, this means that we have to let
$\mu\to 0$ {\it simultaneously} with $L\to \infty$ to fix the density
at some number $>\rho_c$. In this case, we have to be more careful in
replacing the sum in (\ref{sumn}) by an integral. It turns out to be
sufficient to separate the contribution from the lowest energy level,
and approximate the contribution from the remaining terms by an
integral.  The result is that, for $\rho>\rho_c$,
\beq\label{rhow}
\rho = \rho_c(\beta) + \rho_0, \eeq where \beq \rho_0 =
\lim_{L\to\infty} \frac 1{V} \frac
1{\exp\left(\beta(\eps_0-\mu)\right)-1}
\eeq
is the density of the
\lq\lq condensate\rq\rq. The dependence of $\mu$ on $L$ is determined
by (\ref{sumn}), writing $\langle N\rangle = L^3 \rho$ with fixed
$\rho$.

The phenomenon that a single particle level has a macroscopic
occupation\index{macroscopic occupation},
i.e., a non-zero density in the thermodynamic limit, is
called {\it Bose-Einstein condensation} (BEC)\index{Bose-Einstein condensation}.
Note that in the model
considered there is no condensation into the excited energy levels, and
one always has
\beq
\lim_{L\to\infty} \frac 1{V} \frac 1{\exp\left(\beta(\eps_i-\mu)\right)-1}= 0
\eeq
for $i\geq 1$, since $\eps_i-\mu \geq \eps_i-\eps_0 = \const L^{-2}$.

Note that in the case of zero temperature, i.e., the ground state,
{\it all} the particles are in the condensate, i.e., $\rho=\rho_0$.
In a sector of fixed particle number, the ground state wave function
is simply a product of single particle wave-functions in the lowest
energy state.

\section{The Concept of Bose-Einstein Condensation}\label{defbecsect}

\index{Bose-Einstein condensation}
So far we have merely reproduced the standard textbook discussion of
BEC for non-interacting particles. The situation changes drastically
if one considers interacting systems, however. For particles
interacting via a pair potential $v(|\x_i-\x_j|)$, the Hamiltonian\index{Hamiltonian}
takes the form

\begin{equation}\label{hamnew}
H_{N} = - \frac{\hbar^2}{2m}\sum_{i=1}^{N} \Delta_i +
\sum_{1 \leq i < j \leq N} v(|\x_i - \x_j|).
\end{equation}
(We could also include three- and higher body potentials, but we
exclude them for simplicity. These do exist among real atoms, but for
understanding the basic physics it is presumably sufficient to
consider only pair potentials.)

Even at zero temperature, it is not entirely obvious what is meant by
a macroscopic occupation of a one-particle state, because the
eigenfunctions of $H_N$ are {\it not} products of single particle
states.

The concept of a macroscopic occupation\index{macroscopic occupation}
of a single one-particle state
acquires a precise meaning through the {\it one-particle density
matrix}\index{one-particle density matrix}. Given the normalized ground state wave function
\index{ground state}
of $H_N$ (or any other
wave function, for that matter), $\Psi_0$, this is the
operator on $L^2(\R^3)$ given by the kernel
\begin{equation}\label{defgammanew}
 \gamma(\x,\x')=N\int \Psi_0(\x,\X)
\Psi_0(\x',\X) d\X \ ,
\end{equation}
 where we introduced the short hand
notation
\begin{equation}\label{defX}
\X=(\x_2,\dots,\x_N)\qquad{\rm and}\quad  d\X=\prod\limits_{j=
2}^N d\x_j.
\end{equation}
Then $\int \gamma(\x, \x) d\x =\Tr[\gamma] = N$. {\it BEC in the
  ground state means, by definition, that this operator has an
  eigenvalue of order $N$ in the thermodynamic limit.}
\index{Bose-Einstein condensation} This formulation was first stated
in \cite{penrose} by\index{Penrose}\index{Onsager} Penrose and
Onsager. For the ground state $\Psi_0$ of $H_N$, the kernel $\gamma$
is positive and, hopefully, translation invariant in the thermodynamic
limit, and hence the eigenfunction belonging to the largest eigenvalue must be
the constant function $L^{-3/2}$. Therefore, another way to say that
there is BEC in the ground state is that
\begin{equation}\label{defbecnew}
 \frac 1{V} \int\!\!\!\int \gamma(\x,\, \y) \,d\x\,
d\y = \textrm{O}(N)\
\end{equation}
as $N\to \infty$, $L\to \infty$ with $N/L^3$ fixed; more precisely
Eq.\ \eqref{defbecnew} requires that there is a $c>0$ such that the left
side is $>cN$ for all large $N$.

This concept of BEC as a large eigenvalue of the one-particle reduced
density matrix immediately generalizes to thermal states, both in the
canonical and grand-canonical ensembles (or, more generally, to states
defined by arbitrary density matrices). For interacting systems,
however, the ground state already poses a challenging problem which is
still largely unsolved. In fact, BEC has, so far, never been proved for
many-body Hamiltonians with genuine interactions --- {\it except for one
special case:} hard core bosons\index{hard core interaction}\index{lattice gas}\index{half-filling}
on a lattice at half-filling  (i.e.,
$N=$ half the number of lattice sites). The proof was given in
\cite{DLS} and \cite{KLS}. This, and a generalization to a lattice gas
in a periodic external potential, is described in
Chapter~\ref{xychap}.

There are physical situations where it is natural to consider
generalizations of the concept of BEC just described. In particular,
for trapped
\index{trapped Bose gas}
Bose gases, as considered in
Chapters~\ref{sectgp}--\ref{3dto2d}, the system is inhomogeneous and
the thermodynamic limit at fixed density has to be replaced by an
appropriate scaling
\index{scaling} of the potentials involved.\footnote{Also, the way
  to take a thermodynamic limit is not completely unambiguous. In
  the standard van Hove limit \cite{ruelle} the volume grows
  essentially without changing the shape of the domain $\Lambda$, e.g., the
  growing cubes considered in Section~\ref{ideal}. If instead the
  limit is taken in such a way that the size of $\Lambda$ grows at
  different speed in different directions, one may obtain different
  results (see, e.g., \cite{pule,pule2}).} What remains as the criterion of
BEC is the occurrence of a large eigenvalue (i.e., of the order of the
particle number) of the one-particle density matrix in the limit
considered.

Finally, we comment on the relation between BEC and {\it spontaneous
breaking of gauge symmetry}\index{gauge symmetry!breaking}.\index{symmetry breaking}
Gauge symmetry in the present context is
defined by the one-parameter group of unitary transformations in Fock
space
\index{Fock space}
generated by the particle number operator. The Hamiltonian
(\ref{hamnew}) preserves particle number and this implies gauge
invariance of the grand-canonical
\index{grand-canonical}
equilibrium state on the Fock space,
i.e., the direct sum of all the $N$-particle Hilbert spaces. This
symmetry can be explicitly broken by adding a term $\sqrt{V} \lambda
(\aoo+\aos)$ to the Hamiltonian, where $\aos$ and $\aoo$ are the
creation and annihilation operators of the lowest energy mode (i.e.,
in a box, the constant wave function in the one-particle
space). Formally, this is analogous to adding an external
\index{magnetic field}
magnetic
field to the Hamiltonian of a magnet.  In the grand-canonical state
defined by the so modified Hamiltonian the operator $\aoo$ has a
non-zero expectation value (which goes to zero as $\lambda\to 0$ for
any {\it fixed} volume $V$). Gauge symmetry breaking means that this
expectation value, divided by $\sqrt{V}$, remains non-zero even as
$\lambda\to 0$, {\it after} the thermodynamic limit has been
taken. (For the magnet, this corresponds to spontaneous
magnetization.)

In Appendix~\ref{justapp}, we show that under quite general
assumptions, BEC goes hand in hand with spontaneous gauge symmetry
breaking. Breaking of a continuous symmetry is notoriously difficult
to prove, and in one and two dimensions it is excluded, at least at
positive temperature, by the Hohenberg-Mermin-Wagner Theorem
\cite{Ho,MW}. This partly explains why a rigorous proof of BEC for interacting systems
is still lacking in general.

\section{Overview and Outline}

A central theme of these notes is the evaluation of the ground state
energy\label{ground state energy} in various situations. Already in
1927 W.~Lenz \cite{lenz}\index{Lenz} gave a heuristic argument
indicating that the ground state energy of a dilute hard-core gas was
proportional to the scattering length\index{scattering length}, $a$,
of the potential in three dimensions.\footnote{Lenz calculated the
  ground state energy of a single particle in the presence of $N-1$
  randomly placed particles in a $L\times L\times L$ box. This would
  give the correct energy, $4\pi a/V$, {\it provided\/} the fixed
  particles are uniformly distributed. But in a random arrangement
  there are bound to be holes of arbitrary large size (as $N\to\infty$)
  and the moving particle can always fit in a hole with arbitrary
  small energy. It is the inevitability of such holes that makes it
  impossible to convert Lenz's argument into a rigorous one.}  The 3D
formula for the energy per particle at low density is $e_0(\rho) =
(\hbar^2/2m) 4\pi \rho a$, to leading order in the density $\rho =
N/V$.  Bogoliubov's\index{Bogoliubov} 1947 paper \cite{BO} showed how
a perturbative version of this formula could be derived in a more
systematic and general way, and there was further work on this in the
1950's and 60's \cite{Lee-Huang-YangEtc, Lieb63}, but it was not until
1998 \cite{LY1998} that it was proved rigorously.  The issue here was
to derive a lower bound for the energy.  In 1957 Dyson\index{Dyson}
\cite{dyson} had established an asymptotically correct upper bound for
the energy and also a lower bound, but the latter was 14 times too
small.

In two dimensions a different formula, postulated as late as 1971 by
Schick\index{Schick} \cite{schick}, holds and was rigorously proved to
be correct in \cite{LY2d}.  With the aid of the methodology developed
to prove the lower bound for the homogeneous gas, several other
problems could successfully be addressed. One is the proof that the
Gross-Pitaevskii equation\index{Gross-Pitaevskii!equation} correctly
describes the ground state in the `traps' actually used in the
experiments \cite{LSY1999, LSY2d}. For such systems it is also
possible to prove complete Bose condensation and superfluidity
\cite{LS02, LSYsuper}.  On the frontier of experimental developments
is the possibility that a dilute gas in an elongated trap will behave
like a one-dimensional system and this topic was addressed in a
mathematical way in \cite{LSY}.

Another topic is a proof that Foldy's\index{Foldy} 1961 theory
\cite{FO} of a {\it high density} Bose gas of charged particles, which
is based on Bogoliubov's theory, correctly describes the ground state
energy of this gas; using this it is also possible to prove the
$N^{7/5}$ formula for the ground state energy of the two-component
charged Bose gas\index{charged Bose gas} proposed by
Dyson\index{Dyson} in 1967 \cite{D2}.  All of this is quite recent
work \cite{LS, LSo02, So} and it is hoped that the mathematical
methodology might be useful, ultimately, to solve more complex
problems connected with these interesting systems.

One of the most remarkable recent developments in the study of
ultracold Bose gases is the observation of a reversible transition
from a Bose-Einstein condensate to a state composed of localized atoms
as the strength of a periodic, optical trapping potential is varied
\cite{JBCGZ, G1, G2}.  Together with M. Aizenman we have rigorously
analyzed a model of this phenomenon \cite{ALSSY}.  The gas is a hard
core lattice\index{lattice gas} gas and the optical lattice is modeled by a periodic
potential of strength $\lambda$. For small $\lambda$ and temperature
BEC is proved to occur, while at large $\lambda$ BEC disappears, even
in the ground state, which is a Mott insulator\index{Mott insulator} state with a
characteristic gap. The inter-particle interaction is essential for
this effect.
\bigskip\bigskip

Let us briefly describe the structure of these notes.  The discussion
centers around six main topics:

\begin{itemize}
\item [1.]  The dilute, homogeneous Bose gas with repulsive
interaction (2D and 3D).
\vspace*{-2mm}
\item [2.]  Repulsive bosons in a trap (as used in recent experiments)
  and the \lq\lq Gross-Pitaevskii\rq\rq\ equation.
  \vspace*{-2mm}
\item [3.]  BEC  and superfluidity\index{superfluidity} for dilute trapped
gases.
\vspace*{-2mm}
\item [4.]  Low-dimensional behavior of three-dimensional gases in
elongated or disc-shaped traps.
\vspace*{-2mm}
\item [5.]  Foldy's \lq\lq jellium\rq\rq\ \index{jellium} model of charged particles
  in a neutralizing background and the extension to the two-component
  gas.
  \vspace*{-2mm}
\item[6.] The model of an optical lattice that shows BEC
at weak coupling and no BEC at strong coupling.
\end{itemize}
The discussion below of topic~1 is based on \cite{LY1998} and
\cite{LY2d}, and of topic~2 on \cite{LSY1999} and \cite{LSY2d}. See
also \cite{LYbham,LSYdoeb,S4,LSYnn}.

The original references for topic~3 are \cite{LS02} and
\cite{LSYsuper}, but for transparency we also include here a chapter
on the special case when the trap is a rectangular box. This case
already contains the salient points, but avoids several complications
due the inhomogeneity of the gas in a general trap.  An essential
technical tool for topic~3 is a generalized Poincar\'e
inequality\index{Poincar\'e inequality!generalized}, which is
discussed in a separate chapter.

Topic~4 is a summary of the work in \cite{LSY} and \cite{SY}.

The discussion of topic~5 is based on \cite{LS}, \cite{LSo02}
and \cite{So}.
Topic~6 is based on \cite{ALSSY} and \cite{ALSSY2}.

Topic~1 (3 dimensions) was the starting point and contains essential
ideas. It is explained here in some detail and is taken, with minor
modifications (and corrections), from \cite{LYbham}. In terms of
technical complexity, however, the fifth  topic is the most involved
and can not be treated here in full detail.

The interaction potential between pairs of particles in the
Jellium\index{jellium} model and the two-component gas in topic~5 is
the repulsive, {\it long-range} Coulomb potential,
\index{Coulomb potential}
while in topics~1--4 it is assumed to be repulsive and
{\it short range}. Topic~6 concerns a hard-core lattice gas. For
alkali atoms in the recent experiments on Bose-Einstein condensation
the interaction potential has a repulsive hard core, but also a quite
deep attractive contribution of van der Waals type and there are many
two body bound states\index{bound states} \cite{PS}. The Bose
condensate seen in the experiments is thus not the true ground state
(which would be a solid) but a metastable state\index{metastable
  state}. Nevertheless, it is usual to model this metastable state as
the ground state of a system with a repulsive two body potential
having the same scattering length as the true potential, and this is
what we shall do. In these notes all interaction potentials will be
positive, except in Appendices A, C and D.  \bigskip

There are four appendices to these notes:
\begin{itemize}
\item[{A.}]Elements of Bogoliubov Theory
\item[{B.}]An Exactly Soluble Model
\item[{C.}]Definition and Properties of the Scattering Length
\item[{D.}]$c$-Number Substitutions and Gauge Symmetry Breaking
\end{itemize}
Appendices~\ref{chap2} and~\ref{chap3} are reproductions, with only minor
modifications, of two sections of the survey article \cite{EL2}.
This material is still relevant after 40 years and parts of it were
discussed in the Oberwolfach course. Appendix~\ref{appscatt} is taken from the paper
\cite{LY2d}. Appendix~\ref{justapp} is a slightly extended version of \cite{lsyc}.

\chapter{The Dilute Bose Gas in 3D} \label{sect3d}

We consider the Hamiltonian\index{Hamiltonian} for $N$ bosons of mass
$m$ enclosed in a cubic box $\Lambda$ of side length $L$ and
interacting by a
spherically symmetric pair potential
$v(|\x_i - \x_j|)$:
\begin{equation}\label{ham}
H_{N} = - \mu\sum_{i=1}^{N} \Delta_i +
\sum_{1 \leq i < j \leq N} v(|\x_i - \x_j|).
\end{equation}
Here $\x_i\in\R^3$, $i=1,\dots,N$ are the positions of the particles,
$\Delta_i$ the Laplacian with respect to $\x_{i}$, and we have denoted
${\hbar^2}/{ 2m}$ by $\mu$ for short. (By choosing suitable units
$\mu$ could, of course, be eliminated, but we want to keep track of
the dependence of the energy on Planck's constant and the mass.)  The
interaction potential\index{interaction potential} will be assumed to
be {\it nonnegative} and to decrease faster than $1/r^3$ at infinity.
Note that for potentials that tend to zero at infinity `repulsive' and
`non-negative' are synonymous --- in the quantum mechanical literature
at least.  In classical mechanics, in contrast, a potential that is
positive but not monotonically decreasing is not called repulsive.

The Hamiltonian (\ref{ham}) operates on {\it symmetric} wave functions
in the Hilbert space $L^2(\Lambda^{N}, d\x_1\cdots d\x_N)$ as is
appropriate for bosons.  Let us note an important fact here that will
be useful later. To say that we are in the symmetric tensor product of
$L^2(\R^3)$ is equivalent, in plain language, to the statement that we
consider only wave functions $\Psi(\x_1,\dots , \x_N)$ that are
invariant under permutation of the $N$ coordinates. On the other hand,
we could ask for the bottom of the spectrum of $H_N$ without imposing
this symmetry restriction, i.e., on the whole tensor product. The fact
is that the two are the same. The absolute ground state
\index{absolute ground state}
energy is the boson ground state energy. Moreover, the
ground state is unique, i.e., there is only one and it is symmetric.
This fact will be useful later because it means that we can get an
upper bound to the ground state energy by using any handy function.

The proof of this assertion, very briefly, goes as follows: The
absolute ground state energy is the infimum of $\langle \Psi | H_N|
\Psi\rangle$ over all normalized $\Psi$.  If $\Psi$ is a candidate for
a minimizer, we can consider $\Phi(\X) = |\Psi (\X)|$, which has the
same norm, the same potential energy, and a kinetic energy that can be
lower but not higher \cite{LL01}.  This $\Phi$ will not satisfy the
Schr\"odinger equation unless $\Phi =\Psi$, up to an overall phase.
Thus, the absolute ground state is unique and it has to be either
symmetric or antisymmetric since these are the only one-dimensional
representations of the permutation group.  Since it has only one sign
it is symmetric.

We are interested in the ground state energy $E_{0}(N,L)$ of
(\ref{ham}) in the {\it thermodynamic
  limit}\index{thermodynamic!limit} when $N$ and $L$ tend to infinity
with the density $\rho=N/L^3$ fixed. The energy per particle in this
limit is
\begin{equation}\label{eq:thmlimit}
e_{0}(\rho)=\lim_{L\to\infty}E_{0}(\rho L^3,L)/(\rho
L^3).
\end{equation}
Our results about $e_{0}(\rho)$ are based on estimates on $E_{0}(N,L)$
for finite $N$ and $L$, which are important, e.g., for the
considerations of inhomogeneous systems in
Chapters~\ref{sectgp}--\ref{3dto2d}.  To define $E_{0}(N,L)$ precisely
one must specify the boundary conditions. These should not matter for
the thermodynamic limit.  To be on the safe side we use Neumann
boundary conditions for the lower bound, and Dirichlet boundary
conditions\index{boundary condition} for the upper bound since these
lead, respectively, to the lowest and the highest energies.

For experiments with dilute gases the {\it low density
  asymptotics}\index{low density} of $e_{0}(\rho)$ is of importance.
Low density means here that the mean interparticle distance,
$\rho^{-1/3}$ is much larger than the {\it scattering
  length}\index{scattering length} $a$ of the potential, which is
defined as follows. (See Appendix~\ref{appscatt} for details.) The
zero energy scattering Schr\"odinger equation
\begin{equation}\label{3dscatteq}
-2\mu \Delta \psi + v(r) \psi =0
\end{equation}
has a solution  of the form, asymptotically as $|\x|=r\to \infty$
(or for all $r>R_0$ if $v(r)=0$ for $ r>R_0$),
\begin{equation}\label{3dscattlength}
\psi_0(\x) = 1-a/|\x|\,.
\end{equation}
(The factor $2$ in (\ref{3dscatteq}) comes from the reduced mass of
the
two particle problem.) Writing $\psi_0(\x)=u_0(|\x|)/|\x|$ this is the
same as
\begin{equation}
a=\lim_{r\to\infty}r-\frac{u_{0}(r)}{u_{0}'(r)},
\end{equation}
where $u_{0}$ solves the zero energy (radial) scattering equation,
\begin{equation}\label{scatteq}
-2\mu u_{0}^{\prime\prime}(r)+ v(r) u_{0}(r)=0
\end{equation}
with $u_{0}(0)=0$.

An important special case is the hard core potential $v(r)= \infty$ if
$r<a$ and  $v(r)= 0$ otherwise. Then the scattering length and the
radius $a$ are the same.

Our main result is a
rigorous proof of the formula
\begin{equation} e_{0}(\rho)\approx4\pi\mu\rho a\end{equation}
for $\rho a^3\ll 1$, more precisely of
\begin{thm}[\textbf{Low density limit of the ground state
energy}]\label{3dhomthm}
\begin{equation}\label{basic}
\lim_{\rho a^3\to 0}\frac {e_{0}(\rho)}{4\pi\mu\rho a}=1.
\end{equation}
\end{thm}
This formula is independent of the boundary conditions used for the
definition of $e_{0}(\rho)$ . It holds for every positive radially
symmetric pair
potential such that $\int_R^\infty v(r)r^2 dr<\infty$ for some $R$,
which guarantees a finite scattering length, cf.\ Appendix~\ref{appscatt}.

The genesis of an understanding of $e_{0}(\rho)$ was the pioneering
work \cite{BO,Bog} of Bogoliubov\index{Bogoliubov}, and in the 50's
and early 60's several derivations of (\ref{basic}) were presented
\cite{Lee-Huang-YangEtc}, \cite{Lieb63}, even including higher order
terms:
\begin{equation}\frac{e_{0}(\rho)}{4\pi\mu\rho a}=
1+\mfr{128}/{15\sqrt \pi}(\rho a^3)^{1/2}
+8\left(\mfr{4\pi}/{3}-\sqrt 3\right)(\rho a^3)\log (\rho a^3)
+O(\rho a^3)\,.
\end{equation}
These early developments are reviewed in \cite{EL2}. They all rely
on some special assumptions about the ground state that have never
been
proved, or on the selection of special terms from a perturbation
series
which likely diverges. The only rigorous estimates of this period were
established by Dyson\index{Dyson}, who derived the following bounds in 1957 for a
gas of hard spheres \cite{dyson}:
\begin{equation} \frac1{10\sqrt 2} \leq
    \frac{e_{0}(\rho)}{ 4\pi\mu\rho a}\leq\frac{1+2 Y^{1/3}}{
(1-Y^{1/3})^2}
\end{equation}
with $Y=4\pi\rho a^3/3$. While the upper bound has the asymptotically
correct form, the lower bound is off the mark by a factor of about
1/14.
But for about 40 years this was the best lower bound available!

Under the assumption that (\ref{basic}) is a correct asymptotic
formula for the energy, we see at once that understanding it
physically, much less proving it, is not a simple matter.
Initially, the problem presents us with two lengths, $a \ll
\rho^{-1/3}$ at low density. However, (\ref{basic}) presents us
with another length generated by the solution to the problem. This
length is the de Broglie wavelength, or  `uncertainty principle'
length
(sometimes called `healing length'\index{healing length})
\begin{equation}\label{ellc}\ell_c\sim (\rho a)^{-1/2} .
\end{equation}
 The reason for saying that
$\ell_c$ is the de Broglie wavelength is that in the hard core
case all the energy is kinetic (the hard core just imposes a $\psi
=0$ boundary condition whenever the distance between two particles
is less than $a$). By the uncertainty principle, the kinetic
energy is proportional to an inverse length squared, namely
$\ell_c$. We then have the relation (since $\rho a ^3$ is small)
\begin{equation}\label{scales} a \ll \rho^{-1/3}\ll \ell_c
\end{equation}
which implies, physically, that
it is impossible to localize the particles relative to each other
(even though $\rho$ is small). Bosons in their ground state are
therefore `smeared out' over distances large compared to the mean
particle distance and their individuality is entirely lost. They
cannot be localized with respect to each other without changing
the kinetic energy enormously.

Fermions, on the other hand, prefer to sit in
`private rooms', i.e., $\ell_{c}$ is never bigger than $\rho^{-1/3}$
times a fixed factor.
In this respect the quantum nature of bosons is much more pronounced
than for fermions.

Since (\ref{basic}) is a basic result about the Bose gas it is clearly
important to derive it rigorously and in reasonable generality, in
particular for more general cases than hard spheres.  The question
immediately arises for which interaction potentials one may expect it
to be true. A notable fact is that it {\it is not true for all} $v$ with
$a>0$, since there are two body potentials with positive scattering
length that allow many body bound states. (There are even such
potentials without two body bound states but with three body bound
states \cite{BA}.) For such potentials \eqref{basic} is clearly false.
Our proof, presented in the
sequel,  works for nonnegative $v$, but we conjecture that
(\ref{basic})
holds if $a>0$ and $v$ has no $N$-body bound states for any $N$. The
lower bound is, of course, the hardest part, but the upper bound is
not
altogether trivial either.

Before we start with the estimates a simple computation and some
heuristics may be helpful to make
(\ref{basic}) plausible and motivate the formal proofs.

With  $\psi_{0}$ the zero energy scattering solution,
partial integration, using \eqref{3dscatteq} and
\eqref{3dscattlength},
gives, for $R\geq R_0$,
\begin{equation}\label{partint}
\int_{|\x|\leq R}\{2\mu|\nabla \psi_{0}|^2+v|\psi_{0}|^2\}d\x=
8\pi\mu a\left(1-\frac aR\right)\to 8\pi\mu a\quad\mbox{\rm for
$R\to\infty$}.
\end{equation}
Moreover, for positive interaction potentials the scattering solution
minimizes
the quadratic form in (\ref{partint}) for each $R\geq R_0$ with
the boundary condition $\psi_0(|\x|=R)=(1-a/R)$. Hence the energy
$E_{0}(2,L)$ of two
particles in a large box, i.e., $L\gg
a$, is approximately $8\pi\mu a/L^3$. If the gas is sufficiently
dilute it is not unreasonable to expect that the energy is essentially
a sum of all such two particle contributions. Since there are
$N(N-1)/2$ pairs, we are thus lead to $E_{0}(N,L)\approx 4\pi\mu a
N(N-1)/L^3$, which gives (\ref{basic}) in the thermodynamic limit.

This simple heuristics is far from a rigorous proof, however,
especially for the lower bound. In fact, it is rather remarkable that
the same asymptotic formula holds both for `soft' interaction
potentials, where perturbation theory can be expected to be a good
approximation, and potentials like hard spheres where this is not so.
In the former case the ground state is approximately the constant
function and the energy is {\it mostly potential}: According to
perturbation theory $E_{0}(N,L)\approx N(N-1)/(2 L^3)\int v(|\x|)d\x$.
In particular it is {\it independent of} $\mu$, i.e. of Planck's
constant and mass. Since, however, $\int v(|\x|)d\x$ is the first Born
approximation\index{Born series} to $8\pi\mu a$ (note that $a$
depends on $\mu$!), this is not in conflict with (\ref{basic}).  For
`hard' potentials on the other hand, the ground state is {\it highly
  correlated}, i.e., it is far from being a product of single particle
states. The energy is here {\it mostly kinetic}, because the wave
function is very small where the potential is large. These two quite
different regimes, the potential energy dominated one and the kinetic
energy dominated one, cannot be distinguished by the low density
asymptotics of the energy. Whether they behave differently with
respect to other phenomena, e.g., Bose-Einstein
condensation\index{Bose-Einstein condensation}, is not known at
present.

Bogoliubov's analysis \cite{BO,Bog} presupposes the existence of
Bose-Einstein condensation. Nevertheless, it is correct (for the
energy) for the one-dimensional delta-function Bose gas \cite{LL},
despite the fact that there is (presumably) no condensation in that
case \cite{PiSt}.  It turns out that BE condensation is not really
needed in order to understand the energy.  As we shall see, `global'
condensation can be replaced by a `local' condensation on boxes whose
size is independent of $L$. It is this crucial understanding that
enables us to prove Theorem~\ref{3dhomthm} without having to decide
about BE condensation.

An important idea of Dyson was to transform the hard sphere
potential into a soft potential at the cost of sacrificing the
kinetic energy, i.e., effectively to move from one
regime to the other. We shall make use of this idea in our proof
of the lower bound below. But first we discuss the simpler upper
bound, which relies on other ideas from Dyson's beautiful paper
\cite{dyson}.

\section{Upper Bound}\label{upsec}

The following generalization of Dyson's upper bound holds
\cite{LSY1999}, \cite{S1999}:
\begin{thm}[\textbf{Upper bound}]\label{ub} Let
$\rho_1=(N-1)/L^3$ and $b=(4\pi\rho_1/3)^{-1/3}$. For non-negative potentials $v$ and $b>a$
the ground state energy of \eqref{ham} with periodic boundary
conditions
satisfies
\begin{equation}\label{upperbound}
E_{0}(N,L)/N\leq 4\pi \mu \rho_1 a \left( 1+ \const \frac a b\right) .
\end{equation}
Thus in the thermodynamic limit (and for all boundary conditions)
\begin{equation}
\frac{e_{0}(\rho)}{4\pi\mu\rho a}\leq 1 + \const Y^{1/3}
\end{equation}
provided $Y=4\pi\rho a^3/3<1$.
\end{thm}

\begin{proof}
  We first remark that the expectation value of (\ref{ham}) with any
  trial wave function gives an upper bound to the bosonic ground state
  energy, even if the trial function is not symmetric under
  permutations of the variables. (See the discussion at the beginning
  of this chapter.) The reason is that an absolute ground state of the
  elliptic differential operator (\ref{ham}) (i.e., a ground state
  without symmetry requirement) is a nonnegative function which can be
  symmetrized without changing the energy because (\ref{ham}) is
  symmetric under permutations.  In other words, the absolute ground
  state energy is the same as the bosonic ground state energy.

Following \cite{dyson} we choose a trial function of
the following form
\begin{equation}\label{wave}
\Psi(\x_1,\dots,\x_N)=F_1(\x_{1}) \cdot F_2(\x_{1},\x_{2}) \cdots
F_N(\x_{1},\dots,\x_{N}).
\end{equation}
More specifically, $F_{1}\equiv 1$ and $F_{i}$ depends only on the
distance of $\x_{i}$ to its nearest neighbor among the points
$\x_1,\dots ,\x_{i-1}$ (taking the periodic boundary into
account):
\begin{equation}\label{form}
F_i(\x_1,\dots,\x_i)=f(t_i), \quad t_i=\min\left(\xij,j=1,\dots,
i-1\right),
\end{equation}
with a function $f$ satisfying
\begin{equation}0\leq
f\leq 1, \quad f'\geq 0.
\end{equation}
The intuition behind the ansatz (\ref{wave}) is that the particles
are inserted into the system one at the time, taking into account
the particles previously inserted. While such a wave function
cannot reproduce all correlations present in the true ground
state, it turns out to capture the leading term in the energy for
dilute gases. The form (\ref{wave}) is  computationally easier to
handle than an ansatz of the type $\prod_{i<j}f(|\x_{i}-\x_{j}|)$,
which might appear more natural in view of the heuristic remarks
after Eq.\ \eqref{partint}.

The function $f$ is chosen to be
\begin{equation}\label{deff}
f(r)=\begin{cases}
f_0(r)/f_{0}(b)&\text{for $0\leq r\leq b$},\\
1&\textrm{for $r>b$},
\end{cases}
\end{equation}
with $f_{0}(r)=u_{0}(r)/r$ the zero energy scattering solution defined
by \eqref{scatteq}.

We start by computing the kinetic energy of our trial state defined above.
Define $\eps_{ik}(\x_1,\dots,\x_N)$ by
\begin{equation}
\eps_{ik} = \left\{ \begin{array}{cl}
1 & \text{for $i=k$}\\
-1 &\text{for $t_i=|\x_i-\x_k|$}\\
0 &\text{otherwise.}
\end{array} \right.
\end{equation}
Let $\niv_i$ be the unit vector in the direction of $\x_i-\x_{j(i)}$,
where $\x_{j(i)}$ denotes the nearest neighbor of $\x_i$ among the
points $(\x_1,\dots, \x_{i-1})$. (Note that $j(i)$ really depends on
all the points $\x_1,\dots,\x_i$ and not just on the index $i$. Except
for a set of zero measure, $j(i)$ is unique.) Then
\begin{equation}
F^{-1} \bfnab_k F= \sum_i  F_i^{-1}\eps_{ik} \niv_i f'(t_i) \ ,
\end{equation}
and after summation over $k$ we obtain
\begin{eqnarray}\nonumber
&& F^{-2} \sum_k  |\bfnab_k F|^2 = \sum_{i,j,k}\eps_{ik}\eps_{jk}(\niv_i\cdot
\niv_j)
F_i^{-1}F_j^{-1}f'(t_i)f'(t_j)\\ &&\leq 2 \sum_i
F_i^{-2}f'(t_i)^2+2
\sum_{k\leq i<j}|\eps_{ik}\eps_{jk}|F_i^{-1}F_j^{-1}f'(t_i)f'(t_j)\ .\label{term}
\end{eqnarray}
The expectation value of the Hamiltonian can thus be bounded as follows:
\begin{eqnarray}\nonumber
\frac{\langle\Psi|H_{N}|\Psi\rangle}{\langle\Psi|\Psi\rangle}&\leq&
2 \mu \sum_{i=1}^N\frac{\int |\Psi|^2F_i^{-2}f'(t_i)^2}{\int
|\Psi|^2}+\sum_{j<i}\frac{\int |\Psi|^2v(|\x_i-\x_j|)}{ \int |\Psi|^2}\\
&& +2 \mu \sum_{k\leq
i<j}\frac{\int
|\Psi|^2|\eps_{ik}\eps_{jk}|F_i^{-1}F_j^{-1}f'(t_i)f'(t_j)}{\int
|\Psi|^2}.\label{expu}
\end{eqnarray}
For $i<p$, let $F_{p,i}$ be the value that $F_p$ would take
if the point $\x_i$ were
omitted from consideration as a possible nearest neighbor.
Note that $F_{p,i}$ is independent of $\x_i$.
Analogously we define $F_{p,ij}$ by omitting $\x_i$ and $\x_j$. The functions
$F_i$ occur both in the numerator and the denominator so we need estimates
from below and above. Since $f$ is monotone increasing,
\begin{equation}
F_p=\min\{F_{p,ij},f(|\x_p-\x_j|),f(|\x_p-\x_i|)\} \ ,
\end{equation}
and we have, using $0\leq f\leq 1$,
\begin{equation}
F_{p,ij}^2f(|\x_p-\x_i|)^2f(|\x_p-\x_j|)^2\leq F_p^2\leq F_{p,ij}^2 \ .
\end{equation}
Hence, for $j<i$, we have the upper bound
\begin{equation}
F_{j+1}^2\cdots F_{i-1}^2F_{i+1}^2\cdots F_N^2\leq F_{j+1,j}^2\cdots
F_{i-1,j}^2F_{i+1,ij}^2\cdots F_{N,ij}^2\ , \label{above}
\end{equation}
and the lower bound
\begin{align}  \label{below}
 F_j^2\cdots F_N^2  \geq &\,
F_{j+1,j}^2\cdots F_{i-1,j}^2F_{i+1,ij}^2\cdots F_{N,ij}^2 \\ \notag & \times\Big(1-\!\!\sum_{k=1,\, k\neq i,j}^N\!\!(1- f(|\x_j-\x_k|)^2)\Big) \Big(1-\!\!\sum_{k=1,\, k\neq i}^N\!\!(1-f(|\x_i-\x_k|)^2)\Big)\,.
\end{align}
We now consider the first two terms on the right side of \eqref{expu}.
In the numerator of the first term we use, for each fixed $i$, the estimate
\begin{equation}
f'(t_i)^2\leq\sum_{j=1}^{i-1} f'(\x_i-\x_j)^2\ ,
\end{equation}
and in the second term we use $F_i\leq f(|\x_i-\x_j|)$. For fixed $i$
and $j$ we can eliminate $\x_i$ and $\x_j$ from the rest of the
integrand by using the bound (\ref{above}) in the numerator and
(\ref{below}) in the denominator to do the $\x_i$ and $\x_j$
integrations.  Note that, using partial integration, (\ref{3dscatteq}) implies that
\begin{equation}
\int \left( \mu f'(|\x|)^2+\half v(|\x|)f(|\x|)^2\right)\D\x \leq 4\pi a (1-a/b) f_0(b)^{-2} .
\end{equation}
Moreover, $f_0(b)\geq 1-a/b$. We thus obtain
\begin{equation} \label{trafo}
\int \left(2\mu f'(\x_i-\x_j)^2+v(\x_i-\x_j)f(\x_i-\x_j)^2\right)\D\x_i \D\x_j  \leq 8\pi a \mu L^3 \left[1-a/b\right]^{-1}\ .
\end{equation}
In the denominator, we estimate
\begin{equation}  \label{iuu}
\int\Big(1-\sum_{p=1,\,p\neq i}^N(1-f(|\x_p-\x_i|)^2)\Big)\D\x_i
 \geq L^3 - (N-1) I \ ,
\end{equation}
where we set $I=\int (1-f(|\x|)^2)\D\x$. Using that $f(|\x|)\geq
[1-a/|\x|]_+$ (see Appendix~\ref{appscatt}), we get that $I\leq (4\pi/3) a b^2$.
 The same factor comes from the
$\x_j$-integration. The remaining factors are identical in numerator
and denominator and hence we conclude that the first and second term
in (\ref{expu}) are bounded above by
\begin{equation}
\sum_{i=1}^N \, (i-1)  \frac{8\pi a\mu}{(1-a/b)(1-I \rho_1)^2} \leq  \frac {N (N-1)}{L^3} 4\pi a\mu \left( 1 + O\left(a/b\right)\right) \ . \label{bound1}
\end{equation}

A similar argument is now applied to the third term of (\ref{expu}). We omit the details. The result is an upper bound
\begin{equation}\label{xy}
\frac{2}{3}N (N-1) (N-2) \frac{\mu K^2 }{(L^3-(N-1)I)^2} \ ,
\end{equation}
with $K$ given by $K= \int f(|\x|)f'(|\x|)\D\x$. Using again that
$[1-a/|\x|]_+\leq f(|\x|)\leq 1$ as well as partial integration, we
can estimate $K\leq 4\pi a b(1+O(a/b))$. Hence (\ref{xy}) is, for
bounded $a/b$, bounded by $\const N \mu a^2 \rho_1 /b $. Combining this
estimate with (\ref{bound1}) proves (\ref{upperbound}).
\end{proof}


\section{Lower Bound}\label{subsect22}

It was explained previously in this chapter why the lower bound
for the bosonic ground state energy of (\ref{ham}) is not easy to
obtain.
The three different length scales \eqref{scales} for bosons will play
a role in the
proof below.
\begin{itemize}
\item The scattering length $a$.
\item The mean particle distance $\rho^{-1/3}$.
\item The `uncertainty principle length' $\ell_{c}$, defined by
$\mu\ell_{c}^{-2}=e_{0}(\rho)$, i.e., $\ell_{c}\sim (\rho a)^{-1/2}$.
\end{itemize}

Our lower bound for $e_{0}(\rho)$ is as follows.
\begin{thm}[\textbf{Lower bound in the thermodynamic
limit}]\label{lbth}
For a  positive potential $v$ with finite range and $Y$ small enough
\begin{equation}\label{lowerbound}\frac{e_{0}(\rho)}{4\pi\mu\rho
a}\geq
(1-C\,
Y^{1/17})
\end{equation}
with $C$ a constant. If $v$ does not have finite range,
but decreases faster than
$1/r^{3}$ (more precisely, $\int_R^\infty v(r)r^2 dr<\infty$ for some
$R$)
then an analogous
bound to \eqref{lowerbound}
holds, but with $CY^{1/17}$ replaced by $o(1)$ as $Y\to 0$.
\end{thm}
It should be noted right away that the error term $-C\, Y^{1/17}$ in
(\ref{lowerbound}) is of no fundamental significance and is
not believed to reflect the true state of affairs. Presumably, it
does not even have the right sign. We mention in passing that
$C$ can be taken to be
$8.9$ \cite{S1999}.

As mentioned at the beginning of this chapter after
Eq.\ \eqref{eq:thmlimit}, a lower bound on $E_{0}(N,L)$ for finite $N$
and $L$ is of importance for applications to inhomogeneous gases, and
in fact we derive (\ref{lowerbound}) from such a bound. We state it in
the following way:
\begin{thm}[\textbf{Lower bound in a finite box}] \label{lbthm2}
    For a  positive potential $v$ with finite range there is
a $\delta>0$ such that the ground state energy of \eqref{ham} with
Neumann
boundary conditions satisfies
\begin{equation}\label{lowerbound2}E_{0}(N,L)/N\geq 4\pi\mu\rho
a \left(1-C\,
Y^{1/17}\right)
\end{equation}
for all $N$ and $L$ with $Y<\delta$ and $L/a>C'Y^{-6/17}$. Here $C$
and $C'$ are positive  constants, independent of $N$ and $L$. (Note
that the
condition on $L/a$ requires in particular that $N$ must be large
enough, $N>\hbox \const Y^{-1/17}$.)  As in Theorem~{\rm\ref{lbth}}
such a bound, but possibly with a different error term, holds also for
potentials $v$ of infinite range that decrease sufficiently fast at
infinity.
\end{thm}

The first step in the proof of Theorem \ref{lbthm2} is a
generalization of
a lemma of Dyson\index{Dyson!lemma}, which allows us to replace $v$ by a `soft'
potential,
at the cost of sacrificing kinetic energy and increasing the
effective range.

\begin{lem}\label{dysonl} Let $v(r)\geq 0$ with finite range
$R_{0}$. Let
$U(r)\geq 0$
be any function satisfying $\int U(r)r^2dr\leq 1$ and $U(r)=0$ for
$r<R_{0}$.
Let
${\mathcal B}\subset \R^3$ be star shaped with respect to $0$ (e.g.\
convex with $0\in{\mathcal B}$). Then for all differentiable
functions $\psi$
\begin{equation}\label{dysonlemma}
    \int_{\mathcal B}\left[\mu|\nabla\psi|^2+\mfr1/2
v|\psi|^2\right]
\geq \mu a \int_{\mathcal B} U|\psi|^2.\end{equation}
\end{lem}

\begin{proof}
Actually, (\ref{dysonlemma}) holds with $\mu |\nabla \psi
(\x)|^2$
replaced by the (smaller) radial kinetic energy,
 $\mu |\partial \psi (\x)/ \partial r|^2$, and  it  suffices to
prove
the analog of (\ref{dysonlemma}) for the integral along each radial
line with fixed angular variables. Along such a line we write
$\psi(\x) = u(r)/r$ with $u(0)=0$. We consider first the special case
when $U$ is a delta-function at some radius $R\geq
R_0$,
i.e., \begin{equation}\label{deltaU}U(r)=\frac{1}{
R^2}\delta(r-R).\end{equation}
For such $U$ the analog of (\ref{dysonlemma}) along the radial line is
\begin{equation}\label{radial}\int_{0}^{R_{1}}
    \{\mu[u'(r)-(u(r)/r)]^2+\mfr1/2v(r)|u(r)|^2\}dr\geq
    \begin{cases}
        0&\text{if $R_{1}<R$}\\
            \mu a|u(R)|^2/R^2&\text{if $R\leq R_{1}$}
\end{cases}
\end{equation}
where $R_{1}$ is the length of the radial line segment in ${\mathcal
B}$.
The case $R_{1}<R$ is trivial,
because $\mu|\partial \psi/\partial r|^2+\mfr1/2 v|\psi|^2\geq 0$.
(Note that positivity of $v$ is used here.) If $R\leq R_{1}$ we
consider the integral on the left side of (\ref{radial}) from 0
to $R$
instead of $R_{1}$ and
minimize it under the boundary condition that $u(0)=0$
and $u(R)$ is a fixed constant. Since everything is homogeneous in
$u$ we may
normalize this value to $u(R)=R-a$.
This minimization problem leads to the zero energy
scattering
equation (\ref{scatteq}). Since $v$ is positive, the
solution is a true minimum and not just a
stationary point.

Because $v(r)=0$ for $r>R_{0}$ the solution, $u_{0}$, satisfies
$u_{0}(r)=r-a$
for $r>R_{0}$.  By partial integration,
\begin{equation}\int_{0}^{R}\{\mu[u'_{0}(r)-(u_{0}(r)/r)]^2+
    \mfr1/2v(r)|u_{0}(r)]^2\}dr=\mu a|R-a|/R\geq \mu
a|R-a|^2/R^2.
    \end
{equation}
But $|R-a|^2/R^2$ is precisely
the right side of (\ref{radial}) if $u$ satisfies the normalization
condition.

This derivation of (\ref{dysonlemma}) for the special case
(\ref{deltaU})
implies the
general case, because every $U$ can be written as a
superposition of  $\delta$-functions, i.e.,
$U(r)=\int R^{-2}\delta(r-R)\,U(R)R^2 dR$, and $\int U(R)R^2 dR\leq 1$
by assumption.
\end{proof}

By dividing $\Lambda$ for given points $\x_{1},\dots,\x_{N}$ into
Voronoi cells\index{Voronoi cells} ${\mathcal B}_{i}$ that contain all
points closer to $\x_{i}$ than to $\x_{j}$ with $j\neq i$ (these cells
are star shaped w.r.t. $\x_{i}$, indeed convex), the following
corollary of Lemma \ref{dysonl} can be derived in the same way as the
corresponding Eq.\ (28) in \cite{dyson}.

\begin{corollary}\label{2.6} For any $U$ as in Lemma {\rm\ref{dysonl}}
\begin{equation}\label{corollary}H_{N}\geq \mu a W\end{equation}
with $W$ the multiplication operator
\begin{equation}\label{W}W(\x_{1},\dots,\x_{N})=\sum_{i=1}^{N}U(t_{i}),
\end{equation}
where $t_{i}$ is the distance of $\x_{i}$ to its {\it nearest
neighbor} among the other points $\x_{j}$, $j=1,\dots, N$, i.e.,
\begin{equation}\label{2.29}t_{i}(\x_{1},\dots,\x_{N})=\min_{j,\,j\neq
i}|\x_{i}-\x_{j}|.\end{equation}
\end{corollary}
\noindent
(Note that $t_{i}$ has here a slightly different meaning than in
(\ref{form}), where it denoted the distance to the nearest neighbor
among the $\x_{j}$ with $j\leq i-1$.)

Dyson considers in \cite{dyson} a one parameter family of $U$'s that
is essentially the same as the following choice, which is convenient
for the
present purpose:
\begin{equation}\label{softened}
U_{R}(r)=\begin{cases}3(R^3-R_{0}^3)^{-1}&\text{for
$R_{0}<r<R$ }\\
0&\text{otherwise.}
\end{cases}
\end{equation}
We denote the corresponding interaction (\ref{W}) by $W_R$. For the
hard core
gas one obtains
\begin{equation}\label{infimum}E(N,L)\geq
\sup_{R}\inf_{(\x_{1},\dots,\x_{N})}
\mu a
W_R(\x_{1},\dots,\x_{N})\end{equation}
where the infimum is over $(\x_{1},\dots,x_{N})\in\Lambda^{N}$ with
$|\x_{i}-\x_{j}|\geq R_{0}=a$,
because of the hard core. At fixed $R$ simple geometry gives
\begin{equation}\label{fixedR}\inf_{(\x_{1},\dots,\x_{N})}
W_R(\x_{1},\dots,\x_{N})\geq \left(\frac{A}{R^3}-\frac{B}{ \rho
R^6}\right)\end{equation}
with certain constants $A$ and $B$. An evaluation of these constants
gives Dyson's bound
\begin{equation}E(N,L)/N\geq \frac{1}{10\sqrt 2} 4\pi\mu \rho
a.\end{equation}

The main reason this method does not give a better bound is that
$R$ must be chosen quite big, namely of the order of the mean
particle distance $\rho^{-1/3}$, in order to guarantee that the
spheres of radius $R$ around the $N$ points overlap. Otherwise the
infimum of $W_R$ will be zero. But large $R$ means that $W_R$ is
small. It should also be noted that this method does not work for
potentials other than hard spheres: If $|\x_{i}-\x_{j}|$ is
allowed to be less than $R_{0}$, then the right side of
(\ref{infimum}) is zero because $U(r)=0$ for $r<R_{0}$.

For these reasons we take another route. We still use  Lemma
\ref{dysonl} to get into the soft potential regime, but we do {\it
not} sacrifice {\it all} the kinetic energy as in
(\ref{corollary}). Instead we write, for $\varepsilon>0$
\begin{equation}
    H_{N}=\varepsilon H_{N}+(1-\varepsilon)H_{N}\geq \varepsilon
T_{N}+(1-\varepsilon)H_{N}
\end{equation}
with $T_{N}=-\mu\sum_{i}\Delta_{i}$ and use (\ref{corollary}) only for
the
part $(1-\varepsilon)H_{N}$. This gives
\begin{equation}\label{halfway}H_{N}\geq \varepsilon
T_{N}+(1-\varepsilon)\mu
a W_R.\end{equation} We consider the operator on the right side
from the viewpoint of first order perturbation theory, with
$\varepsilon T_{N}$ as the unperturbed  part, denoted $H_{0}$.

The ground state of $H_{0}$ in a box of side length $L$ is
$\Psi_{0}(\x_{1},\dots,\x_{N})\equiv L^{-3N/2}$ and we denote
expectation values in this state by $\langle\cdot\rangle_{0}$.
A  computation, cf.\ Eq.\ (21) in \cite{LY1998} (see also Eqs.\
\eqref{2dfirstorder}--\eqref{firstorder2}), gives
\begin{eqnarray}\label{firstorder}4\pi\rho\left(1-\mfr1/N\right)&\geq&
\langle W_R\rangle_{0}/N  \nonumber   \\ &\geq& 4\pi\rho
\left(1-\mfr1/N\right)\left(1-\mfr{2R}/L\right)^3
\left(1+4\pi\rho(R^3-R_{0}^3)/3\right)^{-1}.
\end{eqnarray}
The rationale behind the various factors is as follows: $(1-\mfr1/N)$
comes from the fact that the number of pairs is $N(N-1)/2$ and not
$N^2/2$, $(1-{2R}/L)^3$ takes into account the fact that the particles
do not interact beyond the boundary of $\Lambda$, and the last factor
measures the probability to find another particle within the
interaction range of the potential $U_R$ for a given particle.

The estimates (\ref{firstorder}) on the first order term look at first
sight quite promising, for if we let $L\to \infty$, $N\to \infty$ with
$\rho=N/L^3$ fixed, and subsequently take $R\to 0$, then $\langle
W_R\rangle_{0}/N$ converges to $4\pi\rho$, which is just what is
desired.  But the first order result (\ref{firstorder}) is not a
rigorous bound on $E_0(N,L)$, we need {\it error estimates}, and these
will depend on $\varepsilon$, $R$ and $L$.

We now recall {\it Temple's inequality}\index{Temple's inequality} \cite{TE} for the expectation
values of an operator $H=H_{0}+V$ in the ground state
$\langle\cdot\rangle_{0}$ of $H_{0}$. It is a simple
consequence of the operator inequality
\begin{equation}(H-E_{0})(H-E_{1})\geq 0\end{equation}
for the two lowest eigenvalues, $E_{0}<E_{1}$, of
$H$ and reads
\begin{equation}\label{temple}
E_{0}\geq \langle
H\rangle_{0}-\frac{\langle
H^2\rangle_{0}-\langle
H\rangle_{0}^2}{E_{1}-\langle H\rangle_{0}}
\end{equation}
provided $E_{1}-\langle H\rangle_{0}>0$.
Furthermore, if $V\geq 0$ we may use $E_{1}\geq E_{1}^{(0)}$= second
lowest
eigenvalue of $H_{0}$ and replace $E_{1}$ in (\ref{temple}) by
$E_{1}^{(0)}$.

{F}rom (\ref{firstorder}) and (\ref{temple}) we get the estimate
\begin{equation}\label{estimate2}\frac{E_{0}(N,L)}{ N}\geq 4\pi \mu
a\rho
\left(1-{\mathcal
E}(\rho,L,R,\varepsilon)\right)\end{equation}
with
\begin{eqnarray}\nonumber
&& 1-{\mathcal
E}(\rho,L,R,\varepsilon) \\ && =(1-\varepsilon)\left(1-\mfr1/{\rho
L^3}\right)\left(1-\mfr{2R}/L\right)^3
\left(1+\mfr{4\pi}/3\rho(R^3-R_{0}^3)\right)^{-1}\nonumber\\
&&\quad \times \left(1-\frac{\mu a\big(\langle
W_R^2\rangle_0-\langle W_R\rangle_0^2\big)}{\langle
W_R\rangle_0\big(E_{1}^{(0)}-\mu a\langle W_R\rangle_0\big)}\right).
\label{error}
\end{eqnarray}
To evaluate this further one may use the estimates (\ref{firstorder})
and the
bound
\begin{equation}\label{square}
\langle W_R^2\rangle_0\leq 3\frac N{R^3-R_0^3}\langle W_R\rangle_0
\end{equation}
which follows from $U_R^2=3({R^3-R_0^3})^{-1}U_R$ together with the
Schwarz inequality. A glance at the form of the error term reveals,
however, that it is {\it not} possible here to take the thermodynamic
limit $L\to\infty$ with $\rho$ fixed: We have
$E_{1}^{(0)}=\varepsilon\pi^2\mu/L^2$ (this is the kinetic energy of a
{\it single} particle in the first excited state in the box), and the
factor $E_{1}^{(0)}-\mu a\langle W_R\rangle_0$ in the denominator in
(\ref{error}) is, up to unimportant constants and lower order terms,
$\sim (\varepsilon L^{-2}-a\rho^2L^3)$. Hence the denominator
eventually becomes negative and Temple's inequality looses its
validity if $L$ is large enough.

As a way out of this dilemma we divide the big box $\Lambda$ into
cubic {\it
cells} of side length $\ell$ that is kept {\it fixed} as $L\to
\infty$.  The
number of cells, $L^3/\ell^3$, on the other hand, increases with
$L$.  The $N$
particles are distributed among these cells, and we use
(\ref{error}), with
$L$
replaced by $\ell$, $N$ by the particle number, $n$, in a cell and
$\rho$ by
$n/\ell^3$, to estimate the energy in each cell with {\it Neumann}
conditions
on the boundary.
For each distribution of the particles we add the
contributions from the cells, neglecting interactions across
boundaries.
Since
$v\geq 0$ by assumption, this can only lower the energy.  Finally, we
minimize
over all possible choices of the particle numbers for the various
cells
adding up to $N$.  The energy obtained in this way is a lower bound to
$E_0(N,L)$,
because we are effectively allowing discontinuous test functions for
the
quadratic form given by $H_N$.

In mathematical terms, the cell method\index{box method} leads to
\begin{equation}\label{sum}
E_0(N,L)/N\geq(\rho\ell^3)^{-1}\inf \sum_{n\geq 0}c_nE_0(n,\ell)
\end{equation}
where the infimum is over all choices of coefficients $c_n\geq 0$
(relative
number of cells containing exactly $n$ particles), satisfying the
constraints
\begin{equation}\label{constraints}
\sum_{n\geq 0}c_n=1,\qquad \sum_{n\geq 0}c_n n=\rho\ell^3.
\end{equation}

The minimization problem for the distributions of the particles among
the
cells would be easy if we knew that the ground state energy
$E_0(n,\ell)$ (or
a
good
lower bound to it) were convex in $n$.  Then we could immediately
conclude
that
it is best to have the particles as evenly distributed among the
boxes as
possible, i.e., $c_n$ would be zero except for the $n$ equal to the
integer closest to
$\rho\ell^3$. This would give
\begin{equation}\label{estimate3}\frac{E_{0}(N,L)}{ N}\geq 4\pi \mu
a\rho
\left(1-{\mathcal E}(\rho,\ell,R,\varepsilon)\right)\end{equation}
i.e.,
replacement of $L$ in (\ref{estimate2}) by $\ell$, which is
independent of
$L$.
The blow up of ${\mathcal E}$ for $L\to\infty$ would thus be avoided.

Since convexity of $E_0(n,\ell)$ is not known (except in the
thermodynamic
limit)
we must resort to other means to show that $n=O(\rho\ell^3)$ in all
boxes. The rescue
comes from {\it superadditivity}\index{superadditivity} of $E_{0}(n,\ell)$, i.e., the
property
\begin{equation}\label{superadd}
 E_0(n+n',\ell)\geq E_0(n,\ell)+E_0(n',\ell)
\end{equation}
which follows immediately from $v\geq 0$ by dropping the interactions
between
the $n$ particles and the $n'$ particles. The bound (\ref{superadd})
implies
in
particular that for any $n,p\in{\mathbb N}$ with $n\geq p$
\begin{equation}\label{superadd1}
E_{0}(n,\ell)\geq [n/p]\,E_{0}(p,\ell)\geq \frac n{2p}E_{0}(p,\ell)
\end{equation}
since the largest integer $[n/p]$ smaller than $n/p$ is in any case
$\geq
n/(2p)$.

The way (\ref{superadd1}) is used is as follows:
Replacing $L$ by $\ell$, $N$ by $n$ and $\rho$ by $n/\ell^3$ in
(\ref{estimate2})  we have for fixed $R$ and $\varepsilon$
\begin{equation}\label{estimate4}
E_{0}(n,\ell)\geq\frac{ 4\pi \mu a}{\ell^3}n(n-1)K(n,\ell)
\end{equation}
with a certain function $K(n,\ell)$ determined by (\ref{error}).
We shall see that $K$ is monoton\-ously decreasing in $n$, so that
if $p\in{\mathbb N}$  and $n\leq p$ then
\begin{equation}\label{n<p}
E_{0}(n,\ell)\geq\frac{ 4\pi \mu a}{\ell^3}n(n-1)K(p,\ell).
\end{equation}
We now split the sum in (\ref{sum}) into two parts.
For $n<p$ we use (\ref{n<p}), and for $n\geq p$ we use
(\ref{superadd1})
together with (\ref{n<p}) for $n=p$. The task is thus to minimize
\begin{equation}\label{task}
\sum_{n<p}c_n n(n-1)+\mfr1/2\sum_{n\geq p}c_nn(p-1)
\end{equation}
subject to the constraints ({\ref{constraints}).
Putting
\begin{equation}
k:=\rho\ell^3 \quad\text{and}\quad t:=\sum_{n<p}c_n n\leq k
\end{equation}
we have $\sum_{n\geq p}c_n n=k-t$, and since
$n(n-1)$ is convex in $n$ and vanishes for $n=0$, and
$\sum_{n<p}c_n\leq 1$, the expression
(\ref{task})
is
\begin{equation}
\geq t(t-1)+\mfr1/2(k-t)(p-1).
\end{equation}
We have to minimize this for $1\leq t\leq k$. If $p\geq 4k$ the
minimum is
taken
at $t=k$ and is equal to $k(k-1)$. Altogether we have thus shown that
\begin{equation}\label{estimate1}
\frac{E_{0}(N,L)}{ N}\geq 4\pi \mu a\rho\left(1-\frac1{\rho\ell^3}
\right)
K(4\rho\ell^3,\ell).
\end{equation}

What remains is to take a closer look at $K(4\rho\ell^3,\ell)$,
which depends on the parameters $\varepsilon$ and $R$ besides
$\ell$, and choose the parameters in an optimal way.
{F}rom (\ref{error}) and (\ref{square}) we obtain
\begin{align}\notag
K(n,\ell)=&\,(1-\varepsilon) \left(1-\mfr{2R}/\ell\right)^3
\left(1+\mfr{4\pi}/3(R^3-R_{0}^3)\right)^{-1}
\\ &\times \left(1-\frac3\pi
\frac{an}{(R^3-R_{0}^3)(\pi\varepsilon\ell^{-2}-4a\ell^{-3}n(n-1))}\right)\,. \label{Kformula}
\end{align}
The estimate (\ref{estimate4}) with this $K$ is valid as long as the
denominator in the last factor
in (\ref{Kformula}) is $\geq 0$, and in order to have a formula
for
all $n$ we can take 0 as a
trivial lower bound in other cases or when (\ref{estimate4}) is
negative. As required
for (\ref{n<p}), $K$ is monotonously decreasing in $n$. We now insert
$n=4\rho\ell^3$ and obtain
\begin{align}\notag
K(4\rho\ell^3,\ell)\geq&\,(1-\varepsilon)\left(1-\mfr{2R}/\ell\right)^3
\left(1+ \const)Y(\ell/a)^3 (R^3-R_{0}^3)/\ell^3\right)^{-1}
\\ &\times \left(1-
\frac{\ell^3}{(R^3-R_{0}^3)}\frac{\const Y}
{(\varepsilon(a/\ell)^{2}-(\const)Y^2(\ell/a)^3)}\right)\label{Kformula2}
\end{align}
with $Y=4\pi\rho a^3/3$ as before. Also, the factor
\begin{equation}
\left(1-\frac1{\rho\ell^3} \right)=(1-({\rm
const.})Y^{-1}(a/\ell)^{3})
\end{equation}
in (\ref{estimate1})
(which is the ratio between
$n(n-1)$ and $n^2$) must not be forgotten. We now make the ansatz
\begin{equation}\label{ans}
\varepsilon\sim Y^\alpha,\quad a/\ell\sim Y^{\beta},\quad
(R^3-R_{0}^3)/\ell^3\sim Y^{\gamma}
\end{equation}
with exponents $\alpha$, $\beta$ and $\gamma$ that we choose
in an optimal way. The conditions to be met are as follows:
\begin{itemize}
\item $\varepsilon(a/\ell)^{2}-({\rm const.})Y^2(\ell/a)^3>0$. This
holds for all small enough $Y$, provided
$\alpha+5\beta<2$ which follows from the conditions below.
\item $\alpha>0$ in order that $\varepsilon\to 0$ for $Y\to 0$.
\item $3\beta-1>0$ in order that  $Y^{-1}(a/\ell)^{3}\to 0$ for
$Y\to
0$.
\item $1-3\beta+\gamma>0$ in order that
$Y(\ell/a)^{3}(R^3-R_{0}^3)/\ell^3\to 0$ for $Y\to 0$.
\item $1-\alpha-2\beta-\gamma>0$ to control the last factor in
(\ref{Kformula2}).
\end{itemize}
Taking
\begin{equation}\label{exponents}
\alpha=1/17,\quad \beta=6/17,\quad \gamma=3/17
\end{equation}
all these conditions are satisfied, and
\begin{equation}
\alpha= 3\beta-1=1-3\beta+\gamma=1-\alpha-2\beta-\gamma=1/17.
\end{equation}
It is also clear that
$2R/\ell\sim Y^{\gamma/3}=Y^{1/17}$, up to higher order terms.
This completes the proof of Theorems \ref{lbth} and \ref{lbthm2}, for
the case of potentials with finite range. By optimizing the
proportionality constants in (\ref{ans}) one can show that $C=8.9$ is
possible in Theorem \ref{lbth} \cite{S1999}. The extension to
potentials of infinite range\index{infinite range potential} but
finite scattering length is obtained by approximation by finite range
potentials, controlling the change of the scattering length as the
cut-off is removed. See Appendix~\ref{appscatt} for details. We remark
that a slower decrease of the potential than $1/r^3$ implies infinite
scattering length.  \hfill$\qedsymbol$\bigskip

The exponents (\ref{exponents}) mean in particular that
\begin{equation}a\ll R\ll \rho^{-1/3}\ll \ell \ll(\rho
a)^{-1/2},\end{equation}
whereas Dyson's method required $R\sim \rho^{-1/3}$ as already
explained.
The condition $\rho^{-1/3}\ll \ell$ is required in order to have many
particles in each box and thus $n(n-1)\approx n^2$. The condition
$\ell \ll(\rho a)^{-1/2}$ is necessary for a spectral
gap $\gg e_{0}(\rho)$ in Temple's inequality. It is also clear that
this choice of $\ell$  would lead to a far too big
energy and no bound for $ e_{0}(\rho)$ if we had chosen Dirichlet
instead of
Neumann boundary
conditions for the cells. But with the latter the method works!
\newpage
\
\thispagestyle{empty}

\chapter{The Dilute Bose Gas in 2D} \label{sect2d}

In contrast to the three-dimensional theory, the two-dimensional Bose
gas\index{two-dimensional Bose gas} began to receive attention only
relatively late.  The first derivation of the correct asymptotic
formula was, to our knowledge, done by Schick\index{Schick}
\cite{schick} for a gas of hard discs. He found
\begin{equation}
    e(\rho)  \approx 4\pi \mu \rho |\ln(\rho a^2) |^{-1}.
\label{2den}
\end{equation}
This was accomplished by an infinite summation of `perturbation
series'
diagrams. Subsequently, a corrected modification of \cite{schick} was
given in \cite{hines}. Positive temperature extensions were given in
\cite{popov} and in \cite{fishho}. All this work involved an analysis
in
momentum space, with the exception
of a method due to one of us that works directly in configuration
space
\cite{Lieb63}.  Ovchinnikov \cite{Ovch} derived \eqref{2den} by using,
basically, the method in \cite{Lieb63}. These derivations require
several unproven assumptions and are not rigorous.

In two dimensions the scattering length $a$ is defined using the zero
energy scattering equation (\ref{3dscatteq}) but instead of
$\psi(r)\approx 1-a/r$ we now impose the asymptotic condition
$\psi(r)\approx \ln(r/a)$.
This is explained in Appendix~\ref{appscatt}.

Note that in two dimensions the ground state energy
could not possibly be $e_0(\rho)  \approx 4\pi \mu
\rho a$ as in three dimensions because that would be dimensionally
wrong.
Since $e_0(\rho) $ should essentially be proportional to $\rho$,
there is apparently no room for an $a$ dependence --- which is
ridiculous! It turns out that this dependence comes about in the
$\ln(\rho a^2)$ factor.

One of the intriguing facts about \eqref{2den} is that the energy for
$N$
particles is {\it not equal} to $N(N-1)/2$  times the energy for two
particles in the low density limit --- as is the case in
three dimensions.  The latter quantity,  $E_0(2,L)$, is,
asymptotically
for large $L$, equal to $8\pi \mu L^{-2} \left[ \ln(L^2/a^2)
\right]^{-1}$.  (This is seen in an analogous way as \eqref{partint}.
The
three-dimensional boundary condition $\psi_0(|\x|=R)=1-a/R$ is
replaced by $\psi_0(|\x|=R)=\ln (R/a)$ and moreover it has to be taken
into account that with this normalization $\|\psi_0\|^2={\rm
(volume)}(\ln (R/a))^2$ (to leading order), instead of just the volume
in the three-dimensional case.)  Thus, if the $N(N-1)/2$ rule were to
apply, \eqref{2den} would have to be replaced by the much smaller
quantity $4\pi \mu
\rho\left[ \ln(L^2/a^2) \right]^{-1}$. In other words, $L$, which
tends
to $\infty$ in the thermodynamic limit, has to be replaced by the mean
particle separation, $\rho^{-1/2}$ in the logarithmic factor. Various
poetic formulations of this curious fact have been given, but the fact
remains that the non-linearity is something that does not  occur in
more
than two dimensions and its precise nature is hardly obvious,
physically. This anomaly is the main reason that the two-dimensional
case is
not a trivial extension of the three-dimensional one.

Eq.\ \eqref{2den} was proved in \cite{LY2d} for nonnegative, finite
range two-body potentials by finding upper and lower bounds of the
correct form, using similar ideas as in the previous chapter for the
three-dimensional case. We discuss below the modifications that have
to
be made in the present two-dimensional case.  The restriction to
finite range can be relaxed as in three dimensions, but the
restriction to nonnegative $v$ cannot be removed in the current state
of our methodology.  The upper bounds will have relative remainder
terms O($|\ln(\rho a^2)|^{-1}$) while the lower bound will have
remainder O($|\ln(\rho a^2)|^{-1/5}$).  It is claimed in \cite{hines}
that the relative error for a hard core gas is negative and O$(\ln
|\ln(\rho a^2)||\ln(\rho a^2)|^{-1})$, which is consistent with our
bounds.

The upper bound is derived in complete analogy with the three
dimensional case. The function $f_0$ in the variational ansatz
\eqref{deff} is in
two dimensions also the zero energy scattering solution --- but for
2D, of course. The result is
\begin{equation}\label{upperbound3}
E_{0}(N,L)/N\leq \frac{2\pi\mu\rho}{\ln(b/a)-\pi\rho
b^{2}}\left(1+{\rm
O}([\ln(b/a)]^{{-1}})\right).
\end{equation}
The minimum over $b$ of the leading term is obtained for
$b=(2\pi\rho)^{{-1/2}}$. Inserting this in \eqref{upperbound3} we
thus
obtain
\begin{equation}\label{upperbound1}
E_{0}(N,L)/N\leq \frac{4\pi\mu\rho}{|\ln(\rho a^{2})|}\left(1+{\rm
O}(|\ln(\rho a^{2})|^{{-1}})\right).
\end{equation}

To prove the lower bound the essential new step is to modify Dyson's
lemma\index{Dyson!lemma}
for 2D. The 2D version of
Lemma \ref{dysonl} is:

\begin{lem}\label{dyson2d}
Let $v(r)\geq0$ and $v(r)=0$ for $r>R_0$.
 Let $U(r)\geq 0$ be any function satisfying
\begin{equation}\label{1dyson}
 \int_0^\infty U(r)\ln(r/a)rdr \leq 1~~~~~{\rm and}~~~~~ U(r)=0
~~~{\rm
for}~
 r<R_0.
\end{equation}
Let ${\mathcal B}\subset \R^2$ be star-shaped  with respect
to $0$ (e.g.\ convex
with $0\in{\mathcal B}$).
Then, for all functions $\psi$ in the Sobolev space
$H^1(\mathcal{B})$,
\begin{equation}
\int_{\mathcal B} \left(\mu|\nabla \psi (\x)|^2 + \half v(|\x|)
|\psi (\x)|^2\right)~d\x
\geq  \mu  \int_{\mathcal B} U(|\x|)
|\psi (\x)|^2 ~d\x.
\label{dysonlem2d}
\end{equation}
\end{lem}

\begin{proof}
In polar coordinates, $r,\theta$, one has
$|\nabla \psi|^2 \geq |\partial \psi /\partial r|^2$. Therefore, it
suffices to
prove that for each angle $\theta \in [0,2\pi)$, and with
$\psi (r,\theta)$ denoted simply by $f(r)$,
\begin{equation} \label{radial2}
\int_0^{R(\theta)}\left( \mu |\partial f(r) /\partial r|^2 +
\half v(r)|f(r)|^2 \right)rdr \geq
 \mu  \int_0^{R(\theta)}  U(r)|f(r)|^2 ~rdr,
\end{equation}
where $R(\theta)$ denotes the distance of the origin to the boundary
of $\mathcal{B}$ along the ray~$\theta$.

If $R(\theta) \leq R_0$ then \eqref{radial2} is trivial because the
right side is zero while the left side is evidently nonnegative.
(Here,
$v\geq0$ is used.)

If $R(\theta) > R_0$ for some given
value of $\theta$, consider the disc $\mathcal{D}(\theta)=
\{\x\in \mathbb{R}^2 \  0\leq |\x|\leq R(\theta) \}$ centered at the
origin in $\mathbb{R}^2$ and of radius $R(\theta) $.
Our function $f$ defines a radial function, $\x\mapsto f(\vert \x\vert)$ on
$\mathcal{D}(\theta)$, and \eqref{radial2} is
equivalent to
\begin{equation}\label{disc}
\int_{{\mathcal D}(\theta)} \left(\mu|\nabla f (|\x|)|^2 +
\frac{1}{2}v(|\x|)
|f(|\x|)|^2\right)d\x\geq
\mu\int_{{\mathcal D}(\theta)} U(|\x|)|f(|\x|)|^2 d\x.
\end{equation}

Now choose some $R\in (R_0,\ R(\theta))$ and note that the left side
of
\eqref{disc} is not smaller than the same quantity with
${\mathcal D}(\theta)$ replaced by the smaller disc ${\mathcal D}_R=
\{\x\in \mathbb{R}^2 \  0\leq |\x|\leq R \}$. (Again, $v\geq 0$ is
used.)
We now minimize this integral over ${\mathcal D}_R$, fixing $f(R)$.
This minimization problem leads to the zero energy scattering
equation.
Plugging in the solution and integrating by parts leads to
\begin{equation} \label{pointwise}
2\pi \int_0^{R(\theta)}\left( \mu |\partial f(r) /\partial r|^2 +
\frac{1}{2}v(r)|f(r)|^2 \right)rdr \geq    \frac{2\pi \mu}{\ln
(R/a)}|f(R)|^2 .
\end{equation}
The proof is completed
by multiplying  both sides of \eqref{pointwise} by $U(R)R\ln(R/a)$
and
integrating with respect to $R$ from $R_0$ to $R(\theta)$.
\end{proof}

As in Corollary \ref{2.6}, Lemma \ref{dyson2d} can be used to bound
the
many body Hamiltonian $H_N$ from below, as follows:
\begin{corollary}
For any $U$ as in Lemma {\rm\ref{dyson2d}} and any $0<\varepsilon<1$
\begin{equation}\label{epsilonbd}
H_N \geq \varepsilon T_N +(1-\varepsilon)\mu W
\end{equation}
with $T_N=-\mu\sum_{i=1}^{N}\Delta_{i}$ and
\begin{equation}
\label{W2}W(\x_{1},\dots,\x_{N})=\sum_{i=1}^{N}U\left(\min_{j,\,j\neq
i}|\x_{i}-\x_{j}|.\right).
\end{equation}
\end{corollary}

For $U$ we choose the following functions, parameterized by $R>R_{0}$:
\begin{equation}U_{R}(r)=\begin{cases}\nu(R)^{-1}&\text{for
$R_{0}<r<R$ }\\
0&\text{otherwise}
\end{cases}
\end{equation}
with $\nu(R)$ chosen so that
\begin{equation}
\int _{R_{0}}^{R}U_{R}(r)\ln(r/a)r\,dr=1
\end{equation}
for all $R>R_{0},$
i.e.,
\begin{equation}\label{nu}
\nu(R)=\int_{R_{0}}^{R}\ln(r/a)r\,dr=\mfr1/4 \left\{R^{2}
\left(\ln(R^{2}/a^{2})-1\right)-R_{0}^{2}
\left(\ln(R_{0}^{2}/a^{2})-1\right)\right\}.
\end{equation}
The nearest neighbor
interaction\index{nearest neighbor interaction} \eqref{W2} corresponding to $U_{R}$ will be denoted
$W_{R}$.

As in Section~\ref{subsect22} we shall need estimates  on the
expectation
value, $\langle W_R\rangle_{0}$,  of $W_{R}$ in the ground state of
$\varepsilon T_N$ of \eqref{epsilonbd} with
Neumann boundary conditions\index{boundary condition}. This is just the average value of $W_{R}$
in a hypercube in $\R^{2N}$.  Besides the
normalization factor $\nu(R)$, the computation involves
the volume (area) of the support of
$U_{R}$, which is
 \begin{equation} A(R)=\pi(R^{2}-R_{0}^{2}).
\end{equation}

In contrast to the three-dimensional situation the normalization
factor
$\nu(R)$ is not just a constant ($R$ independent) multiple of $A(R)$;
 the factor $\ln(r/a)$ in \eqref{1dyson} accounts for the more
complicated expressions in the two-dimensional case.   Taking into
account that $U_{R}$ is proportional to the characteristic function
of
a disc of radius $R$ with a hole of radius $R_{0}$, the following
inequalities for $n$ particles in a box of side length
$\ell$ are obtained by the same geometric reasoning as lead to
\eqref{firstorder}, cf.\
\cite{LY1998}:
\begin{eqnarray}\label{2dfirstorder}
\langle
W_R\rangle_{0}&\geq&\frac {n}{\nu(R)}
\left(1-\mfr {2R}/{\ell}\right)^2\left[1-(1-Q)^{(n-1)}\right]\\
\langle
W_R\rangle_{0}&\leq&
\frac
{n}{\nu(R)}\left[1-(1-Q)^{(n-1)}\right]
\end{eqnarray}
with
\begin{equation}
Q=A(R)/\ell^{2}
\end{equation}
  being   the relative volume occupied by the
support of the potential $U_{R}$.
Since $U_{R}^{2}=\nu(R)^{-1}U_{R}$ we also have
\begin{equation}
\langle
W_R^{2}\rangle_{0}\leq \frac n{\nu(R)}\langle
W_R\rangle_{0}.
\end{equation}

As in \cite{LY1998} we estimate $[1-(1-Q)^{(n-1)}]$ by
\begin{equation}
(n-1)Q\geq \left[1-(1-Q)^{(n-1)}\right]\geq \frac{(n-1)Q}{1+(n-1)Q}
\end{equation}
This gives
\begin{eqnarray}\label{firstorder2}
\langle
W_R\rangle_{0}&\geq&\frac {n(n-1)}{\nu(R)}
\cdot\frac{Q}{1+(n-1)Q},\\
\langle
W_R\rangle_{0}&\leq&
\frac {n(n-1)}{\nu(R)}
\cdot Q\ .
\end{eqnarray}

{F}rom  Temple's inequality\index{Temple's inequality} \cite{TE} we
obtain like in \eqref{temple} the estimate
\begin{equation}\label{temple2d}E_{0}(n,\ell)
\geq (1-\varepsilon)\langle
W_R\rangle_{0}\left(1-\frac{\mu \big(\langle
W_R^2\rangle_0-\langle W_R\rangle_0^2\big)}{\langle
W_R\rangle_0\big(E_{1}^{(0)}-\mu \langle W_R\rangle_0\big)}\right)
\end{equation}
where
\begin{equation}
E_{1}^{(0)}=\frac{\varepsilon\mu}{\ell^{2}}
\end{equation}
is the energy of the lowest excited state of $\varepsilon T_n$.
This estimate is valid for $E_{1}^{(0)}/\mu > \langle
W_R\rangle_0$, i.e., it is important that $\ell$ is not too big.

Putting \eqref{firstorder2}--\eqref{temple2d} together we obtain
the estimate
\begin{equation}\label{alltogether}E_{0}(n,\ell)\geq
\frac{n(n-1)}{\ell^{2}}\,\frac {A(R)}{\nu(R)}\,
K(n)
\end{equation}
with
\begin{equation}\label{k}
K(n)=(1-\varepsilon)\cdot
\frac{(1-\mfr{2R}/{\ell})^{2}}{1+(n-1)Q}\cdot
\left(1-\frac n{(\varepsilon\,\nu(R)/\ell^{2})-n(n-1)\,Q}\right)
\end{equation}
Note that $Q$ depends on $\ell$ and $R$, and $K$ depends on
$\ell$, $R$ and $\varepsilon$ besides $n$.  We have here dropped the
term  $\langle W_R\rangle_0^2$ in the numerator in \eqref{temple2d},
which is  appropriate  for the purpose of a lower bound.

We note that $K$ is monotonically decreasing in $n$, so for a given
$n$ we may replace $K(n)$ by $K(p)$ provided $p\geq n$.  As
explained in the previous chapter,
\eqref{superadd}--\eqref{estimate1},
convexity of $n\mapsto n(n-1)$ together with
superadditivity\index{superadditivity} of $E_{0}(n,\ell)$ in $n$
leads, for $p=4\rho\ell^{2}$,  to an estimate for the energy of $N$
particles in the
large box when the  side length $L$ is  an integer multiple of
$\ell$:
\begin{equation}\label{alltogether2}E_{0}(N,L)/N\geq \frac
{\rho A(R)}{\nu(R)}\left (1-\frac 1{\rho\ell^{2}}\right)
K(4\rho\ell^{2})
\end{equation}
with $\rho=N/L^2$.

Let us now look at the conditions on the parameters $\varepsilon$,
$R$
and $\ell$ that have to be met in order to
obtain a lower bound with the same leading term as the upper bound
\eqref{upperbound1}.

{F}rom \eqref{nu} we have
\begin{equation}\label{alltogether3}
\frac{A(R)}{\nu(R)}=\frac{4\pi}
{\left(\ln(R^{2}/a^{2})-1\right)}\left(1-{\rm
O}((R_{0}^{2}/R^{2})\ln(R/R_{0})\right)
\end{equation}
We thus see that as long as $a<R<\rho^{-1/2}$ the logarithmic factor
in the denominator in \eqref{alltogether3} has the right form for a
lower
bound. Moreover, for
Temple's inequality the denominator in the third factor in \eqref{k}
must be positive. With $n=4\rho\ell^2$ and
        $\nu(R)\geq {\rm(const.)} R^2\ln(R^2/a^2)\ {\rm for}\ R\gg
R_{0}$,
this condition amounts to
\begin{equation}\label{templecond}
{\rm (const.)}\varepsilon \ln(R^2/a^2) /\ell^{2}>\rho^{2}\ell^{4}.
\end{equation}
The relative error terms in \eqref{alltogether2} that have to be $\ll
1$
are
\begin{equation}\label{errors}
        \varepsilon,\quad \frac{1}{\rho\ell^{2}},\quad
\frac{R}{\ell},\quad\rho R^2,\quad
\frac{\rho\ell^4}{\varepsilon R^2\ln(R^2/a^2)}.
\end{equation}
We now choose
\begin{equation}
\varepsilon\sim|\ln(\rho a^2)|^{-1/5},
\quad \ell\sim \rho^{-1/2}|\ln(\rho a^2)|^{1/10},
\quad R\sim \rho^{-1/2}|\ln(\rho a^2)|^{-1/10}
\end{equation}

Condition \eqref{templecond} is satisfied since the left side is
$>{\rm
(const.)}|\ln(\rho a^2)|^{3/5}$ and the right side is $\sim |\ln(\rho
a^2)|^{2/5}$. The first three error terms in \eqref{errors} are all of
the same order, $|\ln(\rho a^2)|^{-1/5}$, the last is $\sim
|\ln(\rho a^2)|^{-1/5}(\ln|\ln(\rho a^2)|)^{-1}$. With these
choices, \eqref{alltogether2} thus leads to the following:

\begin{thm}[{\bf Lower bound}]
For all $N$ and $L$ large enough, such that $L>{\rm (\const )}
\rho^{-1/2}|\ln(\rho a^2)|^{1/10}$ and
$N>{\rm (\const )}|\ln(\rho a^2)|^{1/5}$ with $\rho=N/L^2$, the
ground state energy with Neumann boundary condition satisfies
\begin{equation} \label{lower}
E_{0}(N,L)/N\geq \frac{4\pi\mu\rho}{|\ln(\rho a^2)|}
\left(1-{\rm O}(|\ln(\rho a^2)|^{-1/5})\right).
\end{equation}
\end{thm}

In combination with the upper bound \eqref{upperbound1} this also
proves
\begin{thm}[{\bf Energy at low density in the thermodynamic
limit}]
\begin{equation}
\lim_{\rho a^2\to 0}\frac{e_0(\rho)}{4\pi\mu\rho|\ln(\rho
a^2)|^{-1}}=1
\end{equation}
where $e_0(\rho)=\lim_{N\to\infty} E_0(N,\rho^{-1/2}N^{1/2})/N$.
This holds irrespective of boundary conditions.
\label{limitthm}
\end{thm}

As in the three-dimensional case, Theorem \ref{limitthm} is also valid
for an infinite range potential\index{infinite range potential} $v$ provided that $v\geq 0$ and for
some $R$ we have $\int_{R}^{\infty} v(r)r\ dr <\infty$, which
guarantees a finite scattering length.

\chapter[Generalized Poincar\'e Inequalities]{Generalized Poincar\'e\vspace*{-2mm}\newline Inequalities}\label{poincare}

This chapter contains some lemmas that are of independent mathematical
interest, but whose significance for the physics of the Bose gas may
not be obvious at this point. They will, however, turn out to be
important tools for the discussion of Bose-Einstein condensation (BEC)
and superfluidity in the next chapters.

The classic Poincar\'e inequality\index{Poincar\'e inequality}
\cite{LL01} bounds the $L^q$-norm of a function, $f$, orthogonal to a
given function $g$ in a domain $\K$, in terms of some $L^p$-norm of
its gradient in $\K$.  For the proof of BEC we shall need a
generalization of this inequality where the estimate is in terms of
the gradient of $f$ on a subset $\Omega\subset\K$ and a remainder that
tends to zero with the volume of the complement $\Omega^c=\K\setminus
\Omega$.  For superfluidity it will be necessary to generalize this
further by adding a vector potential\index{vector potential} to the
gradient.  This is the most complex of the lemmas because the other
two can be derived directly from the classical Poincar\'e inequality
using H\"older's inequality.  The first lemma is the simplest variant
and it is sufficient for the discussion of BEC in the case of a
homogeneous gas.  In this case the function $g$ can be taken to be the
constant function.  The same holds for the second lemma, which will be
used for the discussion of superfluidity in a homogeneous gas with
periodic boundary conditions\index{boundary condition}, but the
modification of the gradient requires a more elaborate proof.  The
last lemma, that will be used for the discussion of BEC in the
inhomogeneous case, is again a simple consequence of the classic
Poincar\'e and H\"older inequalities. For a more comprehensive
discussion of generalized Poincar\'e inequalities with further
generalizations we refer to \cite{lsy02}.

\begin{lem}[\textbf{Generalized Poincar{\'e} inequality:
Homogeneous case}]
\label{lem2b}
Let $\K\subset \R^3$ be a cube of side length $L$, and define the
average of a function $f\in L^1(\K)$ by $$ \langle
f\rangle_\K=\frac 1{L^3} \int_\K f(\x)\, d\x \ .$$ There
exists a constant $C$ such that for all measurable sets
$\Omega\subset\K$ and all $f\in H^1(\K)$ the inequality
\begin{equation} \label{poinchom}
 \int_{\K} |f(\x)-\langle f\rangle_\K |^2 d\x \leq C
\left(L^2\int_\Omega |\nabla f(\x)|^2 d\x
+|\Omega^c|^{2/3}\int_\K |\nabla f(\x)|^2 d\x \right)
\end{equation}
holds. Here $\Omega^c=\K\setminus\Omega$, and
$|\cdot|$ denotes the measure of a set.
.
\end{lem}

\begin{proof} By scaling, it suffices to consider the case $L=1$.
Using
the usual Poincar\'e-Sobolev inequality on $\K$ (see
\cite{LL01}, Thm. 8.12), we infer that there exists a $C>0$ such
that
\begin{eqnarray}\nonumber
\|f-\langle f\rangle_\K\|_{L^2(\K)}^2&\leq& \half C
\|\nabla f\|_{L^{6/5}(\K)}^2\\ &\leq&  C\left(\|\nabla
f\|_{L^{6/5}(\Omega)}^2+\|\nabla f\|_{L^{6/5}(\Omega^c)}^2\right)\
.
\end{eqnarray}
Applying H\"older's inequality $$ \|\nabla f\|_{L^{6/5}(\Omega)}
\leq \|\nabla f\|_{L^{2}(\Omega)}|\Omega|^{1/3} $$ (and the
analogue with $\Omega$ replaced by $\Omega^c$), we see that
(\ref{poinchom}) holds.
\end{proof}

In the next lemma $\K$ is again a cube of side length $L$, but we
now replace the gradient $\nabla$ by
\beq\label{nablaphi}
\nabla_{\varphi}:=\nabla+i(0,0,\varphi/L),
\eeq
where $\varphi$ is a real parameter, and require periodic boundary
conditions on $\K$.

\begin{lem}[\textbf{Generalized Poincar\'e inequality with a
vector potential}]\label{L2}
For any $|\varphi|<\pi$ there  are constants
$c>0$ and $C<\infty$  such that for all subsets
$\Omega\subset\K$ and all functions $f\in H^1(\K)$ with periodic
boundary
conditions on $\K$  the
following estimate holds:
\begin{multline}\label{poinc}
\Vert\nabla_\varphi
f\Vert_{L^2(\Omega)}^2\geq \frac{\varphi^2}{L^2}\|f\|_{L^2(\K)}^2
+\frac c{L^2} \Vert
f-\langle f\rangle_{\K}
\Vert_{L^2(\K)}^2\\ -C\left(\|\nabla_\varphi f\|_{L^2(\K)}^2 + \frac
1{L^2}
\|f\|_{L^2(\K)}^2\right) \left(\frac{|\Omega^c|}{L^3}\right)^{1/2} \
.
 \end{multline} Here $\vert\Omega^c\vert$ is the volume of
 $\Omega^c=\K\setminus\Omega$, the complement of $\Omega$ in $\K$.
\end{lem}

\begin{proof}
We shall derive (\ref{poinc}) from a special form of this inequality
that holds for all functions that are orthogonal to the constant
function. Namely, for any positive $\alpha<2/3$ and some constants
$c>0$ and $\widetilde C<\infty$ (depending only on $\alpha$ and
$|\varphi|<\pi$) we claim that
\begin{equation}\label{poinc2}
\|\nabla_\varphi h\|_{L^2(\Omega)}^2  \geq
\frac{\varphi^2+c}{L^2}
\Vert h\Vert_{L^2(\K)}^2-\widetilde
C\left(\frac{|\Omega^c|}{L^3}\right)^\alpha\Vert\nabla_\varphi
h\Vert_{L^2(\K)}^2 \ ,
\end{equation}
provided
$\langle 1,h\rangle_{\K} =0$. (Remark: Eq.~(\ref{poinc2})
holds also for $\alpha=2/3$, but the proof is slightly more
complicated in that case. See \cite{lsy02}.) If (\ref{poinc2}) is
known
the
derivation of (\ref{poinc}) is easy: For any $f$, the function
$h=f-L^{-3}\langle 1,f\rangle_{\K}$ is orthogonal to
$1$. Moreover,
\begin{eqnarray}\nonumber
\Vert\nabla_\varphi h\Vert^2_{L^2(\Omega)}&=&
\Vert\nabla_\varphi h\Vert^2_{L^2(\K)}-\Vert\nabla_\varphi
h\Vert^2_{L^2(\Omega^c)}\nonumber\\
&=&
\Vert\nabla_\varphi f\Vert^2_{L^2(\Omega)}-
\frac{\varphi^2}{L^2}\vert\langle L^{-3/2},f\rangle_{\K}\vert^2
\left(1+\frac{|\Omega^c|}{L^3}\right)
\nonumber \\ &&\quad +2\frac{\varphi}{L}{\rm
Re}\,\langle L^{-3/2},f\rangle_{\K}
\langle\nabla_\varphi f,L^{-3/2}
\rangle_{\Omega^c}\nonumber\\ \nonumber
&\leq& \Vert\nabla_\varphi f\Vert^2_{L^2(\Omega)}-
\frac{\varphi^2}{L^2} \vert\langle L^{-3/2} ,f\rangle_{\K}\vert^2
\\ \nonumber && \quad
+\frac{|\varphi|}{L}
\left(L \Vert\nabla_\varphi f\Vert_{L^2(\K)}^2+\frac 1L \Vert
f\Vert_{L^2(\K)}^2\right) \left(\frac{|\Omega^c|}{L^3}\right)^{1/2}
\\
\label{quadrat1}
\end{eqnarray}
and
\begin{eqnarray}\nonumber
\frac{\varphi^2+c}{L^2}\Vert
h\Vert_{L^2(\K)}^2&=&\frac{\varphi^2}{L^2}\left(  \Vert
f\Vert_{L^2(\K)}^2- \vert\langle L^{-3/2}
,f\rangle_{\K}\vert^2\right)\\  && +
\frac c{L^2} \Vert f-L^{-3}\langle 1 ,f\rangle_{\K} \label{quadrat2}
\Vert_{L^2(\K)}^2 \ .
\end{eqnarray}
Setting $\alpha=\half$, using $\|\nabla_\varphi h\|_{L^2(\K)}\leq
\|\nabla_\varphi f\|_{L^2(\K)}$ in the last term in (\ref{poinc2})
and combining (\ref{poinc2}), (\ref{quadrat1}) and (\ref{quadrat2})
gives
(\ref{poinc}) with $C=|\varphi|+\widetilde C$.

We now turn to the proof of (\ref{poinc2}). For simplicity we set
$L=1$. The general case  follows by scaling.  Assume that
(\ref{poinc2})
is false. Then there exist sequences of constants $C_{n}\to\infty$,
functions $h_{n}$ with $\Vert h_{n}\Vert_{L^2(\K)}=1$ and
$\langle 1,h_{n}\rangle_{\K} =0$, and domains $\Omega_{n}\subset\K$
such
that
\begin{equation}\label{false}
\lim_{n\to\infty}\left\{
\Vert\nabla_\varphi
h_{n}\Vert_{L^2(\Omega_{n})}^2+C_{n}\vert\Omega_{n}^c\vert^\alpha\Vert\nabla_\varphi
h_{n}\Vert_{L^2(\K)}^2\right\}
\leq \varphi^2 \ .
\end{equation}
We shall show that this leads to a contradiction.
\goodbreak

Since the sequence $h_{n}$ is bounded in $L^2(\K)$, it has a
subsequence, de\-no\-ted again by $h_{n}$, that converges weakly to
some $h\in L^2(\K)$ (i.e., $\langle g,h_{n}\rangle_{\K}\to \langle
g,h\rangle_{\K}$ for all $g\in L^2(\K)$).  Moreover, by H\"older's
inequality the $L^p(\Omega_{n}^c)$ norm $\Vert \nabla_\varphi
h_{n}\Vert_{L^p(\Omega_{n}^c)}=(\int_{\Omega^c_{n}} \vert
\nabla_{\varphi}h(\x)\vert^pd\x)^{1/p}$ is bounded by
$\vert\Omega_{n}^c\vert^{\alpha/2}\Vert\nabla_\varphi
h_{n}\Vert_{L^2(\K)}$ for $p=2/(\alpha+1)$.  From (\ref{false}) we
conclude that $\Vert \nabla_\varphi h_{n}\Vert_{L^p(\Omega_{n}^c)}$ is
bounded and also that $\Vert \nabla_\varphi
h_{n}\Vert_{L^p(\Omega_{n})}\leq\Vert \nabla_\varphi
h_{n}\Vert_{L^2(\Omega_{n})}$ is bounded. Altogether, $\nabla_\varphi
h_{n}$ is bounded in $L^p(\K)$, and by passing to a further
subsequence if necessary, we can therefore assume that $\nabla_\varphi
h_{n}$ converges weakly in $L^p(\K)$. The same applies to $\nabla
h_{n}$. Since $p=2/(\alpha+1)$ with $\alpha<2/3$ the hypotheses of the
Rellich-Kondrashov Theorem\index{Rellich-Kondrashov Theorem}
\cite[Thm.~8.9]{LL01} are fulfilled and consequently $h_{n}$ converges
{\it strongly} in $L^2(\K)$ to $h$ (i.e., $\Vert
h-h_{n}\Vert_{L^2(\K)}\to 0$).  We shall now show that
\begin{equation}\label{lowersemi}
\liminf_{n\to\infty}\Vert\nabla_\varphi
h_{n}\Vert_{L^2(\Omega_{n})}^2\geq \Vert\nabla_\varphi
h\Vert_{L^2(\K)}^2 \ .
\end{equation}
This will complete the proof because the $h_{n}$ are normalized and
orthogonal to $1$ and the same holds for $h$ by strong
convergence. Hence the right side of (\ref{lowersemi}) is necessarily
$>\varphi^2$, since for $|\varphi|<\pi$ the lowest eigenvalue of
$-\nabla_\varphi^2$, with constant eigenfunction, is
non-degenerate. This contradicts (\ref{false}).

Eq.~(\ref{lowersemi}) is essentially a consequence of the weak lower
semicontinuity of the $L^2$ norm, but the dependence on $\Omega_{n}$
leads to a slight complication.  First, Eq.~(\ref{false}) and
$C_{n}\to
\infty$ clearly imply that $\vert\Omega_{n}^c\vert\to 0$, because
$\Vert\nabla_\varphi
h_{n}\Vert_{L^2(\K)}^2>\varphi^2$.  By choosing
a subsequence we may assume that
$\sum_{n}\vert\Omega_{n}^c\vert<\infty$.  For some fixed $N$ let
$\widetilde\Omega_{N}=\K\setminus\cup_{n\geq N}\Omega_{n}^c$. Then
$\tilde\Omega_{N}\subset\Omega_{n}$ for $n\geq N$.
Since $\Vert\nabla_\varphi
h_{n}\Vert_{L^2(\Omega_{n})}^2$ is bounded, $\nabla_\varphi
h_{n}$ is also bounded in $L^2(\widetilde\Omega_{N})$ and a
subsequence
of it converges weakly in $L^2(\widetilde\Omega_{N})$ to
$\nabla_\varphi
h$. Hence
\begin{equation}\label{lowersemi2}
\liminf_{n\to\infty}\Vert\nabla_\varphi
h_{n}\Vert_{L^2(\Omega_{n})}^2  \geq
\liminf_{n\to\infty}\Vert\nabla_\varphi
h_{n}\Vert_{L^2(\widetilde \Omega_{N})}^2\geq\Vert\nabla_\varphi
h\Vert_{L^2(\widetilde\Omega_{N})}^2  \ .
\end{equation}
Since
$\widetilde\Omega_{N}\subset \widetilde\Omega_{N+1}$ and
$\cup_{N}\widetilde\Omega_{N}=\K$ (up to a set of measure zero), we
can
now let $N\to\infty$ on the right side of (\ref{lowersemi2}). By
monotone convergence this converges to $\Vert\nabla_\varphi
h\Vert_{L^2(\K)}^2$. This proves (\ref{lowersemi}) which, as remarked
above, contradicts
(\ref{false}).
\end{proof}

The last lemma is a simple generalization of Lemma \ref{lem2b} with
$\K\subset\R^m$ a bounded and connected set that is sufficiently nice
so that the Poincar\'e-Sobolev inequality\index{Poincar\'e-Sobolev
  inequality} (see \cite[Thm.~8.12]{LL01}) holds on $\K$. In
particular, this is the case if $\K$ satisfies the cone
property\index{cone property} \cite{LL01} (e.g., if $\K$ is a
rectangular box or a cube).  Moreover, the constant function on $\K$
is here replaced by a more general bounded function.

\begin{lem}[\textbf{Generalized Poincar{\'e} inequality:
Inhomog. case}]\label{lem2}
For $d\geq 2$ let $\K\subset\R^d$ be as explained above, and let
$h$ be a bounded function with $\int_\K h=1$. There exists a
constant $C$ (depending only on $\K$ and $h$) such that for all
measurable sets $\Omega\subset\K$ and all $f\in H^1(\K)$  with
$\int_\K f h\, d\x=0$, the inequality
\begin{equation} \label{poinc3}
 \int_{\K} |f(\x)|^2 d\x \leq C \left(\int_\Omega |\nabla f(\x)|^2 d\x
+\left(\frac{|\Omega^c|}{|\K|}\right)^{2/d}\int_\K |\nabla
f(\x)|^2 d\x \right)
\end{equation}
holds. Here $|\cdot|$ denotes the measure of a set, and
$\Omega^c=\K\setminus\Omega$.
\end{lem}

\begin{proof} By the usual Poincar\'e-Sobolev inequality on $\K$ (see
\cite[Thm.~8.12]{LL01}),
\begin{eqnarray}\nonumber
\|f\|_{L^2(\K)}^2&\leq& \tilde C \|\nabla
f\|_{L^{2d/(d+2)}(\K)}^2\\ &\leq& 2\tilde C\left(\|\nabla
f\|_{L^{2d/(d+2)}(\Omega)}^2+\|\nabla
f\|_{L^{2d/(d+2)}(\Omega^c)}^2\right),
\end{eqnarray}
if $d\geq 2$
and $\int_\K f h=0$. Applying H\"older's inequality $$ \|\nabla
f\|_{L^{2d/(d+2)}(\Omega)} \leq \|\nabla
f\|_{L^{2}(\Omega)}|\Omega|^{1/d} $$ (and the analogue with
$\Omega$ replaced by $\Omega^c$), we see that (\ref{poinc}) holds
with $C=2|\K|^{2/d}\tilde C$.
\end{proof}
\newpage
\
\thispagestyle{empty}

\chapter[BEC and Superfluidity for Homogeneous
Gases]{Bose-Einstein Condensation and\vspace*{-2mm}\newline Superfluidity for Homogeneous\vspace*{-2mm}\newline
Gases}\label{bec}

\section{Bose-Einstein Condensation}
\index{Bose-Einstein condensation}

Bose-Einstein condensation (BEC) is the phenomenon of a macroscopic
occupation of a single one-particle quantum state, discovered by
Einstein for thermal equilibrium states of an ideal Bose gas at
sufficiently low temperatures \cite{Einstein}. We are here concerned
with interacting Bose gases, where the question of the existence of
BEC is highly nontrivial even for the ground state. Due to the
interaction the many body ground state is not a product of
one-particle states but the concept of a macroscopic occupation of a
single state acquires a precise meaning through the {\it one-particle
  density matrix}\index{one-particle density matrix}, as discussed in
Section~\ref{defbecsect}. Namely, this is the statement that the
operator on $L^2(\R^d)$ ($d=2$ or $3$) given by the kernel
\begin{equation}\label{defgamma}
 \gamma(\x,\x')=N\int \Psi_0(\x,\X)
\Psi_0(\x',\X) d\X
\end{equation}
has an eigenvalue of
order $N$. Here, $\Psi_0$ denotes the normalized ground state wave function.
In case the eigenfunction corresponding to the largest
eigenvalue of $\gamma$ is constant (or, at least, not orthogonal to the constant
function), this means that
\begin{equation}\label{defbec}
 \frac 1{L^d} \int\!\!\!\int \gamma(\x,\, \y) d\x
d\y \geq c N
\end{equation}
for all large $N$, with $c>0$ depending only on the density $N/L^d$.

The problem remains open after more than 75 years since the first
investigations on
the Bose gas \cite{Bose,Einstein}.
Our
construction in Chapter~\ref{sect3d} shows that (in 3D) BEC exists on
a length scale of order $(\rho^{-1} Y^{ -1/17})^{1/3}$ which, unfortunately,
is not a `thermodynamic' length like $\textrm{volume}^{1/3}$. The same
remark applies to the 2D case of Chapter~\ref{sect2d}, where BEC is proved over a
length scale $\rho^{-1/2}|\ln(\rho a^2)|^{1/10}$.

In a certain limit, however, one can prove (\ref{defbec}), as has been
shown in \cite{LS02}.  In this limit the interaction
potential $v$ is varied with $N$ so that the ratio $a/L$ of the
scattering length to the box length is of order $1/N$, i.e., the
parameter $Na/L$ is kept fixed.
Changing $a$
with $N$ can be done by scaling, i.e., we write
\begin{equation}\label{v1}
v(|\x|)=\frac 1{a^2} v_1(|\x|/a)
\end{equation}
for some $v_1$ having scattering length $1$, and vary $a$ while
keeping $v_1$ fixed.\footnote{By scaling, this is mathematically
  equivalent to fixing the interaction potential $v$ (and therefore
  fixing $a$) but taking $L\sim N$, i.e., $\rho=N/L^3\sim N^{-2}$.} It
is easily checked that the $v$ so defined has scattering length $a$.
It is important to note that, in the limit considered, $a$ tends to
zero (as $N^{-2/3}$ since $L=(N/\rho)^{1/3}\sim N^{1/3}$ for $\rho$
fixed), and $v$ becomes a {\it hard} potential of {\it short} range.
This is the {\it opposite} of the usual mean field limit where the
strength of the potential goes to zero while its range tends to
infinity.

We shall refer to this as the {\it Gross-Pitaevskii (GP)
  limit}\index{Gross-Pitaevskii!limit} since $Na/L$ will turn out to
be the natural interaction parameter for inhomogeneous Bose gases
confined in traps, that are described by the Gross-Pitaevskii equation
discussed in Chapters~\ref{sectgp} and~\ref{becsect}. Its significance
for a homogeneous gas can also be seen by noting that $Na/L$ is the
ratio of $\rho a$ to $1/L^2$, i.e., in the GP limit the interaction
energy per particle is of the same order of magnitude as the energy
gap in the box, so that the interaction is still clearly visible, even
though $a\to 0$. Note that $\rho a^3\sim N^{-2}$ in the GP limit, so
letting $N\to\infty$ with $\rho$ fixed and $Na/L$ fixed can be
regarded as a {\it simultaneous thermodynamic and low density limit}.
For simplicity, we shall here treat only the 3D case.

\begin{thm}[\textbf{BEC in a dilute limit}]\label{hombecthm}
Assume that, as $N\to\infty$, $\rho=N/L^3$ and $\g=Na/L$ stay
fixed, and impose either periodic or Neumann boundary conditions
for $H_N$. Then
\begin{equation}\label{xxyy}
\lim_{N\to\infty} \frac 1N \frac 1{L^3} \int\!\!\!\int \gamma(\x,\, \y)
d\x d\y = 1\ .
\end{equation}
\end{thm}

The reason  we do not
deal with Dirichlet boundary conditions at this point should be clear
from the discussion preceding the theorem: There would be an
additional contribution $\sim 1/L^2$ to the energy, of the same order as the
interaction energy, and the
system would not be homogeneous any more. Dirichlet boundary
conditions can, however, be treated with the methods of Chapter~\ref{becsect}.

By scaling, the limit in Theorem \ref{hombecthm} is equivalent to
considering a Bose gas in a {\it fixed box} of side length $L=1$, and
keeping $Na$ fixed as $N\to\infty$, i.e., $a\sim 1/N$.  The ground
state energy of the system is then, asymptotically, $N\times 4\pi Na$,
and Theorem \ref{hombecthm} implies that the one-particle reduced
density matrix $\gamma$ of the ground state converges, after division
by $N$, to the projection onto the constant function.  An analogous
result holds true for inhomogeneous systems as will be discussed
in Chapter~\ref{becsect}.

The proof of Theorem \ref{hombecthm} has two main ingredients. One is
{\it localization of the energy}\index{energy localization} that is
stated as Lemma \ref{L1} below.  This lemma is a refinement of the
energy estimates of Section~\ref{subsect22} and says essentially that
the kinetic energy of the ground state is concentrated in a subset of
configuration space where at least one pair of particles is close
together and whose volume tends to zero as $a\to 0$. The other is the
generalized Poincar\'e inequality\index{Poincar\'e inequality!generalized},
Lemma~\ref{lem2b}, from which one deduces
that the one-particle density matrix is approximately constant if the
kinetic energy is localized in a small set.

The localization lemma will be proved in a slightly more general
version than is necessary for Theorem~\ref{hombecthm}, namely with the
gradient $\nabla$ replaced by
$\nabla_{\varphi}=\nabla+i(0,0,\varphi/L)$, cf.\ Eq.\
\eqref{nablaphi}.  We denote by $H_{N}'$ the corresponding many-body
Hamiltonian \eqref{ham} with $\nabla_{\varphi}$ in place of $\nabla$.
This generalization will be used in the subsequent discussion of
superfluidity, but a reader who wishes to focus on Theorem
\ref{hombecthm} only can simply ignore the $\varphi$ and the reference
to the diamagnetic inequality\index{diamagnetic inequality} in the
proof.  We denote the gradient with respect to $\x_{1}$ by $\nabla_1$,
and the corresponding modified operator by $\nabla_{1,\varphi}$.

\begin{lem}[\textbf{Localization of energy}]\label{L1}
Let $\K$ be a box of side length $L$.
    For all symmetric, normalized wave functions
$\Psi(\x_{1},\dots,\x_{N})$ with periodic boundary conditions on
$\K$, and for $N\geq Y^{-1/17}$,
\begin{eqnarray} \nonumber
\frac1N\langle\Psi, H_{N}'\Psi\rangle &\geq& \big(1-\const
Y^{1/17}\big) \\ && \times \Big(4\pi\mu\rho
a+
\mu \int _{\K^{N-1}}
d\X \int_{\Omega_{\X}}d\x_{1}\big|
\nabla_{1,\varphi}\Psi(\x_{1},\X)\big\vert^2\Big) \ , \nonumber \\
\label{lowerbd}
\end{eqnarray}
\vskip-2mm\noindent
where $\X=(\x_{2},\dots,\x_{N})$, $d\X=\prod_{j=2}^N d\x_j$, and
\begin{equation}
\Omega_{\X}=\left\{\x_{1}: \min_{j\geq 2}\vert\x_{1}-
\x_{j}\vert\geq R \right\}
    \end{equation}
    \vspace*{-1mm}with $R=a Y^{-5/17}$.
\end{lem}
\goodbreak

\begin{proof}
Since $\Psi$ is symmetric, the left side of
(\ref{lowerbd}) can be written as
\begin{equation}
\int_{\K^{N-1}} d\X \int_\K d\x_{1}
\Big[\mu\big|\nabla_{1,\varphi}\Psi(\x_{1},\X)\big\vert^2
 +\half\sum_{j\geq 2}
v(\vert\x_{1}-\x_{j}\vert)|\Psi(\x_{1},\X)\vert^2\Big]\ .
\end{equation}
For any $\varepsilon>0$ and $R>0$ this is
\begin{equation}
\geq \varepsilon T+(1-\varepsilon)(T^{\rm in}+I)+(1-\varepsilon)
T_{\varphi}^{\rm out} \ ,
\end{equation}
with
\begin{equation}\label{15}
T=\mu\int_{\K^{N-1}} d\X\int_\K
d\x_{1}\big|\nabla_1\vert\Psi(\x_{1},\X)
\vert\big|^2 \ ,
\end{equation}
\begin{equation}\label{16}
\quad T^{\rm in}=\mu
\int_{\K^{N-1}} d\X\int_{\Omega^c_{\X}} d\x_{1}\big|\nabla_1
\vert\Psi(\x_{1},\X)\vert\big|^2\ ,
\end{equation}
\begin{equation}
T_{\varphi}^{\rm out}=\mu\int_{\K^{N-1}} d\X\int_{\Omega_{\X}} d\x_{1}
\big\vert\nabla_{1,\varphi}
\Psi(\x_{1},\X)\big\vert^2\ ,
\end{equation}
and
\begin{equation}
I=\half\int_{\K^{N-1}}
d\X\int_\K d\x_{1}\sum_{j\geq 2}
v(\vert\x_{1}-\x_{j}\vert)|\Psi(\x_{1},\X)\vert^2 \ .
\end{equation}
Here
\begin{equation}\Omega^c_{\X}=\left\{\x_{1}: \vert\x_{1}-
\x_{j}\vert<R\ \text{for some }j\geq 2\right\}
\end{equation}
is the complement of $\Omega_{\X}$, and the diamagnetic inequality
\cite{LL01}
$\vert\nabla_{\varphi}f(\x)\vert^2\geq
\left|\nabla|f(\x)|\right|^2$
has been used.
The proof is completed by using the estimates used for the proof of
Theorem 2.4,
in particular (\ref{estimate1}) and
(\ref{Kformula2})--(\ref{exponents}),
which tell us that for
$\varepsilon=Y^{1/17}$ and $R=aY^{-5/17}$
\begin{equation}\label{abs}
\varepsilon T+(1-\varepsilon)(T^{\rm in}+I)\geq\big(1-\const
Y^{1/17}\big)
4\pi\mu\rho a
\end{equation}
as long as $N\geq Y^{-1/17}$.
\end{proof}

\begin{proof}[Proof of Theorem {\rm\ref{hombecthm}}]
We combine Lemma \ref{L1}
(with $\varphi=0$ and hence $H'_{N}=H_{N}$) with
Lemma \ref{lem2b} that gives a lower bound to the second term on the
right side of \eqref{lowerbd}. We thus infer that, for any symmetric
$\Psi$
with
$\langle \Psi,\Psi\rangle=1$ and for $N$ large enough,
\begin{eqnarray} \nonumber
&&\frac1N\langle\Psi,H_{N}\Psi\rangle  \big(1-\const
Y^{1/17}\big)^{-1}\\ \nonumber && \geq  4\pi\mu\rho a  - C Y^{1/17}
\Big(\frac 1{L^2} - \frac 1N
\big\langle\Psi,\mbox{$ \sum_j$} \nabla_j^2
\Psi\big\rangle\Big)
 \\ \nonumber && \quad + \frac c{L^2}
\int_{\K^{N-1}}
d\X \int_{\K} d\x_{1}\Big|
\Psi(\x_1,\X) -L^{-3}\big[\mbox{$ \int_\K$} d\x
\Psi(\x,\X)\big]
\Big|^2 \ ,\\ \label{lowerbd2}
\end{eqnarray}
where we used that $|\Omega^c|\leq \mbox{$\frac{4\pi}3$} N R^3=
\const L^3 Y^{2/17}$. For the ground state wave
function
$\Psi_0$ the energy per particle, $N^{-1}\big\langle\Psi_0,
H_N
\Psi_0\big\rangle$,  is bounded from above by
$4\pi\mu\rho a(1+{\rm const}\, Y^{1/3})$ according to
 \eqref{upperbound}. The same holds for the kinetic energy per particle,
$-N^{-1}\big\langle\Psi_0,
\mbox{$ \sum_j$} \nabla_j^2
\Psi_0\big\rangle$ that appears on the right side of (\ref{lowerbd2}). We
now multiply
the inequality (\ref{lowerbd2}) by $L^2$ and use
the upper bound \eqref{upperbound}. The terms  $ 4\pi\mu\rho a L^2$
on both sides of the resulting inequality cancel. For the remaining terms we note that
in the limit considered $\rho a L^2=\const$ while $Y\to 0$. Hence the positive last term has to vanish in the limit and because $c>0$ this means
\begin{equation}
\lim_{N\to\infty} \int_{\K^{N-1}}
d\X \int_{\K} d\x_{1}\Big|
\Psi_0(\x_1,\X)\\ -L^{-3}\big[\mbox{$ \int_\K$} d\x
\Psi_0(\x,\X)\big]
\Big|^2 = 0 \ .
\end{equation}
This proves (\ref{xxyy}), since
\begin{eqnarray}\nonumber
 &&\int_{\K^{N-1}}
d\X \int_{\K} d\x_{1}\Big|
\Psi_0(\x_1,\X)-L^{-3}\big[\mbox{$ \int_\K$} d\x \Psi_0(\x,\X)\big]
\Big|^2 \\ && = 1- \frac 1{NL^3} \int_{\K\times\K} \gamma(\x,\x') d\x
d\x'
\ . \label{ahaha}
\end{eqnarray}
\vskip-3mm
\end{proof}

\section{Superfluidity}
\index{superfluidity}

The phenomenological two-fluid model of superfluidity (see, e.g.,
\cite{TT})
is based on the
idea that the particle density $\rho$ is composed of two parts, the
density $\rho_{\rm s}$ of the inviscid superfluid and the normal
fluid density
$\rho_{\rm n}$.  If an external velocity field is imposed on the fluid
(for instance by moving the walls of the container) only the viscous
normal
component responds to the velocity field, while the superfluid
component stays at rest.  In accord
with these ideas the superfluid density in the ground state is
often defined as follows \cite{HM}: Let $E_0$ denote the ground
state energy of the system in the rest frame and $E_0'$ the ground
state energy, measured in the
moving frame, when a velocity field ${\bf v}$ is imposed.
Then for small ${\bf v}$
\begin{equation}\label{rhos}
    \frac{E_0'}N=\frac {E_0}N+({\rho_{\rm s}}/\rho)\half{ m} {\bf
    v}^2+O(|{\bf v}|^4)
\end{equation}
where $N$ is the particle number and $m$ the particle mass.  At
positive temperatures the ground state energy should be replaced by
the free energy.  (Remark: It is important here that (\ref{rhos})
holds uniformly for all large $N$; i.e., that the error term $O(|{\bf
  v}|^4)$ can be bounded independently of $N$.  For fixed $N$ and a
finite box, Eq.\ (\ref{rhos}) with $\rho_{\rm s}/\rho=1$ always holds
for a Bose gas with an arbitrary interaction if ${\bf v}$ is small
enough, owing to the discreteness of the energy spectrum.\footnote{The
  ground state with ${\bf v}=0$ remains an eigenstate of the
  Hamiltonian with arbitrary ${\bf v}$ (but not necessarily a ground
  state) since its total momentum is zero.  Its energy is $\half m
  N{\bf v}^2$ above the ground state energy for ${\bf v}=0$.  Since in
  a finite box the spectrum of the Hamiltonian for arbitrary ${\bf v}$
  is discrete and the energy gap above the ground state is bounded
  away from zero for ${\bf v}$ small, the ground state for ${\bf v}=0$
  is at the same time the ground state of the Hamiltonian with ${\bf
    v}$ if $\half m N{\bf v}^2$ is smaller than the gap.}) There are
other definitions of the superfluid density that may lead to different
results \cite{PrSv}, but this is the one we shall use here. We shall
not dwell on this issue since it is not clear that there is a \lq\lq
one-size-fits-all\rq\rq\ definition of superfluidity.  For instance,
in the definition we use here the ideal Bose gas is a perfect
superfluid in its ground state, whereas the definition of
Landau\index{Landau} in terms of a linear dispersion relation of
elementary excitations\index{elementary excitations} would indicate
otherwise.  Our main result is that with the definition adopted here
there is 100\% superfluidity in the ground state of a 3D Bose gas in
the GP limit explained in the previous section.

One of the unresolved issues in the theory of superfluidity is its
relation to Bose-Einstein condensation (BEC).  It has been argued that
in general neither condition is necessary for the other (c.f., e.g.,
\cite{huang,giorgini,KT}), but in the case considered here, i.e., the
GP limit of a 3D gas, we show that 100\% BEC into the constant wave
function (in the rest frame) prevails even if an external velocity
field is imposed.  A simple example illustrating the fact that BEC is
not necessary for superfluidity is the 1D hard-core Bose gas.  This
system is well known to have a spectrum like that of an ideal Fermi
gas \cite{gir} (see also Chapter~\ref{1dsect} and
Appendix~\ref{chap3}), and it is easy to see that it is superfluid in
its ground state in the sense of (\ref{rhos}).  On the other hand, it
has no BEC \cite{Lenard,PiSt}.  The definition of the superfluid
velocity as the gradient of the phase of the condensate wave function
\cite{HM,baym} is clearly not applicable in such cases.

We consider a Bose gas with the Hamiltonian (\ref{ham}) in a box $\K$
of side length $L$, assuming periodic boundary
conditions\index{boundary condition} in all three coordinate
directions.  Imposing an external velocity field ${\bf
  v}=(0,0,\pm|{\bf v}|)$ means that the momentum operator ${\bf
  p}={-{\rm i}\hbar \nabla}$ is replaced by ${\bf p}-m{\bf v}$,
retaining the periodic boundary conditions.  The Hamiltonian in the
moving frame is thus
\begin{equation}\label{hamprime}
H_N'  = -\mu\sum_{j=1}^N \nabla_{j,\varphi}^2+
\sum_{1\leq i<j\leq N}v(\vert\x_{i}-\x_{j}\vert) \ ,
\end{equation}
where  $\nabla_{j,\varphi}=\nabla_{j}+{\rm i}(0,0,\varphi/L)$ and
the dimensionless phase $\varphi$ is connected to
the velocity ${\bf v}$ by
\begin{equation}
\label{varphi}\varphi=\frac{\pm|{\bf v}|Lm}\hbar\ .
\end{equation}

Let $E_0(N,a,\varphi)$ denote the ground state energy of
(\ref{hamprime}) with periodic boundary conditions. Obviously it is no
restriction to consider only the case $-\pi\leq \varphi\leq \pi$,
since $E_0$ is periodic in $\varphi$ with period $2\pi$ (see Remark 1
below).  For $\Psi_0$ the ground state of $H_N'$, let $\gamma_N$ be
its one-particle reduced density matrix
\index{one-particle density matrix}.  We are interested in the {\it Gross-Pitaevskii}
(GP) limit\index{Gross-Pitaevskii!limit} $N\to\infty$ with $Na/L$
fixed. We also fix the box size $L$. This means that $a$ should vary
like $1/N$ which, as explained in the previous section, can be
achieved by writing $v(r)=a^{-2}v_1(r/a)$, where $v_1$ is a fixed
potential with scattering length 1, while $a$ changes with $N$.

\begin{thm}[{\bf Superfluidity and BEC of homogeneous gas}]\label{T1a}
For $|\varphi| \leq \pi$
\begin{equation}\label{i}
\lim_{N\to\infty} \frac{E_0(N,a,\varphi)}N = 4\pi\mu  a\rho +
\mu\frac{\varphi^2}{L^2}
\end{equation}
in the limit $N\to \infty$ with $Na/L$ and $L$ fixed. Here
$\rho=N/L^3$,
so $a\rho$ is fixed too.
In the same
limit,
for $|\varphi| < \pi$,
\begin{equation}\label{ii}
\lim_{N\to\infty} \frac 1N\, \gamma_N(\x,\x')=
\frac 1{L^3}
\end{equation}
in trace class norm, i.e., $\lim_{N\to\infty} \Tr
\big[\,\big|\gamma_N/N - |L^{-3/2}\rangle\langle L^{-3/2}|\,
\big|\,\big]=0$.
\end{thm}

Note that, by the definition (\ref{rhos}) of $\rho_{s}$ and Eq.\
(\ref{varphi}), Eq.\ (\ref{i}) means that $\rho_{s}=\rho$, i.e., there
is 100\% superfluidity. For $\varphi=0$, Eq.\ (\ref{i}) follows from
Eq.\ (\ref{basic}), while (\ref{ii}) for $\varphi=0$ is the BEC
of Theorem \ref{hombecthm}.\footnote{The convention in
Theorem \ref{hombecthm}, where $\rho$ and $Na/L$ stay fixed,
is different from the one employed here, where $L$ and $Na/L$ are
fixed, but these two conventions are clearly equivalent by scaling.}

\bigskip
\noindent {\it Remarks:}
1. By a unitary gauge transformation\index{gauge transformation},
\begin{equation}\label{gauge}
\big(U\Psi\big)(\x_1,\dots,\x_N)= e^{{\rm i}\varphi(\sum_i z_i)/L}\,
\Psi(\x_1,\dots,\x_N) \ ,
\end{equation}
the passage from (\ref{ham}) to (\ref{hamprime})  is equivalent to
replacing periodic boundary
conditions in a box by the
{\it twisted
boundary condition}\index{boundary condition}
\begin{equation}\label{twist}
\Psi(\x_1 + (0,0,L), \x_2, \dots, \x_N)= e^{{\rm i}\varphi}
\Psi(\x_1,
\x_2, \dots, \x_N)
\end{equation}
in the direction of
the velocity field, while retaining the original Hamiltonian
(\ref{ham}).

2. The criterion $|\varphi|\leq\pi$ means that $|{\bf v}|\leq
\pi\hbar/(mL)$. The corresponding energy $\half m ( \pi\hbar/(mL)
)^2$ is the gap in the excitation spectrum of the one-particle
Hamiltonian in the finite-size system.

3. The reason that we have to restrict ourselves to $|\varphi|<\pi$
in the
second part of Theorem~\ref{T1a} is that for $|\varphi|=\pi$ there are
two ground states of the operator $(\nabla+{\rm i}\vp/L)^2$ with
periodic
boundary conditions.  All we can say in this case is
that there is a subsequence of $\gamma_N$ that converges to a density
matrix of rank $\leq 2$, whose range is spanned by these two
functions

\begin{proof}[Proof of Theorem~{\rm\ref{T1a}}]

  As in the proof of Theorem \ref{hombecthm} we combine the
  localization Lemma \ref{L1}, this time with $\varphi\neq 0$, and a
  generalized Poincar\'e inequality
\index{Poincar\'e inequality!generalized}, this time Lemma \ref{L2}.  We thus infer that, for
  any symmetric $\Psi$ with $\langle \Psi,\Psi\rangle=1$ and for $N$
  large enough,
\begin{eqnarray} \nonumber
&&\frac1N\langle\Psi,H_{N}'\Psi\rangle  \big(1-\const
Y^{1/17}\big)^{-1}\\ \nonumber && \geq  4\pi\mu\rho a + \mu
\frac{\varphi^2}{L^2} - C Y^{1/17} \Big(\frac 1{L^2} - \frac 1N
\big\langle\Psi,\mbox{$ \sum_j$} \nabla_{j,\varphi}^2
\Psi\big\rangle\Big)
 \\ \nonumber && \quad + \frac c{L^2}
\int_{\K^{N-1}}
d\X \int_{\K} d\x_{1}\Big|
\Psi(\x_1,\X) -L^{-3}\big[\mbox{$ \int_\K$} d\x
\Psi(\x,\X)\big]
\Big|^2 \ ,\label{lowerbd2a}
\end{eqnarray}
where we used that $|\Omega^c|\leq \mbox{$\frac{4\pi}3$} N R^3=
\const L^3 Y^{2/17}$. From this
we can infer two things. First, since the kinetic energy, divided by
$N$, is certainly bounded independently of $N$, as the upper bound
shows, we get that
\begin{equation}
\liminf_{N\to\infty} \frac{E_0(N,a,\varphi)}N \geq 4\pi \mu \rho a +
\mu \frac{\varphi^2}{L^2}
\end{equation}
for any $|\varphi|<\pi$. By continuity this holds also for
$|\varphi|=\pi$, proving (\ref{i}). (To be precise,
$E_0/N-\mu\varphi^2L^{-2}$ is concave in $\varphi$, and therefore
stays
concave, and in particular continuous, in the limit $N\to\infty$.)
Secondly, since the upper and the lower bounds to $E_0$ agree in the
limit considered, the positive last term in (\ref{lowerbd2}) has to
vanish in the limit. I.e., we get that for the ground state wave
function $\Psi_0$ of $ H_N'$
\begin{equation}
\lim_{N\to\infty} \int_{\K^{N-1}}
d\X \int_{\K} d\x_{1}\Big|
\Psi_0(\x_1,\X)  -L^{-3}\big[\mbox{$ \int_\K$} d\x
\Psi_0(\x,\X)\big]
\Big|^2 = 0 \ .
\end{equation}
Using again (\ref{ahaha}), this proves (\ref{ii}) in a weak sense.
As explained in \cite{LS02,LSSY}, this suffices for  the convergence
$N^{-1}\gamma_N \to |L^{-3/2}\rangle \langle L^{-3/2}|$  in trace
class norm.
\end{proof}

Theorem~\ref{T1a} can be generalized in various ways to a physically
more realistic setting, for example replacing the periodic box by a
cylinder centered at the origin. We shall comment on such extensions
at the end of Chapter~\ref{becsect}.

\chapter[Gross-Pitaevskii Equation for Trapped Bosons]{Gross-Pitaevskii Equation for\vspace*{-2mm}\newline Trapped Bosons} \label{sectgp}
\index{Gross-Pitaevskii!equation}

In the recent experiments on Bose condensation (see, e.g.,
\cite{TRAP}), the particles are confined at very low temperatures in a
`trap'\index{trapped Bose gas} where the particle density is {\em inhomogeneous},
contrary to the case of a large `box', where the density is
essentially uniform.  We model the trap by a slowly varying confining
potential $ V$, with $V(\x)\to \infty $ as $|\x|\to \infty$.  The
Hamiltonian\index{Hamiltonian} becomes
\begin{equation}\label{trapham}
H =  \sum_{i=1}^{N}\left\{ -\mu \Delta_i +V(\x_i)\right\} +
 \sum_{1 \leq i < j \leq N} v(|\x_i - \x_j|) \ .
\end{equation}
Shifting the energy scale if necessary
we can assume that $V$ is nonnegative.
The ground state energy, $\hbar\omega$, of
$- \mu \Delta + V(\x)$ is
a natural energy unit and the corresponding
length unit, $\sqrt{\hbar/(m\omega)}=\sqrt{2\mu/(\hbar\omega)}
\equiv L_{\rm osc}$, is a measure of the extension
of the trap.

In the sequel we shall be considering a limit
where $a/L_{\rm osc}$ tends to zero while $N\to\infty$.
Experimentally
$a/L_{\rm osc}$ can be changed in two ways:  One can either vary
$L_{\rm
osc}$ or $a$.  The first alternative is usually simpler in practice
but
very recently a direct tuning of the scattering length itself has also
been shown to be feasible \cite{Cornish}.  Mathematically, both
alternatives
are equivalent, of course.  The first corresponds to writing
$V(\x)=L_{\rm osc}^{-2}  V_1(\x/L_{\rm osc})$ and keeping $V_1$ and
$v$ fixed.  The second corresponds to writing the interaction
potential
as $v(|\x|)=a^{-2}v_1(|\x|/a)$ like
in \eqref{v1}, where $v_1$ has unit scattering length,
and keeping $V$ and $v_1$ fixed. This is equivalent to the first,
since for given $V_1$ and $v_1$ the ground state energy of
(\ref{trapham}),
measured
in units of $\hbar\omega$, depends only on $N$ and $a/L_{\rm osc}$.
In the dilute limit when $a$ is much smaller than the mean particle
distance, the energy becomes independent of $v_1$.

We choose $L_{\rm osc}$ as a length unit. The energy unit is
$\hbar\omega=2\mu L_{\rm osc}^{-2}=2\mu$.  Moreover, we find it
convenient to regard $V$ and $v_1$ as fixed. This justifies the notion
$E_0(N,a)$ for the quantum mechanical ground state energy.

The idea is now to use the information about the thermodynamic
limiting energy of the dilute Bose gas in a box to find the ground
state energy of (\ref{trapham}) in an appropriate limit. This has been
done in \cite{LSY1999,LSY2d} and in this chapter we give an account of
this work.  As we saw in Chapters~\ref{sect3d} and~\ref{sect2d} there
is a difference in the $\rho$ dependence between two and three
dimensions, so we can expect a related difference now.  We discuss 3D
first.

\section{Three Dimensions}

Associated with the quantum mechanical ground state energy problem
is the Gross-Pitaevskii (GP) energy functional\index{Gross-Pitaevskii!functional}
\cite{G1961,G1963,P1961}
\begin{equation}\label{gpfunc3d}
\E^{\rm
GP}[\phi]=\int_{\R^3}\left(\mu|\nabla\phi|^2+V|\phi|^2+4\pi \mu
a|\phi|^4\right)d\x
\end{equation}
with the subsidiary condition \begin{equation}\label{norm}
\int_{\R^3}|\phi|^2=N.\end{equation}
As before, $a>0$ is the scattering length\index{scattering length} of $v$.
The corresponding energy is
\begin{equation}\label{gpen3d}
E^{\rm GP}(N,a)=\inf_{\int|\phi|^2=N}\E^{\rm GP}[\phi]= \E^{\rm
GP}[\phi^{\rm GP}],\end{equation} with a unique, positive $\phi^{\rm
GP}$. The existence of the minimizer $\phi^{\rm GP}$ is proved by
standard techniques and it can be shown to be continuously
differentiable, see \cite{LSY1999}, Sect.~2 and Appendix A. The
minimizer depends on $N$ and $a$, of course, and when this is
important we denote it by $\phi^{\rm GP}_{N,a}$.

The variational equation satisfied by the minimizer is the
{\it GP equation}\index{Gross-Pitaevskii!equation}
\begin{equation}\label{gpeq}
-\mu\Delta\phi^{\rm GP}(\x)+ V(\x)\phi^{\rm GP}(\x)+8\pi\mu a
\phi^{\rm
GP}(\x)^3 = \mu^{\rm GP} \phi^{\rm GP}(\x),
\end{equation}
where $\mu^{\rm GP}$ is the chemical potential\index{chemical potential}, given by
\begin{equation}\label{mugp}
\mu^{\rm GP}=dE^{\rm GP}(N,a)/dN=E^{\rm GP}(N,a)/N+
(4\pi \mu a/N)\int |\phi^{\rm GP}(\x)|^4 d\x.
        \end{equation}

The GP theory has the following scaling property:\index{scaling property}
\begin{equation}\label{scalen}
E^{\rm GP}(N,a)=N E^{\rm GP}(1,Na),
\end{equation}
and
\begin{equation}\label{scalphi}
\phi^{\rm GP}_{N,a}(\x)= N^{1/2} \phi^{\rm GP}_{1,Na}(\x).
\end{equation}
Hence we see that the relevant parameter in GP theory is the
combination $Na$.

We now turn to the relation of $E^{\rm GP}$ and
$\phi^{\rm GP}$ to the quantum mechanical ground state.
If $v=0$, then the ground state of \eqref{trapham} is
$$\Psi_{0}(\x_{1},\dots,\x_{N})=\hbox{$\prod_{i=1}^{N}$}\phi_{0}(\x_{i})
$$
with $\phi_{0}$ the normalized ground state of $-\mu \Delta + V(\x)$.
In this case
clearly $\phi^{\rm GP}=\sqrt{N}\ \phi_{0}$, and then
$E^{\rm GP}=N\hbar\omega = E_0$.  In the other extreme, if $V(\x)=0$
for
$\x$ inside a large box of volume $L^3$ and $V(\x)= \infty$ otherwise,
then $\phi^{\rm GP} \approx \sqrt{N/L^3}$ and we get $E^{\rm
GP}(N,a) = 4\pi \mu a N^2/L^3$, which is the previously considered
energy
$E_0$ for the homogeneous gas in the low
density regime. (In this case, the gradient term in $\E^{\rm GP}$
plays no role.)

In general, we expect that for {\it dilute} gases\index{dilute gas} in a suitable limit
\begin{equation}\label{approx}E_0
        \approx E^{\rm GP}\quad{\rm and}\quad \rho^{\rm QM}(\x)\approx
\left|\phi^{\rm GP}(\x)\right|^2\equiv \rho^{\rm
GP}(\x),\end{equation}
where the quantum mechanical particle density in the ground state\index{ground state!density} is
defined by \begin{equation} \rho^{\rm
QM}(\x)=N\int|\Psi_{0}(\x,\x_{2},\dots,\x_{N})|^2d\x_{2}\cdots
d\x_{N}.  \end{equation} {\it Dilute} means here that
\begin{equation}\bar\rho a^3\ll 1,\end{equation} where
        \begin{equation}\label{rhobar}
\bar\rho=\frac 1N\int|\rho^{\rm GP}(\x)|^2 d\x
\end{equation}
is the {\it mean density}.\index{mean density}

The limit in which \eqref{approx} can be expected to be true
should be chosen so that {\it all three} terms in
$\E^{\rm GP}$ make a contribution. The scaling relations
\eqref{scalen} and \eqref{scalphi}
indicate that fixing
$Na$ as $N\to\infty$ is the right thing to do (and this is quite
relevant since
experimentally $N$ can be quite large, $10^6$ and more,  and
$Na$ can range from about 1 to $10^4$ \cite{DGPS}).
 Fixing $Na$
(which we refer to as the GP
case) also means that
we really are dealing with a dilute limit, because the mean density
$\bar \rho$ is then of the order $N$ (since
$\bar\rho_{N,a}=N\bar\rho_{1,Na}$) and hence
\begin{equation}
a^3\bar\rho\sim N^{-2}.
\end{equation}

The precise statement of \eqref{approx} is:

\begin{thm}[\textbf{GP limit of the QM ground state energy and
density}]\label{thmgp3}
If $N\to\infty$ with $Na$ fixed, then
\begin{equation}\label{econv}
\lim_{N\to\infty}\frac{{E_0(N,a)}}{ {E^{\rm GP}(N,a)}}=1,
\end{equation}
and
\begin{equation}\label{dconv}
\lim_{N\to\infty}\frac{1}{ N}\rho^{\rm QM}_{N,a}(\x)= \left
|{\phi^{\rm GP}_{1,Na}}(\x)\right|^2
\end{equation}
in the weak $L^1$-sense.
\end{thm}

Convergence can not only be proved for the ground state energy and
density, but also for the individual energy components:

\begin{thm}[\textbf{Asymptotics of the energy
components}]\label{compthm}
Let $\psi_0$ denote the solution to the zero-energy
scattering equation for
$v$ (under the boundary condition
$\lim_{|\x|\to\infty}\psi_0(\x)=1$) and
$s=\int|\nabla\psi_0|^2/(4\pi a)$. Then $0<s\leq 1$ and, in the same
limit as in Theorem~{\rm\ref{thmgp3}} above,
\begin{subequations}\label{parttwo}
\begin{eqnarray}\nonumber
&&\!\!\!\!\!\!\!\!\!\! \lim_{N\to\infty} \int  |\nabla_{\x_1}
\Psi_0(\x_1,\X)|^2 d\x_1\,
d\X
\\ \label{3a}
 &&\qquad= \int|\nabla\phi^{\rm GP}_{1,Na}(\x)|^2d\x + 4\pi Na s
\int|\phi^{\rm GP}_{1,Na}(\x)|^4
d\x,\\  &&\!\!\!\!\!\!\!\!\!\!\lim_{N\to\infty} \int
V(\x_1)|\Psi_0(\x_1,\X)|^2 d\x_1\,
d\X = \int V(\x) |\phi^{\rm GP}_{1,Na}(\x)|^2 d\x, \\ \nonumber
&&\!\!\!\!\!\!\!\!\!\!\lim_{N\to\infty} \half\sum_{j=2}^N \int
v(|\x_1-\x_j|)|\Psi_0(\x_1,\X)|^2 d\x_1\, d\X
\\ \label{part2} &&\qquad=(1-s) 4\pi Na \int|\phi^{\rm
GP}_{1,Na}(\x)|^4
d\x.
\end{eqnarray}
\end{subequations}
\end{thm}

Here we introduced again the short hand notation (\ref{defX}).
Theorem~\ref{compthm} is a simple consequence of Theorem~\ref{thmgp3}
by variation with respect to the different components of the energy,
as was also noted in \cite{CS01a}. More precisely, Eq.\ \eqref{econv}
can be written as
\begin{equation}\label{econv22}
\lim_{N\to \infty}\frac1N E_{0}(N,a)=E^{\rm GP}(1,Na).
\end{equation}
The ground state
energy is a concave function of the mass parameter $\mu$, so it is
legitimate to differentiate both sides of \eqref{econv22} with respect
to $\mu$. In doing so, it has to be noted that $Na$ depends on $\mu$
through the scattering length. Using \eqref{partint} one sees that
\begin{equation}\label{dmua}
\frac{d(\mu a)}{d\mu}=\frac1 {4\pi}\int |\nabla\psi_0|^2d\x
\end{equation}
by the Feynman-Hellmann principle\index{Feynman-Hellmann principle},
since $\psi_0$ minimizes the left side of \eqref{partint}.

We remark that in the case of a
two-dimensional Bose gas, where the relevant parameter to be kept
fixed in
the GP limit is $N/|\ln (a^2 \bar\rho_N)|$ (c.f.\
Chapter~\ref{sect2d} and Section~\ref{sub2d}), the parameter
$s$ in Theorem \ref{compthm} can be shown to be always equal to
$1$. I.e., in 2D the interaction energy is purely kinetic in the GP
limit (see \cite{CS01b}).

\bigskip

To describe situations where $Na$ is very large, it is appropriate
to consider a limit where, as $N\to\infty$,  $a\gg
N^{-1}$, i.e. $Na\to\infty$, but still $\bar\rho a^3\to 0$.
 In
this case, the gradient term in the GP functional becomes
negligible compared to the other terms and the so-called {\it
Thomas-Fermi (TF) functional}\index{Thomas-Fermi functional}
\begin{equation}\label{gtf}
\E^{\rm TF}[\rho]=\int_{\R^3}\left(V\rho+4\pi \mu a\rho^2\right)d\x
\end{equation}
arises. (Note that this functional has nothing to do with the
fermionic theory invented by Thomas and Fermi in 1927, except
for a certain formal analogy.) It is
defined for nonnegative functions $\rho$ on $\R^3$. Its ground
state energy $E^{\rm TF}$ and density $\rho^{\rm TF}$ are defined
analogously to the GP case. (The TF functional is especially
relevant for the two-dimensional Bose gas. There  $a$ has to
decrease exponentially with $N$ in the GP limit, so the TF limit
is more adequate; see Section~\ref{sub2d} below).

Our second main result of this chapter is that minimization of
(\ref{gtf}) reproduces correctly the ground state energy and density
of the many-body Hamiltonian in the limit when $N\to\infty$, $a^3\bar
\rho\to 0$, but $Na\to \infty$ (which we refer to as the TF case),
provided the external potential is reasonably well behaved. We will
assume that $V$ is asymptotically equal to some function $W$ that is
homogeneous\index{asymptotically homogeneous} of some order $s>0$,
i.e., $W(\lambda\x)=\lambda^s W(\x)$ for all $\lambda>0$, and locally
H\"older continuous (see \cite{LSY2d} for a precise definition). This
condition can be relaxed, but it seems adequate for most practical
applications and simplifies things considerably.

\begin{thm}[\textbf{TF limit of the QM ground state energy
and   density}]\label{thm2} Assume that $V$ satisfies
the conditions
stated above. If $\g\equiv Na\to\infty$ as $N\to\infty$, but still
$a^3\bar\rho\to 0$, then
\begin{equation}\label{econftf}
\lim_{N\to\infty}\frac{E_0(N,a)} {E^{\rm TF}(N,a)}=1,
\end{equation}
and
\begin{equation}\label{dconvtf}
\lim_{N\to\infty}\frac{\g^{3/(s+3)}}{N}\rho^{\rm
QM}_{N,a}(\g^{1/(s+3)}\x)= \tilde\rho^{\rm TF}_{1,1}(\x)
\end{equation}
in the weak $L^1$-sense, where $\tilde\rho^{\rm TF}_{1,1}$ is the
minimizer of the TF functional under the condition $\int\rho=1$,
$a=1$, and with $V$ replaced by $W$.
\end{thm}

In the following, we will present the essentials of the proofs
Theorems \ref{thmgp3} and \ref{thm2}.
We will derive appropriate upper and lower bounds on the ground
state energy $E_0$.

The proof of the lower bound in Theorem \ref{thmgp3}
presented here is a modified version of (and partly simpler than)
the original proof in \cite{LSY1999}. For an alternative method
see~\cite{SeirNorway}.

The convergence of the densities follows from
the convergence of the energies in the usual way by variation with
respect to the external potential. For simplicity, we set $\mu\equiv
1$ in
the following.

\begin{proof}[Proof of Theorems {\rm\ref{thmgp3}} and {\rm\ref{thm2}}] {\it
    Part $1$: Upper bound to the QM energy.} To derive an upper bound on
  $E_0$ we use a generalization of a trial wave function of
  Dyson\index{Dyson!wave function} \cite{dyson}, who used this
  function to give an upper bound on the ground state energy of the
  homogeneous hard core Bose gas (c.f.\ Section~\ref{upsec}).  It is
  of the form
\begin{equation}\label{ansatz}
\Psi(\x_{1},\dots,\x_{N})
    =\prod_{i=1}^N\phi^{\rm
GP}(\x_{i})F(\x_{1},\dots,\x_{N}),
\end{equation}
where $F$ is constructed in the following way:
\begin{equation}F(\x_1,\dots,\x_N)=\prod_{i=1}^N
f(t_i(\x_1,\dots,\x_i)),\end{equation} where $t_i =
\min\{|\x_i-\x_j|, 1\leq j\leq i-1\}$ is the distance of $\x_{i}$
to its {\it nearest neighbor} among the points
$\x_1,\dots,\x_{i-1}$, and $f$ is a  function of $r\geq 0$. As in
\eqref{deff} we
choose it to be
\begin{equation}
f(r)=\left\{\begin{array}{cl} f_{0}(r)/f_0(b) \quad
&\mbox{for}\quad r<b\\ 1 &\mbox{for}\quad r\geq b,
\end{array}\right.
\end{equation}
where $f_0$ is the solution of the zero energy scattering equation
(\ref{3dscatteq}) and $b$ is some cut-off parameter of order
$b\sim \bar\rho^{-1/3}$. The function (\ref{ansatz}) is not
totally symmetric, but for an upper bound it is nevertheless an
acceptable test wave function since the bosonic ground state
energy is equal to the {\it absolute} ground state energy.\index{absolute ground state}

The result of a somewhat lengthy computation, similar to the one given
in Section~\ref{upsec} (see \cite{LSY1999} for details), is the upper bound
\begin{equation}\label{ubd}
E_0(N,a)\leq E^{\rm GP}(N,a) \left( 1+O(a\bar\rho^{1/3})\right).
\end{equation}

\bigskip\noindent {\it Part $2$: Lower bound to the QM energy, GP case.}
To obtain a lower bound for the QM ground state energy the strategy is
to divide space into boxes\index{box method} and use the estimate on
the homogeneous gas, given in Theorem \ref{lbthm2}, in each box with
{\it Neumann} boundary conditions. One then minimizes over all
possible divisions of the particles among the different boxes.  This
gives a lower bound to the energy because discontinuous wave functions
for the quadratic form defined by the Hamiltonian are now allowed. We
can neglect interactions among particles in different boxes because
$v\geq 0$. Finally, one lets the box size tend to zero. However, it is
not possible to simply approximate $V$ by a constant potential in each
box. To see this consider the case of noninteracting particles, i.e.,
$v=0$ and hence $a=0$.  Here $E_0=N\hbar\omega$, but a `naive' box
method gives only $\min_\x V(\x)$ as lower bound, since it clearly
pays to put all the particles with a constant wave function in the box
with the lowest value of $V$.

For this reason we start by separating out the GP wave function in
each variable and write a general wave function $\Psi$ as
\begin{equation}\label{5.23}
\Psi(\x_{1},\dots,\x_{N})=\prod_{i=1}^N\phi^{\rm
GP}(\x_{i})F(\x_{1},\dots,\x_{N}).
\end{equation}
Here $\phi^{\rm GP}=\phi^{\rm GP}_{N,a}$ is normalized so that
$\int|\phi^{\rm GP}|^2=N$.
Eq.\ \eqref{5.23} defines $F$ for a given $\Psi$ because $\phi^{\rm
GP}$ is
everywhere strictly positive, being the ground state of the
operator $- \Delta + V+8\pi a|\phi^{\rm GP}|^2$. We now compute
the expectation value of $H$ in the state $\Psi$. Using partial
integration and the variational equation (\ref{gpeq}) for
$\phi^{\rm GP}$, we see that
\begin{equation}\label{ener2}
\frac{\langle\Psi|H\Psi\rangle}{\langle\Psi|\Psi\rangle}-E^{\rm
GP}(N,a)=4\pi a
\int |\rho^{\rm GP}|^2 +Q(F),
\end{equation}
with
\begin{multline}
Q(F)=\\ \sum_{i=1}^{N} \frac{\int\prod_{k=1}^{N}\rho^{\rm GP}(\x_k)
\left(|\nabla_i F|^2+\left[\half\sum_{j\neq i} v(|\x_i-\x_j|)-8\pi a
\rho^{\rm GP}(\x_i)\right]|F|^2\right)} {\int\prod_{k=1}^{N}\rho^{\rm
GP}(\x_k)|F|^2}. \label{ener3}
\end{multline}
We recall that $\rho^{\rm GP}(\x)=|\phi^{\rm
GP}_{N,a}(\x)|^2$. For computing the ground state energy
of $H$ we have to minimize the normalized quadratic form $Q$.
 Compared to the expression for the energy involving
$\Psi$ itself we have thus obtained the replacements
\begin{equation}\label{repl}
V(\x)\to -8\pi a\rho^{\rm GP}(\x) \quad\mbox{and}\quad
\prod_{i=1}^Nd\x_i \to \prod_{i=1}^N\rho^{\rm GP}(\x_{i})d\x_{i}\ .
\end{equation}
We now use the box method on {\it this} problem. More precisely,
labeling the boxes by an index $\alpha$, we have
\begin{equation}\label{5.26}
\inf_F Q(F)\geq \inf_{\{n_\al\}} \sum_\al \inf_{F_\al}Q_\al
(F_\al),
\end{equation}
where $Q_\al$ is defined by the same formula as $Q$  but with the
integrations limited to the box $\alpha$,  $F_{\alpha}$ is a wave
function with particle number $n_\alpha$, and the infimum is taken
over all distributions of the particles with $\sum n_\al=N$.

We now fix some $M>0$, that will eventually tend to $\infty$, and
restrict ourselves to boxes inside a cube $\Lambda_M$ of side length
$M$. Since $v\geq 0$ the contribution to
\eqref{5.26} of boxes outside this cube is easily estimated from
below by
$-8\pi Na \sup_{\x\notin \Lambda_M}\rho^{\rm GP}(\x)$, which, divided
by $N$, is arbitrarily small for $M$ large, since $Na$ is fixed and
$\pgp/N^{1/2}=\pgp_{1,Na}$ decreases faster than exponentially at
infinity (\cite{LSY1999}, Lemma A.5).

For the boxes inside the cube $\Lambda_M$ we want to use Lemma
\ref{dysonl} and therefore we must approximate $\rho^{\rm GP}$ by
constants in each box. Let $\rmax$ and $\rmin$, respectively, denote
the maximal and minimal values of $\rho^{\rm GP}$ in box $\al$. Define
\begin{equation}
\Psi_\alpha(\x_1,\dots, \x_{n_\al})=F_\alpha(\x_1,\dots, \x_{n_\al})
\prod_{k=1}^{n_\al}\phi^{\rm GP}(\x_k),
\end{equation}
and
\begin{equation}
\Psi^{(i)}_\alpha(\x_1,\dots, \x_{n_\al})=F_\alpha(\x_1,\dots,
\x_{n_\al})
\prod_{\substack{k=1 \\ k\neq
i}}^{n_\al}\phi^{\rm GP}(\x_k).
\end{equation}
We have, for all $1\leq i\leq n_\al$,
\begin{equation}\label{5.29}
\begin{split}
&\int\prod_{k=1}^{n_\alpha}\rho^{\rm GP}(\x_k)\left(|\nabla_i
F_\alpha|^2\right.+\half\sum_{j\neq i}\left.
v(|\x_i-\x_j|)|F_\alpha|^2\right)
\\&\geq
\rmin\int \left(|\nabla_i\Psi^{(i)}_\alpha|^2\right.
+\half\sum_{j\neq i}\left.
v(|\x_i-\x_j|)|\Psi^{(i)}_\alpha|^2\right).
\end{split}
\end{equation}
We now use
Lemma \ref{dysonl} to get, for all $0\leq \eps\leq 1$,
\begin{equation}\label{5.30}
(\ref{5.29})\geq \rmin\int \left(\varepsilon
|\nabla_i\Psi^{(i)}_\alpha|^2
+a(1-\eps)U(t_i)|\Psi^{(i)}_\alpha|^2\right)
\end{equation}
where $t_i$ is the distance to the nearest neighbor of $\x_i$, c.f.,
\eqref{2.29}, and $U$ the potential \eqref{softened}.

Since $\Psi_\alpha=\pgp(\x_i)\Psi^{(i)}_\al$ we can estimate
\begin{equation}\label{tpsial}
|\nabla_i\Psi_\alpha|^2 \leq 2\rmax |\nabla_i\Psi^{(i)}_\alpha|^2 +
2|\Psi^{(i)}_\alpha|^2 N C_M
\end{equation}
with
\begin{equation}
C_M=\frac1N\sup_{\x\in\Lambda_M}|\nabla\pgp(\x)|^2=\sup_{\x\in\Lambda_M}
|\nabla\pgp_{1,Na}(\x)|^2.  \end{equation} Since $Na$ is fixed, $C_M$
is independent of $N$. Inserting \eqref{tpsial} into
\eqref{5.30}, summing over $i$ and using $\rho^{\rm GP}(\x_i)\leq
\rmax$ in
the
last term of \eqref{ener3} (in the box $\al$), we get
\begin{equation}\label{qalfal}
Q_\al(F_\alpha)\geq \frac{\rmin}{\rmax}E^{U}_\eps(n_\al,L)-8\pi
a\rmax n_\al -
\eps C_M n_\al,
\end{equation}
where $L$ is the side length of the box and $E^{U}_\eps(n_\al,L)$ is
the ground state energy of
\begin{equation}\label{eueps}
\sum_{i=1}^{n_\al}(-\half\eps \Delta_i+(1-\eps)aU(t_i))
\end{equation}
in the box (c.f.\ \eqref{halfway}). We want to minimize \eqref{qalfal}
with respect to $n_\al$ and drop the subsidiary condition
$\sum_\al{n_\al}=N$ in \eqref{5.26}. This can only lower the minimum.
For the time being we also ignore the last term in
\eqref{qalfal}. (The total contribution of this term for all boxes is
bounded by $\eps C_M N$ and will be shown to be negligible compared to
the other terms.)

Since the lower bound for the energy of Theorem
\ref{lbthm2} was obtained precisely from a lower bound to the operator
\eqref{eueps}, we can use the statement and proof of Theorem
\ref{lbthm2}.
{F}rom this we see that
\begin{equation}\label{basicx}
E^{U}_\eps(n_\al,L)\geq (1-\varepsilon)\frac{4\pi
an_\al^2}{L^3}(1-CY_\al^{1/17})
\end{equation}
with $Y_\al=a^3n_\al/L^3$, provided  $Y_\al$ is small enough, and that
$\eps\geq Y_\al^{1/17}$ and
$n_\al\geq {\rm (\const )} Y_\al^{-1/17}$. The condition on $\eps$ is
certainly fulfilled if we choose $\eps=Y^{1/17}$ with
$Y=a^3N/L^3$. We now want to show that the $n_\alpha$ minimizing
the right side of \eqref{qalfal} is large enough for \eqref{basicx}
to apply.

If the minimum of the right side of \eqref{qalfal}
(without the last term) is taken for some $\bar n_\al$, we have
\begin{equation}\label{minnal}
\frac{\rmin}{\rmax}
\left(E^{U}_\eps(\bar n_\al+1,L)-E^{U}_\eps(\bar n_\al,L)\right)\geq
8\pi a\rmax.
\end{equation}
On the other hand, we  claim that
\begin{lem} For any $n$
\begin{equation}\label{chempot}
E^{U}_\eps( n+1,L)-E^{U}_\eps(n,L)\leq 8\pi a\frac{
n}{L^3}.
\end{equation}
\end{lem}
\begin{proof}
Denote the operator \eqref{eueps} by $\tilde H_n$,  with
$n_\alpha=n$, and
let $\tilde
\Psi_n$
be its ground state. Let $t_i'$ be the distance to the nearest
neighbor
of $\x_i$ among the $n+1$ points $\x_1,\dots,\x_{n+1}$ (without
$\x_i$) and $t_i$ the
corresponding distance excluding $\x_{n+1}$. Obviously, for $1\leq
i\leq n$,
\begin{equation}
U(t_i')\leq U(t_i)+U(|\x_i-\x_{n+1}|)
\end{equation}
and
\begin{equation}
U(t_{n+1}')\leq \sum_{i=1}^nU(|\x_i-\x_{n+1}|).
\end{equation}
Therefore
\begin{equation}
\tilde H_{n+1}\leq \tilde
H_{n}-\half\eps\Delta_{n+1}+2a\sum_{i=1}^nU(|\x_i-\x_{n+1}|).
\end{equation}
Using $\tilde\Psi_n/L^{3/2}$  as trial function for $\tilde H_{n+1}$
we arrive at
\eqref{chempot}.
\end{proof}
Eq.\ \eqref{chempot} together with \eqref{minnal} shows that
 $\bar n_\al$ is at least $\sim \rmax L^3$.
We shall choose
$L\sim N^{-1/10}$, so
the conditions needed for (\ref{basicx}) are fulfilled for $N$ large
enough,
since $\rmax\sim N$ and hence
$\bar n_\al\sim N^{7/10}$ and
$Y_\al\sim N^{-2}$.

In order to obtain a lower bound on $Q_\al$ we therefore have to
minimize
\begin{equation}\label{qalpha}
 4\pi
a\left(\frac{\rmin}{\rmax}\frac{n_\al^2}{L^3}\left(1-CY^{1/17}\right)
-2n_\al\rmax\right).
\end{equation}
We can drop the
requirement that $n_\al$ has to
be an integer. The minimum of (\ref{qalpha}) is obtained for
\begin{equation}
n_\al= \frac{\rmax^2}{\rmin}\frac{L^3}{(1-CY^{1/17})}.
\end{equation}
By Eq.\ (\ref{ener2}) this gives the following lower bound,
including now the last term in \eqref{qalfal} as well as the
contributions from the
boxes outside $\Lambda_M$,
\begin{equation}\label{almostthere}
\begin{split}
&E_0(N,a)-E^{\rm GP}(N,a)\geq \\
&4\pi a\int|\rho^{\rm GP}|^2-4\pi a\sum_{\al\subset\Lambda_M}
\rmin^2
L^3\left(\frac{\rmax^3}{\rmin^3}\frac{1}{(1-CY^{1/17})}\right)\\
&-Y^{1/17}NC_M-4\pi
aN\sup_{\x\notin\Lambda_M}\rho^{\rm GP}(\x).
\end{split}
\end{equation}
Now $\rho^{\rm GP}$ is differentiable and strictly
positive. Since all the boxes are in the fixed cube $\Lambda_M$ there
are
constants
$C'<\infty$, $C''>0$,
such that
\begin{equation}
\rmax-\rmin\leq NC'L,\quad \rmin\geq NC''.
\end{equation}
Since $L\sim N^{-1/10}$ and $Y\sim N^{-17/10}$ we therefore have, for
large $N$,
\begin{equation}
\frac{\rmax^3}{\rmin^3}\frac{1}{(1-CY^{1/17})}\leq
1+{\rm (const.)}N^{-1/10}
\end{equation}
Also,
\begin{equation}
4\pi a\sum_{\al\subset\Lambda_M} \rmin^2 L^3\leq 4\pi a\int
|\rho^{\rm GP}|^2\leq E^{\rm
GP}(N,a).
\end{equation}
Hence, noting that $ E^{\rm GP}(N,a)=N E^{\rm GP}(1,Na)\sim N$ since
$Na$ is fixed,
\begin{equation}\label{there}
\frac{E_0(N,a)}{E^{\rm GP}(N,a)}\geq 1-{\rm
(const.)}(1+C_M)N^{-1/10}-{\rm (const.)}
\sup_{\x\notin \Lambda_M}|\pgp_{1,Na}|^2,
\end{equation}
where the constants depend on $Na$. We can now take $N\to\infty$ and
then $M\to\infty$.

\bigskip
\noindent {\it Part $3$: Lower bound to the QM energy, TF case.}
In the above proof of the lower bound in the GP case we did not
attempt to
keep track of the dependence of the constants on $Na$. In the TF case
$Na\to\infty$, so one would need to take a closer look at this
dependence if one wanted to carry the proof directly over to this
case. But we don't have to do so, because there is a simpler direct
proof. Using the explicit form of the TF minimizer, namely
\begin{equation}\label{tfminim}
\rho^{\rm TF}_{N,a}(\x)=\frac 1{8\pi a}[\mu^{\rm TF}-V(\x)]_+,
\end{equation}
where  $[t]_+\equiv\max\{t,0\}$ and $\mu^{\rm TF}$ is chosen so
that the normalization condition $\int \rho^{\rm TF}_{N,a}=N$
holds, we can use
\begin{equation}\label{vbound}
V(\x)\geq \mu^{\rm TF}-8\pi a \rho^{\rm
TF}(\x)
\end{equation}
to get a replacement as in (\ref{repl}), but without
changing the measure. Moreover, $\rho^{\rm TF}$ has compact
support, so, applying again the box method described above, the
boxes far out do not contribute to the energy. However, $\mu^{\rm
TF}$ (which depends only on the combination $Na$) tends to
infinity as $Na\to\infty$. We need to control the
asymptotic behavior of $\mtf$, and this leads to
the restrictions on $V$ described in the paragraph preceding
Theorem \ref{thm2}. For simplicity, we shall here only consider the
case when $V$
itself is homogeneous, i.e., $V(\lambda\x)=\lambda^sV(\x)$ for all
$\lambda>0$ with
some $s>0$.

In the same way as in \eqref{mugp} we have, with $g=Na$,
\begin{equation}\label{mutf}
\mu^{\rm TF}(g)=dE^{\rm TF}(N,a)/dN=E^{\rm TF}(1,g)+
4\pi g\int |\rho^{\rm TF}_{1,g}(\x)|^2 d\x.
        \end{equation}
The TF energy, chemical potential and minimizer
satisfy the scaling relations
\begin{equation}
E^{\rm TF}(1,g)=g^{s/(s+3)}E^{\rm TF}(1,1),
\end{equation}
\begin{equation}
\mu^{\rm TF}(g)=g^{s/(s+3)} \mu^{\rm TF}(1)  ,
\end{equation}
and
\begin{equation}
g^{3/(s+3)}\rho^{\rm TF}_{1,g}(g^{1/(s+3)}\x)= \rho^{\rm
TF}_{1,g}(\x)  .
\end{equation}
We also introduce the scaled interaction potential, $\widehat v$, by
\begin{equation}
\widehat v(\x)  =g^{2/(s+3)}v(g^{1/(s+3)}\x)
\end{equation}
with scattering length
\begin{equation}
\widehat a=g^{-1/(s+3)}a.
\end{equation}
 Using \eqref{vbound}, \eqref{mutf} and
the scaling relations we obtain
\begin{equation}
E_0(N,a)\geq E^{\rm TF}(N,a)+4\pi N g^{s/(s+3)}\int |\rho^{\rm
TF}_{1,1}|^2 +
g^{-2/(s+3)}Q
\end{equation}
with
\begin{equation}
Q=\inf_{\int|\Psi|^2=1}\sum_{i}\int\left(|\nabla_i\Psi|^2\right.+\half
\sum_{j\neq i}
\left.\widehat v(\x_i-\x_j)|\Psi|^2-8\pi
N\widehat a \rtf_{1,1}(\x_i)|\Psi|^2\right).
\end{equation}
We can now proceed exactly as in Part 2 to arrive at the analogy
of
Eq.\ \eqref{almostthere}, which in the present case becomes
\begin{equation}\label{almosttherex}
\begin{split}
&E_0(N,a)-E^{\rm TF}(N,a)\geq \\
&4\pi N g^{s/(s+3)}\int |\rho^{\rm TF}_{1,1}|^2-4\pi N\widehat
a\sum_{\al}
\rmax^2
L^3(1-C\widehat Y^{1/17})^{-1}.
\end{split}
\end{equation}
Here $\rmax$ is the maximum of $\rho^{\rm TF}_{1,1}$ in the box
$\alpha$, and $\widehat Y=\widehat a^3 N/L^3$. This holds as long as
$L$ does not
decrease too fast with $N$. In particular, if $L$ is simply fixed,
this
holds for all large enough $N$. Note that
\begin{equation}
\bar\rho=N\bar\rho_{1,g}\sim N g^{-3/(s+3)} \bar\rho_{1,1},
\end{equation}
so that $\widehat a^3 N\sim a^3 \bar
\rho$ goes to zero as $N\to\infty$ by assumption. Hence, if we first
let
$N\to\infty$ (which implies $\widehat Y\to 0$) and then take $L$ to
zero, we
arrive at the desired
result
\begin{equation}\label{lowertf}
\liminf_{N\to\infty}\frac{E_0(N,a)}{E^{\rm TF}(N,a)}\geq 1
\end{equation}
in the limit $N\to\infty$, $a^3\bar\rho\to 0$. Here
we used the fact that (because $V\!$, and hence $\rtf$, is continuous by
assumption) the Riemann sum $\sum_\al\rmax^2 L^3$ converges to
$\int|\rtf_{1,1}|^2$ as $L\to 0$.  Together with the upper bound
(\ref{ubd})
and
the fact that $E^{\rm GP}(N,a)/E^{\rm TF}(N,a)=E^{\rm GP}(1,Na)/E^{\rm
TF}(1,Na)\to 1$ as $Na\to\infty$, which holds under our regularity
assumption on $V$ (c.f.\ Lemma 2.3 in \cite{LSY2d}), this proves
(\ref{econv}) and (\ref{econftf}).

\bigskip
\noindent {\it Part $4$: Convergence of the densities.} The
convergence of the energies implies the convergence of the
densities in the usual way by variation of the external potential.
We consider the TF case here; the GP case is analogous. Set again
$\g=Na$. Making the replacement
\begin{equation}
V(\x)\longrightarrow V(\x)+\delta\g^{s/(s+3)}Z(\g^{-1/(s+3)}\x)
\end{equation}
for some positive $Z\in C_0^\infty$ and redoing the upper and
lower bounds we see that (\ref{econftf}) holds with $W$ replaced
by $W+\delta Z$. Differentiating with respect to $\delta$ at
$\delta=0$ yields
\begin{equation}
\lim_{N\to\infty}\frac{\g^{3/(s+3)}}N\rho^{\rm
QM}_{N,a}(\g^{1/(s+3)}\x) =\tilde\rho^{\rm TF}_{1,1}(\x)
\end{equation}
in the sense of distributions. Since the functions all have
$L^1$-norm 1, we can conclude that there is even weak
$L^1$-convergence.
\end{proof}

\section{Two Dimensions}\label{sub2d}

In contrast to the three-dimensional case the energy per particle for
a dilute gas in two dimensions is {\it nonlinear} in $\rho$. In view
of Schick's\index{Schick} formula \eqref{2den} for the energy of the homogeneous gas
it would appear natural to take the interaction into account in two
dimensional GP theory by a term
\begin{equation}
4\pi\int_{\R^2} |\ln(|\phi(\x)|^2 a^2)|^{-1}|\phi(\x)|^4{
d}\x,\end{equation}
and such a term has, indeed, been suggested in
\cite{Shev} and \cite{KoSt2000}.  However, since the nonlinearity
appears only in a logarithm, this term is unnecessarily complicated
as far as leading order computations are concerned.  For dilute gases
it turns out to be sufficient, to leading order, to use an interaction
term of the same form as in the three-dimensional case, i.e, define
the
GP functional\index{Gross-Pitaevskii!functional} as (for simplicity we put $\mu=1$ in this section)
\begin{equation}\label{2dgpfunc}
\E^{\rm
GP}[\phi]=\int_{\R^2}\left(|\nabla\phi|^2+V|\phi|^2+4\pi
\alpha|\phi|^4\right)d\x,
\end{equation}
where, instead of $a$, the coupling constant is now
\begin{equation}\label{alpha}\alpha=|\ln(\bar\rho_N
a^2)|^{-1}\end{equation}
with $\bar\rho_N$ the {\em mean density}\index{mean density}
for the GP functional
at coupling constant
$1$ and particle number $N$. This is defined analogously to
\eqref{rhobar}
as
\begin{equation}\label{meandens2d}
\bar\rho_N=\frac1N\int|\phi^{\rm GP}_{N,1}|^4d\x
\end{equation}
where $\phi^{\rm GP}_{N,1}$ is the minimizer of \eqref{2dgpfunc} with
$\alpha=1$ and subsidiary condition $\int|\phi|^2=N$.
Note that $\alpha$ in \eqref{alpha} depends on
$N$ through the mean density.

Let us denote the GP energy
for a given $N$ and coupling constant
$\alpha$ by $E^{\rm GP}(N,\alpha)$ and the corresponding minimizer by
$\phi^{\rm GP}_{N,\alpha}$.
As in three dimensions the scaling relations
\begin{equation}E^{\rm GP}(N,\alpha)=NE^{\rm
GP}(1,N\alpha)\end{equation}
and
    \begin{equation}N^{-1/2}\phi^{\rm GP}_{N,\alpha}=\phi^{\rm
GP}_{1,N\alpha}
\end{equation}
hold, and the relevant parameter is
\begin{equation}g\equiv N\alpha.\end{equation}

In three dimensions, where $\alpha=a$,
it is natural to consider the limit $N\to\infty$ with $g=Na$= const.
The analogue of Theorem \ref{thmgp3} in two dimensions is
\begin{thm}[{\bf Two-dimensional GP limit
    theorem}]
\label{2dlimit}
If, for $N\to\infty$, $a^2\brtf_N\to 0$ with
$g=N/|\ln(a^2\brtf_N)|$ fixed, then
\begin{equation}\label{econv2}
\lim_{N\to\infty}\frac{E_{0}(N,a)}{\Egp(N,1/|\ln(a^2\brtf_N)|)}=
1
\end{equation}
and
\begin{equation}\label{dconv2}
\lim_{N\to\infty}\frac{1}{ N}\rho^{\rm QM}_{N,a}(\x)= \left
|{\phi^{\rm GP}_{1,g}}(\x)\right|^2
\end{equation}
in the weak $L^1$-sense.
\end{thm}

This result, however, is of rather limited use in practice.  The
reason is
that in two dimensions the scattering length has to
decrease exponentially with $N$ if $g$ is fixed.
The parameter $g$ is
typically {\it very large} in two dimensions
so it is more appropriate to consider the
limit $N\to\infty$ and $g\to\infty$ (but still $\bar\rho_N a^2\to
0$).

For  potentials $V$ that are {\it homogeneous} functions of $\x$,
i.e.,
\begin{equation}\label{homog}V(\lambda
\x)=\lambda^sV(\x)\end{equation}
for some $s>0$, this limit can be described by the a
`Thomas-Fermi' energy functional\index{Thomas-Fermi functional} like \eqref{gtf} with coupling
constant unity:
\begin{equation}\label{tffunct}
\E^{\rm TF}[\rho]=\int_{\R^2}\left(V(\x)\rho(\x)+4\pi
\rho(\x)^2\right)
{ d}\x.
\end{equation}
This is just the GP functional without the gradient term and
$\alpha=1$.
Here $\rho$ is a nonnegative function on $\R^2$ and the normalization
condition is
\begin{equation}\label{norm2}\int\rho(\x)d\x=1.\end{equation}

The minimizer of \eqref{tffunct} can be given explicitly.  It is
\begin{equation}\label{tfminim2}\rho^{\rm
TF}_{1,1}(\x)=(8\pi)^{-1}[\mu^{\rm TF}-V(\x)]_+\end{equation}
where the chemical potential
$\mu^{\rm TF}$ is determined by the normalization condition
\eqref{norm2}
and $[t]_{+}=t$
for $t\geq 0$ and zero otherwise.
We denote the corresponding energy by $E^{\rm TF}(1,1)$.
By scaling one obtains
\begin{equation}\lim_{g\to\infty}
    E^{\rm GP}(1,g)/g^{s/(s+2)}=E^{\rm TF}(1,1),\end{equation}
 \begin{equation}\label{gptotf}\lim_{g\to\infty}g^{2/(s+2)}
\rho^{\rm
GP}_{1,g}(g^{1/(s+2)}\x)=\rho^{\rm TF}_{1,1}(\x),\end{equation}
with the latter limit in the  strong $L^2$ sense.

Our main result about two-dimensional
Bose gases in  external potentials satisfying \eqref{homog}
is that analogous limits also hold for the many-particle quantum
mechanical
ground state at
low densities:
\begin{thm}[{\bf Two-dimensional TF limit theorem}]\label{thm22}
In $2${\rm D}, if
$a^2\bar\rho_N\to 0$, but $g=N/|\ln(\bar\rho_N a^2)|\to \infty$ as
$N\to\infty$,
then
\begin{equation}\lim_{N\to \infty}\frac{E_0(N,a)}{g^{s/s+2}}=
E^{\rm TF}(1,1)\end{equation}
and, in the weak $L^1$ sense,
\begin{equation}\label{conv}\lim_{N\to\infty}\frac{g^{2/(s+2)}}N
\rho^{\rm
QM}_{N,a}(g^{1/(s+2)}\x)=\rho^{\rm TF}_{1,1}(\x).\end{equation}
\end{thm}

\noindent {\it Remarks:} 1. As in Theorem \ref{thm2}, it is
sufficient that $V$ is
asymptotically equal to some homogeneous potential, $W$. In this case,
$E^{\rm TF}(1,1)$ and $\rho^{\rm TF}_{1,1}$ in Theorem \ref{thm22}
should be replaced by the corresponding quantities for $W$.

2. From Eq.\ \eqref{gptotf} it follows that
\begin{equation}\bar\rho_N\sim N^{s/(s+2)}\end{equation} for large
$N$.
Hence the low density criterion
$a^2\bar\rho\ll 1$, means that
$a/L_{\rm osc}\ll  N^{-s/2(s+2)}$.

We shall now comment briefly on the proofs of Theorems
\ref{2dlimit} and \ref{thm22},
mainly pointing out the differences from the 3D case considered
previously.

The upper bounds for the energy are obtained exactly in a same way as
in three dimensions. For the lower bound in Theorem \ref{2dlimit} the
point to notice is that the expression
 \eqref{qalpha}, that has to be minimized over $n_\al$, is in 2D
 replaced by
\begin{equation}\label{qalpha2}
 4\pi
\left(\frac{\rmin}{\rmax}\frac{n_\al^2}{L^2}\frac1{|\ln(a^2n_\alpha/L^2)|}
\left(1-\frac C{|\ln(a^2N/L^2)|^{1/5}}\right)
-\frac{2n_\al\rmax}{|\ln(a^2\bar\rho_N)|}\right),
\end{equation}
since Eq.\ \eqref{basicx} has to be
replaced by the analogous inequality for 2D (c.f.\ \eqref{lower}).
To minimize \eqref{qalpha2} we use the following lemma:

\begin{lem}\label{xb}
For $0<x,b<1$ and $k\geq 1$ we have
\begin{equation}
\frac{x^2}{|\ln x|}-2\frac b{|\ln b|}xk\geq -
\frac{b^2}{|\ln b|}\left(1+\frac 1{(2|\ln b|)^2}\right)k^2.
\end{equation}
\end{lem}

\begin{proof} Replacing $x$ by $xk$ and using the monotonicity of
$\ln$ we
see that it suffices to consider $k=1$.
Since $\ln x\geq-\frac 1{de}x^{-d}$ for
all $d>0$ we have
\begin{equation}
\frac{x^2}{b^2}\frac{|\ln b|}{|\ln x|}
-2\frac xb\geq\frac{|\ln b|}{b^2}ed x^{2+d}-\frac{2x}{b}
\geq c(d)(b^ded\,|\ln b|)^{-1/(1+d)}
\end{equation}
with
\begin{equation}
c(d)=2^{(2+d)/(1+d)}\left(\frac 1{(2+d)^{(2+d)/(1+d)}}-\frac 1
{(2+d)^{1/(1+d)}}\right)\geq -1-\frac 14d^2.
\end{equation}
Choosing $d=1/|\ln b|$ gives the desired result.
\end{proof}

Applying this lemma with $x=a^2n_\al/L^2$, $b=a^2\rmax$ and
\begin{equation}k=\frac{\rmax}{\rmin}\,
\left(1-\frac
C{|\ln(a^2N/L^2)|^{1/5}}\right)^{-1}\frac{|\ln(a^2\rmax)|}
{|\ln(a^2\bar\rho_N)|}
\end{equation}
we get the bound
\begin{equation}
\eqref{qalpha2}\geq -4\pi\frac{\rmax^2L^2}{|\ln(a^2\bar\rho_N)|}
\left(1+\frac1{4|\ln(a^2\rmax)|^2}\right) k.
\end{equation}
In the limit considered, $k$ and the factor in parenthesis both tend
to 1 and
the Riemann sum over the boxes $\alpha$ converges to the integral as
$L\to 0$.

The TF case, Thm.\ \ref{thm22}, is treated in the same way as in
three
dimensions, with modifications analogous to those just discussed when
passing
from 3D to 2D in GP theory.

\chapter[BEC and Superfluidity for Dilute
Trapped Gases]{Bose-Einstein Condensation and\vspace*{-2mm}\newline Superfluidity for Dilute\vspace*{-2mm}\newline
Trapped Gases}\label{becsect}
\index{Bose-Einstein condensation}\index{trapped Bose gas}

It was shown in the previous chapter that, for each fixed $Na$, the
minimization of the GP functional correctly reproduces the large $N$
asymptotics of the ground state energy and density of $H$ -- but no
assertion about BEC in this limit was made. We will now extend this
result by showing that in the Gross-Pitaevskii
limit\index{Gross-Pitaevskii!limit} there is indeed 100\% Bose
condensation in the ground state. This is a generalization of the
homogeneous case considered in Theorem \ref{hombecthm} and although it
is not the same as BEC in the thermodynamic limit it is quite relevant
for the actual experiments with Bose gases in traps.  In the
following, we concentrate on the 3D case, but analogous considerations
apply also to the 2D case. We also discuss briefly some extensions of
Theorem \ref{T1a} pertaining to superfluidity in trapped gases.

As in the last chapter we choose to keep the length scale $L_{\rm
osc}$ of the confining potential fixed and thus write $Na$ instead of
$Na/L_{\rm osc}$.  Consequently the powers of $N$ appearing in the
proofs are different from those in the proof Theorem \ref{hombecthm},
where we
kept $Na/L$ {\it and} $N/L^3$ fixed.

For later use, we define the projector
\begin{equation}
P^{\rm GP}= |\phi^{\rm GP}\rangle\langle \phi^{\rm GP}|\ .
\end{equation}
Here (and everywhere else in this chapter) we denote $\phi^{\rm
GP}\equiv\phi^{\rm GP}_{1,Na}$ for simplicity, where $\phi^{\rm
GP}_{1,Na}$ is the minimizer of the GP functional (\ref{gpfunc3d})
with parameter $Na$ and normalization condition $\int|\phi|^2=1$
(compare with (\ref{scalphi})). Moreover, we set $\mu\equiv 1$.

In the following, $\Psi_0$ denotes the (nonnegative and normalized)
ground state\index{ground state} of the Hamiltonian (\ref{trapham}).
BEC refers to the reduced one-particle density
matrix\index{one-particle density matrix} $ \gamma(\x,\x')$ of
$\Psi_0$, defined in (\ref{defgamma}). The precise definition of BEC
 is that for some $c>0$ this integral operator has for all large $N$
an eigenfunction with eigenvalue $\geq cN$.

Complete (or 100\%) BEC is defined to be the property that
$\mbox{$\frac{1}{N}$}\gamma(\x,\x')$ not only has an eigenvalue of
order one, as in the general case of an incomplete
BEC\index{incomplete BEC}, but in the limit it has only one nonzero
eigenvalue (namely 1). Thus, $\mbox{$\frac{1}{N}$}\gamma(\x,\x')$
becomes a simple product $\varphi(\x)\varphi(\x')^*$ as $N\to \infty$,
in which case $\varphi$ is called the {\it condensate wave
  function}\index{condensate wave function}.  In the GP limit, i.e.,
$N\to\infty$ with $N a$ fixed, we can show that this is the case, and
the condensate wave function is, in fact, the GP minimizer $\phi^{\rm
  GP}$.

\begin{thm}[\textbf{Bose-Einstein condensation in a
trap}]\label{becthm}
For each fixed $Na$ $$ \lim_{N\to\infty} \frac 1 N \gamma(\x, \x')
= \phi^{\rm GP}(\x)\phi^{\rm GP}(\x')\ . $$ in trace norm,
i.e., $\Tr \left|\frac 1
N \gamma - P^{\rm GP} \right| \to 0$.
\end{thm}

We remark that Theorem \ref{becthm} implies that there is also 100\%
condensation for all $n$-particle reduced density
matrices\index{n-particle r@$n$-particle reduced density matrices}
\begin{eqnarray}\nonumber
&&\gamma^{(n)}(\x_1,\dots,\x_n;\x_1',\dots,\x_n')\\&&=n!\binom{N}{n}\int
\Psi_0(\x_1,\dots,\x_N)\Psi_0(\x_1',\dots,\x_n',\x_{n+1},
\dots\x_N)d\x_{n+1}\cdots d\x_N\nonumber \\
\end{eqnarray}
of
$\Psi_0$, i.e., they converge, after division by the normalization
factor,
to the one-dimensional projector onto
the $n$-fold tensor product of $\phi^{\rm GP}$. In other words,
for
$n$ fixed particles the probability of finding them all in the same
state
$\phi^{\rm GP}$ tends to 1 in the
limit considered. To see this,
let $a^*, a$ denote the boson creation and annihilation operators
for the state $\phi^{\rm GP}$, and observe that
\begin{equation}
1\geq \lim_{N\to\infty} N^{-n}\langle \Psi_0 | (a^*)^n
a^n|\Psi_0\rangle =
\lim_{N\to\infty} N^{-n} \langle \Psi_0 | (a^*a)^n|\Psi_0\rangle  \  ,
\end{equation}
since the terms coming from the commutators $[a, a^*]=1$ are of
lower order as $N\to \infty$ and vanish in the limit. From
convexity it follows that
\begin{equation}
N^{-n}  \langle \Psi_0 | (a^*a)^n|\Psi_0\rangle \geq N^{-n} \langle
\Psi_0 | a^*a|\Psi_0\rangle ^n \,
\end{equation}
which converges to $1$ as $N\to\infty$, proving our claim.

Another corollary, important for the interpretation of experiments,
concerns the momentum distribution\index{momentum!distribution} of the
ground state.

\begin{corollary}[\textbf{Convergence of momentum distribution}] Let
$$\widehat\rho (\k)=\int \int\gamma(\x, \x') \exp [i \k\cdot (\x
-\x')]
 d\x d\x'$$
denote the one-particle momentum  density of $\Psi_0$. Then, for
fixed $Na$, $$ \lim_{N\to\infty} \frac 1N
\widehat\rho(\k)=|\widehat\phi^{\rm GP}(\k)|^2 $$ strongly in
$L^1(\R^3)$. Here, $\widehat\phi^{\rm GP}$ denotes the Fourier
transform of $\phi^{\rm GP}$.
\end{corollary}

\begin{proof} If ${\mathcal F}$ denotes the (unitary) operator
`Fourier
transform' and if $h$ is an arbitrary $L^\infty$-function,
then
\begin{eqnarray}\nonumber
\left|\frac 1N\int \widehat\rho h-\int |\widehat\phi^{\rm
GP}|^2 h\right|&=&\left|\Tr[{\mathcal F}^{-1}
(\gamma/N-P^{\rm GP}){\mathcal F}h]\right|\\ \nonumber
&\leq& \|h\|_\infty \Tr |\gamma/N-P^{\rm GP}|,
\end{eqnarray}
from which we conclude that $$\|\widehat\rho/N-|\widehat\phi^{\rm
GP}|^2 \|_1\leq \Tr|\gamma/N-P^{\rm GP}|\, .$$
\end{proof}

As already stated, Theorem \ref{becthm} is a generalization of Theorem
\ref{hombecthm}, the latter corresponding to the case that $V$ is a
box potential.  It should be noted, however, that we use different
scaling conventions in these two theorems: In Theorem \ref{hombecthm}
the box size grows as $N^{1/3}$ to keep the density fixed, while in
Theorem \ref{becthm} we choose to keep the confining external
potential fixed.  Both conventions are equivalent, of course, c.f.\
the remarks in the second paragraph of Chapter~\ref{sectgp}, but
when comparing the exponents of $N$ that appear in the proofs of the
two theorems the different conventions should be born in mind.

As in Theorem~\ref{hombecthm} there are two essential components of
our proof of Theorem~\ref{becthm}.  The first is a proof that the part
of the kinetic energy that is associated with the interaction $v$
(namely, the second term in (\ref{3a})) is mostly located
\index{energy localization} in small balls surrounding each particle.  More
precisely, these balls can be taken to have radius roughly $N^{-5/9}$,
which is much smaller than the mean-particle spacing $N^{-1/3}$.  (The
exponents differ from those of Lemma \ref{L1} because of different
scaling conventions.)  This allows us to conclude that the function of
$\x$ defined for each fixed value of $\X$ by
\begin{equation}\label{defff}
f_\X(\x)=\frac 1{\phi^{\rm GP}(\x)} \Psi_0(\x,\X)\geq 0
\end{equation}
has the property that $\nabla_\x f_\X(\x)$ is almost zero outside
the small balls centered at points of $\X$.

The complement of the small balls has a large volume but it can be a
weird set; it need not even be connected. Therefore, the smallness of
$\nabla_\x f_\X(\x)$ in this set does not guarantee that $f_\X(\x)$ is
nearly constant (in $\x$), or even that it is continuous. We need
$f_\X(\x)$ to be nearly constant in order to conclude BEC. What saves
the day is the knowledge that the total kinetic energy of $f_\X(\x)$
(including the balls) is not huge.  The result that allows us to
combine these two pieces of information in order to deduce the almost
constancy of $f_\X(\x)$ is the generalized Poincar\'e
inequality\index{Poincar\'e inequality!generalized} in Lemma
\ref{lem2}.  The important point in this lemma is that there is no
restriction on $\Omega$ concerning regularity or connectivity.

Using the results of Theorem \ref{compthm}, partial integration
and the GP equation (i.e., the variational equation for $\phi^{\rm
GP}$,
see Eq. (\ref{gpeq})) we see that
\begin{equation}\label{bound}
\lim_{N\to\infty} \int  |\phi^{\rm GP}(\x)|^2 |\nabla_\x
f_\X(\x)|^2 d\x\,
d\X
 = 4\pi Na s\int |\phi^{\rm GP}(\x)|^4 d\x\ .
\end{equation}
The following Lemma shows that to leading order all the energy in
(\ref{bound}) is concentrated in small balls.

\begin{lem}[\textbf{Localization of the energy in a trap}]\label{lem1}
For fixed $\X$ let
\begin{equation}\label{defomega} \Omega_\X=\left\{\x\in \R^3
\left| \, \min_{k\geq 2}|\x-\x_k|\geq
N^{-1/3-\delta}\right\}\right.
\end{equation} for some $0<\delta< 2/9$. Then $$ \lim_{N\to\infty}
\int d\X \int_{\Omega_\X} d\x |\phi^{\rm GP}(\x)|^2 |\nabla_\x
f_\X(\x)|^2
= 0\ . $$
\end{lem}

\noindent {\it Remark.} In the proof
of Theorem \ref{hombecthm} we chose $\delta$ to be 4/51, but the
following proof shows that one can extend the range of $\delta$ beyond
this value.

\begin{proof}
We shall show that
\begin{eqnarray} \nonumber &&\int
d\X \int_{\Omega_\X^c} d\x\, |\phi^{\rm GP}(\x)|^2 |\nabla_\x
f_\X(\x)|^2\\
\nonumber &&+\int d\X \int d\x  |\phi^{\rm GP}(\x)|^2  |f_\X(\x)|^2
\left[
\half \sum_{k\geq 2}
v(|\x-\x_k|) - 8\pi Na |\phi^{\rm GP}(\x)|^2\right]  \\
\label{lowbound}&& \geq
-4\pi Na
\int|\phi^{\rm GP}(\x)|^4 d\x - o(1)
\end{eqnarray}
as $N\to \infty$. We claim that this implies the assertion of the
Lemma. To see this, note that the left side of (\ref{lowbound}) can be
written as
\begin{equation}
\frac 1N E_0 - \mu^{\rm GP} - \int d\X \int_{\Omega_\X} d\x
|\phi^{\rm GP}(\x)|^2 |\nabla_\x
f_\X(\x)|^2 \ ,
\end{equation}
where we used partial integration and the GP equation (\ref{gpeq}),
and
also the symmetry of $\Psi_0$. The convergence of the energy in
Theorem~\ref{thmgp3} and the relation~(\ref{mugp}) now imply the
desired result.

The proof of
(\ref{lowbound}) is actually just a detailed examination of the
lower bounds to the energy derived in \cite{LSY1999} and
\cite{LY1998} and described in Chapters~\ref{sect3d} and~\ref{sectgp}.
We use the same methods as there,
just describing the differences from the case considered here.

Writing
\begin{equation}
f_\X(\x)=\prod_{k\geq 2}\phi^{\rm GP}(\x_k)F(\x,\X)
\end{equation}
and using that $F$ is symmetric in the particle coordinates, we
see that (\ref{lowbound}) is equivalent to
\begin{equation}\label{qf}
\frac 1N Q_\delta(F)\geq -4\pi Na \int|\phi^{\rm GP}|^4 - o(1),
\end{equation}
where $Q_\delta$ is the quadratic form
\begin{eqnarray}\nonumber Q_\delta(F)&=&\sum_{i=1}^{N}
\int_{\Omega_i^c} |\nabla_i
F|^2\prod_{k=1}^{N}|\phi^{\rm GP}(\x_k)|^2d\x_k\\ \nonumber
&&+\sum_{1\leq
i<j\leq N} \int
v(|\x_i-\x_j|)|F|^2\prod_{k=1}^{N}|\phi^{\rm GP}(\x_k)|^2d\x_k\\
\label{qf2} &&-8\pi Na\sum_{i=1}^{N} \int
|\phi^{\rm GP}(\x_i)|^2|F|^2\prod_{k=1}^{N}|\phi^{\rm
GP}(\x_k)|^2d\x_k.
\end{eqnarray}
Here $\Omega_i^c$ denotes the set
$$\Omega_i^c=\{(\x_1,\X)\in\R^{3N}| \, \min_{k\neq
i}|\x_i-\x_k|\leq N^{-1/3-\delta}\}.$$

While  (\ref{qf}) is not true for all conceivable $F$'s satisfying
the normalization condition $$\int
|F(\x,\X)|^2\prod_{k=1}^{N}|\phi^{\rm GP}(\x_k)|^2d\x_k=1,$$ it
{\it is}
true for an $F$, such as ours, that has bounded kinetic energy
(\ref{bound}). Looking at Chapter~\ref{sectgp}, we see that Eqs.
\eqref{ener2}--\eqref{ener3}, \eqref{almostthere}--\eqref{there}
are similar to (\ref{qf}),
(\ref{qf2}) and almost establish (\ref{qf}), but there are
differences which we now explain.

In our case, the kinetic energy of particle $i$ is restricted
to the subset of $\R^{3N}$ in which $\min_{k\neq i}|\x_i-\x_k|\leq
N^{-1/3-\delta}$. However, looking at the proof of the lower bound
to the ground state energy of a homogeneous Bose gas discussed in
Chapter~\ref{sect3d},
which enters the proof of Theorem \ref{thmgp3}, we
see that if we choose $\delta\leq 4/51$ only this part of the
kinetic energy is needed for the lower bound, except for
some part with a relative magnitude of the order
$\eps=O(N^{-2\alpha})$ with $\alpha=1/17$. (Here we use the a
priori knowledge that the kinetic energy is bounded by
\eqref{bound}.)
We can even do better and choose some
$4/51<\delta<2/9$, if $\alpha$ is chosen small enough. (To be
precise, we choose $\beta=1/3+\alpha$ and $\gamma=1/3-4\alpha$ in
the notation of  (\ref{ans}), and $\alpha$ small enough). The
choice of $\alpha$ only affects the magnitude of the error term,
however, which is still $o(1)$ as $N\to\infty$.
\end{proof}

\begin{proof}[Proof of Theorem {\rm\ref{becthm}}]
For some $R>0$ let $\K=\{\x\in\R^3, |\x|\leq R\}$, and define $$
\langle f_\X\rangle_\K=\frac 1{\int_\K |\phi^{\rm GP}(\x)|^2 d\x}
\int_\K
|\phi^{\rm GP}(\x)|^2 f_\X(\x)\, d\x \  . $$ We shall use Lemma
\ref{lem2},
with $d=3$, $h(\x)=|\phi^{\rm GP}(\x)|^2/\int_\K|\phi^{\rm
GP}|^2$,
$\Omega=\Omega_\X\cap\K$ and $f(\x)= f_\X(\x)-\langle f_\X
\rangle_\K$ (see (\ref{defomega}) and (\ref{defff})). Since
$\phi^{\rm GP}$
is bounded on $\K$ above and below by some positive constants,
this Lemma also holds (with a different constant $C'$) with $d\x$
replaced by $|\phi^{\rm GP}(\x)|^2d\x$ in (\ref{poinc}). Therefore,
\begin{eqnarray}\nonumber
&& \int d\X \int_\K d\x |\phi^{\rm GP}(\x)|^2
\left[f_\X(\x)-\langle
f_\X\rangle_\K\right]^2
\\ \nonumber && \leq C'\int d\X\left[\int_{\Omega_\X\cap \K}
|\phi^{\rm GP}(\x)|^2|\nabla_{\x} f_\X(\x)|^2 d\x\right. \\
&&\left.
\qquad\quad\qquad + \frac {N^{-2\delta}}{R^2} \int_\K
|\phi^{\rm GP}(\x)|^2|\nabla_{\x} f_\X(\x)|^2 d\x \right],
\label{21}
\end{eqnarray}
where we used that $|\Omega_\X^c\cap\K|\leq (4\pi/3)
N^{-3\delta}$. The first integral on the right side of (\ref{21})
tends to zero as $N\to\infty$ by Lemma \ref{lem1}, and the second
is bounded by (\ref{bound}). We conclude, since $$\int_\K
|\phi^{\rm GP}(\x)|^2 f_\X(\x) d\x\leq \int_{\R^3} |\phi^{\rm
GP}(\x)|^2
f_\X(\x)d\x$$ because of the positivity of $f_\X$, that
\begin{eqnarray}\nonumber \liminf_{N\to\infty} \frac 1N \langle
\phi^{\rm GP}|\gamma|\phi^{\rm GP}\rangle &\geq& \int_\K
|\phi^{\rm GP}(\x)|^2 d\x \,
\lim_{N\to\infty}\int d\X \int_\K d\x |\Psi_0(\x,\X)|^2
\\ \nonumber &=&\left[\int_\K |\phi^{\rm GP}(\x)|^2 d\x\right]^2,
\end{eqnarray}
where the last equality follows from (\ref{dconv}). Since the
radius of $\K$ was arbitrary, we conclude that
$$\lim_{N\to\infty}\frac 1 N \langle\phi^{\rm
GP}|\gamma|\phi^{\rm GP}\rangle= 1,$$
implying convergence of $\gamma/N$ to $P^{\rm GP}$ in
Hilbert-Schmidt norm. Since the traces are equal, convergence even
holds in trace norm  (cf. \cite{S79}, Thm.~2.20), and Theorem~\ref{becthm} is proved.
\end{proof}

We remark that the method presented here also works in the case of a
two-dimensional Bose gas. The relevant parameter to be kept fixed in
the GP limit is $N/|\ln (a^2 \bar\rho_N)|$, all other considerations
carry over without essential change, using the results in
\cite{LSY2d,LY2d}, c.f.\ Chapter~\ref{sect2d} and Section~\ref{sub2d}. It
should be noted that the existence of BEC in the ground state in 2D is
not in conflict with its absence at positive temperatures \cite{Ho,MW,M}.
In the hard core lattice gas at half filling precisely this phenomenon
occurs \cite{KLS}.

\bigskip Finally, we remark on generalizations of Theorem~\ref{T1a} on
superfluidity\index{superfluidity} from a torus to some physically
more realistic settings \cite{LSYsuper}.  As an example, let
${\mathcal C}$ be a finite cylinder based on an annulus centered at
the origin. Given a bounded, real function $a(r,z)$ let $A$ be the
vector field\index{vector field} (in polar coordinates)
$A(r,\theta,z)=\varphi a(r,z) \widehat e_\theta$, where $\widehat
e_\theta $ is the unit vector in the $\theta$ direction. We also allow
for a bounded external potential $V(r,z)$ that does not depend
on~$\theta$.

Using the methods of Appendix~A in \cite{LSY1999}, it is not
difficult to
see that there exists a $\varphi_0>0$, depending only on ${\mathcal
C}$ and
$a(r,z)$, such that for all $|\varphi|<\varphi_0$ there is a unique
minimizer $\phi^{\rm GP}$ of the Gross-Pitaevskii functional\index{Gross-Pitaevskii!functional}
\begin{equation}\label{defgp}
\E^{\rm
GP}[\phi]=\int_{\mathcal C}\Big(\big|\big(\nabla+{\rm i}A(\x)\big)
\phi(\x)\big|^2  + V(\x)
|\phi(\x)|^2 + 4\pi\mu N a
|\phi(\x)|^4\Big)d\x
\end{equation}
under the normalization condition $\int|\phi|^2=1$. This minimizer
does not depend on $\theta$, and can be chosen to be positive, for the
following reason: The relevant term in the kinetic energy is $T=
-r^{-2}[\partial/\partial \theta + {\rm i}\varphi\, r\, a(r,z)]^2$. If
$|\varphi\, r\, a(r,z)| < 1/2$, it is easy to see that $T\geq
\varphi^2
a(r,z)^2$, in which case, without raising the energy, we can replace
$\phi$ by the square root of the $\theta$-average of $|\phi|^2$.  This
can only lower the kinetic energy \cite{LL01} and, by convexity of
$x\to x^2$, this also lowers the $\phi^4$ term.

We denote the ground state energy of $\E^{\rm GP}$ by
$E^{\rm GP}$, depending on $Na$ and $\varphi$.
The following Theorem \ref{T2}
concerns the ground state energy $E_0$ of
\begin{equation}
H_N^{A}=\sum_{j=1}^N\Big[- \big(\nabla_j+{\rm i}A(\x_j)\big)^2 +
V(\x_j)\Big]
 +\sum_{1\leq i<j\leq N}v(\vert\x_{i}-\x_{j}\vert) \ ,
\end{equation}
with Neumann boundary conditions on ${\mathcal C}$, and the
one-particle reduced
density matrix $\gamma_N$ of the ground state, respectively. Different
boundary conditions can be treated in the same manner, if they are
also used in (\ref{defgp}).

\medskip
\noindent {\it Remark.}  As a special case, consider a uniformly
rotating\index{rotating system} system. In this case $A(\x)=\varphi r
\widehat e_\theta$, where $2 \varphi$ is the angular velocity. $H_N^A$
is the Hamiltonian in the rotating frame, but with external potential
$V(\x)+ A(\x)^2$ (see e.g. \cite[p.~131]{baym}).

\begin{thm}[{\bf Superfluidity in a cylinder}]\label{T2}
For $|\varphi|<\varphi_0$
\begin{equation}\label{gpone}
\lim_{N\to\infty} \frac{E_0(N,a,\varphi)}N = E^{\rm GP}(Na,\varphi)
\end{equation}
in the limit $N\to \infty$ with $Na$ fixed. In the same limit,
\begin{equation}\label{gpone2}
\lim_{N\to\infty} \frac 1N\, \gamma_N(\x,\x')=
\phi^{\rm GP}(\x)\phi^{\rm GP}(\x')
\end{equation}
in trace class norm, i.e., $\lim_{N\to\infty} \Tr
\big[\,\big|\gamma_N/N - |\phi^{\rm GP}\rangle\langle \phi^{\rm
GP}|\, \big|\,\big]=0$.
\end{thm}

\noindent{\it Remark.}
In the special case of
the curl-free vector potential $A(r,\theta)=\varphi r^{-1} \widehat
e_\theta$, i.e., $a(r,z)=r^{-1}$, one can say more about the role of
$\varphi_0$.  In this case, there is a unique GP
minimizer for all $\varphi\not\in \Z+\half$, whereas there are two
minimizers for $\varphi\in \Z+\half$. Part two of Theorem \ref{T2}
holds in this special case for all $\varphi\not\in \Z+\half$, and
(\ref{gpone}) is true even for all $\varphi$.

\medskip In the case of a uniformly rotating system, where $2\varphi$
is the angular velocity, the condition $|\varphi|< \varphi_0$ in
particular means that the angular velocity is smaller than the
critical velocity for creating vortices\index{vortices}
\cite{rot1,rot2,fetter}. For rapidly rotating gases, the appearance of
these vortices cause spontaneous breaking of the axial symmetry. The
GP minimizer is then no longer unique, and Theorem~\ref{T2} does not
apply to this case. It {\it is}, however, possible to show that the GP
equation still correctly describes the ground state of a dilute Bose
gas in the rapidly rotating case, as was recently shown in
\cite{LS05}, using very different techniques than in the proof of
Theorem~\ref{T2}.

\chapter[One-Dimensional Behavior of Dilute Bose Gases in
Traps]{One-Dimensional Behavior of\vspace*{-2mm}\newline Dilute Bose Gases
in Traps}
\label{1dsect}
\index{one-dimensional behavior}
\index{trapped Bose gas}

Recently it has become possible to do experiments in highly elongated
traps on ultra-cold Bose gases that are effectively one-dimensional
\cite{bongs,goerlitz,greiner,schreck,esslinger}. These experiments
show peculiar features predicted by a model of a one-dimensional Bose
gas with repulsive $\delta$-function pair
interaction\index{delta-f@$\delta$-function Bose gas}, analyzed long ago by
Lieb and Liniger \cite{LL,LL2}.\footnote{This model is discussed in
  Appendix~\ref{chap3}.}  These include quasi-fermionic behavior
\cite{gir}, the absence of Bose-Einstein condensation (BEC) in a
dilute limit \cite{Lenard,PiSt,girardeau}, and an excitation spectrum
different from that predicted by Bogoliubov's\index{Bogoliubov} theory
\cite{LL2,jackson,komineas}.  The theoretical work on the dimensional
cross-over for the ground state in elongated traps has so far been
based either on variational calculations, starting from a 3D
delta-potential \cite{olshanii,das2,girardeau2}, or on numerical
Quantum Monte Carlo studies \cite{blume,astra} with more realistic,
genuine 3D potentials, but particle numbers limited to the order of
100.  This work is important and has led to valuable insights, in
particular about different parameter regions \cite{petrov,dunjko}, but
a more thorough theoretical understanding is clearly desirable since
this is not a simple problem.  In fact, it is evident that for a
potential with a hard core the true 3D wave functions do not
approximately factorize in the longitudinal and transverse variables
(otherwise the energy would be infinite) and the effective 1D
potential can not be obtained by simply integrating out the transverse
variables of the 3D potential (that would immediately create an
impenetrable barrier in 1D).  It is important to be able to
demonstrate rigorously, and therefore unambiguously, that the 1D
behavior really follows from the fundamental Schr\"odinger
equation\index{Schr\"odinger equation}.  It is also important to
delineate, as we do here, precisely what can be seen in the different
parameter regions.  The full proofs of our assertions are long and are
given in \cite{LSY}. Here we state our main results and outline the
basic ideas for the proofs.

We start by describing the setting more precisely.
It is convenient to write the
Hamiltonian in the following way (in units where $\hbar=2m =1$):
\begin{equation} \label{3dham}
H_{N,L,r,a}=\sum_{j=1}^N \left( -\Delta_j +
V^{\perp}_{r}(\x^\perp_j)  + V_{L} (z_j) \right) + \sum_{1\leq i<j\leq
N} v_{a}(|\x_i-\x_j|)
\end{equation}
with $\x=(x,y,z)=(\x^\perp,z)$ and with
\begin{align}
V^{\perp}_{r}(\x^\perp)&=\frac 1{r^2}
 V^{\perp}(\x^\perp/r)\ , \notag \\
V_L(z)=\frac 1{L^2} V (z/L)&\ , \quad v_{a}(|\x|)=\frac
1{a^2}v(|\x|/a)\ .
\end{align}
Here, $r, L, a$ are variable scaling parameters while $V^{\perp}$, $V$
and $v$ are fixed.

We shall be concerned with the ground state of this Hamiltonian for
large particle number $N$, which is appropriate for the consideration
of actual experiments. The other parameters of the problem are the
scattering length\index{scattering length}, $a$, of the two-body
interaction potential, $v$, and two lengths, $r$ and $L$, describing
the transverse and the longitudinal extension of the trap potential,
respectively.

The interaction potential $v$ is supposed to be
nonnegative, of finite range and have scattering length 1; the scaled
potential $v_{a}$ then has scattering length $a$.  The external trap
potentials $V$ and $V^\perp$ confine the motion in the longitudinal
($z$) and the transversal ($\x^{\perp}$) directions, respectively, and
are assumed to be continuous and tend to $\infty$ as $|z|$ and $|
\x^{\perp}|$ tend to $\infty$.  To simplify the discussion we find it
also convenient to assume that $V$ is homogeneous of some order
$s>0$, namely
$V(z)=|z|^s$, but weaker assumptions, e.g. asymptotic homogeneity
(cf. Chapter~\ref{sectgp}), would in fact suffice.  The case of a
simple box with
hard walls is realized by taking $s=\infty$, while the usual harmonic
approximation is $s=2$. It is understood that the
lengths associated with the ground states of $-d^2/dz^2+V(z)$ and
$-\Delta^\perp+V^\perp(\x^\perp)$ are both of the order $1$ so that
$L$ and $r$ measure, respectively, the longitudinal and the transverse
extensions of the trap.  We denote the ground state energy of
(\ref{3dham}) by $E^{\rm QM}(N,L,r,a)$ and the ground state particle
density by $\rho^{\rm QM}_{N,L,r,a}(\x)$.
On the average, this 3D density will always be low in the parameter
range considered here (in the sense that
distance between particles is large compared to the 3D
scattering length). The effective 1D density can be either high or
low, however.

In parallel with the 3D Hamiltonian we consider the
Hamiltonian\index{Hamiltonian} for $n$  bosons in 1D
with delta interaction
and coupling constant $g\geq 0$ , i.e.,
\beq\label{13}
H_{n,g}^{\rm 1D}=\sum_{j=1}^n-\partial^2/\partial z_{j}^2
+ g \sum_{1\leq i<j\leq n}
\delta(z_i-z_j)\ .
\eeq
We consider this
Hamiltonian
for the $z_{j}$ in
an interval of length $\ell$ in the
thermodynamic limit, $\ell\to\infty$, $n\to\infty$ with $\rho=n/\ell$
fixed.
The ground state energy per particle in this limit is independent of
boundary conditions and can, according to \cite{LL} (see
Appendix~\ref{chap3}), be written as
\beq \label{1dendens}
e_{0}^{\rm 1D}(\rho)=\rho^2e(g/\rho) \ ,
\eeq
with a function $e(t)$ determined by a certain
integral equation. Its asymptotic form is $e(t)\approx
\half t$
for $t\ll 1$ and $e(t)\to \pi^2/3$ for $t\to \infty$. Thus
\beq\label{e0limhigh}
e_{0}^{\rm 1D}(\rho)\approx \half g\rho\ \ \hbox{\rm for}\ \
g/\rho\ll
1
\eeq
and
\beq\label{e0limlow}
e_{0}^{\rm 1D}(\rho)\approx (\pi^2/3)\rho^2\ \
\hbox{\rm for}\ \
g/\rho\gg
1\ .
\eeq
This latter energy is the same as for non-interacting fermions in 1D,
which can be understood from the fact that (\ref{13}) with $g=\infty$
is equivalent to a Hamiltonian describing free fermions.\index{free
fermions}

Taking $\rho e_{0}^{\rm 1D}(\rho)$ as a local energy density for an
inhomogeneous 1D system we can form the energy functional
\beq\label{genfunc}
\E[\rho]=\int_{-\infty}^{\infty} \!\!\!\! \left(
|\nabla\sqrt\rho(z)|^2 +
V_{L}(z)\rho(z) + \rho(z)^3 e(g/\rho(z)) \right) dz \ .
\eeq
Its ground state energy is obtained by minimizing over all normalized
densities, i.e.,
\beq\label{genfuncen}
E^{\rm 1D}(N,L,g)=\inf \left\{ \E[\rho] \, : \, \rho(z)\geq 0 , \,
\int_{-\infty}^{\infty}
\rho(z)dz = N \right\} .
\eeq
Using convexity of the map $\rho\mapsto \rho^3 e(g/\rho)$, it is
standard
to show that there exists a unique minimizer of (\ref{genfunc}) (see,
e.g., \cite{LSY1999}). It will be denoted by $\rho_{N,L,g}$. We also
define the {\it mean 1D density}\index{mean density} of this minimizer to be
\beq
\bar\rho= \frac 1N\int_{-\infty}^{\infty}
\left(\rho_{N,L,g}(z)\right)^2 dz \ .
\eeq
In a rigid box, i.e., for $s=\infty$, $\bar \rho$ is simply $N/L$
(except for boundary corrections), but in more general traps it
depends also on $g$ besides $N$ and $L$.  The order of magnitude of
$\bar\rho$ in the various parameter regions will be described below.

Our main result relates the 3D ground state energy of (\ref{3dham})
to the 1D density functional energy $E^{\rm
1D}(N,L,g)$ in the large $N$ limit with $g\sim a/r^2$ provided $r/L$
and $a/r$ are sufficiently small. To state this precisely, let
$e^\perp$ and $b(\x^\perp)$, respectively, denote the ground state
energy and the normalized ground state wave function of
$-\Delta^\perp+V^\perp(\x^\perp)$. The corresponding quantities
for $-\Delta^\perp+V^\perp_{r}(\x^\perp)$ are $e^\perp/r^2$ and
$b_{r}(\x^\perp)=(1/r)b(\x^\perp/r)$. In the case that the trap is a
cylinder with hard walls $b$ is a Bessel function; for a quadratic
$V^\perp$ it is a Gaussian.

Define $g$ by
\beq\label{defg}
g=\frac {8\pi a}{r^2} \int |b(\x^\perp)|^4 d\x^\perp={8\pi a}
\int |b_{r}(\x^\perp)|^4d\x^\perp.
\eeq
Our main result of this chapter is:

\begin{thm}[{\bf From 3D to 1D}]\label{T1}
Let $N\to\infty$ and simultaneously $r/L\to 0$ and $a/r\to 0$ in such
a way that
$r^2\bar\rho\cdot\min\{\bar\rho,g\}\to 0$. Then
\beq\label{lim}
\lim    \frac {E^{\rm QM}(N,L,r,a)-Ne^\perp /r^2 }{E^{\rm 1D}(N,L,g)}
= 1.
\eeq
\end{thm}

An analogous result hold for the ground state density. Define the 1D
QM density by averaging
over the transverse variables, i.e.,
\beq
\hat\rho^{\rm QM}_{N,L,r,a}(z)\equiv \int \rho^{\rm QM}_{N,L,r,a}
(\x^\perp,z)d\x^\perp \ .
\eeq
Let $\bar L:= N/\bar\rho$ denote the extension of the system in
$z$-direction, and define the rescaled density $\widetilde\rho$ by
\beq
\rho_{N,L,g}^{\rm 1D}(z)=\frac N{\bar L} \widetilde \rho (z/\bar L) \ .
\eeq
Note that, although $\widetilde \rho$ depends on $N$, $L$ and $g$,
$\|\widetilde \rho\|_1=\|\widetilde \rho\|_2=1$, which shows in
particular that $\bar L$ is the relevant scale in $z$-direction. The
result for the ground state density\index{ground state!density} is:

\begin{thm}[{\bf 1D limit for density}]\label{T1dens}
In the same limit as considered in Theorem~{\rm\ref{T1}},
\beq
\lim \left( \frac {\bar L}N \hat\rho^{\rm QM}_{N,L,r,a}(z\bar L)  -
\widetilde \rho(z)\right) = 0
\eeq
in weak $L^1$ sense.
\end{thm}

Note that because of (\ref{e0limhigh}) and (\ref{e0limlow}) the
condition
$r^2\bar\rho\cdot
\min\{\bar\rho,g\}\to 0$ is the same as
\beq \label{condition}
e_{0}^{\rm 1D}(\bar \rho)\ll 1/r^2 \ ,
\eeq
i.e., the average energy per particle associated with the longitudinal
motion should be much smaller than the energy gap\index{energy gap} between
the ground
and first excited state of the confining Hamiltonian in the transverse
directions. Thus, the basic physics is highly quantum-mechanical and
has no
classical counterpart. The system can be described by a
1D functional (\ref{genfunc}), {\it even though the
transverse trap dimension is much larger than the range of the atomic
forces.}

\section{Discussion of the Results}

We will now give a discussion of the various parameter regions that
are included in the limit considered in Theorems~\ref{T1}
and~\ref{T1dens}
above. We begin by describing the division of the space of parameters
into two basic regions.  This decomposition will eventually be refined
into five regions, but for the moment let us concentrate on the basic
dichotomy.

In Chapter~\ref{sectgp} we proved that the 3D
Gross-Pitaevskii formula for the energy is correct to leading order in
situations in which $N$ is large but $a$ is small compared to the mean
particle distance. This energy has two parts: The energy necessary to
confine the particles in the trap, plus the internal energy of
interaction, which is $N 4\pi a
\rho^{\rm 3D}$.  This formula was proved to be correct for a {\it
fixed} confining potential in the limit $N\to \infty$ with $a^3
\rho^{\rm 3D}\to 0$. However, this limit does not hold uniformly
if $r/L$ gets small as $N$ gets large.  In other words, new physics
can come into play as $r/L\to 0$ and it turns out that this depends on
the ratio of $a/r^2$ to the 1D density, or, in other words, on
$g/\bar\rho$.  There are two basic regimes to consider in highly
elongated traps, i.e., when $r \ll L$. They are
\begin{itemize}
\item The 1D limit of the
3D Gross-Pitaevskii regime
\item The `true' 1D regime.
\end{itemize}
The former is characterized by $g/\bar\rho\ll 1$, while in the latter
regime $g/\bar\rho$ is of the order one or even tends to infinity. (If
$g/\bar\rho\to\infty$ the particles are effectively impenetrable; this
is usually referred to as the Girardeau-Tonks\index{Girardeau-Tonks gas}
region.) These two
situations correspond to high 1D density (weak interaction) and low 1D
density (strong interaction), respectively. Physically, the main
difference is that in the strong interaction regime the motion of the
particles in the longitudinal direction is highly correlated, while in
the weak interaction regime it is not. Mathematically, this
distinction also shows up in our proofs. The first region is correctly
described by both the 3D and 1D theories because the two give the same
predictions there. That's why we call the second region the `true' 1D
regime.

In both regions the internal energy of the gas is small compared to
the energy of confinement.  However,
this in itself does not imply a specifically 1D behavior. (If $a$ is
sufficiently small it is satisfied in a trap of any shape.)  1D
behavior, when it occurs, manifests itself by the fact that the
transverse motion of the atoms is uncorrelated while the longitudinal
motion is correlated (very roughly speaking) in the same way as pearls
on a necklace.  Thus, the true criterion for 1D behavior is that
$g/\bar\rho$ is of order unity or larger and not merely the condition
that the energy of confinement dominates the internal energy.

We shall now briefly describe the finer division of these two regimes
into five regions altogether. Three of them (Regions 1--3) belong to
the weak interaction regime and two (Regions 4--5) to the strong
interaction regime. They are characterized by the behavior of
$g/\bar\rho$ as $N\to \infty$. In each of these regions the general
functional (\ref{genfunc}) can be replaced by a different, simpler
functional, and the energy $E^{\rm 1D}(N,L,g)$ in Theorem~\ref{T1} by the
ground state energy of that functional. Analogously, the density in
Theorem~\ref{T1dens} can be replaced by the minimizer of the
functional
corresponding to the region considered.

The five regions are
\medskip

\noindent $\bullet$
{\bf Region 1, the Ideal Gas case:} In the trivial case where the
interaction is so weak that it effectively vanishes in the large $N$
limit and everything collapses to the ground state of $-d^2/dz^2+V(z)$
with ground state energy $e^{\parallel}$, the energy $E^{\rm 1D}$ in
(\ref{lim}) can be replaced by $N e^{\parallel} /L^2 $. This is the
case if $g/\bar\rho\ll N^{-2}$, and the mean density is just
$\bar\rho\sim N/L$. Note that $g/\bar\rho\ll N^{-2}$ means that the 3D
interaction energy per particle $\sim a
\rho^{\rm 3D}\ll 1/L^2$.
\medskip

\noindent $\bullet$
{\bf Region 2, the 1D GP case:} In this region $g/\bar\rho\sim
N^{-2}$,
with $\bar\rho\sim N/L$. This case is described by a 1D
Gross-Pitaevskii energy functional\index{Gross-Pitaevskii!functional} of the
form
\beq\label{GPfunct}
\E^{\rm GP}_{\rm 1D}[\rho]=\int_{-\infty}^\infty \left(
|\nabla\sqrt\rho(z)|^2+
V_L(z)\rho(z) + \half g\rho(z)^2 \right) dz \ ,
\eeq
corresponding to the high density approximation (\ref{e0limhigh}) of
the
interaction energy in (\ref{genfunc}). Its ground state energy,
$E^{\rm GP}_{\rm 1D}$, fulfills the scaling relation $E^{\rm GP}_{\rm
1D}(N,L,g) =
NL^{-2} E^{\rm GP}_{\rm 1D}(1,1,NgL)$.
\medskip

\noindent $\bullet$
{\bf Region 3, the 1D TF case:}  $N^{-2}\ll g/\bar\rho \ll 1$,
with $\bar\rho$ being of the order $\bar\rho\sim (N/L)
(NgL)^{-1/(s+1)}$, where $s$ is the degree of
homogeneity of the longitudinal confining potential $V$. This region
is described by a Thomas-Fermi\index{Thomas-Fermi functional} type
functional
\beq\label{TFfunct}
\E^{\rm TF}_{\rm 1D}[\rho]=\int_{-\infty}^\infty \left(
V_L(z)\rho(z) + \half g\rho(z)^2 \right) dz \ .
\eeq
It is a limiting case of Region 2 in the sense that $NgL\gg 1$, but
$a/r$ is sufficiently small so that $g/\bar\rho \ll 1$, i.e., the high
density approximation in (\ref{e0limhigh}) is still valid.  The
explanation of the factor $(NgL)^{1/(s+1)}$ is as follows: The linear
extension $\bar L$ of the minimizing density of (\ref{GPfunct}) is for
large values of $NgL$ determined by $V_{L}(\bar L)\sim g(N/\bar L)$,
which gives $\bar L\sim (NgL)^{1/(s+1)} L$.  In addition condition
(\ref{condition}) requires $g\bar\rho \ll r^{-2}$, which means that
$Na/L(NgL)^{1/(s+1)}\ll 1$. The minimum energy of (\ref{TFfunct}) has
the scaling property $E^{\rm TF}_{\rm 1D}(N,L,g) = NL^{-2}(NgL)^{s/(s+1)}
E^{\rm TF}_{\rm 1D}(1,1,1)$.
\medskip

\noindent $\bullet$
{\bf Region 4, the LL case:}  $g/\bar\rho\sim 1$, with
$\bar\rho\sim  (N/L )
N^{-2/(s+2)}$, described by an energy functional
\beq\label{llfunct}
\E^{\rm LL}[\rho]=\int_{-\infty}^\infty \left( V_{L}(z)\rho(z)
+ \rho(z)^3
e(g/\rho(z)) \right) dz \ .
\eeq
This region corresponds to the case $g/\bar\rho\sim 1$, so that
neither the high density (\ref{e0limhigh}) nor the low density
approximation (\ref{e0limlow}) is valid and the full LL energy
(\ref{1dendens}) has to be used. The extension $\bar L$ of the system
is now determined by $V_L(\bar L)\sim (N/\bar L)^2$ which leads to
$\bar L\sim L N^{2/(s+2)}$. Condition (\ref{condition}) means in
this region that $Nr/\bar L\sim N^{s/(s+2)}r/L\to 0$. Since
$Nr/\bar L\sim(\bar\rho/g)(a/r)$, this condition is
automatically fulfilled if $g/\bar\rho$ is bounded away from zero and
$a/r\to 0$.  The ground state energy of (\ref{llfunct}), $E^{\rm
LL}(N,L,g)$, is equal to $N\gamma^2 E^{\rm LL}(1,1,g/\gamma)$, where
we introduced the density parameter $\gamma := (N/L) N^{-2/(s+2)}$.

\medskip
\noindent $\bullet$
{\bf Region 5, the GT case:} $g/\bar\rho\gg 1$,
with $\bar\rho\sim  (N/L) N^{-2/(s+2)}$, described by a functional
with energy density $\sim \rho^3$, corresponding to the
Girardeau-Tonks\index{Girardeau-Tonks gas} limit of the  LL energy density.
It corresponds to impenetrable particles, i.e, the limiting
case $g/\bar\rho\to\infty$ and hence formula
(\ref{e0limlow}) for the energy density. As in Region 4, the mean
density is here
 $\bar\rho\sim  \gamma$.
The energy functional is
\beq
\E^{\rm GT}[\rho]=\int_{-\infty}^\infty \left( V_{L}(z)\rho(z) +
({\pi^2}/3 )\rho(z)^3  \right) dz \ ,
\eeq
with minimum energy  $E^{\rm GT}(N,L)= N \gamma^2 E^{\rm GT}(1,1)$.

\medskip As already mentioned above, Regions 1--3 can be reached as
limiting cases of a 3D Gross-Pitaevskii theory.  In this sense, the
behavior in these regions contains remnants of the 3D theory, which
also shows up in the fact that BEC\index{Bose-Einstein condensation}
prevails in Regions 1 and 2 (See \cite{LSY} for details.)
Heuristically, these traces of 3D can be understood from the fact that
in Regions 1--3 the 1D formula for energy per particle, $g\rho\sim
aN/(r^2 L)$, gives the same result as the 3D formula, i.e., scattering
length times 3D density.  This is no longer so in Regions 4 and 5 and
different methods are required.

\section{The 1D Limit of 3D GP Theory}\label{1dgpsect}

Before discussing the many-body problem we treat the much simpler
problem of taking the $r/L\to 0$ limit of the 3D GP ground state
energy $E_{\rm 3D}^{\rm GP}(N,L,r,a)$, defined by (\ref{gpen3d}) with the
potential $V(\x)= V_r^\perp(\x^\perp)+V_L(z)$. The result is, apart
from the confining energy, the 1D GP energy with coupling constant
$g\sim a/r^2$.  In particular we see that Regions 4 and 5 cannot be
reached as a limit of 3D GP theory.

\begin{thm}[1D limit of 3D GP energy]\label{1dgplim}
Let $g$ be given by \eqref{defg}. If
$r/L\to 0$, then
\beq\frac{E_{\rm 3D}^{\rm
GP}(N,L,r,a)-Nr^{-2}e^\perp}{E^{\rm
GP}_{\rm 1D}(N,L,g)}\to 1\eeq
uniformly in the parameters, as long as $\bar \rho a\to 0$.
\end{thm}

\noindent {\it Remark.} Since $E^{\rm GP}(1,L,Ng)_{\rm 1D}\sim L^{-2}+
\bar\rho a/r^2$, the
condition $\bar\rho a\to 0$ is equivalent to $r^2 E^{\rm
  GP}_{\rm 1D}(1,L,Ng)\to 0$, which means simply that the 1D GP energy per
particle is much less than the confining energy, $\sim 1/r^2$.

\begin{proof}
  Because of the scaling relation \eqref{GPfunct} and the
  corresponding relation for $E^{\rm GP}_{\rm 3D}$ it suffices to
  consider the case $N=1$ and $L=1$.

We denote the (positive) minimizer of the one-dimensional GP functional
(\ref{GPfunct}) with $N=1$, $L=1$ and $g$ fixed by $\phi(z)$.
Taking
$b_{r}(\x^\perp)\phi(z)$ as trial function for the 3D functional
(\ref{gpfunc3d}) and using the definition (\ref{defg}) of $g$ we obtain
without further ado the upper bound
\beq
E^{\rm GP}_{\rm
3D}(1,1,r,a)\leq e^\perp/r^2+ E^{\rm GP}_{\rm 1D}(1,1,g)
\eeq
for all $r$ and
$a$.

For a lower bound we consider the one-particle Hamiltonian
\beq
H_{r,a}=-\Delta^\perp+V_{r}^\perp(\x^\perp)-\partial_{z}^2+V(z)+8\pi
ab_{r}(\x^\perp)^2\phi(z)^2\ .
\eeq
Taking the 3D Gross-Pitaevskii
minimizer $\Phi(\x)$ for $N=1$, $L=1$, as trial function we get
\beqa\label{hupper} \inf{\,\rm spec}\, H_{r,a}&\leq& E^{\rm GP}_{\rm
  3D}(1,1,r,a)-4\pi a\int \Phi^4 +8\pi a\int
b_{r}^2\phi^2\Phi^2\nonumber\\ &\leq&E^{\rm GP}_{\rm 3D}(1,1,r,a)+4\pi
a\int b_{r}^4\phi^4\nonumber\\ &=&E^{\rm GP}_{\rm 3D}(1,1,r,a)+\frac
g2\int\phi^4 \ .
\eeqa
To bound $H_{r,a}$ from below we consider
first for fixed $z\in\R$ the Hamiltonian (in 2D)
\beq
H_{r,a,z}=-\Delta^\perp+V_{r}(\x^\perp)+8\pi a \phi(z)^2
b_{r}(\x^\perp)^2.
\eeq
We regard $-\Delta^\perp+V_{r}(\x^\perp)$ as
its ``free'' part and $8\pi a \phi(z)^2 b_{r}(\x^\perp)^2$ as a
perturbation.  Since the perturbation is positive all the eigenvalues of
$H_{r,a,z}$ are at least as large as those of
$-\Delta^\perp+V_{r}(\x^\perp)$; in particular, the first excited eigenvalue
is $\sim 1/r^2$.
The expectation
value in the ground state $b_{r}$ of the free part is
\beq
\label{hx}\langle H_{r,a,z}\rangle=e^\perp/r^2+ g |\phi(z)|^2.
\eeq
Temple's inequality\index{Temple's inequality} (\ref{temple}) gives
\beq\label{templegp}
H_{r,a,z}\geq \left(e^\perp/r^2+ g
  |\phi(z)|^2\right)\left(1-\frac{\langle(H_{r,a,z}-\langle
    H_{r,a,z}\rangle)^2\rangle}{\langle H_{r,a,z}\rangle(\tilde e^\perp/r^2
- \langle H_{r,a,z}\rangle)}\right)
\eeq
where $\tilde e^\perp/r^2$ is the
lowest eigenvalue above the ground state energy of
$-\Delta^\perp+V_{r}(\x^\perp)$.  Since
\beq
H_{r,a,z}b_{r}=(e^\perp/r^2) b_{r}+8\pi a \phi(z)^2 b_{r}^3
\eeq
we have $(H_{r,a,z}-\langle
H_{r,a,z}\rangle)b_{r}=8\pi\phi(z)^2ab_{r}^3-g\phi(z)^2 b_{r})$ and
hence, using that $g=8\pi a\int b_{r}^4=8\pi(a/r^2)\int b^4$,
\begin{eqnarray} \nonumber
\langle(H_{r,a,z}-\langle
  H_{r,a,z}\rangle)^2\rangle&=&\phi(z)^4\int\left(8\pi
    ab_r(\x^\perp)^3-
    gb_{r}(\x^\perp)\right)^2d\x^\perp\\ \nonumber
  &=&\phi(z)^4 \int\left[(8\pi a)^2 b_{r}(\x^\perp)^6-16\pi ag
    b_{r}(\x^\perp)^4\right.\\ \nonumber &&+
\left.g^2b_{r}(\x^\perp)^2\right]d\x^\perp\\
  &\leq&\hbox{const.}\,\Vert\phi\Vert_{\infty}^4 g^2\leq
  \hbox{const.}\,E^{\rm GP}_{\rm 1D}(1,1,g)^2,
\end{eqnarray}
where we have used Lemma 2.1 in \cite{LSY2d} to bound $g\Vert
\phi\Vert_{\infty}^2$ by
$\hbox{const.}\,E^{\rm GP}_{\rm 1D}(1,1,g)$. We thus see from
\eqref{hx} and the assumption $r^2 E^{\rm GP}_{\rm 1D}(1,1,g)\to 0$
that the error term in the Temple inequality \eqref{templegp} is
$o(1)$.

Now
$H_{r,a}=-\partial_{z}^2+V(z)+H_{r,a,z}$, so from \eqref{templegp} we
conclude that \beq H_{r,a}\geq \left((e^\perp/r^2)-\partial_{z}^2+V(z)+ g
|\phi(x)|^2\right)(1-o(1)).\eeq On the other hand, the
lowest energy of $-\partial_{z}^2+V(z)+ g |\phi(z)|^2$ is just
$E^{\rm GP}_{\rm 1D}(1,1,g)+ (g/2)\int_{\R}|\phi(z)|^4dz$.
Combining \eqref{hupper} and \eqref{templegp} we
thus get \beq E_{\rm 3D}^{\rm GP}(1,1,r,a)- e^\perp/r^2\geq
E^{\rm GP}_{\rm 1D}(1,1,g)(1-o(1)).\eeq
\end{proof}

\section{Outline of Proof}

We now outline the main steps in the proof of Theorems~\ref{T1}
and~\ref{T1dens}, referring to \cite{LSY} for full details.
To prove (\ref{lim}) one has to establish upper and lower bounds, with
controlled errors, on the QM many-body energy in terms of the energies
obtained by minimizing the energy functionals appropriate for the
various regions.  The limit theorem for the densities can be derived
from the energy estimates in a standard way by variation with respect
to the external potential~$V_{L}$.

The different parameter regions have to be treated by different
methods, a watershed lying between Regions 1--3 on the one hand
and Regions 4--5 on the other. In Regions 1--3,
similar methods as in the proof of the 3D Gross-Pitaevskii limit
theorem discussed in Chapter~\ref{sectgp}  can be used.
This 3D proof needs some
modifications, however,  because there the external
potential was fixed and the estimates are not uniform in the ratio
$r/L$. We will not go into the details here, but mainly focus on
Regions 4 and 5, where new methods are needed. It turns out to be
necessary to localize the particles by dividing the trap into finite
`boxes'\index{box method} (finite in $z$-direction), with a controllable
particle
number in
each box. The particles
are then distributed optimally among the boxes to minimize the
energy, in a similar way as Eq. (\ref{estimate1}) was derived from
Eq. (\ref{estimate4}).

A core lemma for Regions 4--5 is an estimate of the 3D ground state
energy in a finite box in
terms of the 1D energy of the Hamiltonian (\ref {13}).
I.e., we will consider the ground state energy of (\ref{3dham})
with the external potential $V_L(z)$ replaced by a finite box (in
$z$-direction) with length $\ell$. Let $E_{\rm D}^{\rm
QM}(n,\ell,r,a)$ and $E_{\rm N}^{\rm QM}(n,\ell,r,a)$ denote its
ground state energy with Dirichlet and Neumann
boundary conditions, respectively.\index{boundary condition}

\begin{lem} \label{finthm}
Let $E_{\rm D}^{\rm 1D}(n,\ell,g)$ and $E_{\rm N}^{\rm
1D}(n,\ell,g)$ denote the ground state energy of \eqref{13} on
$L^2([0,\ell]^n)$, with Dirichlet and Neumann boundary conditions,
respectively, and let $g$ be given by \eqref{defg}. Then there is a
finite number $C>0$ such that
\beq\label{lbthm}
E_{\rm N}^{\rm QM}(n,\ell,r,a)-\frac{ne^\perp}{r^2} \geq E_{\rm
N}^{\rm
1D}(n,\ell,g)
\left( 1 -C n
\left(\frac{a}{r}\right)^{1/8}\left[1+\frac {nr}{\ell}
\left(\frac{a}{r}\right)^{1/8}   \right]\right)  \ .
\eeq
Moreover,
\beq\label{ubthm}
E_{\rm D}^{\rm QM}(n,\ell,r,a)-\frac {ne^\perp}{r^2} \leq E_{\rm
D}^{\rm
1D}(n,\ell,g)
\left(1+ C \left[ \left(\frac{n
a}{r}\right)^{2}\left( 1+ \frac {a\ell}{r^2}\right)
\right]^{1/3}\right) \ ,
\eeq
provided the term in square brackets is less than $1$.
\end{lem}

This Lemma is the key to the proof of Theorems~\ref{T1}
and~\ref{T1dens}. The reader interested in the details is referred to
\cite{LSY}. Here we only give a sketch of the proof of
Lemma~\ref{finthm}.

\begin{proof}[Proof of Lemma~$\ref{finthm}$]

We start with the upper bound (\ref{ubthm}).
Let $\psi$ denote the ground state of (\ref{13}) with Dirichlet
boundary conditions, normalized by $\langle\psi|\psi\rangle=1$, and
let $\rho^{(2)}_\psi$ denote its two-particle density, normalized by
$\int \rho_\psi^{(2)}(z,z')dzdz'=1$. Let $G$ and $F$ be given by
$G(\x_1,\dots,\x_n)=\psi(z_1,\dots,z_n)\prod_{j=1}^n
b_r(\x^\perp_j)$ and by $F(\x_1,\dots,\x_n)=\prod_{i<j} f(|\x_i-\x_j|)$.
Here $f$ is a monotone increasing function, with $0\leq f\leq 1$
and $f(t)=1$ for $t\geq R$ for some $R\geq R_0$. For $t\leq R$ we
shall choose $f(t)=f_0(t)/f_0(R)$, where $f_0$ is the solution to
the zero-energy scattering equation for $v_a$ (\ref{3dscatteq}).
Note that
$f_0(R)=1-a/R$ for $R\geq R_0$, and $f'_0(t)\leq
t^{-1}\min\{1,a/t\}$. We use as a trial wave function
\beq
\Psi(\x_1,\dots,\x_n)=G(\x_1,\dots,\x_n)F(\x_1,\dots,\x_n) \ .
\eeq

We first have to estimate the norm of $\Psi$. Using the fact that $F$
is $1$
whenever no pair of particles is closer together than a distance
$R$, we obtain
\begin{equation}\label{normbd}
\langle\Psi|\Psi\rangle \geq 1- \frac{n(n-1)}2
\frac {\pi R^2}{r^2} \|b\|_4^4 \ .
\end{equation}
To evaluate the expectation value of the Hamiltonian, we use
\beq
\langle\Psi|-\Delta_j|\Psi\rangle=-\int F^2 G\Delta_j G + \int G^2
|\nabla_j F|^2
\eeq
and the Schr\"odinger equation $H_{n,g}\psi=E_{\rm D}^{\rm 1D}\psi$.
This gives
\begin{eqnarray}\nonumber
\langle\Psi|H|\Psi\rangle &=&\left(E_{\rm D}^{\rm 1D}+
\frac{n}{r^2}e^\perp\right)\langle\Psi|\Psi\rangle-g\langle\Psi|
\sum_{i<j}\delta(z_i-z_j)|\Psi\rangle
\\ \label{exp} && + \int G^2 \left( \sum_{j=1}^n |\nabla_j F|^2 +
\sum_{i<j} v_a(|\x_i-\x_j|)|F|^2\right) \ .
\end{eqnarray}
Now, since $0\leq f\leq 1$ and $f'\geq 0$ by assumption, $F^2\leq
f(|\x_i-\x_j|)^2$, and
\beq\label{fpp}
\sum_{j=1}^n |\nabla_j F|^2 \leq 2 \sum_{i<j} f'(|\x_i-\x_j|)^2 +
4 \sum_{k<i<j} f'(|\x_k-\x_i|)f'(|\x_k-\x_j|)  \ .
\eeq
Consider the first term on the right side of (\ref{fpp}), together
with the last term in (\ref{exp}). These terms are bounded above by
\begin{equation}
n(n-1) \int
b_r(\x^\perp)^2b_r(\y^\perp)^2\rho^{(2)}_\psi(z,z')\left(
f'(|\x-\y|)^2+\half v_a(|\x-\y|) f(|\x-\y|)^2\right) .
\label{cha}
\end{equation}
Let
\beq
h(z)=\int \left( f'(|\x|)^2+\half v_a(|\x|)
f(|\x|)^2\right)d\x^\perp \ .
\eeq
Using Young's
inequality for the integration over the $\perp$-variables, we get
\beq\label{31}
(\ref{cha})\leq \frac{n(n-1)}{r^2} \|b\|_4^4 \int_{\R^2}
\rho^{(2)}_\psi(z,z') h(z-z') dz dz'\ .
\eeq
By similar methods, one can show that the contribution from the last
term in (\ref{fpp}) is bounded by
\beq\label{ahaa}
 \frac 23
n(n-1)(n-2)\frac{\|b\|_\infty^2}{r^2}\frac{\|b\|_4^4}{r^2}\|k\|_\infty
\int_{\R^2} \rho^{(2)}_\psi(z,z') k(z-z') dz
dz' \ ,
\eeq
where
\beq
k(z)=\int f'(|\x|)d\x^\perp \ .
\eeq
Note that both $h$ and $k$ are  supported in $[-R,R]$.

Now, for any $\phi\in H^1(\R)$,
\begin{equation}
\left| |\phi(z)|^2-|\phi(z')|^2\right| \leq  2 |z-z'|^{1/2}
\left(\int_\R
|\phi|^2\right)^{1/4}\left(\int_\R \left|\frac
{d\phi}{dz}\right|^2\right)^{3/4} \ .
\end{equation}
Applying this to $\rho_\psi^{(2)}(z,z')$, considered as a function of
$z$ only, we  get
\begin{eqnarray}\nonumber
&&\int_{\R^2} \rho^{(2)}_\psi(z,z') h(z-z') dz
dz'- \int_\R h(z) dz \int \rho_\psi^{(2)}(z,z)dz \\
&& \leq 2
R^{1/2} \int_\R h(z) dz
\left\langle\psi\left|-\frac{d^2}{dz_1^2}\right|\psi\right
\rangle^{3/4} \ , \label{cha2}
\end{eqnarray}
where we used Schwarz's inequality, the normalization of
$\rho^{(2)}_\psi$ and the symmetry of $\psi$. The same argument is
used for (\ref{ahaa}) with $h$ replaced by $k$.

It remains to bound the second term in (\ref{exp}). As in the
estimate for the norm of $\Psi$, we use again the fact
that $F$ is equal to 1 as long as the particles are not within a
distance $R$ from each other. We obtain
\begin{equation}
\langle\Psi|\sum_{i<j}\delta(z_i-z_j)|\Psi\rangle
\geq \frac{n(n-1)}{2} \int \rho_\psi^{(2)}(z,z) dz
\left(1-\frac{n(n-1)}2 \frac{\pi
R^2}{r^2} \|b\|_4^4\right) \label{cha3} \, .
\end{equation}
We also estimate  $g\half n(n-1)\int
\rho^{(2)}_\psi(z,z)dz\leq E_{\rm D}^{\rm 1D}$ and
$\langle\psi|-d^2/dz_1^2|\psi\rangle\leq E_{\rm D}^{\rm 1D}/n$. We
have
$\int h(z)dz=
4\pi a (1 - a/R)^{-1}$, and the terms containing $k$ can be bounded
by $\|k\|_\infty \leq 2\pi a (1+\ln(R/a))/(1-a/r)$ and $\int k(z) dz
\leq  2\pi aR (1-a/(2R))/(1-a/r)$. Putting together all the bounds,
and
choosing
\beq\label{Rin3.1}
R^3 = \frac {a r^2} {n^2(1+ g \ell )} \ ,
\eeq
this proves the desired result.

We are left with the lower bound (\ref{lbthm}).  We write a general
wave function $\Psi$ as \beq
\Psi(\x_1,\dots,\x_n)=f(\x_1,\dots,\x_n)\prod_{k=1}^n b_r(\x^\perp_k)
\ , \eeq which can always be done, since $b_r$ is a strictly positive
function.  Partial integration gives \beq\label{329}
\langle\Psi|H|\Psi\rangle=\frac{n e^\perp}{r^2}+ \sum_{i=1}^n \int
\left[ |\nabla_i f|^2 +\half \sum_{j,\, j\neq i} v_a(|\x_i-\x_j|)|f|^2
\right] \prod_{k=1}^n b_r(\x^\perp_k)^2 d \x_k.  \eeq Choose some
$R>R_0$, fix $i$ and $\x_j$, $j\neq i$, and consider the Voronoi
cell\index{Voronoi cells} $\Omega_j$ around particle $j$, i.e.,
$\Omega_j=\{\x:\, |\x-\x_{j}|\leq |\x-\x_{k}| \hbox{ for all } k\neq
j\}$. If ${\mathcal B}_j$ denotes the ball of radius $R$ around
$\x_j$, we can estimate with the aid of Lemma~\ref{dysonl}
\begin{eqnarray}\nonumber
&&\int_{\Omega_j\cap {\mathcal B}_j} b_r(\x^\perp_i)^2\left(
|\nabla_i f|^2+
\half v_a(|\x_i-\x_j|)|f|^2\right) d\x_i \\ && \geq \frac{\min_{\x\in
{\mathcal B}_j}b_r(\x^\perp)^2}{\max_{\x\in {\mathcal
B}_j}b_r(\x^\perp)^2} a
\int_{\Omega_j\cap {\mathcal B}_j} b_r(\x^\perp_i)^2
U(|\x_i-\x_j|)|f|^2 \ .
\end{eqnarray}
Here $U$ is given in (\ref{softened}).  For some $\delta>0$ let
${\mathcal B}_\delta$ be the subset of $\R^2$ where $b(\x^\perp)^2\geq
\delta$, and let $\chi_{{\mathcal B}_\delta}$ denote its
characteristic function.  Estimating $\max_{\x\in {\mathcal
    B}_j}b_r(\x^\perp)^2\leq \min_{\x\in {\mathcal
    B}_j}b_r(\x^\perp)^2 + 2(R/r^3) \|\nabla b^2\|_\infty$, we obtain
\beq\label{334} \frac{\min_{\x\in {\mathcal
      B}_j}b_r(\x^\perp)^2}{\max_{\x\in {\mathcal
      B}_j}b_r(\x^\perp)^2}\geq \chi_{{\mathcal
    B}_\delta}(\x^\perp_j/r)\left(1-2\frac R{r} \frac{\|\nabla
    b^2\|_\infty}{\delta}\right) \ .  \eeq Denoting $k(i)$ the nearest
neighbor\index{nearest neighbor interaction} to particle $i$, we conclude
that,
for $0\leq \eps\leq 1$,
\begin{eqnarray}\nonumber
&&\sum_{i=1}^n \int \left[ |\nabla_i f|^2 +\half \sum_{j,\, j\neq
i} v_a(|\x_i-\x_j|)|f|^2 \right] \prod_{k=1}^n b_r(\x^\perp_k)^2
d \x_k \\ \nonumber && \geq \sum_{i=1}^n \int \Big[ \eps
|\nabla_i f|^2 +(1-\eps)|\nabla_i f|^2 \chi_{\min_k |z_i-z_k|\geq
R}(z_i)\\ \label{negl} &&  \qquad + a'
U(|\x_i-\x_{k(i)}|)\chi_{{\mathcal B}_\delta}(\x^\perp_{k(i)}/r)|f|^2
\Big]
\prod_{k=1}^n b_r(\x^\perp_k)^2 d \x_k \ ,
\end{eqnarray}
where $a'=a(1-\eps)(1-2 R \|\nabla b^2\|_\infty/ r \delta)$.

Define $F(z_1,\dots,z_n)\geq 0$ by
\beq\label{defF}
|F(z_1,\dots,z_n)|^2=\int |f(\x_1,\dots,\x_n)|^2\prod_{k=1}^n
b_r(\x_k^\perp)^2 d\x_k^\perp \ .
\eeq
Neglecting the kinetic energy in $\perp$-direction in the second term
in (\ref{negl}) and using the Schwarz inequality to bound the
longitudinal kinetic energy of $f$ by the one of $F$, we get the
estimate
\begin{eqnarray}\nonumber
&&\langle\Psi|H|\Psi\rangle- \frac{n e^\perp}{r^2}\geq \\ \nonumber
&& \sum_{i=1}^n \int\Big[\eps |\partial_i F|^2 +
(1-\eps)|\partial_i F|^2 \chi_{\min_k |z_i-z_k|\geq R}(z_i)
\Big]\prod_{k=1}^n dz_k \\ \nonumber &&+\sum_{i=1}^n \int \left[
\eps|\nabla^\perp_i f|^2 + a'
U(|\x_i-\x_{k(i)}|)\chi_{{\mathcal B}_\delta}(\x^\perp_{k(i)}/r)|f|^2
\right] \prod_{k=1}^n b_r(\x^\perp_k)^2 d \x_k , \\ \label{57}
\end{eqnarray}
where $\partial_j=d/dz_j$, and $\nabla^\perp$ denotes the gradient in
$\perp$-direction. We now investigate the last term in (\ref{57}).
Consider, for fixed $z_1,\dots,z_n$, the expression \beq\label{44}
\sum_{i=1}^n \int \left[ \eps|\nabla^\perp_i f|^2 + a'
  U(|\x_i-\x_{k(i)}|)\chi_{{\mathcal
      B}_\delta}(\x^\perp_{k(i)}/r)|f|^2 \right] \prod_{k=1}^n
b_r(\x^\perp_k)^2 d \x_k^\perp \ .  \eeq To estimate this term from
below, we use Temple's inequality\index{Temple's inequality}, as in
Sect.~\ref{subsect22}. Let $\widetilde e^\perp$ denote the gap
above zero in the spectrum of $-\Delta^\perp+V^\perp-e^\perp$, i.e.,
the lowest non-zero eigenvalue. By scaling, $\widetilde e^\perp/r^2$
is the gap in the spectrum of $-\Delta^\perp+V^\perp_r-e^\perp/r^2$.
Note that under the transformation $\phi \mapsto b_r^{-1} \phi$ this
latter operator is unitarily equivalent to $\nabla^{\perp *}\cdot
\nabla^\perp$ as an operator on $L^2(\R^2,b_r(\x^\perp)^2 d\x^\perp)$,
as considered in (\ref{44}). Hence also this operator has $\widetilde
e^\perp/r^2$ as its energy gap\index{energy gap}. Denoting \beq \label{8.55}\langle
U^k\rangle= \int
\left(\sum_{i=1}^n U(|\x_i-\x_{k(i)}|)\chi_{{\mathcal
      B}_\delta}(\x^\perp_{k(i)}/r)\right)^k \prod_{k=1}^n
b_r(\x^\perp_k)^2 d \x_k^\perp \ , \eeq Temple's inequality implies
\beq\label{60} (\ref{44})\geq |F|^2 a'\langle U \rangle \left(1-
  a'\frac{\langle U^2\rangle}{\langle U\rangle}\frac{1}{\eps
    \widetilde e^\perp/r^2-a'\langle U\rangle}\right) \ .  \eeq Now,
using (\ref{softened}) and Schwarz's inequality, $\langle
U^2\rangle\leq 3n(R^3-R_0^3)^{-1}\langle U\rangle$, and
\begin{equation}\label{8.57}
\langle U\rangle \leq
n(n-1)\frac{\|b\|_4^4}{r^2}\frac{3\pi R^2}{R^3-R_0^3} \ .
\end{equation}
Therefore
\beq
(\ref{60})\geq |F|^2 a''\langle U \rangle \ ,
\eeq
where we put all the error terms into the modified coupling constant
$a''$. It remains to derive a lower bound on $\langle U\rangle$. Let
\beq
d(z-z')=\int_{\R^4} b_r(\x^\perp)^2 b_r(\y^\perp)^2 U(|\x-\y|)
\chi_{{\mathcal B}_\delta}(\y^\perp/r) d\x^\perp d\y^\perp \ .
\eeq
Note that $d(z)=0$ if $|z|\geq R$. An estimate similar to
(\ref{firstorder}) gives
\begin{equation}\label{860}
\langle U\rangle \geq
\sum_{i\neq j}d(z_i-z_j) \left(1-(n-2)
\frac{\pi R^2}{r^2} \|b\|_\infty^2\right) \ .
\end{equation}
Note that, for an appropriate choice of $R$, $d$ is close to a
$\delta$-function with the desired coefficient. To make the
connection with the $\delta$-function, we can use a bit of the
kinetic energy saved in (\ref{57}) to obtain
\begin{eqnarray}\nonumber
&&\int \left[ \frac\eps{n-1} |\partial_i F|^2 + a'''  d(z_i-z_j)
|F|^2\right] dz_i \\ && \geq \half g' \max_{|z_i-z_j|\leq R}
|F|^2 \chi_{[R,\ell-R]}(z_j)
\left(1-\left(\frac{2(n-1)}{\eps}g'R\right)^{1/2}\right) \ .
\end{eqnarray}

Putting all the previous estimates together, we arrive at
\begin{eqnarray}\nonumber
\langle\Psi|H|\Psi\rangle- \frac{ne^\perp}{r^2} &\geq& \sum_{i=1}^n
\int\Big[(1-\eps)|\partial_i F|^2 \chi_{\min_k
|z_i-z_k|\geq R}(z_i) \Big] \prod_{k=1}^n dz_k \\ \nonumber
\label{putt} & & +\sum_{i\neq j} \half g'' \int
\max_{|z_i-z_j|\leq R}|F|^2 \chi_{[R,\ell-R]}(z_j) \prod_{k,\, k\neq
i}dz_k \\
\end{eqnarray}
for an appropriate coupling constant $g''$ that contains all the
error terms.
Now assume that $(n+1)R<\ell$. Given an $F$ with $\int |F|^2
dz_1\cdots
dz_n=1$, define, for
$0\leq z_1\leq z_2\leq \dots \leq z_n\leq \ell-(n+1)R$,
\beq
\psi(z_1,\dots,z_n)=F(z_1+R,z_2+2 R,z_3+3R,\dots,z_n+n R) \ ,
\eeq
and extend the function to all of $[0,\ell-(n+1)R]^n$ by symmetry. A
simple calculation shows that
\begin{eqnarray}\nonumber
(\ref{putt})\geq \langle\psi |H' |\psi\rangle &\geq& (1-\eps)
E_{\rm N}^{\rm 1D}(n,\ell-(n+1)R,g'') \langle\psi|\psi\rangle  \\
\label{puutt}
&\geq&
(1-\eps) E_{\rm N}^{\rm 1D}(n,\ell,g'') \langle\psi|\psi\rangle\ ,
\end{eqnarray}
where $H'$ is the Hamiltonian (\ref{13}) with a factor $(1-\eps)$ in
front of the kinetic energy term.

It remains to estimate $\langle\psi|\psi\rangle$. Using that $F$ is
related to the true ground state $\Psi$ by (\ref{defF}), we can
estimate it in terms of the total QM energy, namely
\begin{eqnarray}\nonumber
\langle\psi|\psi\rangle &\geq& 1- \frac {2R}{g''} \left(E^{\rm
  QM}_{\rm N}(N,\ell,r,a)-\frac {ne^\perp}{r^2}\right) \\ && -2 n
\frac
R\ell - 4 n R \left(\frac 1n E^{\rm QM}_{\rm N}(n,\ell,r,a)-\frac
{e^\perp}{r^2} \right)^{1/2} \ . \label{ddss}
\end{eqnarray}
Collecting all the error terms and choosing
\beq\label{chopara}
R=r\left(\frac{a}{r}\right)^{1/4}\ , \quad
\eps=\left(\frac{a}{r}\right)^{1/8}\ , \quad
\delta=\left(\frac{a}{r}\right)^{1/8} \ ,
\eeq
(\ref{puutt}) and (\ref{ddss}) lead to the desired lower bound.
\end{proof}

As already noted above, Lemma~\ref{finthm} is the key to the proof of
Theorems~\ref{T1} and~\ref{T1dens}. The estimates are used in each
box, and the particles are distributed optimally among the boxes. For
the global lower bound, superadditivity\index{superadditivity} of the
energy and convexity of the energy density $\rho^3 e(g/\rho)$ are
used, generalizing corresponding arguments in Chapter~\ref{sect3d}. We
refer to \cite{LSY} for details.


 \chapter[Two-Dimensional Behavior in Disc-Shaped Traps]{Two-Dimensional
Behavior in\vspace*{-2mm}\newline Disc-Shaped Traps}\label{3dto2d}

In this chapter, which is based on \cite{SY},  we discuss  the dimensional
reduction of a Bose
gas in a trap that confines the particles strongly in one direction so
that two-dimensional behavior\index{two-dimensional behavior} is expected.
There are many similarities
with the emergence of one-dimensional behavior in cigar-shaped traps
discussed in the previous chapter but also some notable differences.
As in the case of cigar-shaped traps, there is a basic division of the
parameter domain into two regions (this is also noted in \cite{Petrov}
and \cite{PiSt2003}): one where a limit of a three-dimensional
Gross-Pitaevskii theory applies, and a complementary region
described by a ``truly'' low dimensional theory.  In the case
discussed in Chapter \ref{1dsect} the latter is a density functional
theory based on the exact Lieb-Liniger solution for the energy of a
strongly interacting (and highly correlated) one-dimensional gas with
delta interactions.  (Note that in 1D strong interactions means low
density.)  In the present case, on the other hand, the gas is {\it
  weakly} interacting in all parameter regions.  In the region not
accessible from 3D GP theory the energy formula (\ref{2den}) for a
dilute two-dimensional Bose gas with a logarithmic dependence on the
density applies.  To enter this region extreme dilution is required.
The Lieb-Liniger region in the 1D case demands also quite dilute
systems, but the requirement is even more stringent in 2D. This will
be explained further below.

  We recall from Chapter~\ref{sect2d} that the
energy per particle of a dilute,
  homogeneous, two-dimensional Bose gas with density $\rho_{\rm 2D}$
  and scattering length $a_{\rm 2D}$ of the interaction potential is
  (in units
such that $\hbar=2m=1$)
  \be\label{SCHNEE2den} e_{\rm 2D}\approx 4\pi\rho_{\rm 2D}|\ln(\rho_{\rm
  2D}a_{\rm 2D}^2)|^{-1}.  \ee The corresponding result in three
  dimensions is { \be\label{SCHNEE3den} e_{\rm 3D}\approx
  4\pi\rho_{\rm 3D}a_{\rm 3D}  \ee
as discussed in Chapter~\ref{sect3d}. In the following we shall denote
  the two-dimensional density, $\rho_{\rm 2D}$, simply by $\rho$ and
  the three dimensional scattering length, $a_{\rm 3D}$, by $a$.
The basic message of this chapter is that when the thickness, $h$, of the
trap tends to zero then Eq.\ (\ref{SCHNEE2den}) can be used with an
effective
two-dimensional scattering length
$a_{\rm 2D}= h\exp(-({\rm const.})h/a)$.
A more precise formula is given in Eq.\ \eqref{SCHNEEatd} below.  If
  $|\ln \rho h^2|\ll h/a$, then $|\ln(\rho a_{\rm 2D}^2)|\approx h/a$,
  and the two-dimensional formula \eqref{SCHNEE2den} leads to the same
  result as the three dimensional formula \eqref{SCHNEE3den}, because
  $\rho_{\rm 3D}\sim \rho/h$.  The ``true'' two dimensional region
  requires $|\ln \rho h^2|\gtrsim h/a$ and hence the condition
  $\rho^{-1/2}\gtrsim h e^{h/a}$ for the interparticle distance,
  $\rho^{-1/2}$.  This should be compared with the corresponding
  condition for the 1D Lieb-Liniger region of the previous chapter
where the
  interparticle distance is ``only'' required to be of the order or
  larger than $h^2/a$.

  The formula $a_{\rm 2D}= h\exp(-({\rm const.})h/a)$ for the
  scattering length appeared first in \cite{Petrov}.  It can be
  motivated by considering a weak, bounded potential, where
  perturbation theory can be used to compute the energy \eqref{emin}
  in Appendix C that is directly related to the scattering length.
  This perturbative calculation is carried out in Section
  \ref{pertscatt} as a step in the proof of a lower bound for the
  many-body energy; its relation to the formula for $a_{\rm 2D}$ is
  explained in the remark after Corollary \ref{SCHNEEscattcorr}.

We now define the setting and state the results more precisely.  We
consider $N$ identical, spinless bosons in a confining,
three-dimensional trap potential and with a repulsive, rotationally
symmetric pair interaction.  We take the direction of strong
confinement as the $z$-direction and write the points $\x\in\R^3$ as
$(\xp,z)$, $\xp\in\R^2$, $z\in\R$.  The Hamiltonian is
\begin{equation}\label{SCHNEEdef:ham}
H_{N,L,h,a} = \sum_{i=1}^N \left(-\Delta_i + V_{L,h}(\x_i) \right)
+ \sum_{1\le
i <j \le N} v_a(|\x_i-\x_j|)
\end{equation}
with
\begin{eqnarray}\label{SCHNEEpot}
V_{L,h}(\x) &=& V_L(\xp)+V^\perp_h(z) =
\frac{1}{L^2}
V(L^{-1}\xp)+ \frac{1}{h^2}V^{\perp
}(h^{-1}z),\\
v_a(|\x|) &=& \frac{1}{a^2}v(a^{-1}|\x|). \label{va}
\end{eqnarray}
The confining potentials $V$ and $V^\perp$ are
assumed to be locally
bounded and tend to $\infty$ as $|\xp|$ and $|z|$
tend to $\infty$.
The interaction potential $v$ is assumed to be nonnegative, of finite range
and
with
scattering length 1; the scaled potential
$v_a$ then has scattering length $a$. We regard $v$,
$V^{\perp}$ and $V$ as fixed and $L,h,a$ as scaling
parameters.  The Hamiltonian (\ref{SCHNEEdef:ham}) acts on symmetric
wave functions in $L^2(\R^{3N},d\x_1\cdots d\x_N)$. Its
ground state energy,  $E^{\rm QM}({N,L,h,a})$, scales with $L$ as
\be\label{SCHNEEqmscal} E^{\rm QM}(N,L,h,a)=\frac1{L^2} E^{\rm
QM}(N,1,h/L,a/L).\ee
\\
Taking $N\to\infty$ but keeping $h/L$ and $Na/L$ fixed leads to a
three dimensional Gross-Pitaevskii description of the ground state as
proved in Chapter~\ref{sectgp}.  The corresponding energy functional
is (cf.\ (\ref{gpfunc3d}))
\be\label{SCHNEE3dgp}{\mathcal E}^{\rm GP}_{\rm
3D}[\phi]=\int_{\R^3}\left\{|\nabla\phi(\x)|^2+V_{L,h}(\x)|\phi(\x)|^2+4\pi
Na |\phi(\x)|^4\right\}d^3\x\ee and the energy per particle is
\beqa \nonumber
E^{\rm GP}_{\rm 3D}(N,L,h,a)/N &=&\inf\{{\mathcal
E}^{\rm GP}_{\rm 3D}[\phi],\, \hbox{$\int|\phi(\x)|^2
d^2\x$}=1\}\\ \label{SCHNEE3dgpen}&=&(1/L^2) E^{\rm GP}_{\rm
3D}(1,1,h/L,Na/L).\eeqa By the GP limit
theorem, Thm.\ \ref{thmgp3}, we have, for fixed $h/L$ and $Na/L$,
\be\label{SCHNEE3dgplim}\lim_{N\to\infty}\frac {E^{\rm
QM}({N,L,h,a})}{E^{\rm GP}_{\rm 3D}(N,L,h,a)}=1.  \ee It is important
to note, however, that the estimates in Chapter~\ref{sectgp} are not uniform
in
the ratio $h/L$ and the question what happens if $h/L\to 0$ is not
addressed in Ch.\ \ref{sectgp}.  It will be shown in the next section that a
{\it part} of the parameter range for thin traps can be treated by
considering, at fixed $Na/h$, the $h/L\to 0$ limit of $E^{\rm GP}_{\rm
3D}(1,1,h/L,Na/L)$ with the ground state energy for the transverse
motion, $\sim 1/h^2$, subtracted.  But this limit can evidently never
lead to a logarithmic dependence on the density and it does not give
the correct limit formula for the energy in the whole parameter range.

To cover all cases we have to consider a two-dimensional
Gross-Pitaevskii theory of the type studied in Section \ref{sub2d}, i.e.,
\be\label{SCHNEE2dgp}{\mathcal E}^{\rm GP}_{\rm
2D}[\varphi]=\int_{\R^2}\left\{|\nabla\varphi(x)|^2+V_{L}(x)|\varphi(x)|^2+4
\pi
Ng |\varphi(x)|^4\right\}d^2x\ee with \be\label{SCHNEEcoupl}
g=|\ln(\bar\rho\atd^2|^{-1}.\ee\\
Here $\bar\rho$ is the mean density, defined by Eq.\ (\ref{meandens2d}).
An explicit formula that is valid in the case $Ng\gg 1$ can be states as follows. Let
\be \rho^{\rm TF}_N(x)=\frac1{8\pi}\left[\mu^{\rm TF}_N-V_L(x)\right]_+ \ee
be the 'Thomas Fermi' density for $N$ particles at coupling constant 1 
in the potential $V_L$, where $\mu^{\rm TF}_N$ is chosen so that
$\int \rho^{\rm TF}_N=N$. Then
\be\label{barrho1} \bar\rho=N^{-1}\int_{\R^2} \rho^{\rm TF}_N(x)^2 dx.\ee
For simplicity we shall assume that $V$ is homogeneous
of some degree $s>0$, i.e., $V(\lambda x)=\lambda^s V(x)$, and in this
case \be\label{SCHNEErhobar}\bar\rho\sim N^{s/(s+2)}/L^2 =N/\bar
L^2\quad\hbox{with}\quad \bar L=N^{1/(s+2)}L.\ee The length $\bar L$
measures the effective extension of the gas cloud of the $N$ particles in the
two-dimensional trap.  A box potential corresponds to $L=\bar L$,
i.e., $s=\infty$ and hence $\bar\rho\sim N/L^2$. 

The case  $Ng=O(1)$ requires a closer look at the definition of $\bar\rho$. First, for any value of  $Ng$ we can consider the minimizer $\varphi^{\rm GP}_{Ng}$ of  \eqref{SCHNEE2dgp} with normalization $\Vert \varphi^{\rm GP}_{Ng}\Vert_2=1$. The corresponding mean density is
\be \bar\rho_{Ng}=N\int |\varphi^{\rm GP}_{Ng}|^4.\ee
A general definition of $\bar \rho$ amounts to solving the equation $\bar\rho=\bar\rho_{Ng}$ with $g$ as in \eqref{SCHNEEcoupl}. This gives the same result as \eqref{barrho1} to leading order in $g$ when $Ng\gg 1$. In the case $|\ln Nh/L|\ll h/a$ (referred to as 'Region I' below) the coupling constant is simply $a/h$ and thus independent of $N$. Moreover, in  a homogeneous potential of degree $s$ the effective length scale $\bar L$ is $\sim (Ng+1)^{1/(s+2)} L$ and thus of  order $L$ if $Ng=O(1)$.

The energy per
particle corresponding to \eqref{SCHNEE2dgp} is \be\label{SCHNEE2dgpen}
E^{\rm
GP}_{\rm 2D}(N,L,g)/N=\inf\{{\mathcal E}^{\rm GP}_{\rm
2D}[\varphi],\, \hbox{$\int|\varphi(x)|^2 d^2x$}=1\}=(1/L^2) E^{\rm
GP}_{\rm 2D}(1,1,Ng).\ee

Let $s_{h}$ be the normalized ground state wave function of the one-particle
Hamiltonian $-d^2/dz^2+V^\perp_h(z)$. It can be written as
$s_{h}(z)=h^{-1/2}s(h^{-1}z)$ and the ground state energy as
$e^\perp_{h}=h^{-2}e^\perp$, where $s(z)$ and $e^\perp$ are,
respectively, the ground state wave function and ground state energy of
$-d^2/dz^2+V^\perp(z)$. We {\it define} the effective two dimensional
scattering length by the formula
\be \label{SCHNEEatd}\atd=h\exp\left(-(\hbox{$\int
s(z)^4dz$})^{-1}h/2a\right).\ee
Then, using (\ref{SCHNEEcoupl}),  \be\label{SCHNEEgformula}
g=|-\ln(\bar\rho h^2)+\hbox{($\int s(z)^4dz$})^{-1}h/a|^{-1}.\ee
The justification of the definition (\ref{SCHNEEatd}) is
Theorem \ref{SCHNEEthm:main} below.\\

\noindent{\it Remark.} Since $\atd$ appears only under a logarithm,
and $a/h\to 0$, one could, at least as far as leading order
computations are concerned, equally well define the two dimensional
scattering length as $\atd'=b\,\exp\left(-(\hbox{$\int
s(z)^4dz$})^{-1}h/2a\right)$ with $b$ satisfying $c\, a\leq b\leq C\,
h$ for some constants $c>0$, $C<\infty$.  In fact, if
$g'=|\ln(\bar\rho(\atd')^2)|^{-1}$, then \be\label{ggprime} \frac
{g}{g'}=1+\frac{2\ln(b/h)} {|-\ln(\bar \rho
h^2)+\hbox{(const.)}h/a|}\to 1\ee because $(a/h)\ln(b/h)\to 0$.

\medskip

We can now state the main result of this chapter:
\begin{thm}[From 3D to 2D]\label{SCHNEEthm:main}
Let $N\to\infty$ and at the same time $h/L\to 0$ and $a/h\to 0$ in such a way that
$h^2\bar \rho g\to 0$ (with $g$ given by Eq.\
\eqref{SCHNEEgformula}). Then
\begin{equation}\label{SCHNEElimit}
\lim  \frac{E^{\rm QM}(N,L,h,a) - N h^{-2} e^{\perp}}{E^{\rm GP}_{\rm
2D}(N,L,g)} =1.
\end{equation}
\end{thm}

\noindent{\it Remarks:} 1.  The condition $h^2\bar \rho g\to 0$ means
that the ground state energy $h^{-2}e^\perp$ associated with the
confining potential in the $z$-direction is much larger than the
energy $\bar \rho g$.  This is the condition of {\it strong
confinement} \index{strong confinement} in the $z$-direction.  In the
case that $h/a\gg |\ln(\bar\rho h^2)|$ we have $g\sim a/h$ and hence
the condition in that region is equivalent to \be
\label{SCHNEEahll}\bar\rho ah\ll 1.\ee On the other hand, if
$h/a\lesssim |\ln(\bar\rho h^2)|$ the strong confinement condition is
equivalent to $h^2\bar\rho|\ln( h^2\bar\rho)|^{-1}\ll 1$, which means
simply that
\be\label{SCHNEEhhll}
\bar\rho h^2\ll 1\,.
\ee
Both \eqref{SCHNEEahll} and \eqref{SCHNEEhhll} clearly imply $\bar
\rho\atd^2\ll
1$, i.e., the gas is dilute in the 2D sense (and also in the 3D sense,
$\rho_{\rm 3D}a^3\ll 1$, because $\rho_{\rm 3D}=\rho/h$).  This is
different from the situation in cigar-shaped traps considered in
Chapter \ref{1dsect} where the
gas can be either dilute or dense in the 1D sense, depending on the
parameters (although it is always dilute in the 3D sense).

\medskip
\noindent 2.  It is, in fact, not necessary to demand $h/L\to 0$
explicitly in Theorem \ref{SCHNEEthm:main}.  The reason is as follows.
In the region where $h/a\lesssim |\ln(\bar\rho h^2)|$, the strong
confinement condition $\bar\rho h^2\ll 1$ immediately implies $h/L\ll
1$ because $\bar\rho\gg 1/L^2$, cf.\ Eq.\ (\ref{SCHNEErhobar}).  If
$h/a\gg |\ln(\bar\rho h^2)|$, then at least $\bar\rho ah\ll 1$ holds
true.  This leaves only the alternatives $h/L\to 0$, or, if $h/L$
stays bounded away from zero, $Na/L\to 0$.  But the latter alternative
means, by the three dimensional Gross-Pitaevskii limit theorem, that
the energy converges to the energy of a noninteracting, trapped gas,
for which \eqref{SCHNEElimit} obviously holds true.

\medskip

We shall refer to the parameter region where $h/a\gg |\ln(\bar\rho
h^2)|$ as {\bf Region I}, and the one where $h/a\lesssim |\ln(\bar\rho
h^2)|$ as {\bf Region II}.  In Region I we can take \be\label{SCHNEEgreg1}
g=(\hbox{$\int s(z)^4dz$})a/h.\ee In Region II $g\sim |\ln(\bar\rho
h^2)|^{-1}$, and in the extreme case that $h/a\ll |\ln(\bar\rho
h^2)|$, \be\label{SCHNEEgreg2} g=|\ln(\bar\rho h^2)|^{-1}.\ee In particular
$g$ is then independent of $a$ (but dependent on $\bar\rho$).  As
remarked earlier, Region II is only relevant for very dilute gases since it
requires interparticle distances $\bar\rho^{-1/2}\gtrsim h e^{h/a}$.

By Eq.\ \eqref{SCHNEE2dgpen} the relevant coupling parameter is $Ng$ rather
than $g$ itself, and both Region I and Region II can be divided
further, according to $Ng\ll 1$, $Ng\sim 1$, or $Ng\gg 1$.  The case
$Ng\ll 1$ corresponds simply to an ideal gas in the external trap
potential.  Note that this limit can both be reached from Region I by
taking $a/h\to 0$ at fixed $\bar\rho h^2$, or from Region II by
letting $\bar\rho h^2$ tend more rapidly to zero than $e^{-h/a}$.
The
case $Ng\sim 1$ in Region I corresponds to a GP theory with coupling
parameter $\sim Na/h$ as was already explained, in particular
after Eq.\ \eqref{SCHNEE3dgplim}.  The case $Ng\gg 1$ is the
`Thomas-Fermi'  case where the gradient term in the energy
functional \eqref{SCHNEE2dgp} can be ignored.  In Region II, the case
$Ng\lesssim 1$ requires $\bar\rho^{-1/2}\gtrsim h e^N$ and is thus
only of academic interest, while
$\bar\rho^{-1/2}\ll h e^N$ (but still $he^{h/a}\lesssim
\bar\rho^{-1/2}$) corresponds to the TF case.

The subdivision of the parameter range just described is somewhat different
from the situation described in Chapter \ref{1dsect}. This is due to
the different form of the energy per particle of the low dimensional gas as
function of the density.

\section{The 2D Limit of 3D GP Theory}

As in Section \ref{1dgpsect} certain aspects of the dimensional
reduction of the many-body system
can be seen already in the much simpler context of GP theory.
In this section we consider the $h/L\to 0$ limit
of the 3D GP ground state energy. The result is, apart from the
confining energy, the 2D GP energy with coupling constant $g\sim a/h$.
This shows in particular that Region II, where $g\sim |\ln(\bar\rho
h^2)|^{-1}$, cannot be reached as a limit of 3D GP theory.

\begin{thm}[2D limit of 3D GP energy] Define $g=\left(\int s(z)^4
dz\right)a/h$. If
$h/L\to 0$, then
\be\frac{E_{\rm 3D}^{\rm
GP}(N,L,h,a)-Nh^{-2}e^\perp}{E_{\rm 2D}^{\rm
GP}(N,L,g)}\to 1\ee
uniformly in the parameters, as long as
$\bar\rho ah\to 0$.
\end{thm}

\noindent {\it Remark.} Since $E_{\rm 2D}^{\rm GP}(1,L,Ng)\sim L^{-2}+
\bar\rho a/h$,
 the condition $\bar\rho ah\to 0$ is equivalent to $h^2 E_{\rm
 2D}^{\rm GP}(1,L,Ng)\to 0$, which  means simply
 that the 2D GP energy per particle
 is much less than the confining energy, $\sim 1/h^2$.

\begin{proof} The proof is analogous to that of Thm.\ \ref{1dgplim}.
Because of the scaling relation \eqref{SCHNEE3dgpen}
it suffices to consider the case $N=1$ and $L=1$.

For an upper bound to the 3D GP ground state energy we make the ansatz
\be\label{SCHNEEans}\phi(\x)= \varphi_{\rm GP}(x)s_{h}(z), \ee where
$\varphi_{\rm GP}$ is the minimizer of the 2D GP functional with
coupling constant $g$.  Then
\be{\mathcal E}^{\rm GP}_{\rm 3D}[\phi]=e^\perp/h^2+E_{\rm 2D}^{\rm
GP}(1,1,g)\ee and hence \be E_{\rm 3D}^{\rm GP}(1,1,h,a)-
e^\perp/h^2\leq E_{\rm 2D}^{\rm GP}(1,1,g).\ee

For the lower bound we consider the one-particle Hamiltonian (in 3D)
\be\label{SCHNEEhha} H_{h,a}=-\Delta+V_{1,h}(\x)+8\pi a|\varphi_{\rm
GP}(x)|^2 s_{h}(z)^2.  \ee Taking the 3D GP minimizer $\Phi$
as a test state gives
\bea\nonumber \hbox{inf spec}\, H_{h,a}&\leq& E_{\rm 3D}^{\rm
GP}(1,1,h,a)-4\pi a\int_{\R^3}|\Phi(\x)|^4 d^3\x\\&& +8\pi a\int_{\R^3}
|\varphi_{\rm
GP}(x)|^2s_{h}(z)^2|\Phi(\x)|^2d^3\x\nonumber
\\ &\leq& E_{\rm 3D}^{\rm
GP}(1,1,h,a)+4\pi a\int_{\R^3}|\varphi_{\rm GP}(x)|^4
s_{h}(z)^4d^3\x\nonumber \\ &=& E_{\rm 3D}^{\rm
GP}(1,1,h,a)+4\pi g\int_{\R^2}|\varphi_{\rm GP}(x)|^4
d^2x.\label{SCHNEEinfspec} \eea To
bound $H_{h,a}$ from below we consider first for fixed $x\in\R^2$ the
Hamiltonian (in 1D) \be H_{h,a,x}=-\partial_{z}^2+V_{h}^\perp(z)+8\pi a
|\varphi_{\rm GP}(x)|^2 s_{h}(z)^2.  \ee We regard
$-\partial_{z}^2+V_{h}^\perp(z)$ as its ``free'' part and $8\pi a
|\varphi_{\rm GP}(x)|^2 s_{h}(z)^2$ as a perturbation.  Since the
perturbation is positive all eigenvalues of $H_{h,a,x}$ are at least
as large as those of $-\partial_{z}^2+V_{h}(z)$; in particular the
first excited eigenvalue is $\sim 1/h^2$.
The expectation value in the ground state $s_{h}$ of the free part is
\be\label{SCHNEEhx}\langle H_{h,a,x}\rangle=e^\perp/h^2+8\pi g |\varphi_{\rm
GP}(x)|^2.\ee Temple's inequality (\ref{temple}) gives
\be\label{SCHNEEtemplegp} H_{h,a,x}\geq \left(e^\perp/h^2+8\pi g
|\varphi_{\rm GP}(x)|^2\right)\left(1-\frac{\langle(H_{h,a,x}-\langle
H_{h,a,x}\rangle)^2\rangle}{\langle H_{h,a,x}\rangle(\tilde
e^\perp-e^\perp)/h^2}\right)\,,\ee where $\tilde e^\perp/h^2$ is the
lowest eigenvalue above the ground state energy of
$-\partial_{z}^2+V_{h}^\perp(z)$.  Since \be H_{h,a,x}s_{h}=(e^\perp/h^2)
s_{h}+8\pi a |\varphi_{\rm GP}(x)|^2 s_{h}^3\ee we have
$(H_{h,a,x}-\langle H_{h,a,x}\rangle)s_{h}=8\pi|\varphi_{\rm
GP}(x)|^2(as_{h}^3-gs_{h})$ and hence, using $g=a\int
s_{h}^4=(a/h)\int s^4$, \begin{eqnarray} \langle(H_{h,a,x}-\langle
H_{h,a,x}\rangle)^2\rangle&=&(8\pi)^2|\varphi_{\rm
GP}(x)|^4\int\left(as_{h}(z)^3-gs_{h}(z)\right)^2dz\nonumber\\ &\leq&
(8\pi)^2\Vert \varphi_{\rm GP}\Vert_{\infty}^4(a/h)^2\left[\hbox{$\int$}
s^6-\left(\hbox{$\int$} s^4\right)^2\right]\nonumber\\&\leq &
\hbox{const.}\,E_{\rm
2D}^{\rm GP}(1,1,g)^2\end{eqnarray} where we have used Lemma 2.1 in
\cite{LSY2d} to bound the term $g\Vert \varphi_{\rm GP}\Vert_{\infty}^2$ by
$\hbox{const.}\,E_{\rm 2D}^{\rm GP}(1,1,g)$.  We thus see from
\eqref{SCHNEEhx} and the assumption $h^2 E_{\rm 2D}^{\rm GP}(1,1,g)\to 0$
that the error term in the Temple inequality \eqref{SCHNEEtemplegp} is
$o(1)$.

Now $H_{h,a}=-\Delta_{x}+V(x)+H_{h,a,x}$, so from
\eqref{SCHNEEtemplegp} we
conclude that \be \label{SCHNEEtemplegp2}H_{h,a}\geq
\left((e^\perp/h^2)-\Delta_{x}+V(x)+8\pi g
|\varphi_{\rm GP}(x)|^2\right)(1-o(1)).\ee On the other hand, the
lowest energy of $-\Delta_{x}+V(x)+8\pi g |\varphi_{\rm GP}(x)|^2$ is just
$E_{\rm 2D}^{\rm GP}(1,1,g)+4\pi g\int_{\R^2}|\varphi_{\rm
GP}(x)|^4d^2x$.  Combining \eqref{SCHNEEinfspec} and \eqref{SCHNEEtemplegp2}
we
thus get \be E_{\rm 3D}^{\rm GP}(1,1,h,a)- e^\perp/h^2\geq E_{\rm
2D}^{\rm GP}(1,1,g)(1-o(1)).\ee
\vskip-3mm
\end{proof}

\noindent{\it Remark.} This theorem holds also for the Gross-Pitaevskii
functional for rotating gases, i.e., if a rotational term, $-\langle
\phi,{\vec \Omega}\cdot{\vec L}\phi\rangle$ is added to the
functional.  Here $\vec \Omega$ is the angular velocity, assumed to
point in the $z$-direction, and ${\vec L}$ the angular momentum
operator.  The minimizer $\varphi_{\rm GP}$ is in this case complex
valued in general and may not be unique \cite{rot2}.

\section{Upper Bound}\label{sect9.3}

We now turn to the many-body problem, i.e., the proof of Theorem
\ref{SCHNEEthm:main}.  For simplicity we shall here only discuss the
situation where the system is homogeneous in the 2D variables $x$,
i.e., where the confining potential $V_{L}(x)$ is replaced by a large
box whose side length $L$ is taken to infinity in a thermodynamic
limit.  An inhomogeneous system in the $x$ directions can be treated
by analogous methods if one in a first step separates out a factor
$\prod_{i}\varphi_{\rm GP}(x_{i})$ in the wave function with
$\varphi_{\rm GP}$ the minimizer of the 2D GP functional.  (This is
the same technique as used in Chapter~\ref{sectgp}.) An alternative
method for dealing with inhomogeneities, based on an additive
splitting of the Hamiltonian, is described in \cite{SeirNorway}. It is
applied in \cite{SY} to the 2D limit of the Hamiltonian
\ref{SCHNEEdef:ham}.

As in the problem discussed in Chapter \ref{1dsect} the key lemmas are
energy bounds in  boxes with {\it finite} particle number.
The bounds for the total system are obtained by distributing the
particles optimally among the boxes.  We shall here focus on the
estimates for the individual boxes, starting with the upper bound.

Consider the Hamiltonian \be \label{SCHNEEboxham}H =  \sum_{i=1}^n
\left(- \Delta_i + V_{h}^\perp(z_i) \right) + \sum_{1\le i <j \le n}
v_a(|\x_i-\x_j|)\ee in a region $\Lambda=\Lambda_{2}\times\R$ where
$\Lambda_{2}$ denotes a box of side length $\ell$ in the 2D $x$
variables.  For the upper bound on the ground state energy of
(\ref{SCHNEEboxham}) we impose {\it Dirichlet} boundary conditions on
the 2D Laplacian.  The goal is to prove, for a given 2D density $\rho$
and parameters $a$ and $h$, that for a suitable choice of $\ell$ and corresponding particle number $n=\rho \ell^2$ the
energy per particle is bounded above by \be\label{boxubound} 4\pi
\rho|\ln(\rho
a_{\rm 2D}^2)|^{-1}(1+o(1)) \ee where $a_{\rm 2D}$ is
given by Eq.\ (\ref{SCHNEEatd}).  Moreover, the Dirichlet localization
energy per particle, $\sim 1/\ell^2$, should be small compared to
(\ref{boxubound}).  The relative error, $o(1)$, in (\ref{boxubound})
tends to zero with the small parameters $a/h$ and $\rho a h$ (Region
I), or $a/h$ and $\rho h^2$ (Region II).

The choice of variational functions depends on the
parameter regions and we are first concerned with the Region II, i.e.,
the case $|\ln(\rho h^2)|\gtrsim h/a$.

Let $f_{0}(r)$ be the solution of the zero energy scattering equation
\be\label{SCHNEEscatteq} -\Delta f_{0}+\half v_{a}f_{0}=0,\ee
normalized so that $f_{0}(r)=(1-a/r)$ for $r\geq R_{0}$ with $R_0$ the
range of $v_a$. Note that $R_0=({\rm const.}) a$ by the scaling
\eqref{va}.  The function $f_0$ satisfies $0\leq f_{0}(r)\leq 1$ and
$0\leq f_{0}'(r)\leq \min\{1/r,a/r^2\}$.  For $R>R_{0}$ we define
$f(r)=f_{0}(r)/(1-a/R)$ for $0\leq r\leq R$, and $f(r)=1$ for $r>R$.
Define a two-dimensional potential by
\be\label{SCHNEEwpot}W(x)=\frac{2\Vert s
  \Vert_4^4}h\int_{\R}\left[f'(|\x|)^2+\half
  v_{a}(|\x|)f(|\x|)^2\right]dz.\ee Clearly, $W(x)\geq 0$, and $W$ is
rotationally symmetric with $W(x)=0$ for $|x|\geq R$.  Moreover, by
partial integration, using \eqref{SCHNEEscatteq}, it follows that
$W\in L^1(\R^2)$ with \be\label{SCHNEEintw}\int_{\R^2}W(x)dx=\frac
{8\pi a\Vert s \Vert_4^4}h (1-a/R)^{-1}.\ee

Define, for $b>R$,
\be\varphi(r)=\left\{ \begin{array}{lcr}
{\ln(R/\atd)}/
{\ln(b/\atd)}&{\rm if} & 0\leq r\leq R\\ \\ {\ln(r/\atd)}/
{\ln(b/\atd)}&{\rm if}& R\leq r\leq b\\ \\ 1 &{\rm if} & b\leq r\end{array}
\right.\ee
As
test function for the three dimensional Hamiltonian \eqref{SCHNEEboxham} we
shall
take
\be\label{SCHNEEFG}\Psi(\x_{1},\ldots,\x_{n})=F(\x_{1},\ldots,\x_{n})G(\x_{1
},\ldots,\x_{n})\ee
with \be F(\x_1,\dots,\x_n)=\prod_{i<j} f(|\x_i-\x_j|) \quad
\hbox{and}\quad
G(\x_1,\dots,\x_n)=\prod_{i<j} \varphi(|x_i-x_j|)\prod_{k=1}^n s_h(z_k).\ee
The parameters $R$, $b$ and also $\ell$ will eventually be chosen so that
the
errors compared to the expected leading term in the energy are small.

As it stands, the function (\ref{SCHNEEFG}) does not satisfy
Dirichlet boundary conditions but this can be taken care of by
multiplying the function with additional
factors at energy cost $\sim 1/\ell^2$ per particle, that will turn
out to be small compared to the energy of (\ref{SCHNEEFG}).

Since  $f(|\x_i-\x_j|) \varphi(|x_i-x_j|)=1$ for $|x_i-x_j|\geq b$
and $s_{h}$ is normalized, the
norm of $\Psi$ can be estimated as (cf.\ the analogous Eq.\
(\ref{normbd}))
\be\label{SCHNEEnormbd}
\langle\Psi|\Psi\rangle \geq
\ell^{2n}\left[1-\frac{\pi n(n-1)}2\frac {b^2}{\ell^2}\right]\,.\ee

Next we consider the expectation value of $H$ with the wave function
$\Psi$. By partial integration we have, for every $j$,
\bea\label{SCHNEEupp1}
\!\!\!\!&&\!\!\!\! \int|\nabla_{j}(FG)|^2\nonumber \\
\!\!\!\!&&\!\!\!\!=\int G^2|\nabla_{j}F|^2-\int
F^2G\Delta_{j}G=\int G^2|\nabla_{j}F|^2-\int
F^2G\,\partial_{z_{j}}^2G-\int F^2 G(\Delta^\parallel_{j} G)\nonumber \\
\!\!\!\!&&\!\!\!\!
=\int G^2|\nabla_{j}F|^2-\int
F^2G\,\partial_{z_{j}}^2G+\int F^2|\nabla_{j}^\parallel G|^2+
2\int FG(\nabla_{j}^\parallel F)\cdot(\nabla_{j}^\parallel G)
\eea
where $\Delta^\parallel_{j}$ and $\nabla_{j}^\parallel$ are, respectively,
the two
dimensional Laplace operator and gradient.
The term $-\int
F^2G\,\partial_{z_{j}}^2G$ together with $\int V^\perp_{h} F^2G^2$ gives
the confinement energy,
$(e^\perp/h^2) \Vert\Psi\Vert^2$.

Next we consider the first and the third term in \eqref{SCHNEEupp1}.
Since $0\leq f\leq 1$, $f'\geq 0$ and $s_{h}$ is normalized, we have
\begin{align}\notag
\sum_{j}\int F^2|\nabla^\parallel_{j}G|^2+\sum_{j}\int
|\nabla_{j}F|^2 G\hfill \nonumber  \leq&
\sum_{j}\int |\nabla_{j}^\parallel \Phi|^2+
 2 \sum_{i<j} f'(|\x_i-\x_j|)^2G^2 \\ \notag &  +
4 \sum_{k<i<j} \int f'(|\x_k-\x_i|)f'(|\x_k-\x_j|) G^2 \\ \label{SCHNEEfpp}
\end{align}
where we have denoted $\prod_{i<j}\varphi(|x_{i}-x_{j}|)$ by $\Phi$
for short. Moreover, since $0\leq \varphi\leq1$, 
\be 2 \sum_{i<j} f'(|\x_i-\x_j|)^2G^2\leq 2 \sum_{i<j} f'(|\x_i-\x_j|)^2s_{h}(z_{i})^2s_{j}(z_{j})^2.\ee

By Young's inequality
\be
2\int_{\R^2} f'(|\x_{i}-\x_{j}|)^2s_{h}(z_{i})^2s_{j}(z_{j})^2
dz_{i}dz_{j}\leq \frac{2\Vert s\Vert^4_4}h\int_{\R} f'(|(x_{i}-x_{j},z)|)^2
dz.
\ee
The right side  gives rise to the first of the two terms in
the formula \eqref{SCHNEEwpot} for the two dimensional
potential $W$. The other part is provided by $\int
F^2G^2 v_{a}(\x_{i}-\x_{j})$, 
using  that $0\leq f\leq1$, $0\leq \varphi\leq1$  and Young's inequality.

Altogether we obtain
\be\label{SCHNEE412}\langle \Psi|H|\Psi\rangle-(ne^\perp/h^2)\langle
\Psi|\Psi\rangle
\leq \sum_{j}\int_{\Lambda^n_{2}}
|\nabla_{j}^\parallel \Phi|^2+
\sum_{i<j}\int_{\Lambda^n_{2}}  W(x_{i}-x_{j})\Phi^2 +\mathcal
R_{1}+\mathcal R_{2}
\ee
with
\be
\mathcal R_{1}=2\int_{\Lambda^n} FG(\nabla_{j}^\parallel F)
\cdot(\nabla_{j}^\parallel G)
\ee
and
\begin{align}
\mathcal R_{2}&= 4 \sum_{k<i<j} \int_{\Lambda^n}
f'(|\x_k-\x_i|)f'(|\x_k-\x_j|)
G^2\\ &\leq \frac 23n(n-1)(n-2)\ell^{2(n-3)}\!\int_{\Lambda^3} \!
f'(|\x_1-\x_2|)f'(|\x_2-\x_3|)s_{h}(z_{1})^2
s_{h}(z_{2})^2s_{h}(z_{3})^2, \notag
\end{align}
where $0\leq \varphi\leq 1$ has been used for the last inequality.

The error term $\mathcal R_{1}$ is easily dealt with: It is zero
because $\varphi(r)$ is constant for $r\leq R$ and $f(r)$ is constant
for $r\geq R$.

The other term, $\mathcal R_{2}$, is estimated as follows. Since
$f'(r)=0$ for $r\geq R$ we can use the Cauchy Schwarz inequality for
the integration over $\x_{1}$ at fixed $\x_{2}$ to obtain
\begin{align}\notag
&\int f'(|\x_{1}-\x_{2}|)s_{h}(z_{1})^2d\x_{1} \\  \notag &\leq
\left(\int f'(|\x_{1}-\x_{2}|)^2
d\x_{1}\right)^{1/2}\left(\int_{|\x_{1}-\x_{2}|\leq R}
s_{h}(z_{1})^4d\x_{1}\right)^{1/2} \\ &\leq (4\pi \Vert s\Vert_{\infty}
a'R^3/3h^2)^{1/2}\label{SCHNEE415}
\end{align}
 with
$a'=a(1-a/R)^{-1}$.  The same estimate for the integration over
$\x_{3}$ and a subsequent integration over $\x_{2}$ gives \be \mathcal
R_{2}\leq {\rm (const.)}\ell^{2n} n^3\frac{a'R^3}{\ell^4 h^2}.
\ee
\goodbreak

We
need $\mathcal R_{2}/\langle\Psi|\Psi\rangle$ to be small compared to
the leading term in the energy, $\sim n^2\ell^{-2}|\ln(\rho
h^2)|^{-1}$ with $\rho=n/\ell^2$.  (Recall that we are in Region II
where $|\ln(\rho h^2)|\gtrsim h/a$.)  Moreover, the leading term
should be large compared to the Dirichlet localization energy, which
is $\sim n/\ell^2$.  We are thus lead to the conditions (the first
comes from \eqref{SCHNEEnormbd}): \be\label{SCHNEEr1cond}\frac
{n^2b^2}{\ell^2}\ll
1,\quad \frac {na'R^3 |\ln(\rho h^2)|} {\ell^2 h^2}\ll 1,\quad
\frac{n}{|\ln(\rho h^2)|}\gg 1, \ee which can also be written
\be\label{SCHNEEr1cond'}{\rho^2\ell^2 b^2}\ll 1,\quad \frac {\rho a'R^3
|\ln(\rho h^2)|} {h^2}\ll 1,\quad \frac{|\ln(\rho
h^2)|}{\rho\ell^2}\ll 1.  \ee These conditions are fulfilled if we
choose \be \label{SCHNEEb} R=h, \quad b=\rho^{-1/2}|\ln(\rho h^2)|^{-\alpha}
\ee with $\alpha>1/2$ and \be\label{SCHNEE419}\rho^{-1/2} |\ln(\rho
h^2)|^{1/2}\ll \ell \ll \rho^{-1/2} |\ln(\rho h^2)|^{\alpha}.\ee Note
also that $n=\rho\ell^2\gg 1$.

It remains to compare
\be\label{SCHNEEphien}
\langle\Psi|\Psi\rangle^{-1}\left(\sum_{j}\int_{\R^{2n}}
|\nabla_{j}^\parallel \Phi|^2+
\sum_{i<j}\int_{\R^{2n}}  W(x_{i}-x_{j})\Phi^2\right) \ee
with the expected leading term of the energy, i.e.,
$4\pi (n^2/\ell^2) |\ln(n\atd^2/\ell^2)|^{-1}$.

We consider first the simplest case,
i.e., $n=2$. We have
\be \int_{\R^2}|\nabla^\parallel \varphi|^2=
(\ln(b/\atd))^{-2}2\pi\int_{R}^b\frac
{dr}r=(\ln(b/\atd))^{-2}2\pi\ln(b/R),\ee
\be\half\int_{\R^2}W\varphi^2=\frac{4\pi a\Vert s \Vert_4^4}
h\left(\frac{\ln(R/\atd)}
{\ln(b/\atd)}\right)^2.\ee
Inserting the formula \eqref{SCHNEEatd} for $\atd$ and using
$R=h$, $b=\rho^{-1/2}|\ln(\rho h^2)|^{-\alpha}$ and
$a'=a(1+o(1))$ we have
\begin{multline}\label{SCHNEE423}
\int_{\R^2}(|\nabla^\parallel \varphi|^2+\half
W\varphi^2)=\\
2\pi(\ln(b/\atd))^{-2}\left[\ln(b/h)+(h/2a'\|s\|_{4}^4)
\right]=
4\pi |\ln(\rho/\atd^2)|^{-1}(1+o(1)).
\end{multline}
For $n>2$ we can use the symmetry of $\Phi$ to write, using
\eqref{SCHNEE423}
as well as $0\leq\varphi(r)\leq 1$,
\bea\label{SCHNEE426}&& \sum_{j}\int_{\Lambda_{2}^n}
|\nabla_{j}^\parallel \Phi|^2+
\sum_{i<j}\int_{\Lambda_{2}^n}  W(x_{i}-x_{j})\Phi^2\nonumber\\&&=n
\left(\int_{\Lambda_{2}^{n}}
|\nabla_{1}^\parallel \Phi|^2+
\half\sum_{i=2}^n\int_{\Lambda_{2}^n}
W(x_{i}-x_{1})\Phi^2\right)\nonumber\\
&&\leq 4\pi n^2 \ell^{2(n-1)}
|\ln(n\atd^2/\ell^2)|^{-1}(1+o(1))+\mathcal R_{3}\eea
with
\be \mathcal R_{3}=n^3\ell^{2(n-3)}\int_{\Lambda_{2}^3}
\varphi'(|x_{2}-x_{1}|)
\varphi'(|x_{3}-x_{1}|).\ee
We estimate  $\mathcal R_{3}$
in the same way as
\eqref{SCHNEE415}, obtaining
\be
\mathcal R_{3}\leq \hbox{\rm (const.)}\ell^{2(n-2)}
n^3 b^2 (\ln(b/\atd))^{-2}2\pi\ln(b/R).
\ee
The condition that $\mathcal R_{3}$ has to be
much smaller than the leading term, given by $4\pi n^2 \ell^{2(n-1)}
|\ln(n\atd^2/\ell^2)|^{-1}$, is equivalent to
\be
\frac {n b^2}{\ell^2}\ln(b/R)\ll 1.
\ee
With the choice \eqref{SCHNEEb} this holds if $\alpha>1/2$.
\bigskip

In Region I the ansatz \eqref{SCHNEEFG} can still be
used, but this time we take $b=R$, i.e., $\varphi\equiv 1$.
In this region  $(a/h)|\ln(\rho h^2)|=o(1)$ and the leading term
in the energy is $\sim n^2\ell^{-2}a/h$.
Conditions
\eqref{SCHNEEr1cond} are now replaced by
\be\label{SCHNEEr2cond}
\frac  {n^2R^2}{\ell^2}\ll 1,\quad \frac {nR^3}{\ell^2 h}\ll
1,\quad  \frac{na}{h}\gg 1
\ee
where have here used that $a'=a(1+o(1))$, provided $R\gg a$. Note that
the last condition in \eqref{SCHNEEr1cond} means in particular that $n\gg
1$.
Putting again $\rho=n/\ell^2$, \eqref{SCHNEEr2cond} can  be written as
\be\label{SCHNEEr3cond}
 \rho^2\ell^2R^2\ll 1,\quad \frac {\rho R^3}{h}\ll
1,\quad  \frac{h}{\rho \ell^2 a}\ll 1.
\ee
By assumption, $a/h\ll 1$, but also $\rho a h\ll 1$ by the condition
of strong confinement, c.f. \eqref{SCHNEEahll}. We take
\be\label{SCHNEER1}
R=a(\rho a h)^{-\beta}
\ee
with $0<\beta$, so $R\gg a$. Further restrictions come from the conditions
\eqref{SCHNEEr3cond}: The first and the last of these conditions imply
together that
\be\label{SCHNEEell0}
\frac h{a}\ll \rho\ell^2\ll \frac 1{\rho R^2}
\ee
which can be fulfilled if
\be\label{SCHNEEell1}
\rho a h\ll (\rho a h)^{2\beta}.
\ee
i.e., if $\beta<\half$. Note that this implies in particular $R\ll
\rho^{-1/2}$. We can then take
\be\label{ellupper}
\ell=\rho^{-1/2}(h/a)^{1/2}(\rho a h)^{-\gamma}
\ee
with
\be
0<\gamma<\frac{1-2\beta}2.
\ee

The second of the
conditions \eqref{SCHNEEr3cond} requires that
\be
\frac{\rho R^3}h=(\rho ah)(a/h)^2(\rho a h)^{-3\beta}\ll 1,
\ee
which holds in any case if $\beta\leq 1/3$. A possible choice satisfying all
conditions is
\be
\beta=\frac 13,\quad \gamma=\frac 1{12}.
\ee
The error terms \eqref{SCHNEEr3cond} are then bounded by $(\rho a
h)^{1/6}$ (first and third term) and $(a/h)^2$ (second term).

Finally, with $\Phi\equiv 1$, Eqs. \eqref{SCHNEE412}, \eqref{SCHNEEnormbd}
and
\eqref{SCHNEEintw} give
\be \langle\Psi|\Psi\rangle^{-1}
\langle \Psi|H|\Psi\rangle-(ne^\perp/h^2)\leq \frac{4\pi
n^2}{\ell^2}\frac {a\|s\|_{4}^4}h(1+o(1)).
\ee
This completes the proof of the upper bound in boxes with finite $n$.

The upper bound for the energy per particle in the 2D
thermodynamic limit  is
now obtained by dividing $\R^2$ into Dirichlet boxes with side
length $\ell$ satisfying (\ref{SCHNEE419}) in Region II, or
(\ref{SCHNEEell0}) in Region I, and distributing the particles
evenly among the boxes.  In other words, the trial wave function
in a large box of side length $L$ is
\beq\Psi=\sum_{\alpha}\Psi_{\alpha}\eeq where $\alpha$ labels the
boxes of side length $\ell$ contained in the large box, and
$\Psi_{\alpha}$ is the Dirichlet ground state wave function for
$n=\rho\ell^2$ particles in the box $\alpha$, with $\rho=N/L^2$.
The
choice of $\ell$ guarantees in particular that the error
associated with the Dirichlet localization is negligible.  In
order to avoid contributions from the interaction between
particles in different boxes the boxes should be separated by the
range, $R_{0}$ of the interaction potential and in the
``corridors'' between the boxes the wave function is put equal to
zero.  The number of particles in each box is then not
exactly $\rho\ell^2$, but rather the smallest integer larger than
$\rho \ell^2(\ell/(\ell+R_{0}))^2$.  In order that the errors are negligible one needs $R_{0}/\ell=o(1)$, as well as $\rho\ell^2\gg 1$, and both are guaranteed by the choice (\ref{ellupper}) of $\ell$.

\section{Scattering Length}\label{pertscatt}

As a preparation for the lower bound we consider in this section the
perturbative calculation of the 2D scattering length of an integrable
potential.

Consider a 2D, rotationally symmetric potential $W\geq 0$ of finite
range $R_{0}$.  As discussed in Appendix \ref{appscatt} the
scattering length is determined by minimizing, for
$R\geq R_{0}$, the functional \be\label{SCHNEEscattfunct}{\mathcal
E}_{R}[\psi]=\int_{|x|\leq
R}\left\{|\nabla \psi|^2+\half W|\psi|^2\right\}\ee with boundary
condition $\psi=1$ for $|x|=R$.  The Euler equation (zero energy
scattering equation) is \be\label{SCHNEEscatt}-\Delta \psi+\half W \psi=0\ee
and for
$r=|x|\geq R$ the minimizer, $\psi_{0}$, is \be\psi_{0}(r)=\ln(r/a_{\rm
scatt})/\ln(R/a_{\rm scatt})\ee with a constant $a_{\rm scatt}$.  This
is, by definition, the 2D scattering length for the potential $W$.  An
equivalent definition follows by computing the energy, \be
E_{R}={\mathcal E}_{R}[\psi_{0}]=2\pi/\ln(R/a_{\rm scatt})\ee which
means that \be\label{SCHNEE3.5} a_{\rm scatt}=R\exp(-2\pi/E_{R}).\ee
\begin{lem}[Scattering length for soft
potentials]\label{SCHNEEscattlemm}
    Assume $W(x)=\lambda w(x)$ with $\lambda \geq 0$, $w\geq 0$, and
    $w\in L^1(\R^2)$, with $\int w(x)d^2x=1$.  Then, for $R\geq R_{0}$,
    \be \label{SCHNEEscattcorreq1}a_{\rm
    scatt}=R\exp\left(-\frac{4\pi+\eta(\lambda,R)}{\lambda}\right)\ee
    with $\eta(\lambda,R)\to 0$ for $\lambda\to 0$.
    \end{lem}
    \begin{proof}
The statement is, by \eqref{SCHNEE3.5},  equivalent to \be E_{R}=\half
\lambda(1 +o(1))  \ee
where the error term may depend on $R$. The
upper bound is clear by the variational principle, taking $\psi=1$ as
a test function.  For the lower bound note first that $\psi_{0}\leq
1$.  This follows from the variational principle: Since $W\geq 0$ the
function $\tilde\psi_{0}(x)=\min\{1,\psi_{0}\}$ satisfies ${\mathcal
E}_{R}[\tilde\psi_{0}]\leq {\mathcal E}_{R}[\psi_{0}]$.  Hence the
function $\varphi_{0}=1-\psi_{0}$ is nonnegative.  It satisfies
\be\label{SCHNEEphieq}-\Delta \varphi_{0}+\half W \varphi_{0}=\half W\ee and
the Dirichlet boundary condition $\varphi_{0}=0$ for $|x|=R$.

Integration of
 (\ref{SCHNEEscatt}), using that $\psi_{0}(r)=1$ for $r=R$, gives
\be E_{R}=\half\int W\psi_{0}=\half\int
W(1-\varphi_{0}).\ee
Since $\varphi_{0}\geq 0$ we thus need to show
that $\|\varphi_{0}\|_{\infty}=o(1)$.

By \eqref{SCHNEEphieq}, and since $\varphi_{0}$ and $W$ are both
nonnegative, we have $-\Delta\varphi_{0}\leq\half W$ and hence
\be \varphi_{0}(x)\leq
\int K_{0}(x,x')W(x')d^2x'\ee where $K_{0}(x,x')$ is the (nonnegative)
integral kernel
of
$(-\Delta)^{-1}$ with Dirichlet boundary conditions at $|x|=R$.  The
kernel $K_{0}(x,x')$ is integrable (the singularity is
$\sim \ln |x-x'|$) and hence, if $W$ is bounded, we have
$\|\varphi_{0}\|_{\infty}\leq {\rm(const.)}\lambda
\|w\|_{\infty}=O(\lambda)$.

If $w$ is not bounded we can, for every $\varepsilon>0$, find a bounded
$w^\varepsilon\leq w$ with $\int (w-w^\varepsilon)\leq \varepsilon$.
Define $C_{\varepsilon}=\|w^\varepsilon\|_{\infty}$.  Without
restriction we can assume that $C_{\varepsilon}$ is a monotonously
decreasing function of  $\varepsilon$ and continuous.  The function
$g(\varepsilon)=\varepsilon/C_{\varepsilon}$ is then monotonously
increasing and continuous with $g(\varepsilon)\to 0$ if
$\varepsilon\to 0$.  For every (sufficiently small) $\lambda$ there is
an $\varepsilon(\lambda)=o(1)$ such that $g(\varepsilon(\lambda))=\lambda$.
Then \be\|\varphi_{0}\|_{\infty}\leq
{\rm(const.)}(\varepsilon(\lambda) +\lambda
C_{\varepsilon(\lambda)})={\rm(const.)}\varepsilon(\lambda)=o(1).\ee
\vskip-3mm
\end{proof}

\begin{corollary}[Scattering length for
scaled, soft
potentials]\label{SCHNEEscattcorr}
    Assume $W_{R,\lambda}(x)=\lambda R^{-2} w_{1}(x/R)$ with
    $w_{1}\geq 0$ fixed and $\int w_{1}=1$. Then the scattering length
    of $W_{R,\lambda}$ is
    \be \label{scattcorreq} a_{\rm
    scatt}=R\exp\left(-\frac{4\pi+\eta (\lambda)}{\lambda}\right)\ee
    with $\eta (\lambda)\to 0$ for $\lambda\to 0$, uniformly in $R$.
    \end{corollary}
    \begin{proof} This follows from Lemma \ref{SCHNEEscattlemm}
    noting that, by scaling, the scattering length of $W_{R,\lambda}$ is $R$
times
    the scattering length of $\lambda w_{1}$. The latter is
    independent of~$R$.
\end{proof}

\noindent{\it Remark.} If $W$ is obtained by averaging a 3D
integrable potential $v$ over an interval of length $h$ in the $z$
variable, the formula (\ref{scattcorreq}), together with
Eq.\ (\ref{emin}), motivates the exponential dependence of
the effective 2D scattering length
(\ref{SCHNEEatd}) of $v$ on $h/a$: The integral $\lambda=\int W(x)d^2x$ is
$h^{-1}\int
v(\x)d^3\x$, which for weak potentials is $h^{-1}8\pi a$ to lowest
order, by Eq.\ (\ref{emin}).
Inserting this into (\ref{scattcorreq}) gives (\ref{SCHNEEatd})
(apart from the dependence on the shape function $s$). This heuristics is,
of course,
only valid for soft potentials $v$. An essential step in the
lower bound in the next section is the replacement of $v$ by a soft
potential to which this reasoning can indeed be applied.

\section{Lower Bound} In the same way as for the upper bound we
restrict the attention to the homogeneous case and finite boxes in the
2D variables, this time with Neumann boundary conditions.  The optimal
distribution of particles among the boxes is determined by using
sub\-additivity and convexity arguments as in Chapters~\ref{sect3d}
and \ref{sectgp}.  Inhomogeneity in the 2D $x$ variables can be
treated by factorizing out a product of GP minimizers or, alternatively, by the method of \cite{SeirNorway} as mentioned at
the beginning of Section \ref{sect9.3}.

In the treatment of the lower bound there is a natural division line
between the case where the mean particle distance $\rho^{-1/2}$ is
comparable to or larger than $h$ and the case that it is much
smaller than $h$.  The first case includes Region II and a part
(but not all) of Region I. When $\rho^{-1/2}$ is much smaller than $h$
the boxes have finite extension in the $z$ direction as well. The
method is  then a fairly simple modification of the 3D estimates in Chapter
\ref{sectgp} (see also Section 4.4 in \cite{LSY}) and will not be discussed
further here.

The
derivation of a lower bound for the case that $\rho h^2\leq
C<\infty$ proceeds by the following steps:
\begin{itemize}
    \item
     Use Dyson's Lemma \ref{dysonl} to replace $v_a$ by an integrable 3D
potential
$U$, retaining part of the kinetic energy.

\item Average the potential $U$ at fixed $x\in\R^2$ over the
$z$-variable to obtain a 2D potential $W$.  Estimate the error
in this averaging procedure by
using Temple's inequality (\ref{temple}) at each fixed $x$.

\item The result is a 2D many body problem with an integrable
interaction potential $W$
which, by Corollary \ref{SCHNEEscattcorr},
has the right 2D scattering length to lowest order in $a/h$,
but reduced kinetic energy inside the range of the potential.
This problem is treated in the same way as in
Chapter~\ref{sect2d}, introducing a 2D Dyson potential and using
perturbation
theory, again estimating the errors by
Temple's inequality.

\item Choose the parameters (size $\ell$ of box, fraction
$\varepsilon$ of the kinetic energy, range $R$ of potential $U$, as
well as the corresponding parameters for the 2D Dyson potential)
optimally to minimize the errors.
\end{itemize}

The first two steps are analogous to the corresponding
steps in the proof of the lower bound in  Lemma \ref{finthm}, cf.\
Eqs.\ (\ref{329})--(\ref{57}). It is, however, convenient to define
the Dyson potential $U$ in a slightly different manner than in Eq.\
(\ref{softened}). Namely, for $R\geq 2R_{0}$ with $R_{0}$ the range
of $v$ we define
\begin{equation}\label{softened2}
U_{R}(r)=\begin{cases}\frac {24}7 R^{-3}&\text{for
$\half R<r<R$ }\\
0&\text{otherwise.}
\end{cases}
\end{equation}
The reason is that this potential has a simple scaling with $R$ which
is convenient when applying Corollary \ref{SCHNEEscattcorr}.
Proceeding  as in Eqs.\ (\ref{329})--(\ref{57})  we write a
general
wave function as
\beq
\Psi(\x_1,\dots,\x_n)=f(\x_1,\dots,\x_n)\prod_{k=1}^n
s_h(z_k) \ ,
\eeq
and define $F(x_1,\dots,x_n)\geq 0$ by
\beq\label{defF2}
|F(x_1,\dots,x_n)|^2=\int |f(\x_1,\dots,\x_n)|^2\prod_{k=1}^n
s_h(z_{k})^2 dz_k\ .
\eeq
Note that $F$ is normalized if $\Psi$ is normalized.
The analogue of Eq.\ (\ref{57}) is
\begin{eqnarray}\label{xbound}\nonumber
&&\langle\Psi|H|\Psi\rangle- \frac{n e^\perp}{h^2}\geq \\ \nonumber
&& \sum_{i=1}^n \int\Big[\eps |\nabla^\parallel_i F|^2 +
(1-\eps)|\nabla^\parallel_i F|^2 \chi_{\min_k |x_i-x_k|\geq R}(x_i)
\Big]\prod_{k=1}^n dx_k \\ \nonumber &&+\sum_{i=1}^n \int \left[
\eps|\partial_i f|^2 + a'
U_{R}(|\x_i-\x_{k(i)}|)\chi_{{\mathcal B}_\delta}(z_{k(i)}/h)|f|^2
\right] \prod_{k=1}^n s_h(z_k)^2 d \x_k , \\ \label{57a}
\end{eqnarray}
where  $\nabla^\parallel_{i}$ denotes the gradient with respect to
$x_{i}$
and $\partial_j=d/dz_j$. Moreover,
$\chi_{{\mathcal B}_\delta}$ is the characteristic function of the subset
 ${\mathcal B}_\delta\subset \R$
where
$s(z)^2\geq \delta$ for $\delta>0$,
\beq \label{aprime}
a'=a(1-\eps)(1-2 R \|\partial s^2\|_\infty/
h \delta),\eeq and $k(i)$ denotes the index of the nearest neighbor to
$\x_{i}$. When deriving \eqref{57a} the Cauchy Schwarz inequality has been used to bound the longitudinal kinetic energy of $f$ in terms of that of $F$, i.e.,
\be\label{cauchy}
\sum_{i=1}^n \int|\nabla^\parallel_i f|^2\prod_{k=1}^n s_h(z_k)^2 d \x_k \geq
\sum_{i=1}^n \int|\nabla^\parallel_i F|^2\prod_{k=1}^n  d x_k.\ee

We now consider, for fixed $x_1,\dots,x_n$, the term
\beq\label{44a}
\sum_{i=1}^n \int \left[
\eps|\partial_i f|^2 + a'
U_{R}(|\x_i-\x_{k(i)}|)\chi_{{\mathcal B}_\delta}(z_{k(i)}/h)|f|^2
\right] \prod_{k=1}^n s_h(z_k)^2 d z_k \ .
\eeq
This is estimated from below in the same way as in Eqs.\
(\ref{60}), using Temple's inequality. The result is, by a
calculation analogous to Eqs.\ (\ref{60})--(\ref{860}),
\begin{eqnarray}\nonumber
\langle\Psi|H|\Psi\rangle- \frac{ne^\perp}{h^2} &\geq& \int\sum_{i=1}^n
\Big[\eps |\nabla^\parallel_i F|^2 +(1-\eps)|\nabla^\parallel_i F|^2 \chi_{
|x_i-x_k(i)|\geq R}(x_i)\nonumber
 \\
\label{SCHNEE:putt} & & +\half W(x_{i}-x_{k(i)})|F|^2 \Big]\prod_{k=1}^n
dx_k\,,
\end{eqnarray}
where $x_{k(i)}$ denotes here the nearest neighbor to $x_{i}$ among
the points $x_{k}\in\R^2$, $k\neq i$ and $W$ is obtained by
averaging $a'U_{R}$ over $z$:
\beq W(x-x')=2a^{\prime\prime}\int_{\R\times\R} s_h(z)^2 s_h(z')^2
U_{R}(|\x-\x'|)
\chi_{{\mathcal B}_\delta}(z'/h) dz dz' \ .
\eeq
Here, $a^{\prime\prime}=a'(1-\eta)$ with an error term $\eta$
containing the error estimates form the Temple inequality and 
from ignoring other points than the nearest neighbor to $\x_{i}$. Moreover,
since $\int U_{R}(\x)d\x=4\pi$, $U_{R}(|\x-\x'|)=0$ for $|\x-\x'|>R$, and
$|s(z)^2-s(z')^2|\leq R \Vert \partial_{z}s^2\Vert_{\infty}$ for
$|z-z'|\leq R$ we have the
estimate
\beqa \int_{\R^2} W(x) dx &\geq &
\frac{8\pi a^{\prime\prime}}
h \left(\int_{\mathcal B_{\delta}}s(z)^4dz-\frac Rh
\Vert \partial_{z}s^2\Vert_{\infty}\right)\nonumber\\
&\geq& \frac{8\pi a^{\prime\prime}}
h\left(\Vert s\Vert^4_{4}-\delta-\frac Rh
\Vert \partial_{z}s^2\Vert_{\infty}\right).\label{Wlower}
\eeqa

As
we will explain in a moment, the
errors, and the replacement of $n-1$ by $n$, require the following
terms to be small:
\begin{equation}\label{SCHNEEerr:3d}\frac{nh^2a}{\varepsilon R^3},\quad
    \frac{nR}{ h}, \quad
\varepsilon, \quad
 \frac1n,  \quad \delta, \quad \frac R{h\delta}.
\end{equation}
The rationale behind the first term is as follows. The Temple errors in
the averaging procedure at fixed $\x_{1},\ldots,x_{n}$ produces a
factor similar to \eqref{SCHNEEtemplegp}, namely
\beq\label{60a}
\left(1- a'\frac{\langle
U^2\rangle}{\langle U\rangle}\frac{1}{({\rm const.})\varepsilon/h^2-({\rm const.})
a'\langle U\rangle}\right)
\eeq
with
\beq
\langle U^m\rangle=  \int \left(\sum_{i=1}^n
U(|\x_i-\x_{k(i)}|)\chi_{{\mathcal B}_\delta}(\x^\perp_{k(i)}/r)\right)^m
\prod_{j=1}^n s_h(z_{j})^2 dz_{j},
\eeq
cf.\ Eqs. (\ref{8.55})--(\ref{60}). The analogue of Eq. (\ref{8.57})
is
\be
\langle U\rangle \leq ({\rm const.})
n(n-1)\frac{\|s\|_4^4}{h R^2},
\ee
and $\langle
U^2\rangle\leq ({\rm const.}) nR^{-3}\langle U\rangle$ by Schwarz's
inequality. Since the denominator in \eqref{60a} must be positive, we
see in particular that the particle number must satisfy
\be \label{nbound1} n(n-1)<({\rm const.})\frac{\varepsilon R^2}{ah}\ee
and the error is of the order ${nh^2a}/{\varepsilon R^3}$ as claimed
in \eqref{SCHNEEerr:3d}.

The estimate from below on $\langle U\rangle$, obtained in the same
way as Eq.\ (\ref{860}), is
\beqa\nonumber
\langle U\rangle\!\!
&\geq& \!\! \sum_{i\neq j} \int
U(|\x_i-\x_j|)\chi_{\mathcal B_\delta}(z_j/h)\left(1-\!\!\! \sum_{k,
\, k\neq i,j} \theta(R-|\x_k-\x_i|)\right)\prod_{l=1}^n
s_{h}(z_l)^2 dz_{l} \\ &\geq& \!\!
\frac1{2a^{\prime\prime}}\sum_{i\neq j}W(x_i-x_j) \left(1-(n-2)
\frac{R}{h} \|s\|_\infty^2\right) \ .
\eeqa
In particular, the second term in \eqref{SCHNEEerr:3d}, $nR/h$, should be small.
The requirement that $\varepsilon$ and $n^{-1}$ are small
needs no further comments.

The potential $W$ can be written as
\beq\label{Wpot}
W(x)=\lambda R^{-2}w_{1}(x/R)\,,
\eeq
where $w_{1}$ is independent of $R$, with \beq \int w_{1}(x)dx=1\eeq
and \beq\label{lambda}\lambda=\frac{8\pi\hbox{ $a\int
s^4$}}h(1-\eta')\,.\eeq
Here, $\eta'$ is an error term involving $\delta$ and
$R/(h\delta)$ (cf.\ (\ref{aprime}) and \eqref{Wlower}) besides the
other terms in
(\ref{SCHNEEerr:3d}). In particular, $\delta$ and $R/(h\delta)$
should be small.  The 2D
scattering length of (\ref{Wpot}) can be computed by
Corollary \ref{SCHNEEscattcorr} and has the right form (\ref{SCHNEEatd})
to leading order in $\lambda$. (Recall from the remark preceding Eq.\
(\ref{ggprime}) that $R$ in (\ref{scattcorreq}) can be replaced by $h$
as long as $ca<R<Ch$.)

The Hamiltonian on the right side of Eq.\ (\ref{SCHNEE:putt}) can
now be treated with the methods of Chapter \ref{sect2d}.  The only
difference from the Hamiltonian discussed in that paper is the
reduced kinetic energy inside the range of the potential $W$. This
implies that $\lambda$ in the error term $\eta(\lambda)$ in Corr.\
\ref{SCHNEEscattcorr} should be replaced by $\lambda/\varepsilon$,
which requires \be
\frac a{\varepsilon h}\ll 1.\label{ahe}\ee
Otherwise the method is the same as in Chapter \ref{sect2d}: a slight
modification of the 2D Dyson Lemma \ref{dyson2d}  (see Appendix A in \cite{SY}) allows to
substitute for $W$ a potential $\tilde U$ of larger range, $\tilde
R$, to which perturbation theory and Temple's inequality can be
applied.  
The fraction of the kinetic energy borrowed for the application of
Temple's  will be
denoted by $\tilde \varepsilon$. The errors that have now to be
controlled are
\begin{equation}\label{err:2d}
\tilde\varepsilon,   \quad n\tilde R^2/\ell^2, \quad \frac{R}{\tilde R},
\quad \frac{n\ell^2}{\tilde\varepsilon
\tilde R^2\ln(\tilde R^2/\atd^2)} .
\end{equation}

To explain these terms we refer to  the estimates in Chapter \ref{sect2d}.  Substituting $\tilde
\varepsilon$ for $\varepsilon$ and $\tilde R$ for $R$ in these
estimates we see first that $\tilde \varepsilon$ and $n\tilde
R^2/\ell^2$ should be small.  
The smallness of $R/\tilde R$ guarantees that the \lq\lq hole" of radius $R$ in the 2D Dyson potential has negligible effect.
Since the denominator in the Temple error must be positive we see also that the particle
number $n$ in the box should obey the bound \be n(n-1)<({\rm
const.})\,\tilde\varepsilon\ln(\tilde R^2/\atd^2)\label{nbound2},\ee
and the Temple error is bounded by $({\rm const.})
{n\ell^2}/({\tilde\varepsilon \tilde R^2\ln(\tilde R^2/\atd^2)})$.

Using superadditivity of the energy (which follows from $W\geq 0$)
and convexity in the same way as in Eqs.\
(\ref{sum})--(\ref{estimate3}) one sees that if $\rho=N/L^2$ is the
density in the thermodynamically
large box of side length $L$ the optimal
choice of $n$ in the box of fixed side length $\ell$ is $n\sim \rho
\ell^2$. We thus have to show that it is possible to choose the
parameters $\varepsilon$, $R$, $\delta$, $\ell$, $\tilde\varepsilon$ and
$\tilde
R$ in such a way that all the error terms (\ref{SCHNEEerr:3d}) and
(\ref{err:2d}), as
well as $\delta$ and $R/h\delta$  are small. We note that the conditions
$a/h\ll 1$ and
$\rho|{\ln(\rho\atd^2)}|^{-1}\ll 1/h^2$ imply $\rho\atd^2\to 0$ and hence
 $|\ln(\rho\atd^2)|^{-1}\to 0$.

We make the ansatz
\be \label{2dparam}\varepsilon=\left(\frac ah\right)^\alpha, \quad \delta=
\left(\frac ah\right)^{\alpha'}, \quad
R=h\left(\frac
ah\right)^\beta,
\ee
and choose $\ell$ such that $L$ is a multiple of $\ell$ with
\be \rho^{-1/2}\ll \ell\lesssim\rho^{-1/2}\left(\frac
ah\right)^{-\gamma}.\label{ell}\ee
Then $n=\rho\ell^2\gg 1$ and
$R/(h\delta)=(a/h)^{\beta-\alpha'}$. The error terms (\ref{SCHNEEerr:3d}) are also
powers of $a/h$ and we have to ensure that all
exponents are positive, in particular
\be\label{errexponents} \beta-\alpha'>0,
\quad\beta-2\gamma>0,\quad{1-\alpha-3\beta-2\gamma}>0.
\ee
This is fulfilled, e.g., for
\be \label{2dexpo}\alpha=\alpha'=\frac19,\quad \beta=\frac29,\quad\gamma=\frac1{18}\,,\ee
with all the exponents (\ref{errexponents}) equal to 1/9. Note also
that with this choice Eq.\ \eqref{ahe} is fulfilled.

Next we write \be \tilde\varepsilon=|\ln(\rho\atd^2)|^{-\kappa},
\quad \tilde R=\rho^{-1/2}|\ln(\rho\atd^2)|^{-\zeta}. \ee Then
$|\ln (a_{{\rm
    2D}}^2/\tilde R^2)|=|\ln (a_{{\rm
    2D}}^2\rho)| (1+o(1))$.  The error terms \eqref{err:2d} are
\be\label{{errest:2d}}
\tilde\varepsilon=|\ln(\rho\atd^2)|^{-\kappa}, \quad \frac{R}{\tilde R}
=\left(\frac ah\right)^\beta (\rho h^2)^{1/2}|\ln(\rho\atd^2)|^{\zeta},
\quad
\frac{n\tilde R^2}{\ell^2}=
|\ln(\rho\atd^2)|^{-2\zeta},
\ee and
\be\label{2dtemperr} \frac{n\ell^2}{\tilde\varepsilon
\tilde R^2\ln(\tilde R^2/\atd^2)}=
\left(\frac ah\right)^{-4\gamma}|
\ln(\rho\atd^2)|^{-(1-\kappa-2\zeta)}(1+O(\ln|\ln(\rho\atd^2)|)).
\ee
Since $\left(a/h\right)^{-4\gamma}|
\ln(\rho\atd^2)|^{-4\gamma}=O(1)$, the error term \eqref{2dtemperr} can also
be written
as
\be\frac{n\ell^2}{\tilde\varepsilon
\tilde R^2\ln(\tilde R^2/\atd^2)}=
O(1)|\ln(\rho\atd^2)|^{-(1-\kappa-2\zeta-4\gamma)}(1+O(\ln|\ln(\rho\atd^2)|)).
\ee
The condition $\rho h^2<C$ is used to bound $R/\tilde R$ in \eqref{{errest:2d}}.
Namely,  \be(\rho h^2)^{1/2}|\ln(\rho\atd^2)|^{\zeta}\leq {\rm
(const.)}(h/a)^{\zeta},\ee
so
\be
\frac{R}{\tilde R}
=O(1)\left(\frac ah\right)^{\beta-\zeta}.
\ee
We choose now
\be  \zeta=\frac 19,\quad\kappa=\frac 29.\ee
Then
\be
\beta-\zeta=\frac 19\quad\text{and}\quad 1-\kappa-2\zeta-4\gamma=\frac 13.
\ee

This completes our discussion of the lower bound for the case $\rho
h^2\leq C<\infty$ and a homogeneous system in the 2D $x$ variables. As
already mentioned, the case $\rho
h^2\gg 1$ can be treated with the 3D methods of Chapters~\ref{sect3d} and
\ref{sectgp} and inhomogeneity in the $x$ variables by
separating out a product of GP minimizers, or by the method of \cite{SeirNorway} .

\newpage
\
\thispagestyle{empty}

\chapter[The Charged Bose Gas, the One- and Two-Component Cases]{The Charged Bose Gas,\vspace*{-2mm}\newline the One- and Two-Component\vspace*{-2mm}\newline Cases}

The setting now changes abruptly. Instead of particles interacting
with a short-range potential $v(|\x_i-\x_j|)$ they interact via the
Coulomb potential\index{Coulomb potential} $$v(|\x_i-\x_j|) =
|\x_i-\x_j|^{-1} $$
(in 3 dimensions). The unit of electric charge is
1 in our units.

We will here consider both the one- and two-component gases.  In the
one-component gas (also referred to as the one-component plasma or
bosonic jellium\index{jellium}) we consider positively charged particles confined to
a box with a uniformly charged background\index{background charge}. In the two-component gas we
have particles of both positive and negative charges moving in all of
space.

\section{The One-Component Gas}
In the one-component gas there are $N$ positively charged particles in
a large box $\Lambda$ of volume $L^3$ as before, with $\rho =N/L^{3}$.

To offset the huge Coulomb repulsion (which would drive the particles
to the walls of the box) we add a uniform negative background of
precisely the same charge, namely density $\rho$. Our
Hamiltonian\index{Hamiltonian} is thus
\begin{equation}\label{foldyham}
H_N^{(1)}=  \sum_{i=1}^{N} - \mu \Delta_i -V(\x_i) +
 \sum_{1 \leq i < j \leq N} v(|\x_i - \x_j|)  +C
\end{equation}
with $$
V(\x)=\rho \int_{\Lambda} |\x-\y |^{-1}d\y \qquad \qquad {\rm
and}\qquad \qquad C= \frac{1}{2} \rho \int_{\Lambda} V(\x)d\x\ .
$$
We shall use Dirichlet boundary conditions\index{boundary condition}.
As before the
Hamiltonian acts on  symmetric wave functions in $L^2(\Lambda^{N},
d\x_1\cdots d\x_N)$.

Each
particle interacts only with others and not with itself. Thus,
despite
the fact that the Coulomb potential is positive definite, the
ground state energy $E_0$ can
be (and is) negative (just take $\Psi=$const.). This time, {\it
large} $\rho$ is the `weakly interacting' regime.

We know from the work in \cite{LN} that the thermodynamic limit\index{thermodynamic!limit}
$e_0(\rho)$ defined as in (\ref{eq:thmlimit}) exists. It also follows
from this work that we would, in fact, get the same thermodynamic
energy if we did not restrict the number of particles $N$, but
considered the grand-canonical case where we minimize the energy over
all possible particle numbers, but keeping the background charge
$\rho$ fixed.

Another way in which this problem is different from the previous one
is that {\it perturbation theory is correct to leading order}. If
one computes $(\Psi, H \Psi)$ with  $\Psi=$const, one gets the right
first order answer, namely $0$. It is the next order in $1/\rho$ that
is interesting, and this is {\it entirely} due to correlations.
In 1961 Foldy\index{Foldy} \cite{FO} calculated this correlation energy according
to the prescription of Bogoliubov's\index{Bogoliubov} 1947 theory. That theory was not
exact for the dilute Bose gas, as we have seen, even to first order.
We are now looking at {\it second} order, which should be even
worse. Nevertheless, there was good physical intuition that this
calculation should be asymptotically {\it exact}. Indeed it is, as
proved in \cite{LS} and \cite{So}.

The Bogoliubov theory states that the main contribution to the
energy comes from pairing of particles into momenta $\k, -\k$ and
is the bosonic analogue of the BCS theory of superconductivity
which came a decade later. I.e., $\Psi_0$ is a sum of products of
terms of the form $\exp\{i\k \cdot (\x_i-\x_j)\}$.

The following theorem is the main result for the one-component gas.

\begin{thm}[\textbf{Foldy's law for the one-component gas}]\index{Foldy!law}
\label{thm:Foldy}
\begin{equation}\label{foldyen}
\lim_{\rho\to\infty}\rho^{-1/4}e_0(\rho)=
-\frac{2}{5}\frac{\Gamma(3/4)}{\Gamma(5/4)}\left(\frac{2}{\mu\pi}\right)^{1/4}.
\end{equation}
\end{thm}

This is the {\it first example} (in more than 1 dimension) in which
Bogoliubov's pairing theory has been rigorously validated. It has to
be emphasized, however, that Foldy and Bogoliubov rely on the
existence of Bose-Einstein condensation.
\index{Bose-Einstein condensation}
We neither make such a hypothesis nor does our result
for the energy imply the existence of such condensation. As we said
earlier, it is sufficient to prove condensation in small boxes of
fixed size.

Incidentally, the one-dimensional example for which Bogoliubov's
theory is asymptotically exact to the first two orders (high density)
is the repulsive delta-function Bose gas
\index{delta-f@$\delta$-function Bose gas}
\cite{LL}, discussed in Appendix~\ref{chap3}, for which there is no Bose-Einstein condensation.

To appreciate the  $-\rho^{1/4}$ nature of (\ref{foldyen}), it is
useful to
compare it with what one would get if the bosons had infinite mass,
i.e., the first term in (\ref{foldyham}) is dropped. Then the energy
would
be proportional to $-\rho^{1/3}$ as shown in \cite{LN}. Thus, the
effect
of quantum mechanics is to lower $\frac{1}{3}$ to $\frac{1}{4}$.

A problem somewhat related to bosonic jellium is \textit{fermionic}
jellium.  Graf and Solovej \cite{GS} have proved that the first two
terms are what one would expect, namely
\begin{equation}
        e_{0}(\rho)=C_{\rm TF}\rho^{5/3}-C_{\rm
D}\rho^{4/3}+o(\rho^{4/3}),
\end{equation}
where $C_{\rm TF}$ is the usual Thomas-Fermi constant and $C_{\rm D}$
is the
usual Dirac exchange constant.

It is supposedly true, for both bosonic and fermionic particles, that
there is a critical mass above which the ground state should show
crystalline ordering (Wigner crystal)\index{Wigner crystal}, but this
has never been proved and it remains an intriguing open problem, even
for the infinite mass case. A simple scaling shows that large mass is
the same as small $\rho$, and is thus outside the region where a
Bogoliubov approximation\index{Bogoliubov!approximation} can be expected to hold.

As for the dilute Bose gas, there are several relevant length scales
in the problem of the charged Bose gas. For the dilute gas there were
three scales. This time there are just two. Because of the long range
nature of the Coulomb problem there is no scale corresponding to the
scattering length $a$. One relevant length scale is again the
interparticle distance $\rho^{-1/3}$. The other is the correlation
length\index{correlation length} scale $\ell_{\rm cor}\sim
\rho^{-1/4}$ (ignoring the dependence on $\mu$).  The order of the
correlation length scale can be understood heuristically as follows.
Localizing on a scale $\ell_{\rm cor}$ requires kinetic energy of the
order of $\ell_{\rm cor}^{-2}$.  The Coulomb potential from the
particles and background on the scale $\ell_{\rm cor}$ is
$(\rho\ell_{\rm cor}^3)/\ell_{\rm cor}$. Thus the kinetic energy and
the Coulomb energy balance when $\ell_{\rm cor}\sim\rho^{-1/4}$. This
heuristics is however much too simplified and hides the true
complexity of the situation.

Note that in the high density limit $\ell_{\rm cor}$ is long compared
to the interparticle distance.  This is analogous to the dilute gas
where the scale $\ell_c$ is also long compared to the interparticle
distance [see (\ref{scales})]. There is however no real analogy
between the scale $\ell_{\rm cor}$ for the charged gas and the scale
$\ell_c$ for the dilute gas. In particular, whereas $e_0(\rho)$ for
the dilute gas is, up to a constant, of the same order as the kinetic
energy $\sim\mu\ell_c^{-2}$ we have for the charged gas that
$e_0(\rho)\not\sim \ell_{\rm cor}^{-2}=\rho^{1/2}$. The reason for
this difference is that on average only a small fraction of the
particles in the charged gas actually correlate.

\section{The Two-Component Gas}
Now we consider $N$ particles with char\-ges $\pm1$. The
Hamiltonian\index{Hamiltonian} is thus
$$
H^{(2)}_N=\sum_{i=1}^N-\mu\Delta_i+\sum_{1\leq i<j\leq
  N}\frac{e_ie_j}{|\x_i-\x_j|}.
$$
This time we are interested in $E^{(2)}_0(N)$ the ground state energy
of
$H^{(2)}_N$ minimized over all possible combination of charges
$e_i=\pm1$, i.e., we do not necessarily assume that the minimum occurs
for the neutral case. Restricting to the neutral case would however
not change the result we give below.

An equivalent formulation is to say that $E^{(2)}_0(N)$ is the ground
state energy of the Hamiltonian acting on all wave functions of space
and charge, i.e., functions in
$L^2\left((\R^3\times\{-1,1\})^N\right)$.  As mentioned in the
introduction, and explained in the beginning of the proof of
Thm.~\ref{ub}, for the calculation of the ground state energy we may
as usual restrict to symmetric functions in this Hilbert
space\index{Hilbert space}.

For the two-component gas there is no thermodynamic limit.  In fact,
Dyson\index{Dyson} \cite{D2} proved that $E^{(2)}_0(N)$ was at least
as negative as $-(\mathrm{const})N^{7/5}$ as $N\to \infty$.  Thus,
thermodynamic stability\index{thermodynamic!stability} (i.e., a linear
lower bound) fails for this gas. Years later, a lower bound of this
$-N^{7/5}$ form was finally established in \cite{CLY}, thereby proving
that this law is correct.

The connection of this $-N^{7/5}$ law with the jellium $-\rho^{1/4}$
law (for which a corresponding lower bound was also given in
\cite{CLY}) was pointed out by Dyson \cite{D2} in the following way.
Assuming the correctness of the $-\rho^{1/4}$ law, one can treat the
2-component gas by treating each component as a background for the
other. What should the density be? If the gas has a radius $L$ and if
it has $N$ bosons then $\rho = N L^{-3}$. However, the extra kinetic
energy needed to compress the gas to this radius is $N L^{-2}$. The
total energy is then $N L^{-2} - N \rho^{1/4}$, and minimizing this
with respect to $L$ gives $L\sim N^{-1/5}$ and leads to the $-N^{7/5}$
law. The correlation length scale is now $\ell_{\rm
  cor}\sim\rho^{-1/4}\sim N^{-2/5}$.

In \cite{D2} Dyson conjectured an exact asymptotic expression
for $E^{(2)}_0(N)$ for large $N$. That this
asymptotics, as formulated in the next theorem, is indeed correct is
proved in \cite{{LSo02}} and \cite{So}.

\begin{thm}[\textbf{Dyson's law for the two-component gas}]\index{Dyson!law}
\label{thm:Dyson}
\begin{equation}\label{eq:Dyson}
  \lim_{N\to\infty}\frac{E_0^{(2)}(N)}{N^{7/5}}=\inf\biggl\{\mu\int_{\R^3}|\nabla\Phi|^2-
  I_0\int_{\R^3} \Phi^{5/2}\
  \biggr|\ 0\leq \Phi,\ \int_{\R^3}\Phi^2=1\biggr\} ,
\end{equation}
where $I_0$ is the constant from Foldy's law:
$$
I_0=\frac{2}{5}\frac{\Gamma(3/4)}{\Gamma(5/4)}\left(\frac{2}{\mu\pi}\right)^{1/4}.
$$
\end{thm}

This asymptotics can be understood as a mean field theory for the gas
density, very much like the Gross-Pitaevskii functional for dilute
trapped gases, where the local energy described by Foldy's law should
be balanced by the kinetic energy of the gas density. Thus if we let
the gas density be given by $\phi^2$ then the ``mean field'' energy
should be
\begin{equation}\label{eq:Dysonunscaled}
  \mu\int_{\R^3}|\nabla\phi|^2-
  I_0\int_{\R^3} \phi^{5/2}.
\end{equation}
Here $\int\phi^2=N$. If we now define
$\Phi(\x)=N^{-8/5}\phi(N^{-1/5}\x)$ we see that $\int\Phi^2=1$ and
that the above energy is
$$
N^{7/5}\left(\mu\int_{\R^3}|\nabla\Phi|^2-
  I_0\int_{\R^3} \Phi^{5/2}\right).
$$

It may be somewhat surprising that it is exactly the same constant
$I_0$
that appears in both the one- and two-component cases. The reason that
there are no extra factors to account for the difference between one
and two components is, as we shall see below, a simple consequence
of Bogoliubov's method. The origin of this equivalence, while clear
mathematically, does not appear to have a simple physical
interpretation.

\section{The Bogoliubov Approximation}
\index{Bogoliubov!approximation}

In this section we shall briefly explain the Bogoliubov approximation
and how it is applied in the case of the charged Bose gas.  The
Bogoliubov method relies on the exact diagonalization of a
Hamiltonian, which is quadratic in creation and annihilation
operators. For the charged Bose gas one only needs a very simple case
of the general diagonalization procedure.  On the other hand, the
operators that appear are not exact creation\index{creation operator}
and annihilation operators\index{annihilation operator}. A slightly
more general formulation is needed.
\begin{thm}[\textbf{Simple case of Bogoliubov's
method}]\label{thm:FoldyBogoliubov}
  Assume that $\bn_{\pm,\pm}$ are four (possibly unbounded)
  commuting operators satisfying the operator inequality $$
  \quad\left[\bn_{\tau,e},b^*_{\tau,e}\right]\leq1  \quad\hbox{for
    all }e,\tau=\pm.
  $$ Then for all real numbers $\cA,\cB_+,\cB_-\geq0$ we have \begin{eqnarray*}
    \lefteqn{\cA\sum_{\tau,e=\pm1}b_{\tau,e}^*\bn_{\tau,e} }&&\\&&+
    \sum_{e,e'=\pm1}\sqrt{\cB_e\cB_{e'}}ee'(b^*_{+,e}\bn_{+,e'}+b^*_{-,e}
\bn_{-,e'}+b^*_{+,e}b^*_{-,e'}
    +\bn_{+,e}\bn_{-,e'})\\ &\geq&
    -(\cA+\cB_++\cB_-)+\sqrt{(\cA+\cB_++\cB_-)^2-(\cB_++\cB_-)^2}.
  \end{eqnarray*} If $\bn_{\pm,\pm}$ are four annihilation operators
  then the lower bound is sharp.
\end{thm}

\begin{proof} Let us introduce $$
d_\pm^*=(\cB_++\cB_-)^{-1/2}(\cB_+^{1/2}b^*_{\pm
,+}-\cB_-^{1/2}b^*_{\pm
  ,-}),
$$ and $$ c_\pm^*=(\cB_++\cB_-)^{-1/2}(\cB_-^{1/2}b^*_{\pm
,+}+\cB_+^{1/2}b^*_{\pm
  ,-}).
$$ Then these operators satisfy $$
 [\dn_+,d_+^*]\leq1,\quad [\dn_-,d_-^*]\leq1.
$$ The operator that we want to estimate from below may be rewritten
as \begin{eqnarray*}
  &&\cA(d_+^*\dn_++d_-^*\dn_-+c_+^*\cn_++c_-^*\cn_-)\\
&&+(\cB_++\cB_-)\left(d_+^*\dn_++d_-^*\dn_-+d^*_+d^*_- +\dn_+\dn_-
\right).  \end{eqnarray*} We may now complete the squares to write
this as
\begin{eqnarray*}
&&\cA(c_+^*\cn_++c_-^*\cn_-)+D(d_+^*+\lambda \dn_-)(d_+^*+\lambda
\dn_-)^*\\ &&
+D(d^*_-+\lambda\dn_+)(d^*_-+\lambda\dn_+)^*-D\lambda^2([\dn_+,d^*_+]+[\dn_-,d^*_-])
\end{eqnarray*} if $$
D(1+\lambda^2)=\cA+\cB_++\cB_-,\quad 2D\lambda=\cB_++\cB_-.  $$ We
choose the solution $
\lambda=1+\frac{\cA}{\cB_++\cB_-}-\sqrt{\left(1+\frac{\cA}
{(\cB_++\cB_-)}\right)^2-1}.  $ Hence $$ D\lambda^2=
\mfr{1}/{2}\left(\cA+\cB_++\cB_--\sqrt{(\cA+\cB_++\cB_-)^2-(\cB_++\cB_-)^2}\right)\,
.$$
\end{proof}

In the theorem above one could of course also have included linear
terms in $b_{\tau,e}$ in the Hamiltonian. In the technical proofs in
\cite{LS,LSo02} the Bogoliubov diagonalization with linear terms is
indeed being used to control certain error terms.  Here we shall not
discuss the technical details of the proofs. We have therefore stated
the theorem in the simplest form in which we shall need it to derive
the leading contribution.

In our applications to the charged Bose gas the operators $b_{\pm,e}$
will correspond to the annihilation of particles with charge $e=\pm$
and momenta $\pm \k$ for some $\k\in\R^3$. Thus, only equal and
opposite momenta couple. In a translation invariant\index{translation
  invariance} case this would be a simple consequence of momentum
conservation. The one-component gas is not translation invariant, in
our formulation.  The two-component gas is translation invariant, but
it is natural to break translation invariance by going into the center
of mass frame. In both cases it is only in some approximate sense that
equal and opposite momenta couple.

In the case of the one-component gas we only need particles of one
sign. In this case we use the above theorem with $b_{\pm,-}=0$ and
$\cB_-=0$.

We note that the lower bounds in Theorem \ref{thm:FoldyBogoliubov} for
the
one- and two-com\-po\-nent gases are the same except for the replacement
of
$\cB_+$ in the one-component case by $\cB_+ +\cB_-$ in the
two-component
case. In the application to the two-component gas $\cB_+$ and $\cB_-$
will be proportional to the particle densities for respectively the
positive or negatively charged particles. For the one-component gas
$\cB_+$ is proportional to the background density.\index{background charge}

The Bogoliubov diagonalization method cannot be immediately
applied to the operators $H^{(1)}_N$ or $H^{(2)}_N$
since these operators are {\it not} quadratic in creation and
annihilation operators. In fact, they are quartic.  They have
the general form \begin{equation}\label{eq:generalform}
\sum_{\alpha,\beta}t_{\alpha\beta}a^*_\alpha\an_\beta
+\mfr{1}/{2}\sum_{\alpha,\beta,\mu,\nu}w_{\alpha\beta\mu\nu}a^*_\alpha
a^*_\beta\an_{\nu}\an_\mu, \end{equation} with $$
t_{\alpha\beta}=\langle\alpha|T|\beta\rangle,\qquad
w_{\alpha\beta\mu\nu}=\langle\alpha\beta|W|\mu\nu\rangle, $$ where $T$
is the one-body part of the Hamiltonian and $W$ is the two-body-part
of
the Hamiltonian.
\goodbreak

The main step in Bogoliubov's approximation
\index{Bogoliubov!approximation}
is now to assume Bose-Einstein condensation, i.e.,
that almost all particles are in the same one-particle state. In case
of the two-component gas this means that almost half the particles are
positively charged and in the same one-particle state as almost all
the other half of negatively charged particles.  We denote this
condensate state by the index $\alpha=0$ in the sums above.  Based on
the assumption of condensation Bogoliubov now argues that one may
ignore all terms in the quartic Hamiltonian above which contain 3 or 4
non-zero indices and at the same time replace all creation and
annihilation operators of the condensate by their expectation values.
The result is a quadratic Hamiltonian (including linear terms) in the
creation and annihilation with non-zero index.  This Hamiltonian is of
course not particle number preserving, reflecting the simple fact that
particles may be created out of the condensate or annihilated into the
condensate.

In Section~\ref{sec:chargedupper} below it is explained how to
construct trial wave functions for the one- and two-component charged
gases whose expectations agree essentially with the prescription in
the Bogoliubov approximation. The details appear in \cite{So}.
This will imply upper bounds on the energies corresponding to the
asymptotic forms given in Theorems~\ref{thm:Foldy} and
\ref{thm:Dyson}.

In \cite{LS,LSo02} it is proved how to make the steps in the Bogoliubov
approximation rigorous as lower bounds.  The main difficulty is to
control the degree of condensation.  As already explained it is not
necessary to prove condensation in the strong sense described above.
We shall only prove condensation in small boxes. Put differently, we
shall not conclude that most particles are in the same one-particle
state, but rather prove that most particles occupy one-particle states
that look the same on short scales, i.e., that vary slowly.
Here the short scale is the correlation length scale $\ell_{\rm cor}$.

\section{The Rigorous Lower Bounds}

As already mentioned we must localize into small boxes of some fixed
size $\ell$. This time we must require $\ell_{\rm cor}\ll\ell$. For
the one-component gas this choice is made only in order to control the
degree of condensation. For the two-component gas it is required both
to control the order of condensation, and also to make a local
constant density approximation\index{local density approximation}. The
reason we can control the degree of condensation in a small box is
that the localized kinetic energy has a gap above the lowest energy
state. In fact, the gap is of order $\ell^{-2}$. Thus if we require
that $\ell$ is such that $N\ell^{-2}$ is much greater than the energy
we may conclude that most particles are in the lowest eigenvalue state
for the localized kinetic energy. We shall always choose the localized
kinetic energy in such a way that the lowest eigenstate, and hence the
condensate, is simply a constant function.

\subsection{Localizing the interaction}
In contrast to the dilute gas the long range Coulomb potential
prevents us from simply ignoring the interaction
between the small boxes.  To overcome this problem we use a sliding
technique\index{sliding technique} first introduced in \cite{CLY}.
\begin{thm}[\textbf{Controlling interactions by
sliding}]\label{thm:sliding}
  Let $\upchi$ be a smooth approximation to the characteristic
  function of the unit cube centered at the origin. For $\ell>0$
  and $\z\in\R^3$ let $\upchi_{\z}(\x)=\upchi((\x-\z)/\ell)$. There
exists
  an $\omega>0$ depending on $\upchi$ (in such a way that it tends to
  infinity as $\upchi$ approximates the characteristic function) such
  that
$$
\sum_{1\leq i<j\leq
  N}\frac{e_ie_j}{|\x_i-\x_j|}\geq\left(\int\upchi^2\right)^{-1}
\int_{\R^3}\sum_{1\leq i<j\leq
  N}e_ie_jw_{\ell\z}(\x_i,\x_j)d\z-\frac{N\omega}{2\ell},
$$
for all $\x_1,\ldots\in\R^3$ and $e_1,\ldots=\pm1$,
where
$$
w_\z(\x,\y)=\upchi_{\z}(\x)Y_{\omega/\ell}(\x-\y)\upchi_{\z}(\y)
$$
with
$Y_\mu(\x)=|\x|^{-1}\exp(-\mu |\x|)$ being the Yukawa
potential.
\end{thm}

The significance of this result is that the two-body potential $w_\z$
is localized to the cube of size $\ell$ centered at $\ell\z$. The
lower bound above is thus an integral over localized interactions
sliding around with the integration parameter.

We have stated the sliding estimate in the form relevant to the
two-com\-po\-nent problem. There is an equivalent version for the
one-component gas, where the sum of the particle-particle,
particle-background, and back\-ground-back\-ground interactions may be
bounded below by corresponding localized interactions.

Since $\ell\gg\ell_{\rm cor}$ the error in the sliding
estimate is much smaller than $\omega N/\ell_{\rm cor}$, which for
both the one and two-component gases is of order $\omega$ times the
order of the energy. Thus, since $\ell$ is much
bigger than $\ell_{\rm cor}$, we have room to let $\omega$ be very
large, i.e., $\upchi$ is close to the characteristic function.

\subsection{Localizing the kinetic energy}
\index{energy localization}
Having described the technique to control the interaction between
localized regions we turn next to the {\it localization of the kinetic
  energy}.

For the two-component gas this is done in two steps. As already
mentioned it is natural to break the translation invariance of the
two-component gas. We do this by localizing the system into a box of
size $L'\gg N^{-1/5}$ (which as we saw is the expected size of the
gas) as follows. By a partition of unity we can divide space into
boxes of this size paying a localization error due to the kinetic
energy of order $NL'^{-2}\ll N^{7/5}$. We control the interaction
between these boxes using the sliding technique.

We may now argue, as follows, that the energy is smallest if all the
particles are in just one box.  For simplicity we give this argument
for the case of two boxes.  Suppose the two boxes have respective wave
functions $\psi$ and $\widetilde\psi$. The total energy of these two
non-interacting boxes is $E+\widetilde E$. Now put all the particles
in one box with the trial function $\Psi=\psi\widetilde\psi$.  The
fact that this function is not bosonic (i.e., it is not symmetric with
respect to all the variables) is irrelevant because the true bosonic
ground state energy is never greater than that of any trial state
(Perron-Frobenius Theorem\index{Perron-Frobenius Theorem}). The energy of $\Psi$ is
$$E+\widetilde
E+\iint \uprho_\psi(\x)|\x-\y|^{-1}\uprho_{\widetilde\psi}(\y)d\x d\y,
$$
where $\uprho_\psi$ and $\uprho_{\widetilde\psi}$ are the
respective {\it charge} densities of the states $\psi$ and
$\widetilde\psi$.  We claim that the last Coulomb term can be made
non-positive. How? If it is positive then we simply change the state
$\widetilde\psi$ by interchanging positive and negative charges (only
in $\widetilde\psi$ and not in $\psi$). The reader is reminded that we
have not constrained the number of positive and negative particles but
only their sum.  This change in $\widetilde \psi$ reverses the
relative
charge of the states $\psi$ and $\widetilde\psi$ so, by symmetry the
energies $E$ and $\widetilde E$ do not change, whereas the Coulomb
interaction changes sign.

The localization into smaller cubes of size $\ell$ can however not be
done by a crude partition of unity localization. Indeed, this would
cost a localization error of order $N\ell^{-2}$, which as explained is
required to be of much greater order than the energy.

For the one-component charged gas we may instead use a Neumann
localization of the kinetic energy, as for the dilute Bose gas. If we
denote by $\Delta_\ell^{(\z)}$ the Neumann Laplacian for the cube of
size $\ell$ centered at $\z$ we may, in the spirit of the sliding
estimate, write the Neumann localization Laplacian in all of $\R^3$ as
$$
-\Delta=\int -\Delta_\ell^{(\ell\z)} d\z.
$$
In order to write the localized kinetic energy in the same form as
the localized interaction we must introduce the smooth localization
$\upchi$ as in Theorem~\ref{thm:sliding}. This can be achieved
by ignoring the low momentum part of the kinetic energy.

More precisely, there exist $\varepsilon(\upchi)$ and $s(\upchi)$
such that $\varepsilon(\upchi)\to0$ and $s(\upchi)\to0$ as $\upchi$
approaches the characteristic function of the unit cube and such that
(see Lemma 6.1 in \cite{LS})
\begin{equation}\label{eq:1kineticloc}
  -\Delta_\ell^{(\z)}\geq
  (1-\varepsilon(\upchi))\cP_{\z}\upchi_\z(\x)F_{\ell
s(\chi)}(-\Delta)\upchi_\z(\x)\cP_{\z}
\end{equation}
where $\cP_\z$ denotes the projection orthogonal to constants in
the cube of size $\ell$ centered at $z$ and
$$
F_s(u)=\frac{u^2}{u+s^{-2}}.
$$
For $u\ll s^{-2}$ we have that $F_s(u)\ll u$. Hence the
effect of $F$ in the operator estimate above is to ignore the low
momentum part of the Laplacian.

For the two-component gas one cannot use the Neumann localization as
for the one-component gas. Using a Neumann localization ignores the
kinetic energy corresponding to long range variations in the wave
function and one would not get the kinetic energy term
$\int\mu|\nabla\Phi|^2$ in (\ref{eq:Dyson}).  This is the essential
difference between the one- and two-component cases.  This problem is
solved in \cite{{LSo02}} where a new kinetic energy localization
technique is developed. The idea is again to separate the high and low
momentum part of the kinetic energy. The high momentum part is then
localized as before, whereas the low momentum part is used to connect
the localized regions by a term corresponding to a discrete Laplacian.
(For details and the proof the reader is referred to \cite{{LSo02}}.)
\begin{thm}[\textbf{A many body kinetic energy
localization}]\label{thm:kinloc}
  Let $\upchi_\z$, $\cP_\z$ and $F_s$ be as above. There exist
  $\varepsilon(\upchi)$ and $s(\upchi)$ such that
  $\varepsilon(\upchi)\to0$ and $s(\upchi)\to0$ as $\upchi$ approaches
  the characteristic function of the unit cube and such that for all
  normalized symmetric wave functions $\Psi$ in
  $L^2((\R^3\times\{-1,1\})^N)$ and all $\Omega\subset\R^3$ we have
  \begin{eqnarray*}
    (1+\varepsilon(\upchi))\left(\Psi,\sum_{i=1}^N-\Delta_i\Psi\right)
    &\geq& \int_\Omega
\Bigl[(\Psi,\cP_{\ell\z}\upchi_{\ell\z}(\x)F_{\ell
      s(\chi)}(-\Delta)\upchi_{\ell\z(\x)}\cP_{\ell\z}\Psi)\\&&
    +\mfr{1}/{2}\ell^{-2}\sum_{\y\in\Z^3\atop
      |\y|=1}(S_\Psi(\ell(\z+\y))-S_\Psi(\ell\z))^2
\Bigr]d\z\\&&-\const\ell^{-2}{\rm Vol}(\Omega),
  \end{eqnarray*}
  where
  $$
  S_\Psi(\z)=\sqrt{\left(\Psi,(a^*_{0+}(\z)\an_{0+}(\z)+a^*_{0-}(\z)\an_{0-}(\z))\Psi\right)+1}-1
  $$
  with $\an_{0\pm}(z)$ being the annihilation of a particle of
  charge $\pm$ in the state given by the normalized characteristic
  function of the cube of size $\ell$ centered at $\z$.
\end{thm}
The first term in the kinetic energy localization in this theorem is
the same as in (\ref{eq:1kineticloc}). The second term gives rise to a
discrete Laplacian for the function $S_\Psi(\ell\z)$, which is
essentially the number of condensate particles in the cube of size
$\ell$ centered at $\ell\z$. Since we will eventually conclude that
most particles are in the condensate this term will after
approximating the discrete Laplacian by the continuum Laplacian lead
to
the term $\int\mu|\nabla\phi|^2$ in (\ref{eq:Dysonunscaled}). We
shall not
discuss this any further here.

When we apply this theorem to the two-component gas the set
$\ell\Omega$ will be the box of size $L'$ discussed above. Hence the
error term $\ell^{-2}{\rm Vol}(\Omega)$ will be of order
$L'^3/\ell^{-5}\ll (N^{2/5}\ell)^{-5}(N^{1/5}L')^3N^{7/5}$. Thus since
$\ell\gg N^{-2/5}$ we may still choose $L'\gg N^{-1/5}$, as required,
and have this error term be lower order than $N^{7/5}$.

\subsection{Controlling the degree of condensation}
After now having localized the problem into smaller cubes we are ready
to control the degree of condensation.  We recall that the condensate
state is the constant function in each cube. Let us denote by $\hn_\z$
the number of excited (i.e., non-condensed particles) in the box of
size $\ell$ centered at $\z$. Thus for the two-component gas $\hn_\z
+a^*_{0+}(\z)\an_{0+}(\z)+a^*_{0-}(\z)\an_{0-}(\z)$ is the total
number of particles in the box and a similar expression gives the
particle number for the one-component gas.

As discussed above we can use the fact that the kinetic energy
localized to a small box has a gap above its lowest eigenvalue
to control the number of excited particles. Actually, this will
show that the expectation $(\Psi,\hn_\z\Psi)$ is much smaller than the
total number of particles in the box for any state
$\Psi$ with negative energy expectation.

One needs, however, also a good bound on $(\Psi,\hn_\z^2\Psi)$ to
control the Coulomb interaction of the non-condensed particles.  This
is more difficult.  In \cite{LS} this is not achieved directly through
a bound on $(\Psi,\hn_\z\Psi)$ in the ground state.  Rather it is
proved that one may change the ground state without changing its
energy very much, so that it only contains values of $\hn_\z$
localized close to $(\Psi,\hn_\z\Psi)$. The following theorem gives
this very general localization technique. Its proof can be found in
\cite{LS}.
\begin{thm}[\textbf{Localizing large matrices}]\label{local}
  Suppose that ${\cA}$ is \index{localizing large matrices}
  an $N+1\times N+1$ Hermitian matrix and let ${\cA}^k$, with
$k=0,1,...,N$,
  denote the matrix consisting of the $k^{\rm th}$ supra- and
  infra-diagonal of ${\cA}$.  Let $\psi \in {\bf C}^{N+1}$ be a
normalized vector
  and set $d_k = (\psi , {\cA}^k \psi) $ and $\lambda = (\psi , {\cA}
\psi) =
  \sum_{k=0}^{N} d_k$.  \  ($\psi$ need not be an eigenvector of
${\cA}$.) \

  Choose some positive integer $M \leq N+1$.
  Then, with $M$ fixed, there is some $n \in [0, N+1-M]$ and some
normalized
  vector $ \phi \in   {\bf C}^{N+1}$ with the property that
  $\phi_j =0$ unless $n+1 \leq j \leq n+M$ \ (i.e., $\phi $ has
length $M$)
  and such that
  \begin{equation}\label{localerror}
    (\phi , {\cA} \phi) \leq \lambda + \frac{C}{ M^2}\sum_{k=1}^{M-1}
k^2 |d_k|
    +C\sum_{k=M}^{N} |d_k|\ ,
  \end{equation}
  where $C>0 $ is a  universal constant. (Note that the first sum
starts
  with $k=1$.)
\end{thm}
To use this theorem we start with a ground state (or approximate
ground state) $\Psi$ to the many
body problem. We then consider the projections of $\Psi$ onto the
eigenspaces of $\hn_\z$. Since the possible eigenvalues run from $0$
to $N$ these projections span an at most $N+1$ dimensional space.

We use the above theorem with $\cA$ being the many body Hamiltonian
restricted to this $N+1$ dimensional subspace. Since the Hamiltonian
can change the number of excited particles by at most two we see that
$d_k$ vanishes for $k\geq3$. We shall not here discuss the estimates
on $d_1$ and $d_2$ (see \cite{LS,LSo02}). The conclusion is that we
may, without changing the energy expectation of $\Psi$ too much,
assume that the values of $\hn_\z$ run in an interval of length much
smaller than the total number of particles.  We would like to conclude
that this interval is close to zero.  This follows from the fact that
any wave function with energy expectation close to the minimum must
have an expected number of excited particles much smaller than the
total number of particles.

\subsection{The quadratic Hamiltonian}\index{Hamiltonian!quadratic}

Using our control on the degree of condensation it is now possible to
estimate all unwanted terms in the Hamiltonian, i.e., terms that
contain 3 or more creation or annihilation operators corresponding to
excited (non-condensate) states. The proof which is a rather
complicated bootstrapping argument is more or less the same for the
one- and two-component gases.  The result, in fact, shows that we can
ignore other terms too. In fact if we go back to the general form
(\ref{eq:generalform}) of the Hamiltonian it turns out that we can
control all
quartic terms except the ones with the coefficients:
$$
w_{\alpha\beta00},\ w_{00\alpha\beta},\ w_{\alpha00\beta},\hbox{ and
} w_{0\alpha\beta0}.
$$
To be more precise, let $u_\alpha$, $\alpha=1,\ldots$ be an
orthonormal basis of real functions for the subspace of functions on
the cube of size $\ell$ centered at $\z$ orthogonal to constants, i.e,
with vanishing average in the cube. We shall now omit the subscript
$\z$ and let $a_{0\pm}$ be the annihilation of a particle of charge
$\pm1$ in the normalized constant function in the cube (i.e., in the
condensate).  Let $a_{\alpha\pm}$ with $\alpha\ne0$ be the
annihilation operator for a particle of charge $\pm1$ in the state
$u_\alpha$.  We can then show that the main contribution to the
localized energy of the two-component gas comes from the Hamiltonian
\begin{eqnarray*}
H_{\rm local} \!\!\!\!&=&  \!\!\!\!\!\! \sum_{\alpha,\beta=1\atop e=\pm1}^\infty
t_{\alpha\beta}a^*_{\alpha e}\an_{\beta e}\\&&  \!\!\!\!\!\!+
\mfr{1}/{2}\sum_{\alpha,\beta=1\atop
  e,e'=\pm1}ee'w_{\alpha\beta}(2a_{0e}^*a^*_{\alpha
  e'}\an_{0e'}\an_{\beta e} +a_{0e}^*a^*_{0
  e'}\an_{\alpha e'}\an_{\beta e} +a_{\alpha e}^*a^*_{\beta
  e'}\an_{0e'}\an_{0e}),
\end{eqnarray*}
where
$$
t_{\alpha\beta}=\mu(u_\alpha,\cP_\z\upchi_\z(\x)F_{\ell
  s(\chi)}(-\Delta)\upchi_\z(\x)\cP_\z u_\beta)
$$
and
$$
w_{\alpha\beta}=\ell^{-3}\iint
u_\alpha(\x)\upchi_\z(\x)Y_{\omega/\ell}(\x-\y)\upchi_\z(\y)u_\beta(\x)d\x
d\y.
$$
In $H_{\rm local}$ we have
ignored all error terms and hence also $\varepsilon(\upchi)\approx0$
and $\int\upchi^2\approx1$.

In the case of the one-component gas we get exactly the same local
Hamiltonian, except that we have only one type of particles, i.e, we
may set $\an_{\alpha-}=0$ above.

Let $\nu_\pm=\sum_{\alpha=0}^\infty a^*_{\alpha\pm}\an_{\alpha\pm}$
be the total number of particles in the box with charge $\pm1$.
For $\k\in\R^3$ we let
$\upchi_{\k,\z}(\x)=\upchi_\z(\x)e^{i\k\cdot\x}$. We then introduce
the operators
$$
b_{\k\pm}=(\ell^3\nu_\pm)^{-1/2}a_\pm(\cP_\z\upchi_{\k,\z})a^*_{0\pm},
$$
where
$a_\pm(\cP_\z\upchi_{\k,\z})=
\sum_{\alpha=1}^\infty(\upchi_{\k,\z},u_\alpha)\an_{\alpha\pm}$
annihilates a particle in the state $\upchi_{\k,\z}$ with charge
$\pm1$.  It is then clear that the operators $b_{\k\pm}$ all commute
and a straightforward calculation shows that
$$
[\bn_{\k\pm},b_{\k\pm}^*]\leq(\ell^3\nu_\pm)^{-1}\|\cP_\z\upchi_\z\|^2
a^*_{0\pm}\an_{0\pm}\leq 1.
$$
If we observe that
\begin{eqnarray*}
  \sum_{\alpha,\beta=1\atop e=\pm1}^\infty
  t_{\alpha\beta}a^*_{\alpha e}\an_{\beta e}
  &=&(2\pi)^{-3}\int \mu F_{\ell
    s(\chi)}(\k^2)\sum_{e=\pm}a_e(\cP_\z\upchi_{\k,\z})^*a_e(\cP_\z\upchi_{\k,\z})d\k\\
  &\geq&(2\pi)^{-3}\ell^3\int \mu F_{\ell
    s(\chi)}(\k^2)\sum_{e=\pm}b_{\k e}^*\bn_{\k e},
\end{eqnarray*}
we see that
\begin{eqnarray*}
  \lefteqn{H_{\rm local}\geq \mfr{1}/{2}(2\pi)^{-3}\int
\mu\ell^3F_{\ell
    s(\chi)}(\k^2) \sum_{e=\pm}(b_{\k e}^*\bn_{\k e}+b_{-\k
e}^*\bn_{-\k e})}&&\\
  &&+\sum_{ee'=\pm}\widehat{Y}_{\omega/\ell}(\k)\sqrt{\nu_e\nu_{e'}}ee'(b_{\k
e}^*\bn_{\k,e'}
  +b_{-\k e}^*\bn_{-\k,e'}+b_{\k e}^*b_{-\k,e'}^*+\bn_{-\k
e}\bn_{\k,e'})d\k\\
  &&-\sum_{\alpha\beta=1}w_{\alpha\beta}(a_{\alpha+}^*\an_{\beta+}+a_{\alpha-}^*\an_{\beta-}).
\end{eqnarray*}
The last term comes from commuting $a^*_{0\pm}\an_{0\pm}$ to
$\an_{0\pm}a^*_{0\pm}$.
It is easy to see that this last term is a bounded operator with norm
bounded by
$$
\const(\nu_++\nu_-
)\ell^{-3}\|\widehat{Y}_{\omega/\ell}\|_\infty\leq\const\omega^{-2}(\nu_++\nu_-
)\ell^{-1}.
$$
When summing over all boxes we see that the last term above gives a
contribution bounded by $\const
\omega^{-2}N\ell^{-1}=\omega^{-2}(N^{2/5}\ell)^{-1}N^{7/5}$ which is
lower order than the energy.

The integrand in the lower bound on $H_{\rm local}$ is precisely an
operator of the form treated in the Bogoliubov method\index{Bogoliubov!method}
Theorem~\ref{thm:FoldyBogoliubov}.  Thus up to negligible errors we
see that the operator $H_{\rm local}$ is bounded below by
$$
  \mfr{1}/{2}(2\pi)^{-3}\int
-(\cA(\k)+\cB(\k))+\sqrt{(\cA(\k)+\cB(\k))^2-\cB(\k)^2}\, d\k,
$$
where
$$
\cA(\k)=\mu\ell^3F_{\ell s(\chi)}(\k^2)\quad\hbox{and}\quad
\cB(\k)=\nu\widehat{Y}_{\omega/\ell}(\k)
$$
with $\nu=\nu_++\nu_-$ being the total number of particles in the
small box.  A fairly simple analysis of the above integral shows that
we may to leading order replace $\cA$ by $\mu\ell^3 \k^2$ and
$\cB(\k)$ by $4\pi\nu |\k|^{-2}$, i.e., we may ignore the cut-offs.
The final conclusion is that the local energy is given to leading
order by
$$
  \frac {- 1}{2(2\pi)^{3}}\int
 4\pi\nu|\k|^{-2}+\mu\ell^3|\k|^2
    -\sqrt{(4\pi\nu|\k|^{-2}+\mu\ell^3|\k|^2)^2
    -(4\pi\nu|\k|^{-2})^2} \, d\k
$$
$$=-2^{1/2}\pi^{-3/4}\nu\left(\frac{\nu}{\mu\ell^3}\right)^{1/4}
  \int_0^\infty 1+x^4-x^2(2+x^4)^{1/2}\, dx.
$$
If we finally use that
$$
\int_0^\infty1+x^4-x^2(2+x^4)^{1/2}\, dx=\frac{2^{3/4}\sqrt{\pi}\Gamma(3/4)}{5\Gamma(5/4)}
$$
we see that the local energy to leading order is $ -I_0
\nu({\nu}/{\ell^3})^{1/4} $. For the one-component gas we should set
$\nu=\rho\ell^3$ and for the two-component gas we should set~$\nu=\phi^2\ell^3$ (see (\ref{eq:Dysonunscaled})).  After replacing
the sum over boxes by an integral and at the same time replace the
discrete Laplacian by a continuum Laplacian, as described above, we
arrive at asymptotic lower bounds as in Theorems~\ref{thm:Foldy} and
\ref{thm:Dyson}.

There is one issue that we have not discussed at all and which played
an important role in the treatment of the dilute gas.  How do we know
the number of particles in each of the small cubes?  For the dilute
gas a superadditivity\index{superadditivity} argument was used to show that there was an
equipartition of particles among the smaller boxes. Such an argument
cannot be used for the charged gas.  For the one-component gas
one simply minimizes the energy over all possible particle numbers in
each little box.  It turns out that charge neutrality is essentially
required for the energy to be minimized. Since the background charge
in each box is fixed this fixes the particle number.

For the two-component there is a-priori nothing that fixes the
particle
number in each box.  More precisely, if we ignored the kinetic energy
between the small boxes it would be energetically favorable to put all
particles in one small box.  It is the kinetic energy between boxes,
i.e., the discrete Laplacian term in Theorem~\ref{thm:kinloc}, that
prevents this from happening. Thus we could in principle again
minimize over all particle numbers and hope to prove the correct
particle number dependence (i.e., Foldy's law\index{Foldy}) in each small box. This
is essentially what is done except that boxes with very many or very
few particles must be treated somewhat differently from the ``good''
boxes. In the ``bad'' boxes we do not prove Foldy's law, but only
weaker estimates that are adequate for the argument.

  \bigskip

\section{The Rigorous Upper Bounds}\label{sec:chargedupper}

\subsection{The upper bound for the two-component gas}

To prove an upper bound on the energy $E_0^{(2)}(N)$ of the form given
in Dyson's\index{Dyson} formula Theorem~\ref{thm:Dyson} we shall construct a trial
function from the prescription in the Bogoliubov approximation.  We
shall use as an input a minimizer $\Phi$ for the variational problem
on the right side of (\ref{eq:Dyson}). That minimizers exist can be
easily seen using spherical decreasing rearrangements. It is however
not important that a minimizer exists. An approximate minimizer would
also do for the argument given here.  Define
$\phi_0(\x)=N^{3/10}\Phi(N^{1/5}\x)$. Then again $\int\phi_0^2=1$. In
terms of the unscaled function $\phi$ in (\ref{eq:Dysonunscaled}),
$\phi_0(\x)=N^{-1}\phi(\x)$.

Let $\phi_\alpha$, $\alpha=1,\ldots$ be an orthonormal family of real
functions all orthogonal to $\phi_0$.  We choose these functions
below.

We follow Dyson~\cite{D2} and choose a trial function which does not
have a specified particle number, i.e., a state in the bosonic Fock
space.\index{Fock space}

As our trial many-body wave function we now choose
\begin{eqnarray}\label{eq:trialstate}
\Psi&=&\exp\left(-\lambda_0^2+\lambda_0a^*_{0+}+\lambda_0
a^*_{0-}\right)\nonumber\\&&\times\prod_{\alpha\ne0}(1-\lambda_\alpha^2)^{1/4}\exp\Bigl(-\sum_{e,e'=\pm1}\sum_{\alpha\ne
  0}\frac{\lambda_{\alpha}}{4}
ee'a_{\alpha,e}^*a_{\alpha,e'}^*\Bigr)\left|0\right\rangle,
\end{eqnarray}
where $a_{\alpha,e}^*$ is the creation of a particle of charge
$e=\pm1$ in the state $\phi_\alpha$, $|0\rangle$ is the vacuum state,
and the coefficients $\lambda_0,\lambda_1,\ldots$ will be chosen
below satisfying $0<\lambda_\alpha<1$ for $\alpha\ne0$.

It is straightforward to check that $\Psi$ is a normalized function.

Dyson used a very similar trial state in \cite{D2}, but in his case
the exponent was a purely quadratic expression in creation operators,
whereas the one used here is only quadratic in the creation operators
$a^*_{\alpha e}$, with $\alpha\ne0$ and linear in $a^*_{0\pm}$.  As a
consequence our state will be more sharply localized around the mean
of the particle number.

In fact, the above trial state is precisely what is suggested by the
Bogoliubov approximation.
To see this note that one has
$$
(\an_{0\pm}-\lambda_0)\Psi=0,\quad
\hbox{ and }\quad
\left(a^*_{\alpha+}-a^*_{\alpha-}+\lambda_\alpha
(\an_{\alpha+}-\an_{\alpha-})\right)\Psi=0
$$
for all $\alpha\ne0$. Thus the creation operators for the condensed
states can be replaced by their expectation values and an adequate
quadratic expression in the non-condensed creation and annihilation
operators is minimized.

Consider now the operator
\begin{equation}\label{eq:gamma}
  \gamma=\sum_{\alpha=1}^\infty
  \frac{\lambda_\alpha^2}{1-\lambda_\alpha^2}|\phi_\alpha\rangle\langle\phi_\alpha|.
\end{equation}
A straightforward calculation of the energy expectation in the state
$\Psi$ gives that
\begin{eqnarray*}
  \left(\Psi,\sum_{N=0}^\infty H^{(2)}_N \Psi\right)&=&2\lambda_0^2
\mu\int(\nabla \phi_0)^2
  +\hbox{Tr}\left(-\mu\Delta\gamma\right)\nonumber\\&&
  +2\lambda_0^2\hbox{Tr}\left(\cK\left(\gamma-\sqrt{\gamma(\gamma+1)}\right)\right),\label{eq:upper}
\end{eqnarray*}
where $\cK$ is the operator with integral kernel
\begin{equation}\label{eq:cK}
  \cK(\x,\y)=\phi_0(\x)|\x-\y|^{-1}\phi_0(\y).
\end{equation}
Moreover, the expected particle number in the state $\Psi$ is
$2\lambda_0^2+\hbox{Tr}(\gamma)$. In order for $\Psi$ to be well
defined by the formula (\ref{eq:trialstate}) we must require this
expectation to be finite.

Instead of making explicit choices for the individual functions
$\phi_\alpha$ and the coefficients $\lambda_\alpha$, $\alpha\ne0$ we
may equivalently choose the operator $\gamma$.  In defining $\gamma$
we use the method of coherent states.  Let $\upchi$ be a non-negative
real and smooth function supported in the unit ball in $\R^3$, with
$\int\upchi^2=1$.  Let as before $N^{-2/5}\ll\ell\ll N^{-1/5}$ and
define $\upchi_\ell(\x)=\ell^{-3/2}\upchi(\x/\ell)$. We choose
$$
\gamma=(2\pi)^{-3}\int_{\R^3\times\R^3}f(\u,|\p|)\cP_{\phi_0}^\perp|\theta_{\u,\p}\rangle\langle\theta_{\u,\p}|\cP_{\phi_0}^\perp
d\u d\p
$$
where $\cP_{\phi_0}^\perp$ is the projection orthogonal to
$\phi_0$,
$$
\theta_{\u,\p}(x)=\exp(i\p\cdot\x)\upchi_\ell(\x-\u),
$$
and
$$
f(\u,|\p|)=\frac{1}{2}\left(\frac{\p^4+16\pi\lambda_0^2\mu^{-1}\phi_0(\u)^2}{\p^2
    \left(\p^4+32\pi\lambda_0^2\mu^{-1}\phi_0(\u)^2\right)^{1/2}}-1\right).
$$
We note that $\gamma$ is a positive trace class operator,
$\gamma\phi_0=0$, and
that all eigenfunctions of $\gamma$ may be chosen real. These are
precisely the requirements needed in order for $\gamma$ to define the
orthonormal family $\phi_\alpha$ and the coefficients $\lambda_\alpha$
for $\alpha\ne0$.

We use the following version of the Berezin-Lieb inequality\index{Berezin-Lieb inequality}
\cite{Berezin72,Lieb73}.  Assume that $\xi(t)$ is an {\it operator}
concave function of $\R_+\cup\{0\}$ with $\xi(0)\geq0$.  Then if $Y$
is a
positive semi-definite operator we have
\begin{equation}\label{eq:berezinlieb}
  \hbox{Tr} \left(Y\xi(\gamma)\right)\geq (2\pi)^{-3}\int
  \xi(f(\u,|\p|))\left(\theta_{\u,\p},\cP_{\phi_0}^\perp
    Y\cP_{\phi_0}^\perp\theta_{\u,\p}\right) d\u d\p.
\end{equation}
We use this for the function $\xi(t)=\sqrt{t(t+1)}$.  Of course, if
$\xi$ is the identity function then (\ref{eq:berezinlieb}) is an
identity. If $Y=I$ then (\ref{eq:berezinlieb}) holds for all concave
$\xi$ with
$\xi(0)\geq0$.

Proving an upper bound on the energy expectation
(\ref{eq:upper}) is thus reduced to the calculations of explicit
integrals. After
estimating these integrals one arrives at the leading contribution
(for large $\lambda_0$)
\begin{eqnarray*}
  2\lambda_0^2 \mu\int(\nabla \phi_0)^2
  +&\displaystyle\iint&\left(\mu\p^2+2\lambda_0^2\phi_0(\u)^2\frac{4\pi}{\p^2}\right)f(\u,|\p|)\\
  &&-\frac{4\pi}{\p^2}2\lambda_0^2\phi_0(\u)^2\sqrt{f(\u,|\p|)(f(\u,|\p|)+1)}\
d\p d\u\\
&=&2\lambda_0^2\mu \int(\nabla \phi_0)^2-I_0\int
(2\lambda_0^2)^{5/4}\phi_0^{5/2},
\end{eqnarray*}
where $I_0$ is as in Theorem~\ref{thm:Dyson}.

If we choose $\lambda_0=\sqrt{N/2}$ we get after a simple rescaling
that the energy above is $N^{7/5}$ times the right side of
(\ref{eq:Dyson}) (recall that $\Phi$ was chosen as the minimizer).
We also note that the expected number of particles is
$$
2\lambda_0^2+\hbox{Tr}(\gamma)=N+O(N^{3/5}),
$$
as $N\to\infty$.

The only remaining problem is to show how a similar energy could be
achieved with a wave function with a fixed number of particles $N$,
i.e., how to show that we really have an upper bound on
$E^{(2)}_0(N)$.  We
indicate this fairly simple argument here.

We construct a trial function $\Psi'$ as above, but with an expected
particle number $N'$ chosen appropriately close to, but slightly
smaller than $N$. More precisely, $N'$ will be smaller than $N$ by an
appropriate lower order correction.  It is easy to see then that the
mean deviation of the particle number distribution in the state
$\Psi'$ is lower order than $N$. In fact, it is of order
$\sqrt{N'}\sim\sqrt{N}$. Using that we have a good lower bound on the
energy $E^{(2)}_0(n)$ for all $n$ and that $\Psi'$ is sharply
localized around its mean particle number, we may, without changing
the energy expectation significantly, replace $\Psi'$ by a normalized
wave function $\Psi$ that only has particle numbers less than $N$.
Since the function $n\mapsto E^{(2)}_0(n)$ is a decreasing function we
see that the energy expectation in the state $\Psi$ is, in fact, an
upper bound to $E^{(2)}_0(N)$.

\subsection{The upper bound for the one-component gas}
The upper bound for the one-component gas is proved in a very similar
way as for the two-component gas.  We shall simply indicate the main
differences here.  We will again choose a trial state without a fixed
particle number, i.e., a grand-canonical trial state. Since we know
that the one-component gas has a thermodynamic limit and that there is
equivalence of ensembles\index{equivalence of ensembles} \cite{LN}, it
makes no difference whether we choose a canonical or grand-canonical
trial state.

For the state $\phi_0$ we now choose a normalized function with
compact support in $\Lambda$, that is constant on the set
$\{x\in\Lambda\ |\ \hbox{dist}(x,\partial\Lambda)>r\}$. We shall
choose $r>0$ to go to zero as $L\to\infty$. Let us also choose the
constant $n$ such that $n\phi_0^2=\rho$ on the set where $\phi_0$ is
constant. Then $n\approx\uprho L^3$.

Let again $\phi_\alpha$, $\alpha=1,\ldots$ be an orthonormal family
of real
functions  orthogonal to $\phi_0$.
As our trial state we choose, this time,
\begin{eqnarray}\label{eq:trialstate1}
\Psi&=&\prod_{\alpha\ne0}(1-\lambda_\alpha^2)^{1/4}\exp\Bigl(-\lambda_0^2/2+\lambda_0a^*_{0}
-\sum_{\alpha\ne
  0}\frac{\lambda_{\alpha}}{2}a_{\alpha}^*a_{\alpha}^*\Bigr)\left|0\right\rangle,
\end{eqnarray}
where $a^*_\alpha$ is the creation of a particle in the state
$\phi_\alpha$.  We will choose $\Psi$ implicitly by choosing the
operator $\gamma$ defined as in (\ref{eq:gamma}).

This time we obtain
\begin{eqnarray}
  \lefteqn{\left(\Psi,\sum_{N=0}^\infty H^{(1)}_N
\Psi\right)=\lambda_0^2
    \mu\int(\nabla \phi_0)^2
  \nonumber }&&\\&&+\half\iint\frac{|\gamma(\x,\y)|^2}{|\x-\y|}d\x d\y
  +\half\iint\frac{|\sqrt{\gamma(\gamma+1)}(\x,\y)|^2}{|\x-\y|}
  d\x d\y\nonumber\\&&
  +\half\iint\limits_{\Lambda\times\Lambda}
  \left(\uprho-\uprho_\gamma(\x)-\lambda_0^2\phi_0(\x)^2\right)|\x-\y|^{-1}
  \left(\uprho-\uprho_\gamma(\y)-\lambda_0^2\phi_0(\y)^2\right)
  d\x d\y\nonumber\\&&+\hbox{Tr}\left(-\mu\Delta\gamma\right)
  +\lambda_0^2
  \hbox{Tr}
  \left(\cK\left(\gamma-\sqrt{\gamma(\gamma+1)}\right)\right),
  \label{eq:upperen}
\end{eqnarray}
where $\uprho_\gamma(\x)=\gamma(\x,\x)$ and $\cK$ is again given as in
(\ref{eq:cK}).  We must show that we can make choices such that the
first four terms on the right side above are lower order than the
energy, and can therefore be neglected.

We choose
$$
\gamma=\gamma_\varepsilon=(2\pi)^{-3}\int_{|p|>\varepsilon\rho^{1/4}}f(|\p|)\cP_{\phi_0}^\perp
|\theta_{\p}\rangle\langle\theta_{\p}|\cP_{\phi_0}^\perp
d\p,
$$
where $\varepsilon>0$ is a parameter which we will let tend to 0 at
the end of the calculation. Here
$\cP_{\phi_0}^\perp$ as before is the projection orthogonal to
$\phi_0$ and this time
$$
f(|\p|)=\frac{1}{2}\left(\frac{\p^4+8\pi\mu^{-1}\uprho}{\p^2
    \left(\p^4+16\pi\mu^{-1}\uprho\right)^{1/2}}-1\right)
$$
and
$$
\theta_{\p}(\x)=\sqrt{n\uprho^{-1}}\exp(i\p\cdot\x)\phi_0(\x).
$$
Note that $n\uprho^{-1}\phi_0(\x)^2$ is 1
on most of $\Lambda$. We then again have the Berezin-Lieb inequality
as before.
We also find that
\begin{eqnarray*}
  \uprho_\gamma(\x)&=&(2\pi)^{-3}\int_{|p|>\varepsilon\rho^{1/4}}
f(|\p|)d\p
n\uprho^{-1}\phi_0(\x)^2\left(1+O(\varepsilon^{-1}\uprho^{-1/4}L^{-1})\right)\\
  &=&A_\varepsilon(\rho/\mu)^{3/4}n\uprho^{-1}\phi_0(\x)^2\left(1+O(\varepsilon^{-1}\uprho^{-1/4}L^{-1})\right),
\end{eqnarray*}
where $A_\varepsilon$ is an explicit function of $\varepsilon$.
We now choose $\lambda_0$ such that
$\lambda_0^2=n(1-A_\varepsilon\uprho^{-1/4}\mu^{-3/4})$, i.e., such
that
$$
\lambda_0^2\phi_0^2(\x)+\uprho_\gamma(\x)=
n\phi_0(\x)^2(1+O(\varepsilon^{-1}\uprho^{-1/2}L^{-1}))\approx\uprho.
$$
It is easy to see that the first term in (\ref{eq:upperen}) is of
order $\uprho L^3(rL)^{-1}$ and the fourth term in (\ref{eq:upperen})
is of order $\uprho L^3(\varepsilon^{-2}+\uprho r^2)$.  We may choose
$r$, depending on $L$, in such a way that after dividing by $\uprho
L^3$ and letting
$L\to\infty$ only the error $\varepsilon^{-2}$ remains. This
allows choosing $\varepsilon\ll\rho^{-1/8}$.

To estimate the second term in (\ref{eq:upperen}) we use
Hardy's inequality to deduce
$$
\iint\frac{|\gamma(\x,\y)|^2}{|\x-\y|}d\x d\y\leq2(\Tr\gamma^2)^{1/2}
\Tr(-\Delta\gamma^2)^{1/2},
$$
and these terms can be easily estimated using the Berezin-Lieb
inequality\index{Berezin-Lieb inequality} in the direction opposite
from before, since we are interested now in an upper bound. The third
term in (\ref{eq:upperen}) is controlled in exactly the same way as
the second term.  We are then left with the last two terms in
(\ref{eq:upperen}). They are treated in exactly the same way as for
the two-component gas again using the Berezin-Lieb inequality.
\newpage
\
\thispagestyle{empty}

\chapter[BE Quantum Phase Transition in an
Optical Lattice Model]{Bose-Einstein Quantum Phase\vspace*{-2mm}\newline Transition in an
Optical Lattice\vspace*{-2mm}\newline Model}\label{xychap}
\index{quantum phase transition}
\index{optical lattice}

\section{Introduction}
One of the most remarkable recent developments in the study of
ultracold Bose gases is the observation of a reversible transition
from a Bose-Einstein condensate to a state composed of localized atoms
as the strength of a periodic, optical trapping potential is varied
\cite{G1,G2}.  This is an example of a quantum phase transition
\cite{Sa} where quantum fluctuations and correlations rather than
energy-entropy competition is the driving force and its theoretical
understanding is quite challenging.  The model usually considered for
describing this phenomenon is the Bose-Hubbard model\index{Bose-Hubbard model}  and the
transition is interpreted as a transition between a superfluid and a
{\it Mott insulator}\index{Mott insulator} that was studied in \cite{FWGF} with an
application to ${\rm He}^4$ in porous media in mind.  The possibility
of applying this scheme to gases of alkali atoms in optical traps was
first realized in \cite{JBCGZ}.  The article \cite{Z} reviews these
developments and many recent papers, e.g.,
\cite{GCZKSD,Ziegler,NS,G,Zi,DODS,RBREWC,MA,Auer} are devoted to this
topic.  These papers contain also further references to earlier work
along these lines.

The investigations of the phase transition in the Bose-Hubbard model
are mostly based on variational or numerical methods and the signal of
the phase transition is usually taken to be that an ansatz with a
sharp particle number at each lattice site leads to a lower energy
than a delocalized Bogoliubov state.  On the other hand, there exists
no rigorous proof, so far, that the true ground state of the model has
off-diagonal long range order at one end of the parameter regime that
disappears at the other end.  In this chapter, which is based on the
paper \cite{ALSSY}, we study a slightly different model where just
this phenomenon can be rigorously proved and which, at the same time,
captures the salient features of the experimental situation.

Physically, we are dealing with a trapped Bose gas with short range
interaction like in Chapters~\ref{intro}--\ref{1dsect}. The model we
discuss, however, is not a continuum model but rather a lattice
gas\index{lattice gas}, i.e., the particles are confined to move on a
$d$-dimensional, hypercubic lattice and the kinetic energy is given by
the discrete Laplacian. Moreover, when discussing BEC, it is
convenient not to fix the particle number but to work in a
grand-canonical ensemble\index{grand-canonical}. The chemical
potential is fixed in such a way that the average particle number
equals half the number of lattice sites, i.e., we consider {\it half
  filling}\index{half-filling}.  (This restriction is dictated by our
method of proof.)  The optical lattice is modeled by a periodic,
one-body potential. In experiments the gas is enclosed in an
additional trap potential that is slowly varying on the scale of the
optical lattice but we neglect here the inhomogeneity due to such a
potential and consider instead the thermodynamic limit.

In terms of bosonic creation\index{creation operator} and annihilation
operators\index{annihilation operator}, $a^\dagger_\x$ and
$a^{\phantom\dagger}_\x$, our Hamiltonian\index{Hamiltonian} is
expressed as
\begin{equation}
H= - \half \sum_{\langle \vecx\vecy\rangle} ( a^\dagger_\vecx
a^{\phantom\dagger}_\vecy + a^{\phantom\dagger}_\vecx a^\dagger_\vecy
) + \lambda \sum_\vecx (-1)^\vecx a^\dagger_\vecx a^{\phantom\dagger}_\vecx
+ U \sum_\vecx a^\dagger_\vecx a^{\phantom\dagger}_\vecx
(a^\dagger_\vecx a^{\phantom\dagger}_\vecx-1). \label{hamiltonian}
\end{equation}
The sites $\vecx$ are in a cube $\Lambda\subset \Z^d$ with opposite
sides identified (i.e., a $d$-dimensional torus) and $\langle
\vecx\vecy\rangle$ stands for pairs of nearest neighbors. Units are
chosen such that $\hbar^2/m =1$.

The first term in \eqref{hamiltonian} is the discrete Laplacian
$\sum_{\langle
\vecx\vecy\rangle}(a^\dagger_\x-a^\dagger_\y)
(a^{\phantom\dagger}_\x-a^{\phantom\dagger}_\y)$
minus $2d\sum_{\x} a^\dagger_\vecx a^{\phantom\dagger}_\x$, i.e., we
have subtracted a chemical potential\index{chemical potential} that equals $d$.

The optical lattice gives rise to a potential $\lambda (-1)^{\x}$
which alternates in sign between the $A$ and $B$ sublattices of even
and odd sites.  The inter-atomic on-site repulsion is $U$, but we consider
here only the case of a {\it hard core interaction\index{hard core interaction}}, i.e., $U=\infty$. If
$\lambda = 0$ but $U < \infty$ we have the Bose-Hubbard model.  Then
all sites are equivalent and the lattice represents the attractive
sites of the optical lattice. In our case the adjustable parameter is
$\lambda$ instead of $U$ and for large $\lambda$ the atoms will try to
localize on the $B$ sublattice. The Hamiltonian \eqref{hamiltonian}
conserves the particle number $N$ and it can be shown that, for
$U=\infty$, the lowest
energy is obtained uniquely for $N=\half|\Lambda|$, i.e., half the number of
lattice sites.  Because of the periodic potential the unit cell in
this model consists of two lattice sites, so that we have on average
one particle per unit cell. This corresponds, physically, to filling
factor 1 in the Bose-Hubbard model.

In contrast to the previous chapters we no longer restrict our
attention to the ground state of the system but consider more
generally thermal equilibrium states\index{thermal equilibrium states}
at some nonnegative temperature\index{temperature}
$T$. These states are described by the Gibbs density matrices\index{Gibbs density matrix}
$Z^{-1}\exp(-\beta H)$ with $Z$ the normalization factor (partition
function) and $\beta=1/T$ the inverse temperature. Units are chosen so
that Boltzmann's constant equals 1. The thermal expectation value of
some observable $\mathcal{O}$ will be denoted by $\langle{\mathcal
O}\rangle=Z^{-1}\Tr {\mathcal O}\exp(-\beta H)$.

Our main results about this model can be summarized as follows:

\begin{itemize}

\item[1.]\label{item1} If $T$ and $\lambda$ are both small, there is
  Bose-Einstein condensation\index{Bose-Einstein condensation}.  In
  this parameter regime the one-body density matrix
  $\gamma(\x,\y)=\langle a^\dagger_\x a^{\phantom \dagger}_\y \rangle$
  has exactly one large eigenvalue (in the thermodynamic limit), and
  the corresponding condensate wave function\index{condensate wave function} is $\phi(\x)=$constant.

\item[2.]\label{item2} If either $T$ or $\lambda$ is big enough, then
   the one-body density matrix decays exponentially with the distance
   $|\x-\y|$, and hence there is {\it no BEC}.  In particular, this
   applies to the ground state $T=0$ for $\lambda$ big enough, where
   the system is in a Mott insulator\index{Mott insulator} phase.

\item[3.]\label{item3}
The Mott insulator phase is characterized by a gap, i.e., a jump in
the chemical potential. We are able to prove this, at half-filling, in
the region described in item~2 above.  More precisely, there
is a cusp in the dependence of the ground state energy on the number
of particles; adding or removing one particle costs a non-zero amount
of energy.  We also show that there is no such gap whenever there is
BEC.

\item[4.] The interparticle interaction is essential for items~2 and~3.
Non-interacting bosons {\it always display BEC} for low, but
   positive $T$ (depending on $\lambda$, of course).

\item[5.] For all $T\geq 0$ and all $\lambda > 0$ the diagonal part of
   the one-body density matrix $\langle a^\dagger_\x a^{\phantom
     \dagger}_\x \rangle$ (the one-particle density) is {\it not
     constant}. Its value on the A sublattice is constant, but strictly
   less than its constant value on the B sublattice and this
   discrepancy survives in the thermodynamic limit. In contrast, in the
   regime mentioned in item~1, the off-diagonal long-range
   order is constant, i.e., $\langle a^\dagger_\x a^{\phantom
     \dagger}_\y \rangle \approx \phi(\x) \phi(\y)^*$ for large
   $|\x-\y|$ with $\phi(\x)=$constant.
\end{itemize}

\begin{figure}[htf]
\center
\includegraphics[width=9cm, height=6.6cm]{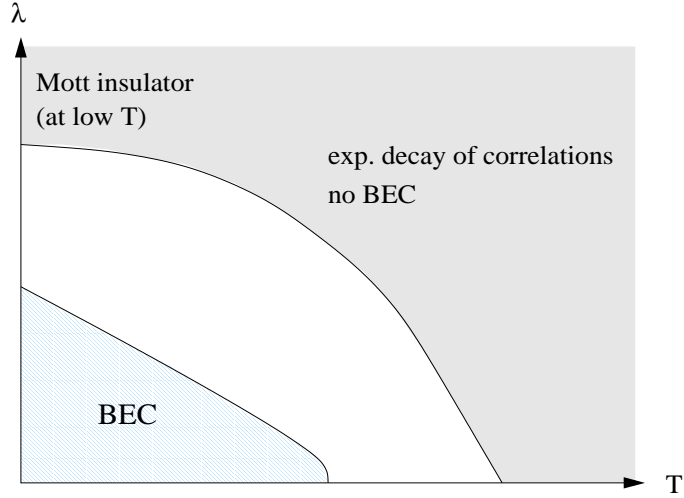}
\caption{Schematic phase diagram at half-filling}
\label{fig1}
\end{figure}

Because of the hard-core interaction between the particles, there is
at most one particle at each site and our Hamiltonian (with
$U=\infty$) thus acts on the Hilbert space ${\mathcal
H}=\bigotimes_{\x\in\Lambda}{\mathbb C}^2$.  The creation and
annihilation operators can be represented as $2\times 2$ matrices with
$$ a^\dagger_{\vecx}\leftrightarrow \left( \begin{array}{cc} 0 & 1 \\
0 & 0
\end{array} \right) , \quad a^{\phantom\dagger}_{\vecx}\leftrightarrow
\left( \begin{array}{cc} 0 & 0 \\ 1 & 0 \end{array} \right) , \quad
a^\dagger_{\vecx}a^{\phantom\dagger}_\vecx\leftrightarrow \left(
\begin{array}{cc} 1 & 0 \\ 0 & 0 \end{array} \right),
$$
for each $\vecx\in\Lambda$. More precisely, these matrices act on
the tensor factor associated with the site $\x$ while
$a^\dagger_{\vecx}$ and $a^{\phantom\dagger}_{\vecx}$ act as the
identity on the other
factors in the Hilbert space ${\mathcal
H}=\bigotimes_{\x\in\Lambda}{\mathbb C}^2$.

The Hamiltonian can alternatively be written in terms of the spin 1/2
operators $$ S^1=\frac12 \left(
\begin{array}{cc} 0 & 1 \\ 1 & 0 \end{array} \right), \quad S^2=\frac12\left(
\begin{array}{cc} 0 & -{\rm i} \\ {\rm i} & 0 \end{array} \right), \quad
S^3=\frac12\left( \begin{array}{cc} 1 & 0 \\ 0 & -1 \end{array}
\right). $$ The correspondence with the creation and annihilation
operators is $$ a^\dagger_{\vecx}=S^1_{\vecx}+{\rm i}S^2_{\vecx}\equiv
S_\vecx^+, \quad a^{\phantom\dagger}_{\vecx}=S^1_{\vecx}-{\rm
i}S^2_{\vecx}\equiv S_\vecx^-, $$ and hence
$a^\dagger_{\vecx}a^{\phantom\dagger}_{\vecx}=S_\vecx^3+\half$.  (This
is known as the Matsubara-Matsuda correspondence \cite{MM}.)  Adding a
convenient constant to make the periodic potential positive, the
Hamiltonian (\ref{hamiltonian}) for $U=\infty$ is thus equivalent to
\begin{eqnarray}\label{hamspin}
H&=&-\half \sum_{\langle \vecx\vecy\rangle} (S^+_\vecx
S^-_\vecy+S^-_\vecx S^+_\vecy)+\lambda\sum_\vecx \left[\half +
(-1)^\vecx S^3_\vecx\right]
\nonumber\\
&=&-\sum_{\langle \vecx\vecy\rangle}(S^1_\vecx S^1_\vecy+S^2_\vecx
S^2_\vecy) +\lambda\sum_\vecx\left[ \half + (-1)^\vecx
S^3_\vecx\right].
\end{eqnarray}
Without loss of generality we may assume $\lambda\geq 0$. This
Hamiltonian is well known as a model for interacting spins, referred
to as the XY model~\cite{DLS}\index{XY model}. The last term has the interpretation of a
staggered magnetic field. We note that BEC for the lattice gas is
equivalent to off-diagonal long range order for the 1- and 2-components
of the spins.

The Hamiltonian \eqref{hamspin} is clearly invariant under
simultaneous rotations of all the spins around the 3-axis. In particle
language this is the $U(1)$ gauge symmetry\index{gauge symmetry} associated with particle
number conservation of the Hamiltonian \eqref{hamiltonian}.
Off-diagonal long range order (or, equivalently, BEC) implies that
this symmetry is spontaneously broken in the state under
consideration.\footnote{See the discussion at the end of
Appendix~\ref{justapp}.} It is notoriously difficult to prove such
symmetry breaking\index{symmetry breaking} for systems with a continuous symmetry. One of the
few available techniques is that of {\it reflection positivity}\index{reflection positivity} (and
the closely related property of {\it Gaussian domination})\index{Gaussian domination} and
fortunately it can be applied to our system. For this, however, the
hard core and half-filling conditions are essential because they imply
a particle-hole symmetry\index{particle-hole symmetry} that is crucial for the proofs to work.
Naturally, BEC is expected to occur at other fillings, but no one has
so far found a way to prove condensation (or, equivalently, long-range
order in an antiferromagnet with continuous symmetry) without using
reflection positivity and infrared bounds, and these require the
additional symmetry.

Reflection positivity was first formulated by K.\ Osterwalder and R.\
Schrader \cite{oster} in the context of relativistic quantum field theory.
Later, J. Fr\"ohlich, B. Simon and T. Spencer used the concept to
prove the existence of a phase transition for a classical spin model
with a continuous symmetry \cite{FSS}, and E.\ Lieb and J.\ Fr\"ohlich
\cite{FL} as well as F.\ Dyson, E.\ Lieb and B.\ Simon \cite{DLS}
applied it for the analysis of quantum spin systems. The proof of
off-diagonal long range order for the Hamiltonian \eqref{hamspin} (for
small $\lambda$) given here is based on appropriate modifications of
the arguments in \cite{DLS}.

\section{Reflection Positivity}
\index{reflection positivity}

In the present context reflection positivity means the following. We
divide the torus $\Lambda$ into two congruent parts, $\Lambda_{\rm L}$
and $\Lambda_{\rm R}$, by cutting it with a hyperplane orthogonal to
one of the $d$ directions. (For this we assume that the side length of
$\Lambda$ is even.) This induces a factorization of the Hilbert space,
${\mathcal H}={\mathcal H}_{\rm L}\otimes {\mathcal H}_{\rm R}$, with
$${\mathcal H}_{\rm L,R}=\bigotimes_{\x\in\Lambda_{\rm L,R}}{\mathbb
C}^2.$$ There is a natural identification between a site
$\x\in\Lambda_{\rm L}$ and its mirror image $\vartheta\x
\in\Lambda_{\rm R}$. If $F$ is an operator on ${\mathcal H}={\mathcal
   H}_{\rm L}$ we define its reflection $\theta F$ as an operator on
   ${\mathcal H}_{\rm R}$ in the following way. If $F=F_\x$ operates
   non-trivially only on one site, $\x\in \Lambda_{\rm L}$, we define
   $\theta F=VF_{\vartheta\x}V^\dagger$ where $V$ denotes the unitary
   particle-hole transformation or, in the spin language, rotation by
   $\pi$ around the 1-axis. This definition extends in an obvious
   way to products of operators on single sites and then, by linearity,
   to arbitrary operators on ${\mathcal H}_{\rm L}$. Reflection
   positivity of a state $\langle\,\cdot\,\rangle$ means that
\begin{equation}\label{reflpos}
\langle
  F\theta \overline F\rangle \geq 0
\end{equation}
  for any $F$ operating on
   ${\mathcal H}_{\rm L}$. Here $\overline F$ is the complex
   conjugate of the operator $F$ in the matrix representation defined
   above, i.e., defined by the basis where the operators $S^3_{\x}$ are
   diagonal.

We now show that reflection positivity holds for any thermal
equilibrium state of our Hamiltonian. We can write the
Hamiltonian~(\ref{hamspin}) as
\beq\label{hamreal}
H=H_{\rm L} + H_{\rm R}-\half \sum_{\langle \vecx\vecy\rangle\in
M}(S^+_\vecx S^-_\vecy +S^-_\vecx
S^+_\vecy),
\eeq
where $H_{\rm L}$ and $H_{\rm R}$ act non-trivially
only on ${\mathcal H}_{\rm L}$ and ${\mathcal H}_{\rm R}$,
respectively. Here, $M$ denotes the set of bonds going from the left
sublattice to the right sublattice. (Because of the periodic
boundary condition these include the bonds that connect the right
boundary with the left boundary.) Note that $H_{\rm R} = \theta H_{\rm L}$, and
$$
\sum_{\langle \vecx\vecy\rangle\in M}(S^+_\vecx S^-_\vecy +S^-_\vecx
S^+_\vecy)= \sum_{\langle \vecx\vecy\rangle\in M}(S^+_\vecx \theta
S^+_\vecx +S^-_\vecx
\theta S^-_\vecx).
$$
For these properties it is essential that we included the unitary
particle-hole transformation $V$ in the definition of the reflection
$\theta$. For reflection positivity it is also important that  all
operators appearing in $H$ (\ref{hamreal}) have a {\it real} matrix
representation. Moreover, the minus sign in (\ref{hamreal}) is
essential.

Using the Trotter product formula, we have
$$
\Tr F \theta \overline F e^{-\beta H} = \\ \lim_{n\to\infty} \Tr F
\theta \overline F \, {\mathcal Z}_n
$$
with
\beq\label{trotter}
{\mathcal Z}_n = \left[ e^{-\frac 1n \beta H_L} \theta e^{-\frac 1n
\beta H_L} \prod_{\langle \vecx\vecy\rangle\in
M}\left(1+\frac{\beta}{2n} \left[ S^+_\vecx \theta S^+_\vecx
+S^-_\vecx
\theta S^-_\vecx)\right] \right) \right]^n.
\eeq
Observe that ${\mathcal Z}_n$ is a sum of terms of the form
\beq\label{aaa}
\mbox{$\prod_i$} A_i \theta A_i,
\eeq
with $A_i$ given by either $e^{-\frac 1n \beta H_L}$ or
$\sqrt{\frac{\beta}{2n}}S^+_\vecx$ or
$\sqrt{\frac{\beta}{2n}}S^-_\vecx$. All the $A_i$ are real matrices,
and therefore
\beq
\Tr_{\mathcal H}\,  F \theta \overline F\, \mbox{$\prod_i $}A_i
\theta  A_i =
\Tr_{\mathcal H}\,  F \mbox{$\prod_i $}A_i \, \theta\left[
\overline F \mbox{$\prod_j$} A_j\right] = \big| \Tr_{{\mathcal
H}_{\rm L}}\, F \mbox{$\prod_i$} A_i \big|^2 \geq 0.
\eeq
Hence $\Tr F\theta\overline F\, {\mathcal Z}_n$  is a sum of
non-negative terms and therefore non-negative. This proves our
assertion.

\section{Proof of BEC for Small $\lambda$ and $T$}

The main tool in our proof of BEC are {\it infrared
  bounds}\index{infrared bounds}. More precisely, for $\p\in
\Lambda^*$ (the dual lattice of $\Lambda$), let $\widetilde
S^\#_\p=|\Lambda|^{-1/2} \sum_\x S_\x^\# \exp({\rm i} \p\cdot \x)$
denote the Fourier transform of the spin operators. We claim that
\beq\label{infb} (\widetilde S_\p^1, \widetilde S^1_{-\p}) \leq
\frac{T}{ 2 E_\vecp}, \eeq with $E_\vecp= \sum_{i=1}^d (1-\cos(p_i))$.
Here, $p_i$ denotes the components of $\vecp$, and $(\, ,\, )$ denotes
the Duhamel two point function\index{Duhamel two-point function} at
temperature $T$, defined by \beq (A,B)=\int _0^1 \Tr\left( A e^{-s
    \beta H} B e^{-(1-s)\beta H} \right) ds / \Tr e^{-\beta H} \eeq
for any pair of operators $A$ and $B$.  Because of invariance under
rotations around the $S^3$ axis, (\ref{infb}) is equally true with
$S^1$ replaced by $S^2$, of course.

The crucial lemma ({\it Gaussian domination})\index{Gaussian domination} is the following.
Define, for a complex valued function $h$ on the bonds $\langle
\vecx\vecy \rangle$ in $\Lambda$,
\beq
Z(h)=\Tr \exp\left[ - \beta K(h) \right],
\eeq
with $K(h)$ the modified Hamiltonian
\beq
K(h)= \frac 14 \sum_{\langle \vecx\vecy\rangle} \left(
\big(S^+_\vecx-S^-_\vecy-h_{\vecx\vecy}\big)^2
+ \big(S^-_\vecx-S^+_\vecy-\overline{h_{\vecx\vecy}}\big)^2\right)+
\lambda\sum_\vecx \big[\half + (-1)^\vecx S^3_\vecx\big].
\eeq
Note that for $h\equiv 0$, $K(h)$ agrees with the Hamiltonian $H$,
because $(S^{\pm})^2=0$. We claim that, for any {\em real valued} $h$,
\beq\label{gauss}
Z(h)\leq Z(0).
\eeq
The infrared bound then follows from $d^2 Z(\varepsilon
h)/d\varepsilon^2|_{\eps=0}\leq 0$, taking $h_{\x\y}=\exp({\rm
i}\p\cdot \x)- \exp({\rm i}\p\cdot \y)$. This is not a real function,
though, but the negativity of the (real!) quadratic form $d^2
Z(\varepsilon
h)/d\varepsilon^2|_{\eps=0}$ for real $h$ implies negativity also for
complex-valued $h$.

The proof of (\ref{gauss}) is very similar to the proof of the
reflection positivity\index{reflection positivity} property (\ref{reflpos}) given above. It follows
along the same lines as in \cite{DLS}, but we repeat it here for
convenience of the reader.

The intuition behind (\ref{gauss}) is the following. First, in
maximizing $Z(h)$ one can restrict to gradients, i.e., $h_{\x\y}= \hat
h_\x-\hat h_\y$ for some function $\hat h_\x$ on $\Lambda$. (This
follows from stationarity of $Z( h)$ at a maximizer $h_{\rm max})$.)
Reflection positivity implies that $\langle A\theta \overline
B\rangle$ defines a scalar product on operators on $\mathcal H_{\rm
   L}$, and hence there is a corresponding Schwarz inequality.
Moreover, since reflection positivity holds for reflections across
{\it any} hyperplane, one arrives at the so-called {\it chessboard
   inequality}, which is simply a version of Schwarz's inequality for
multiple reflections across different hyperplanes. Such a chessboard
estimate implies that in order to maximize $Z(h)$ it is best to choose
the function $\hat h_\vecx$ to be constant. In the case of classical
spin systems \cite{FSS}, this intuition can be turned into a complete
proof of (\ref{gauss}).
Because of non-commutativity of $K(h)$
with $K(0)=H$, this is not possible in the quantum case. However, one
can proceed by using the Trotter formula as follows.

Let $h_{\rm max}$ be a function that maximizes $Z(h)$ for real valued
$h$. If there is more than one maximizer, we choose $h_{\rm max}$ to
be one that vanishes on the largest number of bonds. We then have to
show that actually $h_{\rm max}\equiv 0$. If $h_{\rm max}\not\equiv
0$, we draw a hyperplane such that $h_{\vecx\vecy}\neq 0$ for at
least one pair $\langle \vecx\vecy\rangle$ crossing the plane. We can
again write
\beq
K(h) = K_L(h) + K_R(h)  + \frac 14 \sum_{\langle \vecx\vecy\rangle\in
M} \left(  (S^+_\vecx-S^-_\vecy-h_{\vecx\vecy})^2
+ (S^-_\vecx-S^+_\vecy-h_{\vecx\vecy})^2\right).
\eeq
Using the Trotter formula, we have  $Z(h)=\lim_{n\to\infty} \alpha_n$,
with
\beq\label{tro}
\alpha_n= \Tr \left[ e^{-\beta K_L/n} e^{-\beta K_R/n} \prod_{\langle
\x\y\rangle\in M} e^{-\beta (S^+_\vecx-S^-_\vecy-h_{\vecx\vecy})^2 /4
n} e^{-\beta (S^-_\vecx-S^+_\vecy-h_{\vecx\vecy})^2 /4 n}\right]^n .
\eeq
For any matrix, we can write
\beq
e^{-D^2} = (4\pi)^{-1/2} \int_\R dk\, e^{{\rm i}k D} e^{-k^2/4}.
\eeq
If we apply this to the last two factors in (\ref{tro}), and note
that $S^-_\y = \theta S^+_\x$ if $\langle\x\y\rangle\in M$. Denoting
by $\x_1, \dots, \x_l$ the points on the left side of the bonds in
$M$, we have that
\begin{eqnarray}\nonumber
\alpha_n &=& (4\pi)^{-n l} \int_{R^{2nl}} d^{2nl} k \, \Tr
\left[  e^{-\beta K_L/n} e^{-\beta K_R/n} e^{{\rm i} k_1 (S^+_{\x_1}
- \theta S^+_{\x_1}) \beta^{1/2} /2  n^{1/2}}  \dots \right] \\ &&
\times e^{- k^2/4 } e^{-{\rm i} k_1 h_{\x_1 \vartheta \x_1}
\beta^{1/2}/2n^{1/2} \dots}.
\end{eqnarray}
Here we denote $k^2 =\sum k_i^2$ for short.
Since matrices on the right of $M$ commute with matrices on the left,
and since all matrices in question are {\it real}, we see that the
trace in the integrand above can be written as
\beq
\Tr \left[  e^{-\beta K_L/n} e^{{\rm i} k_1 S^+_{\x_1} \beta^{1/2}/2
n^{1/2}}  \dots \right] \overline{ \Tr \left[  e^{-\beta K_R/n}
e^{{\rm i} k_1 \theta S^+_{\x_1} \beta^{1/2} /2 n^{1/2}}  \dots
\right]}.
\eeq
Using the Schwarz inequality for the $k$ integration, and \lq
undoing\rq\ the above step, we see that
\begin{eqnarray}\nonumber
\!\!\!\!\! |\alpha_n|^2 \!\!\! &\leq& \!\!\!\left( (4\pi)^{-n l}
\int_{R^{2nl}} d^{2nl} k \, e^{- k^2/4} \right.\\ \nonumber && \left.
\quad \times\Tr
\left[  e^{-\beta K_L/n} e^{-\beta \theta K_L/n} e^{{\rm i} k_1
(S^+_{\x_1} - \theta S^+_{\x_1})\beta^{1/2} /2 n^{1/2}}  \dots
\right] \phantom{\int_{R^{2nl}} } \!\!\!\!\! \!\!\!\!\!\!\!\!\!
\right) \\ \nonumber && \!\!\!\!\!\times  \left( (4\pi)^{-n l}
\int_{R^{2nl}} d^{2nl} k \, e^{- k^2/4} \right.\\ && \left. \quad
\times \Tr
\left[  e^{-\beta \theta K_R/n} e^{-\beta K_R/n} e^{{\rm i} k_1
(S^+_{\x_1} - \theta S^+_{\x_1}) \beta^{1/2} /2 n^{1/2}}  \dots
\right] \phantom{\int_{R^{2nl}} } \!\!\!\!\! \!\!\!\!\!\!\!\!\!
\right).
\end{eqnarray}
In terms of the partition function $Z(h)$, this means that
\beq
|Z(h_{\rm max})|^2 \leq Z(h^{(1)}) Z(h^{(2)}),
\eeq
where $h^{(1)}$ and $h^{(2)})$ are obtained from $h_{\rm max}$ by
reflection across $M$ in the following way:
\beq
h^{(1)}_{\x\y} = \left\{ \begin{array}{ll} h_{\x\y} & {\rm if\ } \x,
\y\in \Lambda_L \\ h_{\vartheta\x\vartheta\y} & {\rm if\ } \x, \y\in
\Lambda_R \\ 0 & {\rm if\ } \langle \x \y\rangle \in M \end{array}
\right.
\eeq
and $h^{(2)}$ is given by the same expression, interchanging $L$ and $R$.
Therefore also $h^{(1)}$ and $h^{(2)}$ must be maximizers of $Z(h)$.
However, one of them will contain strictly more zeros than $h_{\rm
max}$, since $h_{\rm max}$ does not vanish identically for bonds
crossing $M$. This contradicts our assumption that $h_{\rm max}$
contains the maximal number of zeros among all maximizers of $Z(h)$.
Hence $h_{\rm max}\equiv 0$ identically. This completes the proof of
(\ref{gauss}).

The next step is to transfer the upper bound on the Duhamel two point
function (\ref{infb}) into an upper bound on the
thermal expectation value. This involves convexity arguments and
estimations of double commutators like in Section~3 in \cite{DLS}.
For this purpose, we have to evaluate the double commutators
\begin{equation}
   [\widetilde S^1_\vecp,[H,\widetilde S^1_{-\vecp}]]+
[\widetilde S^2_\vecp,[H,\widetilde S^2_{-\vecp}]]=-\frac 2 {|\Lambda|}
       \Big(H - \half \lambda|\Lambda| + 2\sum_{\langle
\vecx\vecy\rangle} S^3_\vecx
S^3_\vecy \cos \vecp\cdot(\vecx-\vecy)\Big).
\end{equation}
Let $C_\vecp$ denote the expectation value of this last expression,
$$
C_\vecp= \langle [\widetilde S^1_\vecp,[H,\widetilde
S^1_{-\vecp}]]+[\widetilde S^2_\vecp,[H,\widetilde S^2_{-\vecp}]]
\rangle\geq 0.
$$
The positivity of $C_\vecp$ can be seen from an
eigenfunction-expansion of the trace.  {F}rom \cite[Corollary~3.2 and
Theorem~3.2]{DLS} and (\ref{infb}) we infer that
\begin{equation}\label{dlsb}
\langle \widetilde S_\vecp^1 \widetilde S_{-\vecp}^1 + \widetilde S_\vecp^2
\widetilde S_{-\vecp}^2\rangle\leq \frac 12 \sqrt {
\frac  {C_\vecp}{E_\vecp}} \coth \sqrt{\beta^2 C_\vecp E_\vecp /4}.
\end{equation}
Using $\coth x \leq 1+1/x$ and Schwarz's inequality, we obtain for the
sum over all $\vecp\neq \0$,
\begin{equation}
     \sum_{\vecp\neq \0}\langle \widetilde S_\vecp^1 \widetilde S_{-\vecp}^1 +
      \widetilde S_\vecp^2 \widetilde S_{-\vecp}^2\rangle \leq \frac 1{\beta}
      \sum_{\vecp\neq \0} \frac 1{E_\vecp} + \frac 12 \Big(
\sum_{\vecp\neq \0} \frac
      1{E_\vecp} \Big)^{1/2} \Big( \sum_{\vecp\neq \0} C_\vecp
\Big)^{1/2}. \label{sump}
\end{equation}
We have $\sum_{\vecp\in\Lambda^*} C_\vecp = -2 \langle H
\rangle+\lambda|\Lambda|$, which can
be bounded from above using the following lower bound on the Hamiltonian:
\beq\label{lowH}
H\geq -\frac {|\Lambda|}4\left
[d(d+1)+4\lambda^2\right]^{1/2}+\half\lambda|\Lambda|.
\eeq
This inequality follows from the fact that the lowest eigenvalue of
\begin{equation}\label{sumt}
-\frac 12 S_\vecx^1 \sum_{i=1}^{2d} S_{\vecy_i}^1-\frac 12 S_\vecx^2
\sum_{i=1}^{2d} S_{\vecy_i}^2+ \lambda S_\vecx^3
\end{equation}
is given by $-\mbox{$\frac 14$}[d(d+1)+4\lambda^2]^{1/2}$. This can
be shown exactly in the same way as
\cite[Theorem~C.1]{DLS}. Since the Hamiltonian $H$ can be written as
a sum of terms like
(\ref{sumt}), with $\vecy_i$ the nearest neighbors of $\vecx$, we get from
this fact the lower bound (\ref{lowH}).

With the aid of the sum rule
$$
\sum_{\vecp\in \Lambda^*}\langle \widetilde S_\vecp^1 \widetilde S_{-\vecp}^1 +
\widetilde S_\vecp^2
\widetilde S_{-\vecp}^2\rangle=\frac {|\Lambda|}2
$$
(which follows from $(S^1)^2=(S^2)^2=1/4$), we obtain from
(\ref{sump}) and (\ref{lowH}) the following lower
bound in the thermodynamic limit:
\begin{eqnarray}\nonumber
      &&\lim_{\Lambda\to \infty} \frac 1{|\Lambda|} \langle \widetilde S_\0^1
      \widetilde S_{\0}^1 + \widetilde S_\0^2 \widetilde S_{\0}^2\rangle\\
      &&\geq \frac 12 -\frac12 \left(\half
        \left[d(d+1)+4\lambda^2\right]^{1/2} c_d \right)^{1/2} - \frac
      1{\beta} c_d, \label{frrom}
\end{eqnarray}
with $c_d$ given by
\beq
c_d=\frac 1{(2\pi)^{d}} \int_{ [-\pi,\pi]^d} d\vecp\frac 1{E_\vecp} .
\eeq
This is our final result. Note that $c_d$ is finite for $d\geq 3$.
Hence the right side of (\ref{frrom}) is positive, for large enough
$\beta$, as long as
$$
\lambda^2 < \frac 1{c_d^2}-\frac {d(d+1)}4.
$$
In $d=3$, $c_3\approx 0.505$ \cite{DLS}, and hence this condition is
fulfilled  for
$\lambda \lesssim 0.960$.  In \cite{DLS} it was also shown that $dc_d$
is monotone decreasing in $d$, which implies a similar result for all
$d>3$.

The connection with BEC is as follows.
Since $H$ is real, also $\gamma(\vecx,\vecy)$ is real and we have
$$
\gamma(\vecx,\vecy)= \langle S_\vecx^+ S_\vecy^ - \rangle = \langle
S_\vecx^1 S_\vecy^1+S_\vecx^2 S_\vecy^2
\rangle.
$$
Hence, if
$\varphi_0=|\Lambda|^{-1/2}$ denotes the constant function,
$$
\langle \varphi_0 |\gamma| \varphi_0\rangle =  \langle \widetilde S_\0^1
      \widetilde S_{\0}^1 + \widetilde S_\0^2 \widetilde S_{\0}^2\rangle,
$$
and thus the bound (\ref{frrom}) implies that the largest eigenvalue
of $\gamma(\x,\y)$ is bounded from below by the right side of
(\ref{frrom}). In addition one can show that the infrared
bounds imply that there is at most {\it one} large eigenvalue (of the
order $|\Lambda|$), and that the corresponding eigenvector (the \lq
condensate wave function\rq)\index{condensate wave function} is strictly constant in the
thermodynamic limit \cite{ALSSY}.
The constancy of the condensate wave function is surprising and is
not expected to hold for densities different from $\half$, where
particle-hole symmetry\index{particle-hole symmetry} is absent.
In contrast to the condensate wave function the particle density shows the
staggering of the periodic potential \cite[Thm.~3]{ALSSY}. It also
contrasts with the situation
for zero interparticle interaction, as discussed at the end of this chapter.

\bigskip

In the BEC phase there is {\it no gap} for adding particles beyond half
filling (in the thermodynamic limit): The ground state energy,
$E_{k}$, for $\half|\Lambda|+k$
particles satisfies
\beq\label{enfin}
0\leq E_{k}-E_{0}\leq\frac{(\rm const.)}{|\Lambda|}
\eeq
(with a constant that depends on $k$ but not on $|\Lambda|$.) The
proof of~(\ref{enfin})
is by a variational calculation, with a trial state of the form
$(\widetilde S^+_\0)^k |0\rangle$, where $|0\rangle$ denotes the
absolute ground state, i.e., the ground state for half filling. (This
is the unique ground state of the Hamiltonian, as can be shown using
reflection positivity. See Appendix~A in \cite{ALSSY}.)
Also, in the thermodynamic limit, the energy per site for a given
density, $e(\varrho)$, satisfies
\beq\label{enth}
e(\varrho)-e(\half)\leq \const (\varrho - \half)^2.
\eeq
Thus there is no cusp at $\varrho=1/2$. To show this, one takes a
trial state of the form
\beq
|\psi_\y\rangle= e^{{\rm i}\varepsilon \sum_\x S_\x^2} (S_\y^1+\half)|0\rangle.
\eeq
The motivation is the following: we take
the ground state and first project onto a given direction of $S^1$ on
some site $\vecy$. If there is long-range order, this should imply
that essentially all the spins point in this direction now. Then we
rotate slightly around the $S^2$-axis. The particle number should then
go up by $\eps|\Lambda|$, but the energy only by $\eps^2|\Lambda|$.
We refer to \cite[Sect.~IV]{ALSSY} for the details.

The absence of a gap in the case of BEC is not surprising, since a gap
is characteristic for a Mott insulator state\index{Mott insulator}. We
show the occurrence of a gap, for large enough $\lambda$, in the next
section.

\section{Absence of BEC and Mott Insulator Phase}
\index{Mott insulator}

The main results of this section are the following:
If either
\begin{itemize}
\item  $\lambda \geq 0$ and $T> d/(2 \ln 2)$, or
\item  $T\geq 0$ and $\lambda \geq 0$  such that $\lambda + |e(\lambda)|> d$,
with $e(\lambda)=$ ground state energy per site,
\end{itemize}
then there is exponential decay of correlations:
\beq
\gamma(\x,\y)\leq{\rm (const.)}\exp(-\kappa|\x-\y|)
\eeq
with $\kappa>0$. Moreover, for $T=0$, the ground state energy in a
sector of fixed particle number $N=\half |\Lambda|+k$, denoted by
$E_k$, satisfies
\beq
E_k + E_{-k} -2 E_0 \geq (\lambda + |e(\lambda)|-
d)|k| .
\eeq
I.e, for large enough $\lambda$ the chemical potential  has a jump at
half filling.

The derivation of these two properties is based on a path
integral representation of the equilibrium state at temperature $T$,
and of the
ground state which is obtained in the limit $T\to
\infty$.
density matrix.
The analysis starts from the observation
that the density operator $e^{-\beta H}$ has non-negative
matrix
elements in the basis in which $\{ S_{\x}^3\}$ are diagonal,
i.e. of
states with specified particle occupation numbers.
It is
convenient to focus on the dynamics
of the `quasi-particles'\index{quasi-particles} which
are defined so that the presence
of one at a site $\x$ signifies a
deviation there from the occupation state which minimizes the
potential-energy.
Since the Hamiltonian is $H=H_0+\lambda W$,
with $H_0$ the hopping term in (\ref{hamspin}) and $W$ the staggered
field, we define the quasi-particle number operators $n_\x$ as:
\beq
n_\x \ = \  \half+(-1)^\x S_{\x}^3=\begin{cases}
  a^\dagger_{\x}a_{\x} & \hbox{for $\x$ even} \\
    1-a^\dagger_{\x}a_{\x} & \hbox{for $\x$ odd}\end{cases} \,  .
\eeq
Thus $n_{\x}=1$ means presence of a particle if $\x$ is on the
A sublattice (potential maximum) and absence if $\x$ is on the
B sublattice (potential minimum).

The collection of the joint eigenstates of the occupation numbers, $\{
|\{n_\x\}\rangle \}$, provides a convenient basis for the Hilbert
space.  The functional integral representation of $\langle \{n_\x\}
|\, e^{-\beta (H_0+\lambda W)} \, |\{n_\x\}\rangle$ involves an integral
over configurations of quasi-particle loops in a {\em space $\times$
time} for which the (imaginary) `time' corresponds to a variable with
period $\beta$.  The fact that the integral is over a positive measure
facilitates the applicability of statistical-mechanics intuition and
tools.  One finds that the quasi-particles are suppressed by the
potential energy, but favored by the entropy, which enters this
picture due to the presence of the hopping term in $H$.  At large
$\lambda$, the potential suppression causes localization: long
`quasi-particle' loops are rare, and the amplitude for long paths
decays exponentially in the distance, both for paths which may occur
spontaneously and for paths whose presence is forced through the
insertion of sources, i.e., particle creation and annihilation
operators.  Localization is also caused by high temperature, since the
requirement of periodicity implies that at any site which participates
in a loop there should be at least two jumps during the short
`time' interval $[0,\beta)$ and the amplitude for even a single jump
is small, of order $\beta$.

The path integral described above is obtained through the Dyson
expansion\index{Dyson!expansion}
\begin{equation}\label{dyson}
    e^{t(A+B)}=e^{tA}\sum_{m\geq 0}\int_{0\leq t_{1}\leq
    t_{2}\leq \cdots \leq t_m\leq t}B(t_{m})\cdots B(t_{1})dt_{1}\cdots dt_{m}
\end{equation}
for any matrices $A$ and $B$ and $t>0$, with $B(t)=e^{-tA}Be^{tA}$.
(The $m=0$ term in the sum is interpreted here as $1$.)

In
evaluating the matrix elements of $e^{-\beta H} \ = \
e^{-\beta
(H_0+\lambda W)}$,  in the basis $\{  |\{n_\x\}\rangle \}$,
we note
that $W$ is diagonal and
$\langle \{n_\x\}| H_0 |\{n'_\x\}\rangle$
are non-zero only if the configurations  $\{n_\x\}$ and $\{n'_\x\}$
differ at exactly
one nearest neighbor pair of sites where the
change corresponds
to either a creation of a pair of quasi-particles
or the
annihilation of such a pair.
I.e., the matrix elements is
zero unless $n_\x=n_\x'$ for all $\x$
except for a nearest neighbor
pair $\langle \x\y\rangle$,
where $n_\x=n_\y$, $n'_\x=n'_\y$, and
$n_\x+n'_\x=1$.
In this case, the matrix element equals
$-1/2$.

Introducing intermediate states, the partition function\index{partition function} can
thus be
written as follows:
\begin{eqnarray}\nonumber
\Tr\, e^{-\beta H} \!\!\! &=& \!\!\! \sum_{m=0}^\infty \int_{0\leq t_{1}\leq
    t_{2}\leq \cdots\leq t_m \leq  \beta}
\sum_{|\{n^{(i)}_\x\}\rangle, \, 1\leq i \leq m}
\\ \nonumber && \times \exp\left( -\lambda  \sum_{i=1}^{m}
(t_{i}-t_{i-1}) \sum_\x n_\x^{(i)} \right)dt_{1}\cdots dt_{m}
\\ \nonumber && \times (-1)^m \langle \{n^{(1)}_\x\} | H_0 |
\{n^{(m)}_\x\}\rangle  \langle \{n^{(m)}_\x\} |H_0|
\{n^{(m-1)}_\x\}\rangle \\ &&\quad \times \langle
\{n^{(m-1)}_\x\}|H_0|   \{n^{(m-2)}_\x\}\rangle \cdots  \langle
\{n^{(2)}_\x\}|H_0| |\{n^{(1)}_\x\}\rangle   \label{part}
\end{eqnarray}
with the interpretation $t_0=t_m-\beta$. Note that the factor in the
last two lines of (\ref{part}) equals $(1/2)^m$ if adjacent elements
in the sequence of configurations $\{n^{(i)}_\x\}$ differ by exactly
one quasi-particle pair, otherwise it is zero.

Expansions of this type are explained more fully in \cite{AN}. A
compact way of writing (\ref{part}) is:
\beq \label{pathi}
\Tr\, e^{-\beta H}= \int
v(d\omega)e^{-\lambda|\omega|}.
\eeq
Here the \lq path\rq\ $\omega$
stands for a set of disjoint oriented loops in the \lq space-time\rq\
$\Lambda \times [0,\beta]$, with periodic boundary conditions in \lq
time\rq. Each $\omega$ is parametrized by a number of jumps, $m$,
jumping times $0\leq t_1\leq t_2\leq \dots \leq t_m \leq \beta$, and a
sequence of configurations $\{n^{(i)}_\x\}$, which is determined by
the initial configuration $\{n^{(1)}_\x\}$ plus a sequence of \lq
rungs\rq\ connecting nearest neighbor sites, depicting the creation
or annihilation of a pair of neighboring quasi-particles (see
Fig.~\ref{loopfig}).
As in Feynman's picture of QED,
it is convenient to
regard such an event as a jump of the
quasi-particle, at which its
time-orientation is also reversed.
The length of $\omega$,
denoted by $|\omega|$, is the sum of the vertical lengths of the loops.
The measure $v(d\omega)$ is determined
by (\ref{part}); namely, for a given sequence of configurations
$\{n^{(i)}_\x\}$, $1\leq i\leq m$, the integration takes places over
the times of the jumps, with a measure $(1/2)^m dt_1\cdots dt_m$.
\index{path integrals}

One may note that the measure $v(d\omega)$ corresponds to a
Poisson process of random configurations of oriented `rungs', linking
neighboring sites at random times, and signifying either the
creation
or the annihilation of a pair of quasiparticles.
The
matrix element
$\langle \{n_\x\}| e^{-\beta H} |\{n'_\x\}\rangle$
gets no contribution from
rung configurations that are inconsistent,
either internally or with
the boundary conditions corresponding to
the specified state vectors.
A consistent configuration yields a
family of non-overlapping loops
which describe the motion of the
quasi-particles in the
`space-time' $\Lambda\times [0,\beta)$.
Each such configuration
contributes with weight
$e^{-\lambda|\omega|}$ to the above matrix
element (another positive
factor was absorbed in
the measure $v(d\omega)$).
One may note
that long paths are suppressed in the integral  (\ref{gamma1}) at a
rate which increases with $\lambda$.

\begin{figure}[htf]
\center
\includegraphics[width=11cm, height=6cm]{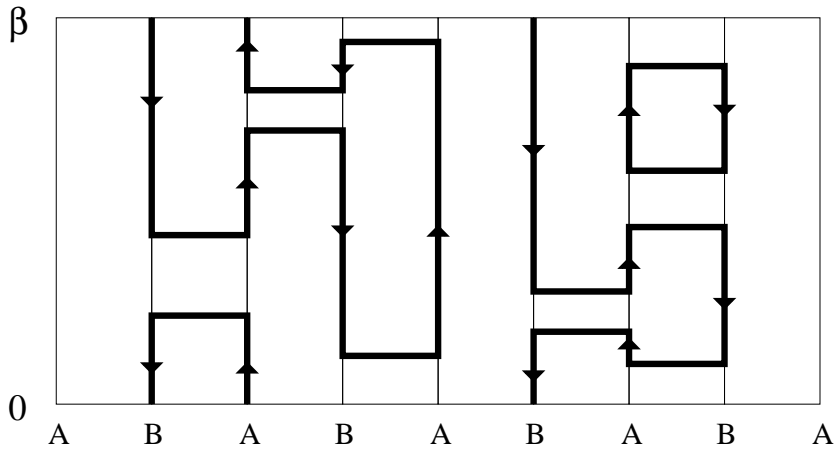}
\caption{Loop gas describing paths of quasi-particles for particle
     number $N=|\Lambda|/2-1$. A line on an A site means presence of a
     particle, while on a B site it means absence. The horizontal rungs
     correspond to hopping of a particle.}
\label{loopfig}
\end{figure}

Likewise, for $\x\neq \y$, we can write \beq \Tr
\,a^\dagger_{\x}a_{\y}e^{-\beta H}=\int_{ {\mathcal A}^{(\x,\y)}}
v(d\omega)e^{-\lambda|\omega|}, \eeq where ${\mathcal A}^{(\x,\y)}$
denotes the set of all loops that, besides disjoint closed loops,
contain one curve which avoids all the loops and connects $\x$ and
$\y$ at time zero.  The one-particle density matrix
\index{one-particle density matrix}
can thus be written \beq\label{gamma1}
\gamma(\x,\y)=\frac{\int_{{\mathcal A}^{(\x,\y)}}
  v(d\omega)e^{-\lambda|\omega|}}{\int
  v(d\omega)e^{-\lambda|\omega|}}.  \eeq

For an upper bound, we can drop the condition in the numerator
that the loops and the curve from $\x$ to $\y$ do not intersect. The
resulting measure space is simply a Cartesian product of the measure
space appearing in the denominator and the space of all curves,
$\zeta$, connecting $\x$ and $\y$, both at time 0. Denoting the
latter by ${\mathcal B}(\x,\y)$, we thus get the upper bound
\beq\label{gamma2}
\gamma(\x,\y)\leq \int_{{\mathcal B}(\x,\y)}
v(d\zeta)e^{-\lambda|\zeta|} .
\eeq

The integral over paths is convergent if either $\lambda$ or
$T$ is
small enough, and away from the convergence threshold the
resulting
amplitude decays exponentially.  A natural random
walk estimate, see
\cite[Lemma~4]{ALSSY}, leads
to the claimed exponential bound provided
\beq
d \left(1-e^{-\beta\lambda}\right) < \lambda.
\eeq
This includes, in particular, the cases $T>d$ for any $\lambda$, and
$\lambda > d$ for any $T$.


Exponential decay actually holds for the larger range of parameters where
\beq\label{condim}
d \left(1-e^{-\beta(\lambda-f)}\right) < \lambda - f,
\eeq
where $f=f(\beta,\lambda)=-(\beta|\Lambda|)^{-1} \ln \Tr e^{-\beta
H}$ is the free energy per site.
Note that $f<0$. This condition can be obtained by a more elaborate
estimate than the one used in obtaining
(\ref{gamma2}) from (\ref{gamma1}), as shown in
\cite[Lemma~3]{ALSSY}. The argument there uses reflection positivity
of the measure $v(d\omega)$. Using simple bounds on $f$ one can then
obtain  from (\ref{condim}) the conditions stated in the beginning of
this section.

\bigskip

The proof of the energy gap is based on an estimate for the ratio
$\frac{\Tr\,{\mathcal P}_{k}e^{-\beta H}}{\Tr\,{\mathcal
P}_{0}e^{-\beta
     H}}$ where ${\mathcal P}_{k}$ projects onto states in Fock space
with particle number $N=\half|\Lambda|+k$, expressing numerator and
denominator in terms of path integrals.
The integral for the numerator is over configurations $\omega$ with a
non-trivial winding number\index{winding number} $k$. Each such configuration includes a
collection of \lq non-con\-tract\-ible\rq\ loops with total length at least
$\beta |k|$. An estimate of the relative weight of such loops yields
the bound
\beq
\frac{\Tr\,{\mathcal P}_{k}e^{-\beta H}}{\Tr\,{\mathcal
P}_{0}e^{-\beta
H}}\leq
{\rm (\const)}(|\Lambda|/|k|)^{|k|}\left(e^{1-{\rm
(const.)}\beta}\right)^{|k|}
\eeq
which gives for $\beta\to \infty$
\beq
E_k-E_{0}\geq \const|k|
\eeq
independently of $|\Lambda|$. We refer to \cite{ALSSY} for details.

\section{The Non-Interacting Gas}\label{sectfree}

The interparticle interaction is essential for the existence of a Mott
insulator phase for large $\lambda$. In case of absence of the
hard-core interaction, there is BEC for any density and any $\lambda$
at low enough temperature (for $d\geq 3$). To see this, we have to
calculate the spectrum of the one-particle Hamiltonian $-\half\Delta +
V(\vecx)$, where $\Delta$ denotes the discrete Laplacian and
$V(\vecx)=\lambda (-1)^\vecx$. The spectrum can be easily obtained by
noting that $V$ anticommutes with the off-diagonal part of the
Laplacian, i.e., $\{ V, \Delta+2d\} = 0$. Hence \beq
\left(-\half\Delta - d + V(\vecx) \right)^2 = \left(-\half \Delta -
   d\right)^2 + \lambda^2, \eeq so the spectrum is given by \beq d\pm
\sqrt{\left(\mbox{$\sum_i$} \cos p_i\right)^2 +\lambda^2}, \eeq where
$\vecp\in\Lambda^*$. In particular, $E(\vecp)-E(0)\sim \half d
(d^2+\lambda^2)^{-1/2} |\vecp|^2$ for small $|\vecp|$, and hence there
is BEC for low enough temperature. Note that the condensate wave
function is of course {\it not} constant in this case, but rather
given by the eigenfunction corresponding to the lowest eigenvalue of
$-\half\Delta+\lambda(-1)^\vecx$.

\section{Conclusion}
In this chapter a lattice model is studied, which is similar to the
usual Bose-Hubbard model and which describes the transition between
Bose-Einstein condensation and a Mott insulator state as the strength
$\lambda$ of an optical lattice potential is increased.  While the
model is not soluble in the usual sense, it is possible to prove
rigorously all the essential features that are observed
experimentally.  These include the existence of BEC for small
$\lambda$ and its suppression for large $\lambda$, which is a
localization phenomenon depending heavily on the fact that the Bose
particles interact with each other.  The Mott insulator regime is
characterized by a gap in the chemical potential, which does not exist
in the BEC phase and for which the interaction is also essential. It
is possible to derive bounds on the critical $\lambda$ as a function
of temperature.

\chapter{New Developments}

In this chapter, we give a brief description of new results that
have been obtained since the publication of these lecture notes. 

\begin{itemize}

\item [$\bullet$] {\it Rotating Bose Gases.} 
As already noted at the end of Chapter~\ref{becsect}, it has recently
been shown in \cite{LS05} that the Gross-Pitaevskii equation also
correctly describes dilute, trapped Bose gases subject to rotation. In
particular, Eq.~(\ref{gpone}) holds for any rotation speed $\varphi$,
not only for small $\varphi$. For $\varphi$ large enough, the GP
minimizer is in general not unique, however, due to the appearance of
quantized vortices. Eq.~(\ref{gpone2}) therefore has to be replaced by
a more complicated expression, stating that the one-particle density matrix
$\gamma$ can, in general, be expressed as a convex combination of
projections onto minimizers of the GP functional.

\item[$\bullet$] {\it Ground State Energy of Dilute Fermi Gases.}
In Chapters~\ref{sect3d} and~\ref{sect2d}, the ground state energy of a dilute
Bose gas in three and two dimensions, respectively, has been
estimated. The analogous result for gases of fermions has now been
obtained in \cite{LSSfermi}. In the three-dimensional case, it says
that
\begin{equation}\label{fermieo}
e_0(\rho) = \frac 35 \left( \frac{6\pi^2}{q}\right)^{2/3}\rho^{2/3} +
4\pi a \left(1-\frac 1q\right)\rho + o(\rho)
\end{equation}
in the fermionic case. Here, $q$ denotes the number of internal
degrees of freedom of the particles; e.g., $q=2$ for spin $1/2$
fermions. The first term in (\ref{fermieo}) is just the ground state
energy of a non-interacting Fermi gas. The first correction term looks
the same as in the bosonic case, except for the factor $(1-1/q)$,
which results from the fact that effectively only fermions in
different spin states interact with each other (to leading order in
the density). 

\item [$\bullet$] {\it Dilute Fermi Gases at Positive Temperature.}
Most of the results discussed in these notes are on the ground state
of various models of quantum gases, i.e., on the zero temperature
states. Actual physical experiments are always carried out at non-zero
temperature, and it is hence desirable to generalize some of the
results to the case of positive temperature. In the fermionic case,
this has been achieved in \cite{SfermiT}. More precisely, an analogue
of (\ref{fermieo}) concerning the free energy of a dilute Fermi gas at
positive temperature $T$ has been obtained. There is now an additional
length scale in the problem, namely the thermal wavelength $\sim
T^{-1/2}$. One
is interested in the case $a \ll \rho^{-1/3} \sim T^{-1/2}$; i.e.,
one considers a dilute gas with $a^3\rho\ll1$, and temperatures of the
order of the Fermi temperature, which is $\sim \rho^{2/3}$.

\item [$\bullet$] {\it Dilute Bose Gases at Positive Temperature.}  In
  the case of a dilute Bose gas, the analogue of Theorem~\ref{lbth} at
  positive temperature $T$ has recently been obtained in \cite{SboseT}.
  I.e., a lower bound was derived on the free energy of a dilute Bose
  gas.  The leading order correction as compared with the expression
  for non-interacting particles is of the form
\begin{equation}\label{fmt}
4\pi a \rho \left( 2 - [1 - \rho_c(T)/\rho]_+^2\right)\,.
\end{equation}
Here, $\rho_c(T) = \const T^{3/2}$ denotes the critical density for
Bose-Einstein condensation (for the non-interacting gas), and $[\,
\cdot \,]_+$ denotes the positive part. Note that below the critical
density (or, equivalently, above the critical temperature), the
leading order correction is $8\pi a \rho$, as compared with $4\pi a
\rho$ at zero temperature. The additional factor 2 is an exchange
effect; heuristically speaking, it is a result of the symmetrization
of the wave functions. This symmetrization only plays a role if the
particles are in different one-particle states, which they mostly are
above the critical temperature. Below the critical temperature,
however, a macroscopic number of particles occupies the zero-momentum
state; there is no exchange effect among these particles, which
explains the subtraction of the square of the condensate density
$[\rho - \rho_c(T)]_+$ in (\ref{fmt}).

\end{itemize}

\appendix

\chapter{Elements of Bogoliubov Theory}\label{chap2}
\index{Bogoliubov!method}

\textit{This appendix is Section $2$ of the $1965$ review article {\rm\cite{EL2}},
  with minor modifications. Unlike the rest of these notes, this
  appendix is not mathematically rigorous, but it is important because
  it shows what is widely believed to be the case about the low end of
  the spectrum of the Hamiltonian $H_N$ --- even if it cannot be
  proved.  It is also a heuristic guide to thinking about Bose gases.}

\textit{In the forty years since the appearance of the review article {\rm\cite{EL2}} the
  Bogoliubov model has been extensively studied; an account of many of
  these developments can be found in the comprehensive
  article~{\rm\cite{zagrebnov}}.}  \bigskip \bigskip

Almost all work in the field of interacting Bose gases has its genesis
in Bogoliubov's\index{Bogoliubov} 1947 paper \cite{BO} and we shall
begin with a brief summary of its main features.  His theory is
important because it set us on the right track.  But of equal
importance is understanding, with a little hindsight, the failures of
the theory, for its success led to a regrettable tendency to take its
predictions as gospel.  Moreover, like many great approximation
schemes in mathematical physics, the first order approximation is
qualitatively correct in a certain regime (weak coupling).  Attempts
to push out of this regime through higher approximations have led to
great difficulties.  Having learned the predictions of the theory,
therefore, we should be prepared to have to seek a new method in order
to understand intermediate and strong coupling.

The basic Hamiltonian\index{Hamiltonian} $H$ of the problem is
\begin{equation}\label{II1}
H = -\frac {\hbar^2}{2m} \sum\limits_{i=1}^N \nabla_i^2 + \frac{1}{2}
    \sum\limits_{i \neq j} v \big(\x_i - \x_j\big)
\end{equation}
wherein $v$ is the two--body potential (no one has yet considered
systems with more--body forces),
$\ulr_i$ is the coordinate of the $i^{th}$
particle, and $N$ is the number of bosons.  The potential $v$ is
necessarily symmetric (i.e., $v( \ulx ) = v(-\ulx )$), but it is not necessary
that it be spherically symmetric, although this assumption is generally
made.  The scene of action is a box of volume $V= L \times L \times L$
and $\rho = N/V$ is the density of particles.  We are interested in the
bulk limit:  $N \to \infty$ with $\rho =$ constant.

The first question that arises is that of boundary
conditions\index{boundary condition}, and this is intimately connected
with the question of extensivity (or saturation in the terminology of
nuclear physics).  The ground state energy, $E_0$, is said to be
extensive if it is of the form $E_0 = N \times$ function of $\rho$.
Similarly, extensivity is defined for the free energy, $F$, at some
non--zero temperature.  The system is said to be extensive if both
$E_0$ and $F$ are.  Unfortunately, necessary and sufficient conditions
on $v$ that the system be extensive are far from known, but one can
prove extensivity for a wide class of potentials.  For some of these,
in turn, one can further show that $E_0$ and $F$ are independent of
boundary conditions (to leading order in $N$), provided they are fixed
homogeneous conditions.  We shall assume that our $v$ is of this kind
and shall use periodic boundary conditions which, while they may be
somewhat unphysical, are mathematically most convenient.  They state
that if $\psi ( \x_1 , \ldots , \x_N)$ be an eigenfunction of $H$ and
if $\x_j$ (for all $j$) be on the wall of the box (so that $\x_j = (x,
y, 0)$ for example), then the value of $\psi$ on the opposite wall
$(\x_j = (x, y, L))$ must be the same as on the first wall.  This is
to be true for all values of $\x_k$ ($k \neq j$), and a similar
condition is imposed on the normal derivative of $\psi$.

Periodic conditions can {\it always} be imposed, but they become
{\it useful} only if $H$ can be periodically extended to all of space.
This in turn requires that $v(\x_1 - \x_2)$ be
periodic in $\x_1$ and $\x_2$, which means, in
effect, that $v$ must depend upon $L$.  Thus, the one--dimensional
potential $v(x_2 - x_1) = \exp (- \gamma|x_2 -x_1|)$ is not periodic for
$x_1$ and $x_2$ in $(0,L)$, but it can be replaced by
$\exp ( - \gamma |x_2 - x_1|) + \exp ( - \gamma L + \gamma |x_2 - x_1|)$
which is periodic in this region.  It can then be periodically
extended to all of space and one can show that the addition to $v$ will
not affect $E_0$ or $F$ in the bulk limit.

The virtue of having $H$ periodic is that one can trivially show that
$\psi (\x + {\bf a})$ is an eigenfunction with the same
energy as $\psi $ (by this we mean that any constant vector
${\bf a}$ is added to all coordinates).  Since all irreducible
representations of the translation group are one--dimensional, every
eigenfunction {\it can be} chosen to have a constant total momentum,
i.e.,
\begin{equation}\label{II2}
 {\bf P} \psi = - i \hbar \biggl( \sum\limits_{i=1}^N
   \nabla_{i} \biggr)
  \, \psi = {\bf p} \psi
\end{equation}
where ${\bf p}$ is a real vector.  Another advantage is that for
$\x_1$ and $\x_2$ in $V$, the potential can be
written as
\begin{equation}\label{II3}
v(\x_2 - \x_1 ) =
   V^{-1} \sum\limits_{\k} \nu (\k)
     e^{{\rm i}\k \cdot (\x_2 - \x_1 )}
\end{equation}
where the $\k$ are vectors of the form $(2\pi/L)(n_1, n_2, n_3)$,
(the $n_j$ are integers).  The point here is that any potential can
always be written as a double Fourier series, but the series is diagonal
as in (\ref{II3}) only when $v$ is periodic.  The Fourier transform is given
by
\begin{equation}\label{II4}
\nu ( \k) = \int_V v(\x)
   e^{-{\rm i}\k \cdot \x}
     d\x \,.
\end{equation}

Now the essence of the Bogoliubov method is that we go into
momentum space.  To do this we define
\begin{subequations}\label{II5}
\begin{equation}\label{II5a}
\varphi (\k_1 , \ldots , \k_N ) =
   \int_V d\x_N \cdots \int_V d\x_1
     \psi (\x_1 , \ldots , \x_N)
       \exp \biggl({\rm i} \sum\limits_{i=1}^N \k_i \cdot \x_i
\biggr)
\end{equation}
\begin{equation}\label{II5b}
\psi ( \ubr_1 , \ldots , \ubr_N) =
V^{-N} \sum\limits_{\ulk_1} \cdots \sum\limits_{\ulk_N}
  \varphi (\ulk_1 , \ldots , \ulk_N )
   \exp \biggl( -{\rm i} \sum\limits_{i=1}^N \ulk_i \cdot \ubr_i
\biggr)
\end{equation}
\end{subequations}
and the equation $H\psi = E\psi$ becomes
\begin{eqnarray}\nonumber
&&\biggl[ \sum\limits_{i=1}^N \epsilon (\ulk_i) \biggr]
  \varphi (\ulk_1 , \ldots , \k_N) + (1/2V)
     \sum\limits_{\ubp} \sum\limits_{i\neq j}
       \varphi (\ulk_1 , \ldots , \ulk_i +
          \ubp , \\
&& \hskip 1in \cdots \ulk_j -\ubp , \ldots , \ulk_N)
     \nu (\ubp) = E \varphi (\ulk_1 , \ldots , \k_N)
 \label{II6}
\end{eqnarray}
where $\epsilon (\k) = (\hbar^2/2m)\k^2$.

Since we are interested in bosons we require solutions,
$\varphi$, to (\ref{II6}) which are symmetric in
$\ulk_1 , \ldots , \ulk_N$.  Since $\varphi$
is symmetric, the sums on $i$ and $i\neq j$ in (\ref{II6}) are to some extent
repetitive and it is convenient to introduce a device to handle
automatically the factors of $N$, $\binom N2$, etc. that will continually
appear on the left hand side of (\ref{II6}).

The device used is to introduce boson creation and annihilation
operators\index{creation operator}\index{annihilation operator}.  But
we wish to emphasize strongly that they are not essential to the
theory.  They are merely a convenient bookkeeping device.  For every
$\ulk$ one introduces a creation operator $a_{\ulk}^\dagger$ and its
Hermitian conjugate (the annihilation operator) $a_{\ulk}$ which
satisfy the following commutation rules:
\begin{alignat}{2}
& \big[ a_{\ulk} , a_{\ulk^\prime} \big] =0=
  \big[ a_{\ulk}^\dagger, a_{\ulk^\prime}^\dagger\big] \nonumber\\
& \big[ a_{\ulk} , a_{\ulk^\prime}^\dagger \big] =
\delta_{{\ulk}  ,  {\ulk^\prime}}\label{II7}
\end{alignat}
The $\delta$ is a Kronecker delta since the $\ulk$'s are discrete.
A vacuum state $|0 \rangle$ in a Hilbert space\index{Hilbert space} is also introduced such
that
\begin{equation}\label{II8}
a_{\ulk} | 0 \rangle = 0 \qquad (\text{for all } \ulk ) \,.
\end{equation}
One can show that the state $|0\rangle$ is essentially unique and that
any state in the Hilbert space is a sum of products of various $a^\dagger$'s
acting on the vacuum.  Next one defines an isomorphism between
the functions $\varphi$ and states $\Psi$ in the Hilbert space as
follows:
\begin{equation}\label{II9}
\Psi = \sum\limits_{\ulk_N} \cdots \sum\limits_{\ulk_1}
  \varphi (\ulk_1 , \ldots , \ulk_N) a_{\ulk_1}^\dagger \cdots
     a_{\ulk_N}^\dagger |0\rangle \,.
\end{equation}
Using (\ref{II7}) it is easy to see that knowing $\Psi$ one can find
$\varphi$, so that the correspondence is indeed one to one.
It is then slightly tedious, but simple, to show that if $\varphi$
satisfied (\ref{II6}) then $\Psi$ satisfies
\begin{equation}\label{II10}
H \Psi \equiv \left\{ \sum\limits_{\ulk} \epsilon (\ulk) a_{\ulk}^\dagger
     a_{\ulk} + (1/2V) \sum\limits_{\ulk , \ulq , \ubp}
      a_{\ulk + \ulp}^\dagger a_{\ulq -\ulp}^\dagger a_{\ulk}
       a_{\ulq} \; \nu (\ulp)  \right\} \Psi = E \Psi \,.
\end{equation}
Equation (\ref{II10}) is the starting point for Bogoliubov's approximation.

Before proceeding it is well to keep in mind certain properties
of the $a_{\ulk}$'s.  We define the total number operator\index{number operator}
\begin{equation}\label{II11}
\eta = \sum\limits_{\ulk} a_{\ulk}^\dagger a_{\ulk}
\end{equation}
and the total momentum\index{momentum} operator
\begin{equation}\label{II12}
{\bf P} = \hbar \sum\limits_{\ulk} \ulk a_{\ulk}^\dagger a_{\ulk} \,.
\end{equation}
These commute with $H$ and when acting on $\Psi$
yield $N$ and $\ulp$ respectively.

Now, consider the state $a_{\ulk}\Psi$.  This state has particle number
$N-1$ and momentum $\ulp - \hbar \ulk$.  We can go back through
(\ref{II9}) and (\ref{II5}) and ask what function this corresponds to in
configuration space.  The result is
\begin{equation}\label{II13}
a_{\ulk} \Psi \Longleftrightarrow \sqrt{\frac N V} \int_V
   \Psi (\ubr_1 , \ldots , \ubr_N)
     e^{-{\rm i}\ulk \cdot \ubr_N}
       d^3 \x_N \,.
\end{equation}
Likewise, the state $a_{\ulk}^\dagger \Psi$ has $N+1$ particles and momentum
$\ulp + \hbar {\ulk}$.  It corresponds to
\begin{equation}\label{II14}
a_{\ulk}^\dagger \Psi \Longleftrightarrow \frac 1{\sqrt{(N+1)V}}
   \sum e^{{\rm i}\ulk \cdot \ubr_{N+1}}
    \Psi (\ubr_1 , \ldots , \ubr_N)
\end{equation}
where the sum is on $N+1$ similar terms.

It is also convenient to define density operators which are
generalizations of (\ref{II11}), namely
\begin{equation}\label{II15}
\rho({\ulk}) = \sum\limits_{\ulq} a_{\ulk + \ulq}^\dagger a_{\ulq} \,;
\end{equation}
it conserves particle number and increases the momentum by an amount
$\hbar \ulk$.  Its effect in configuration space is given by using
(\ref{II13}) and (\ref{II14}) and is
\begin{equation}
\rho({\ulk}) \Psi \Longleftrightarrow
\left( \sum\limits_{j=1}^{N} e^{{\rm i}\ulk \cdot \ubr_j}\right)
 \Psi (\ubr_1 , \ldots , \ubr_N ) \,.
\label{II16}
\end{equation}
We shall have occasion to use the relations (\ref{II13}) to (\ref{II16}) later.

Returning to (\ref{II10}), the reason for going into momentum space is the
following:  If there were no interaction then
$\Psi_0 = (N!)^{1/2} (a_\0^\dagger)^N |0\rangle$ would be the normalized
ground state with energy zero.  The interaction $v$ has the property
that it converts a pair of particles with momenta $\ubp$ and $\ulq$
into a pair with momenta $\ubp +\ulk$ and $\ulq - \ulk$.
The matrix element is $(1/2V) \nu (\ulk )$.
Starting with all $N$ particles having momentum zero
(so-called condensed state), we would first get
$(N-2)$ with momentum zero, together with one pair having momenta $\ulk$
and $-\ulk$.  When the potential is applied again we could get two
possibilities:  one would be two pairs $\ulk , - \ulk$ and $\ulq$, $-\ulq$;
the other would be a genuine triplet $\ulk , \ulq , {\bf r}$, such that
$\ulk + \ulq + {\bf r} = 0$.  But the probability of the former
relative to the latter
would be $(N-2)(N-3)/4$ because there are $(N-1)$ particles with
zero momentum and only
2 with non--zero momentum.  Applying $v$ over and over again
we will ultimately get a
finite fraction of triplets, quartets, etc. as well as pairs, but
hopefully if the
interaction is weak enough we need consider explicitly only pairs in
the ground state wave function.  Stating this more precisely,
we suppose that to a good
approximation $\Psi_0$ is a sum of terms each of which contains several
factors like $a_{\ulk}^\dagger a_{-\ulk}^\dagger$, as well as
$a_\0^\dagger$, acting on $|0\rangle$.

Another way of motivating this {\it ansatz} is to note that in the
non--interacting ground state
$n_{\ulk} = \langle a_{\ulk}^\dagger a_{\ulk} \rangle = N \delta _{\ulk ,0}$,
that is to say all
the particles are condensed.  With a weak interaction we suppose that
$n_\0/N$ is still a number of order unity and that the remaining fractions
are largely grouped into pairs, for it is only pairs that can give rise
to triplets.  The idea that $n_\0/N$ is of order unity is called
the condensation hypothesis.\index{condensation hypothesis}
It need not be true for sufficiently strong interaction and we remark
that Girardeau \cite{ref2boulder} has generalized this concept somewhat.

If $\Psi$ as given by (\ref{II9}) has only pairs (more precisely, for every
$a_{\ulk}^\dagger$ with $\ulk \neq 0$, there is an $a_{-{\ulk}}^\dagger$)
then only certain parts of $H$ result in pair functions when applied to $\Psi$.
There are three possibilities as far as the interaction is concerned:
The first is when all indices are zero, giving $(a_\0^\dagger)^2 (a_{\0})^2$;
the second is when two indices are zero, giving
$(a_\0^\dagger)^2 a_{\ulk}a_{-\ulk}$ or $a_{\ulk}^\dagger a_{-\ulk}^\dagger (a_\0)^2$ or
$a_{\k}^\dagger a_\0^\dagger a_{-\ulk} a_\0$; the third is when no
indices are zero, giving $a_{-\ulq}^\dagger a_{\ulq}^\dagger a_{\ulk} a_{-\ulk}$ or
$a_{\ulk}^\dagger a_{\ulq}^\dagger a_{\ulk}a_{\ulq}$.
Collecting together all such parts of $H$ we derive the pair Hamiltonian\index{Hamiltonian!pair},
\begin{equation}\label{II17}
\begin{aligned}
\quad\quad H_{\text{pair}}=\frac{1}{2}(N-1)
     \rho \nu (\0) &+\sum\limits_{\text{k}}
\hskip -.001cm \raisebox{.9ex}{$^\prime$}
  \bigl[ \epsilon (\ulk) + (1/V)N_0 \nu (\ulk ) \bigr] N_{\ulk} \\
&+ (1/2V) \sum\limits_{\text{k}} \hskip -.001cm \raisebox{.9ex}{$^\prime$}
    \nu (\ulk) \left[\Ap_{\ulk}^\dagger \Ap_\0 +\Ap_\0^\dagger \Ap_{\ulk}
           \right]  \\
&+ (1/2V) \sum\limits_{\ulk,\ulq}
   \hskip -.001cm \raisebox{.9ex}{$^\prime$}
   \hskip .01cm   \nu (\ulk) \left[ \Ap_{\ulq}^\dagger \Ap_{\ulq-\ulk}
        +N_{\ulq - \ulk} N_{\ulq} \right]
\end{aligned}
\end{equation}
where $\Ap_{\ulk} = a_{\ulk} a_{-\ulk}$,
$N_{\ulk} = a_{\ulk}^\dagger a_{\ulk}$ and the prime on the summation means we
delete the terms $\ulk =\0$ and/or $\ulq =\0$ as well as the term
$\ulq=\ulk$ in the double summation (note that $N_\0$ and $\Ap_\0$
are operators).  In deriving (\ref{II17}) we  used the fact that
$\eta \Psi = N \Psi$.

It is important to note that $H_{\text{pair}}$ has a {\it double}
significance.  On the one hand if we can diagonalize it we should have
a good approximation to the ground state and low lying states of the
system for the reasons mentioned above.  On the other hand we have seen
that if we take the expectation value of the total $H$ with respect to
{\it any} state having {\it only} pairs then $H_{\text{pair}}$
is the only part of $H$ that contributes to the final result.
Hence, from the variational theorem, the exact ground state energy,
$E_{0, \text{pair}}$, of $H_{\text{pair}}$
is a true upper bound to the ground state energy, $E_0$, of $H$.
Moreover, any variational upper bound to $E_{0, \text{pair}}$
is thus an upper bound to $E_0$.  It turns out that
$E_{0,\text{pair}}$ can indeed be found if one is prepared to solve a
finite set of non--linear integral equations.  This can be done in
certain limiting cases and has been exploited by
Girardeau and Arnowitt, \cite{b3,b4,b5,b6}.

Basically, what permits us to find the ground state energy of
(\ref{II17}) in the bulk limit, as well as the free energy for non-zero
temperature, is the following observation.  What we have in (\ref{II17}) are
bilinear forms in operators whose expectation values we believe to be
extensive.  Consider, for instance
\begin{equation}\nonumber
\sum\limits_{\ulq }
\hskip -.001cm \raisebox{.9ex}{$^\prime$}
N_{\ulq} \left( \sum\limits_{\ulk\neq \ulq}
\hskip -.001cm \raisebox{.9ex}{$^\prime$}
\nu (\ulk ) N _{\ulq - \ulk} \right) \,.
\end{equation}
The operator in parenthesis (call it $F_{\ulq})$ we believe has an
expectation value of order $N$ (call it $N f_{\ulq}$).
The root-mean-square fluctuation of $F_{\ulq}$ in the ground
state ought to be of order $\sqrt{N}$, and if so, replacing $F$ by $Nf$
in (\ref{II17}) should make no difference to the energy to order $N$.  It is
possible to make this argument more precise \cite{b6,b7,b8,b8b}
by formally expanding the operators about their mean values in a power
series in $N^{-1}$.  The difficulty is that no one has shown that these
series converge and they might well not.  Nevertheless, the energy
(or free energy) obtained under the assumption can be shown to be a
genuine variational energy and so is a true upper bound.

Having replaced $F_{\ulq}$ by $Nf_{\ulq}$ we do the same thing with the
term corresponding to $\Ap_{\ulq - \ulk}$, $N_\0$, $\Ap_\0$ and
$\Ap_\0^\dagger$.  We are then left with a Hamiltonian involving only
quadratic expressions in the $a$'s and $a^\dagger$'s.  This can be
diagonalized in the standard way.  The ground state wave function will
then depend upon the $c$-numbers $f_{\ulq}$, etc. as parameters.
We then adjust these parameters so that the expectation value of
$F_{\ulq}$ is indeed $Nf_{\ulq}$, etc.  This leads to integral equations
and it is clear that what we have really done is a self-consistent field
calculation that we hope is rigorously correct in the bulk limit.

There is, however, still one difficulty which we have glossed over.
If we replace $\Ap_\0$ by a $c$-number then the expression
$\Ap_{\ulk}^\dagger \Ap_\0$ will no longer conserve particles because
$\Ap_{\ulk}^\dagger$ always creates two particles.  Another way of saying
this is that the expectation value of $\Ap_\0$ in the true ground
state is, strictly speaking, zero.  There are two ways around
this difficulty.  The first is the method used by Bogoliubov,
namely to introduce a chemical potential\index{chemical potential}.  We write
\begin{equation}\label{II18}
\begin{aligned}
H_{\text{pair}}^\prime &= H_{\text{pair}}-\mu \eta \,,  \\
(\text{or } H^\prime & = H-\mu \eta) \,.
\end{aligned}
\end{equation}
We then diagonalize $H_{\text{pair}}^\prime$ by replacing $\Ap_\0$,
etc. by $c$-numbers and in addition to the above consistency
conditions we choose $\mu$ by requiring that $\langle N \rangle = N$
in the ground state.  To calculate the free energy we must use a
grand-canonical ensemble\index{grand-canonical!ensemble}.  This method
can be justified to the same extent as in the above discussion.  In
the second method \cite{b6,b9} we redefine operators so that particles
are conserved even after the $c$-number substitution.  Following
Kromminga and Bolsterli we introduce operators $b_{\ulk} =
a_\0^\dagger (N_\0 +1)^{-1/2} a_{\ulk}$ and their conjugates.
It is then easy to show that
$b_{\ulk}^\dagger b_{\ulq} = a_{\ulk}^\dagger a_{\ulq}$ (all $\ulk , \ulq$) and that
$b_{\ulk}b_{\ulq}^\dagger = a_{\ulk} a_{\ulq}^\dagger$
(if $\ulk \neq \0$ and $\ulq \neq \0$
and if this operator does not act on a state with $N_\0 = 0$).
An annoying operator such as $\Ap_{\ulk}^\dagger \Ap_\0$ is then equal
to $b_{\ulk}^\dagger b_{-\ulk}^\dagger (N_\0 (N_\0 -1))^{1/2}$ on any state.
Also $\Ap_{\ulk} \Ap_\0^\dagger = (N_\0 (N_\0 -1))^{1/2} b_{\ulk}b_{-\ulk}$ and
$\Ap_{\ulq}^\dagger \Ap_{\ulk}= b_{\ulq}^\dagger b_{-\ulq}^\dagger b_{\ulk} b_{-\ulk}$.
In addition one easily proves that the $b$'s satisfy the same
commutation relations (\ref{II7}) as the $a$'s (provided $\ulk \neq \0$)
and also the $b$'s conserve particle number (i.e., they commute with $N$).
Thus either $H_{\text{pair}}$ or $H$ may be rewritten in terms of the
$b$'s which behave just like the $a$'s except that now any $c$-number
substitution will automatically preserve particle number.
All the sums explicitly exclude $\ulk = \0$, so the only way in which
$a_\0$ and $a_\0^\dagger$ appear in the new Hamiltonian is through $N_\0$.
But $N_\0$ may be eliminated in favor of $b$'s by the relation
\begin{equation}\nonumber
N_\0 = N - \sum \hskip -.001cm \raisebox{.9ex}{$^\prime$}
a_{\ulk}^\dagger a_{\ulk} = N -
\sum \hskip -.001cm \raisebox{.9ex}{$^\prime$}
b_{\ulk}^\dagger b_{\ulk} \,.
\end{equation}
Now the particle number appears explicitly in $H$, which it did
not before (although it did appear in $H_{\text{pair}}$),
and a term such as $N_\0 N_{\ulk}$ becomes now a quartic form in the $b$'s.

This is a neat trick to overcome the problem of particle conservation
that plagued previous authors (such as Bogoliubov\index{Bogoliubov}).  It obviates the
need for lengthy (and unrigorous) arguments that method (\ref{II18}) gives
the correct answer in the bulk limit.  Needless to say, however, in
every calculation anyone has ever done, the more cumbersome method 2
does indeed give the same result as method 1 and we shall therefore use
the latter.

Bogoliubov did not actually use $H_{\text{pair}}$.
He made a further simplification which consisted in deleting the last
sum in (\ref{II17}) on the grounds that these terms are quadratic pair
operators and so may be expected to be small in comparison with the
first two sums.  This omission unfortunately destroys the bounding
property of the Hamiltonian, but it does turn out that for sufficiently
weak interaction Bogoliubov's ground state energy is indeed an upper
bound to $E_0$ as was to be expected.  To that extent his simplification
is justified.

We begin by replacing $N_\0$, $\Ap_\0$ and $\Ap_\0^\dagger$ by a
(common) $c$-number\index{c-num@$c$-number substitution},
$N_\0$.\footnote{Subsequently, Ginibre \cite{ginibre} proved that one
  can replace $a_\0$ and $a_\0^\dagger$ everywhere in $H$ by the
  c-number $\sqrt{N_\0}$ without making an error in the ground state
  energy/particle in the thermodynamic limit. A simpler and more
  general version of this result can be found in
  Appendix~\ref{justapp}.}  For weak interaction we expect this number
to be close to $N$, and indeed it turns out to be so.  The correction,
$N-N_\0$, (the so-called ground state depletion)\index{depletion} gives a higher order
contribution to $E_0$ --- a correction which is of the same order as
that caused by the neglected quartic terms.  Since the qualitative
results do not depend upon the ground state depletion effect, we shall
simply take $N_\0$ to be $N$.  We thus have Bogoliubov's Hamiltonian
(replacing $N-1$ by $N$ in the bulk limit):
\begin{equation}
H_B = \frac12 N\rho \nu (\0) + \sum\limits_{\ulk} 
\hskip -.001cm \raisebox{.9ex}{$^\prime$} \left(
f(\ulk) a_{\ulk}^\dagger a_{\ulk} + \frac12 g (\ulk)
\big(a_{\ulk} a_{-\ulk} + a_{-\ulk}^\dagger a_{\ulk}^\dagger \big) \right)\,,
\label{II19}
\end{equation}
where
\begin{equation}
\begin{aligned}
f(\ulk) &= \epsilon (\ulk) + \rho \nu (\ulk) \\
g(\ulk) &= \rho \nu (\ulk ) \,.
\end{aligned}
\label{II20}
\end{equation}
This Hamiltonian is a quadratic form in the $a$'s and may be
diagonalized in the usual way.  The transformation that accomplishes
this is $\exp \; ({\rm i}S)$, where
\begin{equation}
{\rm i}S = \frac12 \sum\limits_{\ulk}
\hskip -.001cm \raisebox{.9ex}{$^\prime$}
\left( a_{\ulk}^\dagger a_{-\ulk}^\dagger - a_{\ulk} a_{-\ulk}\right) \psi (\ulk) \,,
\label{II21}
\end{equation}
so that
\begin{equation}
a_{\ulk} \rightarrow b_{\ulk} = e^{{\rm i}S} a_{\ulk} e^{-{\rm i}S} = a_{\ulk}
  \cosh \psi (\ulk) -a_{-\ulk}^\dagger \sinh \psi (\ulk) \,.
\label{II22}
\end{equation}
If we now choose
\begin{equation}
\tanh 2 \psi (\ulk) = g(\ulk ) / f(\ulk)\,,
\label{II23}
\end{equation}
then
\begin{subequations}\label{II24}
\begin{equation}
\begin{aligned}
\qquad\qquad H_B \rightarrow H_B^\prime &
     = \frac12 N \rho \nu (\0)
    - \frac12 \sum\limits_{\ulk}
\hskip -.001cm \raisebox{.9ex}{$^\prime$} f(\ulk) - \big(
f(\ulk)^2 - g(\ulk)^2 \big)^{1/2}  \\
&\;+ \sum\limits_{\ulk}
     \hskip -.001cm \raisebox{.9ex}{$^\prime$}
   b_{\ulk}^\dagger b_{\ulk} \epsilon^\prime (\ulk)  \,,
\end{aligned}
\label{II24a}
\end{equation}
where
\begin{equation}
\epsilon^\prime (\ulk) = \big[ \epsilon (\ulk)^2 +
     2 \epsilon (\ulk) \,\rho \nu
        (\ulk)\big]^{1/2}\,.
\label{II24b}
\end{equation}
\end{subequations}
Notice that this transformation is impossible unless
$|g| < |f|$.  It is also necessary that $f(\ulk) > 0$ for all $\ulk$.
Unless these two conditions are fulfilled the Hamiltonian has no ground
state and it would then be unphysical.  This means that $\nu(\ulk)$
cannot be too negative (attractive), but it can be as repulsive as we
please.

Now let us consider the implications of (\ref{II24}).  Since the
$b$'s are bosons, the ground state wave function is given by
$\Psi_0^\prime$, the vacuum of the $b$'s, i.e.,
\begin{equation}
b_{\ulk} \Psi_0^\prime = 0 \qquad \text{ all } \ulk \neq \0 \,.
\label{II25}
\end{equation}

The ground state energy is then simply
\begin{equation}
E_0 = \half N \rho \nu (\0) -
 N\big(4 \pi^2 \rho \big)^{-1}
    \!\int_0^\infty\! k^2 dk \bigl\{
     \epsilon (\k)+ \rho \nu (\k) - \big( \epsilon (\k)^2
+ 2 \epsilon (\k) \rho \nu (\k)\big)^{1/2} \bigr\},
\label{II26}
\end{equation}
where we have gone to the bulk limit by replacing
\begin{displaymath}
\sum\limits_{\ulk} \quad \text{by} \quad (L/2\pi)^3 \int d^3\k \,,
\end{displaymath}
and have further assumed that the problem is spherically symmetric.
At first sight it appears that the second term in (\ref{II26}) is order
$\rho^{-1}$, thereby violating our intuition that it should be small for
low density.  This is not so because as $\rho \rightarrow 0$ the
integrand itself vanishes.  To see what happens let us assume that
$\nu (\ulk)$ goes to zero for large $\ulk$ faster than $k^{-1}$.
Let us rewrite the integral in (\ref{II26}) as follows:  $I=I_1 + I_2$
where
\begin{displaymath}
I_1 = \int_0^\infty k^2 dk \bigl\{
   \Ap^2k^2 + \rho \nu (k) -
     \big(\Ap^4k^4 + 2 \Ap^2 k^2 \rho \nu (k)\big)^{1/2} -
      \rho^2 \nu(k)^2 /2 \Ap ^2 k^2\bigr\}
\end{displaymath}
and
\begin{equation}
I_2 = \rho^2 / 2\Ap^2 \int_0^\infty \nu (k)^2 dk
\label{II27}
\end{equation}
where $\Ap^2 = \hbar^2/2m$.  The integral $I_2$ certainly converges.
In $I_1$ the integrand goes to zero faster than $k^{-3}$ and is
absolutely convergent.  We see that as $\rho \rightarrow 0$ only
very small $k$ will play a role.  Let us assume that $\nu(k)$ is smooth
for small $k$ and that $\nu(0) \neq 0$.  We may then replace
$\nu (k)$ by $\nu(0)$ everywhere and the integral is then elementary.

Let us define
\begin{displaymath}
a_0 = ( 8 \pi \Ap^2)^{-1} \nu (0)\,,
\end{displaymath}
and
\begin{equation}
a_1 = - (2 \pi^2)^{-1}
  (8 \pi \Ap^2)^{-2}
     \int d^3 k \, \nu(k)^2/k^2 \,.
\label{II28}
\end{equation}
The result for $E_0$ is then
\begin{equation}
E_0 = 4 \pi N \rho \,(\hbar^2/2m)(a_0+a_1) +
  4\pi N\rho \,(\hbar^2/2m)\,
     a_0(128/15\sqrt{\pi}) (\rho \,a_0^3)^{1/2} \, .
\label{II29}
\end{equation}
What is the significance of this result?  If we had neglected the
integral in (\ref{II26}) we would have gotten $E_0 = \frac12 N \, \rho
\, \nu (0) = 4 \pi N \, \rho \hbar^2 (2m)^{-1}a_0$.  We may call this
the zeroth order Bogoliubov approximation.
\index{Bogoliubov!approximation}
But notice that it really does not depend upon Planck's constant and
the mass -- a conclusion that is certainly meaningless because if the
mass were infinite $E_0$ would be equal to the minimum potential
energy which is not necessarily $N \rho \nu (0)$.  The integral term
comes to the rescue, however.  We had naively expected it to
contribute a higher power of the density than $\rho$, but it in fact
contributed a term of the same order as the zeroth approximation --
namely $\rho a_1$.  This term now truly depends on $\Ap^2$.  If we
look a little closer we notice that $a_0 + a_1$ are just the first two
terms in the Born series\index{Born series} for the scattering
length.\index{scattering length} That is to say, if we consider the
zero--energy scattering equation
\begin{equation}
\biggl[ - \Ap^2 \bigl(\nabla_1^2 + \nabla_2^2 \bigr)
   + v\big(\ubr_1 - \ubr_2 \big) \biggr]
     \psi \big(\ubr_1 - \ubr_2 \big) = 0 \,,
\label{II30}
\end{equation}
the asymptotic behavior of $\psi$ is $\psi(\x_1-\x_2) \sim 1 - a/|\x_1 - \x_2|$.
The quantity $a$ is defined to be the scattering length.
(From (\ref{II30}), it is also given by
$a= (8\pi \Ap^2)^{-1} \int v(\x) \psi (\x) d^3\x)$.)
We therefore suspect, despite what we had originally thought,
that Bogoliubov's method is really an expansion in the density
{\it and} in the potential $v$.  It is not truly a low density
expansion unless the potential is very weak.
This idea has been confirmed by doing perturbation theory
on the parts of $H$ not included in (\ref{II19}), using our wave function
previously found as the starting point.
It is indeed true that higher order corrections give contributions
proportional to $\rho$.  They can be recognized as constituting the
full Born series for the scattering length.  Likewise, the second term
in (\ref{II19}) (which came from $I_1$) looks like it is the beginning
of a similar series.  Hence we are tempted to write
\begin{equation}
E_0 = \big(\hbar^2 / 2m\big) 4 \pi N \rho \, a
    \biggl\{ 1+ \big(128/15\sqrt{\pi}\big)\big(\rho a^3 \big)^{1/2}
    + \cdots \biggr\}
\label{II31}
\end{equation}
and presumably we now have the beginning of a genuine series in the
density {\it alone}.

Nevertheless, it is important to note that what started out to be
a very reasonable hypothesis -- the pair approximation -- is invalid as
a density expansion.  Even the full pair Hamiltonian (\ref{II17}) will not
give (\ref{II31}) \cite{b4}.  The trouble was that we had thought we were making
some sort of cluster expansion as one does in classical statistical
mechanics.  This may be a reasonable thing to do, but it is essential
that we treat the two--body interaction fully and completely, and this
cannot be done very easily by perturbation theory.  We shall discuss
this matter more fully later but for the present let us make some
attempt to justify (\ref{II31}).

The expression (\ref{II31}) for $E_0$ has not yet been proved to be
correct.  But it is very reasonable.  For one thing the first term is
simply the number of pairs of particles, $\frac12 N(N-1)$, times the
ground state energy of two particles in a large box, $\Ap^2 8\pi
a/V$.\footnote{This is discussed in Chapter~\ref{sect3d} in connection
  with Eq.~(\ref{partint}).}
The second term is harder to understand.  There is no analogue of it
for two particles and it is clearly some sort of quantum-mechanical
correlation effect.  Nevertheless, if the true second order term in
the density is of the form $\rho^{3/2}$ given by (\ref{II31}) it must,
for dimensional reasons, be proportional to a (length)$^{5/2}$.  But
the only relevant length at low energy is the scattering length.

Having considered the ground state energy let us return to the second
term in (\ref{II24}).  Apart from the fact that $\epsilon^\prime$ is
different from $\epsilon$, this spectrum of $H_B$ is the same as for the
original Hamiltonian without interaction.  Any number, $n$, of bosons
(or phonons) of any momentum $\ulk$ can be excited independently of each
other with energy $n \epsilon^\prime (\ulk)$.  The important difference
is that for low momentum $\epsilon (\ulk)$ is proportional to $\k^2$,
whereas $\epsilon^\prime (\k) = [2\Ap^2 \rho \nu (\0)]^{1/2}|\k|$.
This new spectrum is definitely phonon-like (without an energy gap) and
we expect that it is associated with sound propagation.  If so, the
velocity of sound\index{velocity of sound} would be
\begin{equation}
v_s = \lim_{k\rightarrow 0} (1 / \hbar k)\epsilon^\prime (k) = (\rho \nu (0)/m)^{1/2}
    =(\hbar /m)(4\pi \rho \, a_0)^{1/2} \,.
\label{II32}
\end{equation}
We can check this result by using the fundamental definition of the
velocity of sound in terms of the compressibility,\index{compressibility}
\begin{displaymath}
v_s = \biggl[ m^{-1} \big( \partial /\partial \,\rho \big )
  \,\rho^2 \partial / \partial \, \rho \big(E_0/N \big )\biggr]^{1/2} .
\nonumber
\end{displaymath}
Using (\ref{II29}) this gives
\begin{equation}
v_s = \hbar /m \biggl[ 4\pi \rho(a_0 + a_1 ) + 64 \rho \, a_0
    \big(\pi \, \rho \, a_0 ^3\big)^{1/2} \biggr]^{1/2}  .
\label{II33}
\end{equation}
There is agreement between (\ref{II33}) and (\ref{II32}) and we are led to
surmise as before that the correct expression to the first two orders
in the density is
\begin{equation}
v_s = \hbar/m \biggl[ 4 \pi \, \rho \, a \big(1+ (16/\pi)
(\pi \, \rho \, a^3) \big)^{1/2} \biggr]^{1/2} .
\label{II34}
\end{equation}
It is interesting to note that the expression (\ref{II33}), obtained
from $E_0$, is more accurate than (\ref{II32}) obtained from the
phonon spectrum.\index{phonon spectrum} This is curious since we would
have expected that whatever accuracy was inherent in $H_B$, it should
be the same for $\epsilon^\prime (\k)$ as for $E_0$.

Another important feature of the boson type spectrum that we have
obtained is that while there may be something qualitatively correct
about it, it is much too simple to be taken literally.  It is hardly to
be expected that the spectroscopy of the true spectrum will fall into a
pattern associated with independent normal modes.  It must be more
complicated and indeed higher order perturbation theory indicates that
the phonons interact with one another.  There are two ways to describe
this state of affairs.  The usual way is to say that the interaction
causes the phonons to decay with a finite lifetime.  The second way is
to say that the unitary transformation leading to the $b_{\ulk}$'s was
only a partial diagonalization of the full Hamiltonian which must always
have real eigenvalues since it is Hermitian.  In other words the true
energy spectrum and eigenfunctions are simply more complicated than we
have so far envisaged.  Still, the independent phonon idea may be
justified provided we do not excite too many of them -- this would be
true at low temperature.

We might, however, anticipate a difficulty of another sort.  The
Bogoliubov approximation is essentially perturbation theory, albeit of a
sophisticated sort, because it assumes that the system does not change
drastically when we switch on the interaction.  Were it otherwise $H_B$
could not be justified.  It is generally held that if we imagine
$v(\ubr)$ to be proportional to a coupling constant $\lambda$, then
after passing to the bulk limit there will be an analytic singularity in
$\lambda$ (in $E_0 (\lambda)$ for example) at $\lambda =0$.  A Bose
gas with an interaction, however weak, may be qualitatively different
from the non-interacting gas.  If this is so then we might expect new
types of normal modes (or phonons) that have no counterpart for free
bosons.  Indeed, when we examine a one-dimensional model later we shall
see that to the extent that the spectrum is phonon like, it can best be
described by two separate $\epsilon (\k)$ curves, not one.

Where might this second phonon spectrum come from?  In the Bogoliubov
ground state most of the particles are still condensed at $\ulk = \0$
and there is a small (but non-zero) background of particles with
$\ulk \neq \0$.  The Bogoliubov phonons are qualitatively the same as the
free ``phonons'';  particles are excited out of $\ulk =\0$ sea.
The renormalized energy $\epsilon^\prime$ comes about because the
$\ulk =\0$ background now has to readjust itself.  It is quite possible
that there is another type of excitation which would be associated
directly with excitation of the $\ulk \neq \0$ background.  Similar
suggestions have been made before (see \cite{b10}).

A physical reason for the possibility of another spectrum is the
critical velocity\index{critical velocity} failure of current theory.
Some time ago Landau\index{Landau} gave an argument to account for
superfluidity based on the phonon hypothesis.  Suppose one has a mass,
$M$, of fluid moving with velocity $v$ and momentum $P= Mv$.  Let us
suppose that in the rest system of the fluid it is in its ground state
and hence has momentum equal to zero.  In the laboratory system the
fluid will have energy $E_1=E_0 +P^2/2M$.  If the fluid interacts with
the walls of the channel in which it is moving it must {\it lose}
energy and momentum, so that its energy is now $E_1 - \Delta$ and its
momentum is $P- \delta$.  If we make a Galilean transformation with
velocity $-v$, the energy will be $E_0 - \Delta + \delta v$ and the
momentum will be $- \delta$.  Now consider the $\epsilon ^\prime (k)$
curve.  It is presumably possible to draw a straight line
$\epsilon^{\prime\prime} (k)= sk$ such that $\epsilon^{\prime\prime}$
is just tangent to, and otherwise always under, $\epsilon^\prime (k)$
(we could not do this for free bosons unless $s$ were to be zero).
This line defines a velocity $v_c = s / \hbar$.  In order to impart
momentum $\delta$ to the system the energy must therefore increase at
least by an amount $v_c \delta$.  In the above example, however, the
energy increased by an amount less than $v \delta$.  The conclusion is
that if the velocity of the fluid is less than $v_c$, the fluid will
not be able to lose any momentum to the walls of the channel, and
hence it will display superfluid behavior.

We would expect from the Bogoliubov solution that $v_c$ should be of the
order of magnitude of the sound velocity.  Experimentally, $v_c$ is
found to be very much less than $v_s$, and it is also found to depend
sensitively on the diameter of the channel -- especially for very narrow
channels.  It is clear that another type of excitation, with an energy
much less than that given by (\ref{II24b}), could account for the
discrepancy.  Indeed, we will find such a spectrum in the
one--dimensional model to be analyzed later.

Finally, we wish to emphasize another important feature of
Bogoliubov's analysis, namely that the excitation spectrum is intimately
connected with the ground state wave function.  This fact is important
because it shows that calculating the ground state energy is not merely
an academic exercise.  Although $E_0$ is a rather unimportant number
(it can be measured, however) it is manifestly clear that we cannot
really hope to be able to predict the dynamics of this complicated
system unless we are in a position to calculate the much simpler
quantity, $E_0$, along with some of the important properties of the
ground state, such as the two--particle correlation function which we
shall discuss later.

We shall write down the Bogoliubov wave function
$\Psi_0^\prime = e^{-{\rm i}S} \Psi_0$.  $S$ is given by (\ref{II21}), but that
expression is needlessly complicated because it contains
$a_{\ulk} a_{-\ulk}$ which vanishes when applied to $|0\rangle$.
A little algebra will show that
\begin{equation}
\Psi_0^\prime = \exp \left\{ \sum \hskip -.001cm \raisebox{.9ex}{$^\prime$}
   h(\ulk) a_{\ulk}^\dagger a_{-\ulk}^\dagger \right\} \Psi_0
\label{II35}
\end{equation}
where
\begin{equation}
\begin{aligned}
\qquad\qquad h(\ulk) &=  -\frac{1}{2g(\ulk)} \biggr\{
    f (\ulk) - \big (f(\ulk)^2 - g(\ulk)^2 \big)^{1/2}
             \biggl\}  \\
&= - \frac12 \, -\frac12 (\rho \, \nu (\ulk))^{-1} \big(
     \epsilon (\ulk) - \epsilon^\prime (\ulk) \big)\,.
\end{aligned}
\label{II36}
\end{equation}
In (\ref{II35}) the state $\Psi_0^\prime$ is not normalized, and $\Psi_0$
means the free particle ground state, i.e., $(a_\0^\dagger)^N |0\rangle$.
The difficulty with $\Psi_0^\prime$, of course, is that it does not have
a definite particle number.  One way to fix this would be to replace the
$a^\dagger$'s by the Kromminga, Bolsterli $b^\dagger$'s.  This would not alter
$E_0$.  A simpler procedure would be to multiply each $a_\k^\dagger$ by
$N^{-1/2} a_\0$.  If we then write out the first few terms in
(\ref{II35}) we get
\begin{equation}
\begin{aligned}
\qquad\qquad \Psi_0^\prime = &(a_\0^\dagger)^N |0\rangle +
   N \sum\limits_{\ulk}
       \hskip -.001cm \raisebox{.9ex}{$^\prime$}
     h(\ulk) a_{\ulk}^\dagger a_{-\ulk}^\dagger (a_\0^\dagger)^{N-2} |0\rangle \\
& +N^2/2! \sum\limits_{\ulk , \ulq} h (\ulk) h (\ulq)
  a_\k^\dagger a_{-\k}^\dagger a_{\ulq}^\dagger a_{-\ulq}^\dagger (a_\0^\dagger)^{N-4} |0 \rangle  \\
& + \ldots \,.
\end{aligned}
\label{II37}
\end{equation}
In essence, (\ref{II37}) may be taken as the statement of the
Bogoliubov {\it ansatz}, just as much as (\ref{II19}), for the most general
wave function we could construct would be
\begin{equation}
\begin{aligned}
\qquad\qquad \Psi =  & \; (a_\0^\dagger)^N |0\rangle + N \sum\limits_k
       \hskip -.001cm \raisebox{.9ex}{$^\prime$}
   h^{(2)} (\ulk) a_\k^\dagger a_{-\k}^\dagger (a_\0^\dagger)^{N-2} |0\rangle \\
&+ N \sum\limits_{\ulk , \ulq}
       \hskip -.001cm \raisebox{.9ex}{$^\prime$}
    h^{(3)} (k, q) a_{\ulk}^\dagger a_{\ulq}^\dagger a_{-\ulk -\ulq}^\dagger
       (a_\0^\dagger) ^{N-3} |0\rangle \\
& + N^2/2! \sum\limits_{\ulk , \ulq , \ulp}
       \hskip -.001cm \raisebox{.9ex}{$^\prime$}
  h^{(4)} (\ulk , \ulq , \ulp )
    a_{\ulk}^\dagger a_{\ulq}^\dagger a_{\ulp}^\dagger
    a_{-\ulk -\ulp -\ulq}^\dagger (a_\0^\dagger)^{N-4}|0\rangle\\
& + \ldots \,.
\end{aligned}
\label{II38}
\end{equation}
(Notice that there can be no linear term because of momentum
conservation.)  The quadratic term in (\ref{II37}) is quite general, but
(\ref{II37}) has no triplet or other odd power terms.  Also the quartic term
is of a very special kind.  The Bogoliubov {\it ansatz} is that
\begin{equation}
h^{(4)} (\ulk , \ulq , \ulp ) =
  \delta_{\ulp + \ulk} h^{(2)} (\ulk) h^{(2)} (\q) \,,
\label{II39}
\end{equation}
and so on for succeeding even powers.  We could just as well have
derived the Bogoliubov results by starting with (\ref{II37}) instead of with
$H_B$ in (\ref{II19}).

\newpage
\
\thispagestyle{empty}

\chapter{An Exactly Soluble Model}\label{chap3}

\textit{This appendix is taken from Section $5$ of {\rm\cite{EL2}}, which summarizes
the papers {\rm\cite{LL,LL2}}. Until recently this model was just an
amusing exercise, but it now appears that it is possible to
produce a one-dimensional gas like this in the laboratory and
the experimental results agree, so far, with the rigorous calculations
on the model. In Chapter {\rm\ref{1dsect}} we show how this
one-dimensional model emerges from three-dimensional models when
the trap is long compared to its width.}
\bigskip \bigskip

With various approximation schemes before us, it would certainly be
advantageous to have at least one problem of the type (\ref{II1}) that can be
solved exactly.  We should then be able to verify whether or not the
previously mentioned qualitative ideas are correct.

Such model problem is the one-dimensional Bose gas with a pair-wise
repulsive $\delta$-function potential\index{delta-f@$\delta$-function Bose gas}
(\cite{LL,LL2}).  Using units in which $\hbar^2/2m=1$, the
Schr\"odinger equation\index{Schr\"odinger equation} is
\begin{equation}
\left\{ - \sum\limits_{i=1}^N \partial^2 / \partial x_i^2 + 2c
   \sum_{i < j} \delta(x_i - x_j) \right\} \psi = E \psi \,.
\label{V1}
\end{equation}
Hence $2c \geq 0$ is the amplitude of the $\delta$--function.
If $L$ is the length of the line then $\rho = N/L$.  It is well known
that a $\delta$--function potential is equivalent to the following
boundary conditions\index{boundary condition} whenever any two particles touch each other
(irrespective of the value of the remaining $N-2$ coordinates):
\begin{equation}
(\partial / \partial x_j - \partial/\partial x_k)
       \psi |_{x_j = x_k^+}
    - (\partial / \partial x_j - \partial /\partial x_k ) \psi
        |_{x_j = x_k^-} = 2c \psi |_{x_j = x_k} \,.
\label{V2}
\end{equation}
We also note that we are seeking symmetric solutions to (\ref{V1}) and hence
if we know $\psi$ in $R_1$
\begin{equation}
R_1 : 0 \leq x_1 \leq \cdots \leq x_N \leq L \,,
\label{V3}
\end{equation}
we know $\psi$ everywhere by symmetric extension.
Thus our equations become
\begin{equation}
\left\{ - \sum\limits_{i=1}^N \partial^2 / \partial x_i^2 \right\}
     \psi = E \psi \;\;\underline{\text{inside}} \; R_1
\label{V4}
\end{equation}
\begin{equation}
 \big(\partial / \partial x_{j+1} - \partial/\partial x_j \big)
    \psi |_{x_{j+1} = x_j^+}  = c \psi |_{x_{j+1} = x_j} \,.
\label{V5}
\end{equation}
Moreover, the original periodic boundary conditions can be interpreted
as
\begin{equation}
\begin{aligned}
\psi(0, x_2 , \ldots , x_N) &= \psi(x_2 , \ldots , x_N, L)
      \\
 \partial/\partial x \, \psi(x , x_2 , \ldots , x_N ) |_{x=0}& =
   \partial/\partial x \,\psi(x_2 , \ldots , x_N , x) |_{x=L} \,.
\end{aligned}
\label{V6}
\end{equation}

To solve these equations, consider the function
\begin{equation}
\varphi (x_1, \ldots , x_N) = \text{Det} | \exp ({\rm i} \, \txk_i x_j)|
\label{V7}
\end{equation}
where $\txk_1 , \ldots , \txk_N \equiv \{\txk\}$ are any set
of $N$ distinct numbers.  Now define $\psi$ by
\begin{equation}
\psi = \prod_{j >i} (\partial / \partial x_j - \partial/\partial
       x_i +c)\varphi \,.
\label{V8}
\end{equation}
It is readily verified that $\psi$ satisfies (\ref{V5}) automatically, as
well as (\ref{V4}) with
\begin{equation}
E = \sum\limits_{i=1}^N \txk_i^2 \,.
\label{V9}
\end{equation}
It is (\ref{V6}) which determines the numbers $\{\txk\}$ and it may be shown
that this is equivalent to the $N$ simultaneous equations
\begin{equation}
(-1)^{N-1} \exp (-{\rm i} \, \txk_j L) =
    \exp \left[ {\rm i} \sum\limits_{s=1}^N \theta (\txk_s -\txk_j)\right] \,,
\label{V10}
\end{equation}
where
\begin{equation}
\theta(\txk) = -2 \tan^{-1} (\txk/c) \, , \qquad -\pi < \theta < \pi
    \qquad \text{for real } \, \txk \,.
\label{V11}
\end{equation}
Equation (\ref{V10}) may be rewritten as
\begin{equation}
\delta_j \equiv (\txk_{j+1} - \txk_j) L = \sum\limits_{s=1}^N
 \bigl[ \theta(\txk_s - \txk_j )
     - \theta (\txk_s -\txk_{j+1})\bigr] + 2\pi \,n_j
\label{V12}
\end{equation}
for $j=1, \ldots , N-1$.  Equation (\ref{V12}) is $N-1$ simultaneous
equations for the $\delta_j$; when they are found, the individual
$\txk$'s
may be obtained from (\ref{V10}).  The $n_j$ in (\ref{V12}) are integers, and it
can be shown that for any choice of the $n_j$ such that all $n_j \geq 1$
there is a solution to (\ref{V12}) with real $\delta_j\geq 0$.  Presumably
these solutions are unique and are the only solutions to (\ref{V10}); at
least this is true if $N=2$.  These $n$'s are therefore the quantum
numbers of the system.

If we pass now to the bulk limit, $N \rightarrow \infty$, the
ground state will be obtained when all $n_j = 1$, because this choice
clearly minimizes (\ref{V9}).  It may be verified that all the $\txk$'s must
lie between $- \pi \rho$ and $\pi \rho$, which means that the spacing
between the ${\txk}$'s decreases as $N \rightarrow \infty$.  If we set
\begin{equation}
f(\txk_j) = 1/\delta_j
\label{V13}
\end{equation}
then $Lf(\txk) =$ the number of $\txk$'s between $\txk$
and $\txk+d\txk$.
Furthermore, we denote by $K$ the common value of $\txk_N$ and $-\txk_1$.
Using Poisson's formula, (\ref{V10}) may be converted into an integral
equation:
\begin{equation}
2c \int_{-K}^K f(p) dp \bigg/ \bigl[ c^2 + (p-\txk)^2 \bigr]
    = 2 \pi f(\txk) -1 \,,
\label{V14}
\end{equation}
with
\begin{equation}
\int_{-K}^K f(\txk) \; d\txk = \rho
\label{V15}
\end{equation}
being the condition that the total number of particles be $N$.
This latter condition determines $K$.  The ground state energy is then
\begin{equation}
E_0 = (N/\rho) \int_{-K}^K f(k) \txk^2 \, d\txk \,.
\label{V16}
\end{equation}

At this point it is convenient to introduce the dimensionless coupling
constant
\begin{equation}
\gamma = c/\rho \,,
\label{V17}
\end{equation}
in terms of which we may write
\begin{equation}
E_0 = N\rho^2 e (\gamma) \,.
\label{V18}
\end{equation}
Equations (\ref{V14}) and (\ref{V15}) can easily be solved on a computer
for all values of $\gamma \geq 0$.  Graphs for $K(\gamma)$, $e(\gamma)$
and $f(\txk, \gamma)$ are given in Figures 1, 3 and 2 respectively
of reference \cite{LL}.  The results, briefly, are these:
\begin{equation}\label{V19}
\begin{aligned}
 \underline{\text{small } \gamma} : \qquad
  e(\gamma) &= \gamma - (4/3 \pi ) \gamma^{3/2} \\
K(\gamma)& = 2 \rho \gamma^{1/2} \\
f(\txk, \gamma) &= (2 \pi \rho \gamma )^{-1} (4 \rho^2 \gamma -
\txk^2)^{1/2}
\end{aligned}
\end{equation}
\vskip.3cm
\begin{equation}\label{V20}
\begin{aligned}
\underline{\text{large } \gamma} : \qquad
  e(\gamma) &= (\pi^2 \gamma^2 ) /3) (\gamma +2)^{-2}  \\
K(\gamma) & = \pi \rho \gamma (\gamma +2)^{-1}  \\
f(\txk, \gamma) &= (\gamma +2 )/ 2 \pi \gamma \,.
\end{aligned}
\end{equation}

We may inquire how Bogoliubov's theory\index{Bogoliubov!method} fares for this problem.
This theory yields (\ref{V19}) for $e (\gamma )$ for {\it all} $\gamma$,
a result which is in fair agreement with the correct $e(\gamma)$ up
to $\gamma =2$ and then becomes quite useless.
While the true $e (\gamma)$ is a monotonically
increasing function of $\gamma$, with an asymptotic value of $\pi^2/3$,
(\ref{V19}) is actually negative for $\gamma > (3 \pi /4)^2$.

When $\gamma$ is infinite, the wave functions, (\ref{V8}), are given by
\begin{equation}
\psi = \text{Det } |\exp ({\rm i} \, \txk_i x_j)|
\label{V21}
\end{equation}
where each $k_i$ is of the form:
\begin{equation}\label{V22}
\begin{array}{ll}
\hskip 1.5in \txk = (2\pi/L) \times (\text{integer}) , \;
        & (N \text{ odd})  \\
 \hskip 1.623in  = ( 2\pi /L) \times (\text{integer}) +
        \pi/L, &(N \text{ even}) \,.
\end{array}
\end{equation}
For the ground state the $\txk$'s run between $-\pi \rho$ and $\pi \rho$.
Another way to express the ground state $\psi_0$ (for any $N$) is to recognize (\ref{V21})
as a Vandermonde determinant, whence
\begin{equation}
\psi_0 = \prod_{i<j} \sin (\pi |x_j - x_i |/L)
\label{V23}
\end{equation}
for {\it all} values of the $x_i$ in $(0,L)$.

Turning now to the excitation spectrum, let us first consider the
infinite $\gamma$ case, where the $\txk$'s are given by (\ref{V22}).
The spectrum will be recognized as the same as that of a one--component
Fermi gas.

An elementary excitation\index{elementary excitations} consists in
increasing one of the momenta from $q < K = \pi \rho$ to $\txk > K$.
The energy is then
\begin{equation}
\epsilon (\txk,q) = \txk^2 - q^2 \,,
\label{V24}
\end{equation}
and the momentum of the state is
\begin{equation}
p(\txk,q) = \txk - q \,.
\label{V25}
\end{equation}
The difficulty with this description is that it is completely different
from what we had been led to expect on the basis of the pair Hamiltonian
calculation.  For one thing there is no unique $\epsilon (p)$ curve.  For
another, each excitation may take place only once, whereas a boson type
excitation can be repeated as often as desired.

In order to make this spectrum appear boson-like, let us define
{\it two} types of elementary excitations:  for type I we increase $k_N$
from $K$ to $K+p$ (where $p > 0$).  The momentum of the state is $p$ and
the energy is
\begin{equation}
\epsilon_1 (p) = (K+p)^2 - K^2 = p^2 + 2 \pi  \rho \, p\,.
\label{V26}
\end{equation}
For type II we increase one of the momenta from $K -p +2\pi /L$
(where $0<p<K$) to $K+2\pi /L$.  Here the momentum is again $p$ and
\begin{equation}
\epsilon_2 (p) = 2 \pi \rho \,p - p^2 \,.
\label{V27}
\end{equation}
A type II excitation is defined {\it only} for momentum less than
$\pi \rho$.  With this description we have achieved our aim, but at the
expense of introducing two $\epsilon (p)$ curves.  Any type I excitation
can be repeated as often as desired providing we agree always to take
the last available $k$ less than $K$ and increase it by $p$.
Similarly, with the same proviso, a type II excitation is boson-like.
In addition, both I and II excitations may occur simultaneously.
In fact the excitation in (\ref{V24}) may be thought of as a simultaneous
type I and II excitation with momenta $p-k$ and $K-q$, respectively.
It will be seen that provided we make a {\it finite} number of
excitations by the above rule the energies and momenta will be additive
to order $1/N$.  Thus if we make $n$ type I excitations with momenta
$p_1 , \ldots , p_n$ and $m$ type II excitations with momenta
$q_1 , \ldots , q_m$, the energy as given by (\ref{V26}) and (\ref{V27}) would be
\begin{equation}
E = \sum\limits_{j=1}^n \epsilon_1 (p_j) + \sum\limits_{j=1}^m
     \epsilon_2 (q_j) \,,
\label{V28}
\end{equation}
while the momentum would be
\begin{equation}
P= \sum\limits_{j=1}^n p_j + \sum\limits_{j=1}^m q_j \,.
\label{V29}
\end{equation}
Now if we examine the state we would obtain with these excitations we
will find that the {\it true} energy of the state agrees with (\ref{V28}) to
order $1/N$ while the {\it true} momentum is exactly given by (\ref{V29}).

There is one {\it caveat}, however.  In achieving this boson description
of the excitations we have, in reality, counted each state twice.  If
the rules above are carefully examined it turns out that a state with
$n$ type I excitations with momentum $p=2 \pi m/L$ is identical to the
state with $m$ type II excitations of momentum $p = 2\pi n/L$.
The spectrum is therefore really much more complicated than we had
imagined.  If we give up the double spectrum point of view in order to
avoid the double counting, then we would have to regard a type II
excitation, for example, as essentially an infinite number of type I
excitations with vanishing small momentum, $2 \pi /L$.  Not only is this
unnatural, but the energy would then be given incorrectly.
{F}rom (\ref{V26}) we would conclude that $\epsilon_2 (p) = 2 \pi \rho \, p$,
where as the correct expression is (\ref{V27}).

When $\gamma$ is not infinite the same qualitative conclusions apply.
It will be appreciated that increasing one of the $k$'s is the same as
putting all but one of the $n_j$'s in (\ref{V12}) equal to unity.
For a type I excitation the singular $n$ is $n_{N-1}$ and this is set
equal to $qL/2\pi$.  For a type II excitation the singular $n$ is one of
the $n_j$'s (where $N/2 < j < N-1$) and this is set equal to 2.  The
difficulty with the finite $\gamma$ case is that when the $n$'s
are changed in this way {\it all} the $k$'s are shifted --- not merely
one of them.  This shift can be computed from (\ref{V12}) and one can again
obtain two $\epsilon (p)$ functions.  The details are given in reference
\cite{LL2}.  It is found that as $\gamma \rightarrow 0$,
$\epsilon_1 (p) \rightarrow p^2$ (the free boson function) while
$\epsilon_2 (p) \rightarrow 0$ for all $p$.
But for every $\gamma$ there are always {\it two} $\epsilon (p)$ curves.

Bogoliubov's theory, on the other hand, predicts only {\it one}
$\epsilon (p)$ curve:
\begin{equation}
\epsilon (p) = p(p^2 + 4 \gamma \rho^2)^{1/2} , \quad
          (\text{Bogoliubov}) \,.
\label{V30}
\end{equation}
It turns our that for small $\gamma$, (\ref{V30}) is quite close to the true
$\epsilon_1(p)$ (cf. Fig. 4 of Reference \cite{LL2}).  The $\epsilon_2(p)$
curve is entirely missing from Bogoliubov's theory.

\chapter[Definition and Properties of Scattering Length]{Definition and Properties of\vspace*{-2mm}\newline Scattering Length}\label{appscatt}
\index{scattering length}

\textit{This appendix is from the paper {\rm\cite{LY2d}} and is included
  here because the scattering length plays an important role in these
  notes and because it is not easy to find a rigorous definition in
  the textbooks.}  \bigskip\bigskip

In this appendix we shall define and derive the scattering length and
some of its properties. The reader is referred to \cite{LL01},
especially chapters 9 and 11, for many of the concepts and facts we
shall use here.

We start with a potential $\frac{1}{2}v(\x)$ that depends only on the
radius,
$r=|\x|$, with $\x\in \R^n$. For simplicity, we assume that
$v$ has finite range; this condition can easily be relaxed, but we shall
not do so here, except for a remark at the end that shows how
to extend the concepts to infinite range, nonnegative potentials.
Thus, we assume that
\begin{equation}
v(r) =0  ~~~~~~~~ {\rm for}~ r>R_0.
\end{equation}
We decompose $v$ into its positive
and negative parts,
$v=v_+ - v_-$, with $v_+,\ v_-\geq 0$,  and assume the following for
$v_-$ only (with $\varepsilon >0$):
\begin{equation}
v_- \in \ \ \ \begin{cases} L^1(\R^1) & {\rm for}~ n=1 \\
L^{1+\varepsilon}(\R^2)  & {\rm for}~ n=2 \\
L^{n/2}(\R^n) & {\rm for}~ n\geq3.
\end{cases}
\label{spaces}
\end{equation}
In fact, $v$ can even be a finite, spherically symmetric
measure, e.g., a sum of
delta functions.

We also make the {\it important assumption} that $ \frac{1}{2}v(\x)$
has no negative energy bound states\index{bound states} in
$L^{2}(\R^n)$, which is to say we assume that for all $\phi \in
H^1(\R^n)$ (the space of $L^2$ functions with $L^2$ derivatives)
\begin{equation}
\int_{\R^n} \left( \mu |\nabla \phi (\x)|^2 + \frac{1}{2}v(\x)|\phi(\x)|^2\right)
d^n\x \geq 0 \ .
\label{nobound}
\end{equation}

The scattering length is defined, of course, even when bound states are
present, but it is not defined by the variational principle given below.

\begin{thm}\label{energy}
 Let $R>R_0$ and let $B_R \subset \R^n$
denote the ball $\{\x: 0<|\x| <R\}$ and $S_R$ the sphere $\{\x: |\x| =R\}$.
For $f\in H^1(B_R)$ we set
\begin{equation}
\mathcal{E}_R[\phi] = \int_{B_R} \left( \mu |\nabla \phi(\x)|^2 +
\frac{1}{2}v(\x)|\phi(\x)|^2 \right) d^n\x \,.
\label{E}
\end{equation}
Then, in the subclass of functions such that $\phi(\x) =1$ for all $\x\in
S_R$,
there is a unique function $\phi_0$ that minimizes $\mathcal{E}_R[\phi]$.
This function is nonnegative and spherically symmetric, i.e,
\begin{equation}
\phi_0(\x)=f_0(|\x|)
\end{equation}
with a nonnegative function $f_0$ on the interval $(0,R]$, and it satisfies
the equation
\begin{equation}\label{dist}
-\mu \Delta \phi_0(\x) + \frac{1}{2}v(\x)\phi_0(\x) =0
\end{equation}
in the sense of distributions on $B_R$, with  boundary
condition $f_0(R)=1$.

For $R_0<r<R$
\begin{equation}
f_0(r) =  f_0^{\rm asymp}(r) \equiv
\begin{cases} (r-a)/(R-a)  & {\rm for}\ n=1 \\
\ln (r/a)/\ln(R/a)  & {\rm for}\  n=2 \\
   (1-ar^{2-n})/( 1-aR^{2-n})& {\rm for}\  n\geq 3
\end{cases}
\label{asymp}
\end{equation}
for some number $a$ called the \underline{\em scattering length}.

The minimum  value of $\mathcal{E}_R[\phi]$ is
\begin{equation}
E= \begin{cases} 2\mu /(R-a) & {\rm for}\ n=1 \\
2 \pi \mu /\ln(R/a)  & {\rm for}\  n=2 \\
   2 \pi^{n/2} \mu a/[\Gamma(n/2)( 1-aR^{2-n})]&
{\rm for}\  n\geq 3 .
\end{cases}
\label{emin}
\end{equation}
\label{minimizer}
\end{thm}

\noindent {\it Remarks:}
1. Given that the minimizer is spherically symmetric for every $R$, it
is then easy to see that the $R$ dependence is trivial.  There is really
one function, $F_0$, defined on all of the positive half axis, such that
$f_0(r) =F_0(r)/F_0(R)$. That is why we did not bother to indicate the
explicit
dependence of $f_0$ on $R$. The reason is a simple one: If
$\widetilde{R} >R$, take the minimizer $\widetilde{f}_0$ for
$\widetilde{R} $ and replace its values for $r<R$ by
$f_0(r)\widetilde{f}_0(R)$, where $f_0$ is the minimizer for the $B_R$
problem.  This substitution cannot increase
$\mathcal{E}_{\widetilde R}$.  Thus, by
uniqueness, we must have that $\widetilde{f}_0(r) =
f_0(r)\widetilde{f}_0(R)$ for $r\leq R$.
\medskip

2. {}From \eqref{asymp} we then see that $f_0^{\rm asymp}(r)\geq 0$ for all
$r>R_0$, which implies that $a\leq R_0$ for $n\leq 3$ and
$a\leq R_0^{n-2}$ for $n>3$.
\medskip

3. According to our definition \eqref{asymp}, $a$ has the dimension of a
length only when $n\leq 3$.
\medskip

4. The variational principle \eqref{E}, \eqref{emin} allows us
to  discuss the connection between the
scattering length and $\int v$. We
recall Bogoliubov's perturbation
theory \cite{BO,BZ}, which says that to leading order in the density $\rho$,
the
energy per particle of a Bose gas is $e_0(\rho)\sim 2\pi \rho \int v$,
whereas the correct formula in two-dimensions
is $4 \pi \mu \rho |\ln(\rho a^2)|^{-1}$.
The Bogoliubov formula is an upper bound (for all $\rho$) since it is the
expectation value of $H_{N}$ in the non-interacting ground state
$\Psi \equiv 1$. Thus, we must have
$\frac{1}{2} \int v \geq   |\ln(\rho a^2)|^{-1}$ when
$\rho a^2 \ll 1$, which suggests that
\begin{equation}
\int_{\R^2} v \geq \frac{4\pi \mu }{\ln (R_0/a)}.
\label{2scattineq}
\end{equation}
\bigskip

Indeed, the truth of (\ref{2scattineq}) can be verified by using the
function
$\phi(\x) \equiv 1$ as a trial function in (\ref{E}). Then, using
(\ref{emin}),
$\frac{1}{2} \int v \geq E = 2\pi \mu/ \ln(R/a)$
for all $R \geq R_0$, which proves (\ref{2scattineq}).
As $a \to 0$, (\ref{2scattineq}) becomes an equality, however,
in the sense that $(\int_{\R^2} v) \ln(R_0/a) \to 4\pi \mu$.

In the same way, we can derive the inequality of
Spruch and Rosenberg\index{Spruch--Rosenberg inequality} \cite{SR} for dimension 3 or more:
\begin{equation}\label{3scattineq}
\int_{\R^n} v \geq \frac{4\pi^{n/2} \mu a}{\Gamma(n/2)}.
\end{equation}
(Here, we
take the limit $R\to \infty$ in \eqref{emin}).

In one-dimension we obtain (with $R=R_0$)
\begin{equation}\label{1scattineq}
\int_{\R} v \geq \frac{4\mu}{R_0 - a }.
\end{equation}

\begin{proof}[Proof of Theorem {\rm\ref{energy}}] Given any $\phi\in H^1$ we can
replace it by the square root of the spherical average of $|\phi|^2$. This
preserves the boundary condition at $|\x|=R$, while the   $v$ term in
\eqref{E} is unchanged. It also lowers the gradient term in (\ref{E})
because the map $\rho \mapsto \int (\nabla \sqrt \rho)^2$ is convex
\cite{LL01}.  Indeed,  there is a strict decrease unless $\phi$ is
already
spherically symmetric and nonnegative.

Thus, without loss of generality, we may consider only nonnegative,
spherically symmetric functions. We may also assume that in the annular
region $\mathcal{A} = \{\x \ : \ R_0\leq |\x|\leq R\}$ there is some $a$
such that (\ref{asymp}) is true because these are the only spherically
symmetric, harmonic functions in $\mathcal{A}$.  If we substitute for
$\phi$ the harmonic function in $\mathcal{A}$ that agrees with $\phi$ at
$|\x|=R_0$ and $|\x|=1$ we will lower $\mathcal{E}_R$ unless $\phi$ is
already
harmonic in $\mathcal{A}$.  (We allow the possibility $a=0$ for $n\leq
2$, meaning that $\phi =$ constant.)

Next, we note that $\mathcal{E}_R[\phi]$ is bounded below. If it were
not bounded then (with $R$ fixed) we could find a sequence $\phi^j$
such that $\mathcal{E}_R(\phi^j) \to -\infty$.  However, if $h$ is a
smooth function on ${\mathbb R}_+$ with $h(r) =1 $ for $r<R+1$ and
$h(r) =0$ for $r>2R+1$ then the function $\widehat{\phi^j}(\x) =
\phi^j(\x)$ for $|\x|\leq R$ and $\widehat{\phi^j}(\x) = h (|\x|) $
for $|\x|>R$ is a legitimate variational function for the $L^2(\R^n)$
problem in (\ref{nobound}).  It is easy to see that
$\mathcal{E}_R[\widehat{\phi^j}] \leq \mathcal{E}_R[\phi^j] +{\rm
  (const)} R^{n-2}$, and this contradicts (\ref{nobound}) (recall that
$R$ is fixed).

Now we take a minimizing sequence $\phi^j$ for $\mathcal{E}_R$ and
corresponding $\widehat{\phi^j}$ as above. By the assumptions on $v_-$ we
can see that the kinetic energy $T^j = \int |\nabla \phi^j|^2$ and $\int
|\phi^j|^2$ are  bounded. We can then find a subsequence of the
$\widehat{\phi}^j$
that converges weakly in $H^1$ to some spherically symmetric
$\widehat{\phi}_0(\x)=\widehat f_0(|\x|)$. Correspondingly, $\phi^j(\x)$
converges weakly in $H^1(B_R)$
to $\phi_0(\x)=f_0(|\x|)$.  The important point is that the term -$\int
v_-|\phi^j|^2$ is
weakly continuous while the term $\int v_+|\phi^j|^2$ is weakly lower
continuous \cite{LL01}. We also note that $f_0(R) =1$ since the
functions $\widehat{\phi^j}$ are identically equal to $1$ for $R<|\x|<R+1$
and
the limit $\widehat{\phi}_0$ is continuous away from the origin
since it is spherically symmetric and in $H^1$.

Thus, the limit function $\phi_0$ is a minimizer for  $\mathcal{E}[\phi]$
under the condition $\phi=1$ on $S_R$. Since it is a minimizer, it
must be harmonic in $\mathcal{A}$, so $\eqref{asymp}$ is true.
Eq. \eqref{dist} is standard and is obtained by replacing
$\phi_0$ by $\phi_0+\delta \psi$, where $\psi$ is any infinitely
differentiable
function that is zero for $|\x|\geq R$. The first variation in $\delta$
gives \eqref{dist}.

Eq. \eqref{emin} is obtained by using integration by parts to compute
$\mathcal{E}_R[\phi_0]$.

The uniqueness of the minimizer can be proved in two ways. One way is to
note that if $\phi_0\neq \psi_0$ are two minimizers then, by the convexity
noted above, $\mathcal{E}_R [\sqrt{\phi_0^2 +\psi_0^2}] <
\mathcal{E}_R[\phi_0]
+\mathcal{E}_R[\psi_0]$. The second way is to notice that all minimizers
satisfy \eqref{dist}, which is a linear, ordinary differential equation
 for $f_{0}$  on $(0,R)$ since all minimizers are spherically symmetric, as we noted.
But the solution of such equations, given the value at the end points, is
unique.
\end{proof}

We thus see that if the Schr\"odinger operator on $\R^n$ with potential
$\frac{1}{2}v(\x)$ has no negative energy bound state then  the
scattering length in (\ref{asymp}) is well defined by a variational
principle. Our next task is to find some properties of the minimizer
$\phi_0$. For this purpose we shall henceforth assume that $v$ is
{\it nonnegative}, which guarantees (\ref{nobound}), of course.

\begin{lem} \label{lem1.1}
 If $v$ is nonnegative  then for all $0<r\leq R$ the
minimizer $\phi_0(\x)=f_0(|\x|)$ satisfies

A)
\begin{equation}
f_0(r) \geq f_0^{\rm asymp}(r),
\label{bound2d}
\end{equation}
where $ f_0^{\rm asymp}$ is given in (\ref{asymp})

B)
$f_0(r)$ is a monotonically nondecreasing function of $r$.

C) If $v(r)\geq \widetilde{v}(r) \geq 0 $ for all $r$ then
the corresponding
minimizers satisfy $f_0(r) \leq  \widetilde{f}_0(r)$ for all $r<R$.
Hence, $a >  \widetilde{a} \geq 0$.

\end{lem}

\begin{proof}
Let us define $f_0^{\rm asymp}(r)$ for {\it all} $0<r<\infty$ by
\eqref{asymp}, and let us extend $f_0(r)$ to all $0<r<\infty$ by
setting $f_0(r) = f_0^{\rm asymp}(r)$ when $r\geq R$.

To prove A) Note that $-\Delta \phi_0 = -\frac{1}{2}v\phi_0$, which implies
that $\phi_0$ is subharmonic (we use $v\geq 0$ and $\phi_0 \geq 0$, by
Theorem~\ref{energy}). Set $h_{\varepsilon}(r)= f_0(r) - (1+\varepsilon) f_0^{\rm
asymp}(r)$ with $\varepsilon >0$ and small.  Obviously, $\x\mapsto
h_\varepsilon(|\x|)$ is
subharmonic on the open set $\{\x: 0<|\x|<\infty \}$ because $f_0^{\rm
asymp}(|\x|)$ is harmonic there. Clearly, $h_{\varepsilon} \to -\infty$ as
$r\to \infty$ and $h_{\varepsilon}(R) = -\varepsilon$.  Suppose that
(\ref{bound2d}) is false at some radius $\rho <R$ and that $h_0(\rho) =
-c<0$. In the annulus $\rho < r < \infty$, $h_{\varepsilon}(r) $ has its
maximum on the boundary, i.e., either at $\rho $ or at $\infty$ (since
 $h(|\x|)$  is subharmonic in $\x$ ).  By choosing $\varepsilon$ sufficiently small and
positive we can have that $h_{\varepsilon}(\rho) < -2\varepsilon $
 and this contradicts the fact that the maximum (which is at least
$-{\varepsilon}$ ) is on the boundary.

B) is proved by noting (by subharmonicity again) that the maximum
of $f_0$ in $(0,r)$ occurs on the boundary, i.e., $f_0(r) \geq
f_0(r')$ for any $r'<r$.

C) is proved by studying the function $g=f_0- \widetilde{f}_0$. Since
$f_0$ and $\widetilde{f}_0$ are continuous, the falsity of C) implies
the existence some open subset, $\Omega \subset B_R$ on which $g(|\x|)
>0$.  On $\Omega$ we have that $g(|\x|)$ is subharmonic (because $vf_0
> \widetilde{v} \widetilde{f}_0$). Hence, its maximum occurs on the
boundary, but $g=0$ there. This contradicts $g(|\x|)>0$ on $\Omega$.
\end{proof}

\noindent {\it Remark about infinite range potentials:}
\index{infinite range potential}
If $v(r)$ is infinite
range and nonnegative it is easy to extend the definition of the
scattering length under the assumptions:

1) $v(r) \geq 0$ for all $r$ and

2) For some $R_1$ we have $\int_{R_1}^{\infty} v(r)r^{n-1}\ dr
<\infty$.

If we cut off the potential at some point $R_0 >R_1$ (i.e., set
$v(r)=0$ for $r>R_0$) then the scattering length is well defined but it
will depend on $R_0$, of course. Denote it by $a(R_0)$. By part C of
Lemma (\ref{lem1.1}), $a(R_0) $ is an increasing function of $R_0$.
However, the bounds (\ref{2scattineq}) and (\ref{3scattineq})
and assumption 2) above
guarantee that $a(R_0)$ is bounded above.
(More precisely, we need  a simple modification of
(\ref{2scattineq}) and (\ref{3scattineq}) to the potential
$\widehat{v}(r) \equiv \infty $ for $r \leq R_1 $ and $\widehat{v}(r)
\equiv v(r)$ for $r> R_1$. This is accomplished by replacing the `trial
function' $f(x)=1$ by a smooth radial function that equals $0$ for $r<
R_1$ and equals $1$ for $r>R_2$ for some $R_2>R_1$.)  Thus, $a$
is well defined by
\begin{equation}
a= \lim_{R_0 \to \infty}a(R_0) \ .
\end{equation}

\chapter[$c$-Number Substitutions and Gauge Symmetry Breaking]{$\boldsymbol c$-Number Substitutions and\vspace*{-2mm}\newline Gauge Symmetry Breaking}\label{justapp}
\index{c-num@$c$-number substitution}\index{gauge symmetry!breaking}

\textit{In this appendix, which is a slightly extended version of {\rm\cite{lsyc}},
we give a rigorous justification of part of the Bogoliubov\index{Bogoliubov!approximation}
approximation, discussed in Appendix~{\rm\ref{chap2}} -- namely the
replacement of bosonic creation and annihilation operators by
$c$-numbers. We also discuss the relation between BEC and spontaneous
breaking of gauge symmetry which was mentioned briefly in the
introduction.}
\bigskip \bigskip

One of the key developments in the theory of the Bose gas, especially
the theory of the low density gases currently at the forefront of
experiment, is Bogoliubov's 1947 analysis \cite{BO} of the many-body
Hamiltonian by means of a $c$-number substitution for the most
relevant operators in the problem, the zero-momentum mode operators,
namely $\aoo\to z, \, \aos\to z^*$.  Naturally, the appropriate value
of $z$ has to be determined by some sort of consistency or variational
principle, which might be complicated, but the concern, expressed by
many authors over the years, is whether this sort of substitution is
legitimate, i.e., error free.  We address this latter problem here and
show, by a simple but rigorous analysis, that it is so under very
general circumstances.

The rigorous justification for this substitution, as far as calculating
the pressure is concerned, was done in a classic paper of Ginibre\index{Ginibre}
\cite{ginibre} in 1968, but it does not seem to have percolated into the
general theory community. In textbooks it is often said, for instance,
that
it is tied to the imputed `fact' that the expectation value of the number
operator $\no =\aos \aoo$ is of order $V=$ volume. (This was the argument
in \cite{BO}). That is, Bose-Einstein condensation\index{Bose-Einstein condensation} (BEC)  justifies the
substitution. As Ginibre pointed out, however, BEC has nothing to do with
it. The $z$ substitution still gives the right answer even if $\no$ is
small (but it is a useful calculational tool only if $\no$ is
macroscopic). Thus, despite \cite{ginibre} and the thorough review of these
matters
in \cite{zagrebnov}, there is some confusion in the literature and clarification
could
be useful.

In this appendix  we do three things. 1.) We show how Ginibre's
result can be easily obtained in a few simple lines. While he used
coherent states, he did not use the Berezin-Lieb inequality\index{Berezin-Lieb inequality}
\cite{Berezin72,Lieb73,simon}, derived later, which efficiently gives upper bounds.
This inequality gives explicit error bounds which, typically, are only
order one compared to the total free energy or pressure times volume,
which are order $N=$ particle number. \

2.)  This allows us to go beyond \cite{ginibre} and make $c$-number
substitutions for {\it many} $\kk$-modes at once, provided the number
of modes is lower order than $N$.  \

3.)  We show how the optimum value of $z$ yields, in fact, the
expectation value $\langle\no\rangle$ in the true state when a gauge
breaking term is added to the Hamiltonian. More precisely, in the
thermodynamic limit (TL) the $|z|^2$ that maximizes the partition
function equals $|\langle \aoo \rangle | ^2$ and this equals $\langle
\no \rangle$, which is the amount of condensation --- a point that was
not addressed in full generality in previous work
\cite{ginibre,zagrebnov,BSP,AVZ}.  The second of these equalities has previously
only been treated under some additional assumptions \cite{fannes} or
for some simplified models \cite{zagrebnov,suto}.\footnote
{We note that recently A.\ S\"ut\H o
presented
a different proof of item 3 \cite{suto2}.}

While we work here at positive temperature
$k_{\rm B}T=1/\beta$, our methods also work for the ground state (and
are even simpler in that case).
To keep this note short and, hopefully, readable, we will be a bit
sketchy in places but there is no difficulty filling in the details.

The use of coherent states\index{coherent states} \cite{klauder,feng} to give accurate upper
and lower bounds to energies, and thence to expectation values, is
effective in a wide variety of problems \cite{liebcoh}, e.g., quantum
spin systems in the large $S$ limit \cite{Lieb73}, the Dicke model
\cite{HL}, the strong coupling polaron \cite{LT}, and the proof that
Thomas-Fermi theory is exact in the large atom limit \cite{li2,thir}.
For concreteness and relevance, we concentrate on the Bose gas problem
here, and we discuss only the total, correct Hamiltonian.
Nevertheless, the same conclusions hold also for variants, such as
Bogoliubov's truncated Hamiltonian (the ``weakly imperfect Bose gas''
\cite{zagrebnov,BO}) or other modifications, provided we are in the
stability regime (i.e., the regime in which the models make sense).
We are not
claiming that any particular approximation is valid. That is a completely
different story that has to be decided independently.
The method can also be modified to incorporate inhomogeneous systems.
The message is the same in all cases, namely that the $z$ substitution
causes no errors (in the TL), even if there is no BEC, whenever it is
applied to
physically stable systems.  Conversely, if the system is stable after
the $z$ substitution then so is the original one.

We start with the well-known Hamiltonian for bosons in a large box of
volume $V$, expressed in terms of the second-quantized creation\index{creation operator} and
annihilation operators\index{annihilation operator} $\ak, \aks$ satisfying the canonical
commutation relations,
\begin{equation}
H = \sum_\kk k^2 \aks \ak + \frac{1}{2V}\sum_{\kk, \pp, \qq}
\nu(\pp) a_{\kk + \pp}^* a_{\qq - \pp}^* \ak \aq  ,
\end{equation}
(with $\hbar = 2m =1$). Here, $\nu$ is the Fourier transform of the
two-body potential $v(\x)$.  We assume that there is a bound on the
Fourier coefficients $|\nu(\kk)| \leq \varphi <\infty$.

The case of
hard core potentials\index{hard core interaction} can be taken care of in the following way. First cut
off the hard core potential $v$ at a height $10^{12}$ eV. It is easy
to prove, by standard methods, that this cutoff will have a
negligible effect on the exact answer.
After the cutoff $\varphi$
will be about $10^{12}$ eV \r{A}$^3$, and according to what we prove
below, this substitution will affect the chemical potential only by
about $\varphi /V$, which is truly negligible when $V= 10^{23}$
\r{A}$^3$.

If we replace the operator $\aoo$ by a complex number $z$ and $\aos$
by $z^*$ everywhere in $H$ we obtain a Hamiltonian $H'(z)$ that acts
on the Fock-space\index{Fock space} of all the modes other than the $\aoo$ mode.
Unfortunately, $H'(z)$ does not commute with the particle number
$N^>\equiv\sum_{\kk\neq {\bf 0}} \aks \ak$. It is convenient, therefore,
to work in
the grand-canonical ensemble\index{grand-canonical!ensemble} and consider $H_\mu = H- \mu N = H-\mu (
\aos \aoo + N^>)$ and, correspondingly, $H^\prime_\mu(z)= H'(z) -\mu
(|z|^2 +
N^>)$.

The partition functions\index{partition function} are  given by
\begin{align}\label{partition}
e^{\beta V p(\mu)} \equiv \Xi(\mu) &= \Tr_\hi \exp[ -\beta H_\mu] \\
e^{\beta V p'(\mu)} \equiv\Xi'(\mu) &= \int d^2\!z\, \Tr_\hip
\exp[ -\beta H_\mu^\prime(z)]
\label{partitionint}
\end{align}
where $\hi$ is the full Hilbert (Fock) space, $\hip$ is the Fock space
without
the $\aoo$ mode, and $d^2\!z\,\equiv \pi^{-1} dxdy$ with $z=x+iy$. The
functions  $p(\mu)$
and $p'(\mu)$ are the corresponding finite volume pressures.\index{pressure}

The pressure $p(\mu)$ has a finite TL for all $\mu<
\mu_{\rm critical}$, and it is a convex function of $\mu$. For the
non-interacting gas, $\mu_{\rm critical}=0$, but for any {\it realistic}
system
$\mu_{\rm critical}=+\infty$. In any case, we assume $\mu< \mu_{\rm
  critical}$, in which case both the pressure and the density are finite.

Let $|z\rangle = \exp\{-|z|^2/2 +z \aos \}\, |0\rangle$ be the
coherent state vector in the $\aoo$ Fock space and let $\Pi(z)
=|z\rangle\langle z|$ be the projector onto this vector. There are six
relevant operators containing $\aoo$ in $H_\mu$, which have the following
expectation
values \cite{klauder} (called {\it lower symbols})\index{lower symbol}
\begin{align}\nonumber
\langle z| \aoo  |z \rangle &= z,
& \!\!\!\langle z| \aoo \aoo |z  \rangle &= z^2, &\!\!\!
\langle z| \aos \aoo |z  \rangle &= |z|^2 \\
\langle z| \aos  |z \rangle &= z^*, & \!\!\!
 \langle z| \aos \aos  |z  \rangle & = z^{*2}, & \!\!\!
\langle z|  \aos \aos \aoo \aoo |z  \rangle &= |z|^4. \nonumber
\end{align}
Each also has an {\it upper symbol}\index{upper symbol}, which is a function of $z$ (call
it $u(z)$ generically) such that an operator $F$ is represented as $F
= \int d^2\!z\, u(z) \Pi(z)$. These symbols are
\begin{align}
\aoo &\to z, & \aoo\aoo &\to z^2, &  \aos \aoo &\to |z|^2-1 \nonumber \\
 \aos &\to z^*, & \!\!\!  \aos \aos &\to z^{*2}, &\!\!\!  \aos \aos \aoo
\aoo
 &\to |z|^4 -4|z|^2 +2. \nonumber
\end{align}

It will be noted that the operator $H^\prime_\mu(z)$, defined above,
is obtained from $H_\mu$ by substituting the lower symbols for the six
operators.  If we substitute the upper symbols instead into $H_\mu$ we
obtain a slightly different operator, which we write as $H^{\prime
  \prime}_\mu(z) = H^\prime_\mu(z) + \delta_\mu(z)$ with
\begin{equation}\label{delta}
 \delta_\mu(z) = \mu+ \frac{1}{2V}\Big[ (-4|z|^2 +2) \nu({\bf 0})
- \sum_{\kk \neq {\bf 0}}
a_{\kk}^*a_{\kk}^{\phantom *}\big(2 \nu({\bf 0}) + \nu(\kk)+
\nu(-\kk)\big) \Big] .
\end{equation}

The next step is to mention two inequalities, of which the first is
\begin{equation}\label{clowerbound}
\Xi(\mu) \geq \Xi'(\mu)\,.
\end{equation}
This is a consequence of the following two facts: The completeness
property of coherent states,
$ \int d^2\!z\, \Pi (z) = {\rm Identity}$, and
\begin{equation} \label{jensen}
\langle z\otimes \phi | e^{-\beta H_\mu} | z\otimes \phi \rangle \geq
e^{-\beta \langle z\otimes \phi | H_\mu |z \otimes \phi \rangle } =
 e^ {-\beta \langle \phi |H_\mu^\prime ( z) |\phi \rangle },
\end{equation}
where $\phi$ is any normalized vector in $\hip$. This
 is Jensen's inequality for the expectation value of a
convex function (like the exponential function) of an operator.

To prove (\ref{clowerbound}) we take $\phi$ in (\ref{jensen}) to be one
of the normalized eigenvectors of $H_\mu^\prime (z)$, in which case
$\exp\{\langle \phi |-\beta H_\mu^\prime ( z) |\phi \rangle\} =
\langle \phi | \exp\{-\beta H_\mu^\prime ( z) \}|\phi \rangle $.  We
then sum over all such eigenvectors (for a fixed $z$) and integrate
over $z$.  The left side is then $\Xi(\mu)$, while the right side is
$\Xi'(\mu)$.

The second inequality \cite{Berezin72,Lieb73,simon} is
\begin{equation}\label{upper}
\Xi(\mu) \leq \Xi''(\mu)\equiv e^{\beta V p''(\mu)},
\end{equation}
where $ \Xi''(\mu)$ is similar to $ \Xi'(\mu)$ except that
$H_\mu^\prime(z)$ is replaced by $H_\mu^{\prime \prime}(z)$.  Its
proof is the following. Let $|\Phi_j \rangle \in \hi$ denote the
complete set of normalized eigenfunctions of $H_\mu$. The partial
inner product $|\Psi_j(z)\rangle = \langle z| \Phi_j\rangle $ is a
vector in $\hip$ whose square norm, given by $c_j(z) = \langle \Psi_j (z) |
\Psi_j(z) \rangle_\hip$, satisfies $\int d^2\!z\, c_j(z) =1$. By using the
upper symbols, we can write $\langle \Phi_j | H_\mu | \Phi_j\rangle =\int
d^2\!z\, \langle \Psi_j (z) | H_\mu^{\prime \prime} (z) | \Psi_j
(z)\rangle = \int
d^2\!z\,\langle \Psi_j' (z) | H_\mu^{\prime \prime} (z)|
\Psi_j'(z)\rangle c_j(z) $, where $|\Psi_j'(z) \rangle$ is the
normalized vector $c_j(z)^{-1/2} \Psi_j(z)$.  To compute the trace, we
can exponentiate this to write $\Xi(\mu)$ as
\begin{equation}\nonumber
 \sum_j \exp\left\{-\beta \int d^2\!z\, c_j(z)
\langle \Psi_j' (z) | H_\mu^{\prime \prime} (z)|
\Psi_j'(z)\rangle  \right\}.
\end{equation}
Using Jensen's inequality twice, once for functions and once for
expectations as in (\ref{jensen}), $\Xi(\mu)$ is less than
\begin{align} \nonumber
 &\sum_j \int d^2\!z\, c_j(z) \exp\left\{
\langle \Psi_j' (z) | -\beta H_\mu^{\prime \prime} (z) |
\Psi_j'(z)\rangle \right\} \\ &\leq \sum_j \int d^2\!z\, c_j(z)
\langle \Psi_j' (z) | \exp\left\{-\beta H_\mu^{\prime \prime} (z) \right\}
|
\Psi_j'(z)\rangle .\nonumber
\end{align}
Since $\Tr\, \Pi(z) = 1$, the last
expression can be rewritten
\begin{equation}\nonumber
\int d^2\!z\, \sum_j \langle \Phi_j |  \Pi(z) \otimes
\exp\left\{-\beta H_\mu^{\prime \prime}(z)
\right\}| \Phi_j \rangle = \Xi''(\mu).
\end{equation}

Thus, we have that
\begin{equation}\label{correx0}
\Xi'(\mu) \leq \Xi(\mu) \leq \Xi''(\mu).
\end{equation}
The next step is to try to relate $ \Xi''(\mu)$ to $\Xi'(\mu) $. To
this end we have to bound $\delta_\mu(z)$ in (\ref{delta}). This is
easily done in terms of the total number operator whose lower symbol
is $N^{\prime}(z) = |z|^2 + \sum_{\kk \neq {\bf 0}}\aks \ak$. In
terms of the bound $\varphi$ on $\nu(\pp)$
\begin{equation}\label{bounddelta}
|\delta_\mu(z)|  \leq 2\varphi (N'(z)+\half )/V +|\mu| \ .
\end{equation}
Consequently, $ \Xi''(\mu)$ and
$\Xi'(\mu) $ are related by the inequality
\begin{equation}
\Xi''(\mu) \leq \Xi'(\mu + 2\varphi/V) e^{\beta (|\mu|+\varphi/V)}.
\label{correx1}
\end{equation}
Equality of the pressures $p(\mu)$, $p'(\mu)$ and $p''(\mu)$ in the TL
follows from (\ref{correx0}) and (\ref{correx1}).

Closely related to this point is the question of relating $\Xi(\mu)$
to the maximum value of the integrand in (\ref{partitionint}), which
is $\max_z \Tr_\hip \exp[ -\beta H_\mu^\prime(z)] \equiv e^{\beta V
  p^{\max}}$.  This latter quantity is often used in discussions of
the $z$ substitution problem, e.g., in refs. \cite{ginibre,zagrebnov}.  One
direction is not hard.  It is the inequality (used in ref. \cite{ginibre})
\begin{equation}\label{junk}
\Xi(\mu) \geq \max_z \Tr_\hip  \exp[ -\beta H_\mu^\prime(z)],
\end{equation}
and the proof is the same as the proof of (\ref{clowerbound}), except
that this time we replace the completeness relation for the coherent
states by the simple inequality ${\rm Identity} \geq \Pi(z)$ for
any fixed number $z$.

For the other direction, split
the integral defining $\Xi''(\mu)$ into a part where $|z|^2 <  \xi$ and
$|z|^2\geq \xi$. Thus,
\begin{equation}\label{cheb}
  \Xi''(\mu) \leq  \xi \max_z \Tr_\hip  \exp[ -\beta
H_\mu^{\prime\prime}(z)]
+  \frac 1{\xi} \int_{|z|^2\geq \xi} d^2\!z\,
  |z|^2 \, \Tr_\hip \exp[ -\beta H_\mu^{\prime\prime}(z)].
\end{equation}
Dropping the condition $|z|^2\geq \xi$ in the last integral and using
$|z|^2\leq N'(z)=N''(z)+1$, we see that the last term on the right side of  (\ref{cheb}) is
bounded above by $\xi^{-1} \Xi''(\mu) [V\rho''(\mu)+1]$, where
$\rho''(\mu)$ denotes the density in the $H_\mu''$ ensemble. Optimizing
over $\xi$  leads to
\begin{equation}\label{morejunk}
\Xi''(\mu) \leq 2 [V \rho''(\mu)+1]  \, \max_{z} \Tr_\hip  \exp[ -\beta
H_\mu^{\prime\prime}(z)].
\end{equation}
Note that $\rho''(\mu)$ is order one, since $p''(\mu)$ and $p(\mu)$
agree in the TL (and are convex in $\mu$), and we assumed that the density
in the original
ensemble is finite. By (\ref{bounddelta}), $H_\mu^{\prime\prime}\geq
H_{\mu+2\varphi/V}^\prime-|\mu|-\varphi/V$, and  it follows from
(\ref{upper}), (\ref{morejunk}) and~(\ref{junk}) that $p^{\max}$
agrees with the true pressure $p$ in the TL. Their
difference, in fact, is at most $O(\ln V /V)$. This is the result obtained
by
Ginibre in \cite{ginibre} by more complicated arguments, under the
assumption of superstability\index{superstability} of the interaction, and without the
explicit error estimates obtained here.

To summarize the situation so far, we have four expressions for the
grand-canonical pressure. They are all equal in the TL
limit,
\begin{equation}\label{equality}
p(\mu) = p'( \mu) = p''(\mu) =  p^{\max}(\mu)
\end{equation}
when $\mu$ is not a point at which the density can be infinite.

Our second main point is that not only is the $z$ substitution valid
for $\aoo$ but it can also be done for many modes simultaneously.  As
long as the number of modes treated in this way is much smaller than
$N$ the effect on the pressure will be negligible.  Each such
substitution will result in an error in the chemical potential that is
order $\varphi /V$. The proof of this fact just imitates what was done
above for one mode.  Translation invariance is not important here; one
can replace any mode such as $\sum_{\bf k} g^{\phantom *}_{\bf k} \ak$
by a $c$-number, which can be useful for inhomogeneous systems.

A more delicate point is our third one, and it requires, first, a
discussion of the meaning of `condensate fraction'\index{condensate fraction} that goes beyond
what is usually mentioned in textbooks, but which was brought out in
\cite{BO,griffiths,roepstorff}. The `natural' idea would be to
consider $V^{-1}\langle \no \rangle$. This, however, need not be a
reliable measure of the condensate fraction for the following reason.
If we expand $\exp \{-\beta H\}$ in eigenfunctions of the number
operator $\no$ we would have $\langle \no \rangle = \sum_n n
\gamma (n)$, where $\gamma (n)$ is the probability that $\no=n$. One
would like to think that $\gamma(n) $ is sharply peaked at some
maximum $n$ value, but we do not know if this is the case. $\gamma(n)$
could be flat, up to the maximum value or, worse, it could have a
maximum at $n=0$.
Recall that precisely this happens for the
Heisenberg quantum ferromagnet\index{Heisenberg quantum ferromagnet} \cite{griffiths}; by virtue of
conservation of total spin angular momentum, the distribution of
values of the $z$-component of the total spin, $S^z$, is a strictly
decreasing function of $|S^z|$. Even if it were flat, the expected
value of $S^z$ would be half of the spontaneous magnetization that one
gets by applying a weak magnetic field.

With this example in mind, we see that the only physically reliable
quantity is $\lim_{\lambda \to 0} \lim_{V \to \infty} V^{-1}\langle
\no \rangle_{\mu, \lambda}$, where the expectation is now with respect
to a Hamiltonian $H_{\mu,\lambda}= H_\mu + \sqrt{V}(\lambda \aoo
+\lambda^*\aos)$ \cite{BO}. Without loss of generality, we assume
$\lambda$ to be real.  We will show that for almost every $\lambda$,
the density $\gamma(V\rho_0)$ converges in the TL to a $\delta$-function
at
the point $\widehat\rho_0=\lim_{V\to\infty} |z_{\rm max}|^2/V$, where
$z_{\rm
  max}$ maximizes the partition function $\Tr_{\hip} \exp\{ -\beta
H'_{\mu,\lambda}(z)\}$.  That is,
\begin{equation}\label{19}
\lim_{V\to \infty} \frac 1V \langle \no\rangle_{\mu,\lambda}=\lim_{V\to \infty} \frac 1V
 |\langle \aoo\rangle_{\mu,\lambda}|^2
=\lim_{V\to \infty} \frac 1V  |z_{\max} |^2 \,.
\end{equation}
This holds for those $\lambda$ where the pressure in the TL
is differentiable; since $p(\mu,\lambda)$ is convex (upwards) in
$\lambda$ this is true almost everywhere. The right and left
derivatives exist for every $\lambda$ and hence
$\lim_{\lambda \to 0+} \lim_{V\to\infty} V^{-1} |\langle
\aoo\rangle_{\mu,\lambda}|^2$ exists.

The expectation values $\langle \no\rangle_{\mu,\lambda}$ and $\langle
\aoo\rangle_{\mu,\lambda}$ are obtained by integrating $(|z|^2-1)$ and
$z$, respectively, with the weight $W_{\mu,\lambda}(z)$, given by $W_{\mu,\lambda}(z)\equiv
\Xi(\mu,\lambda)^{-1} \Tr_{\hip} \langle z| \exp\{-\beta
H_{\mu,\lambda}\}| z\rangle$. We will show that this weight converges
to a $\delta$-function at $z_{\max}$ in the TL, implying (\ref{19}).
If we could replace $W_{\mu,\lambda}(z)$ by $W_{\mu,0}(z)e^{-\beta
  \lambda\sqrt{V}(z+z^*)}$, this would follow from Griffiths'
argument\index{Griffiths' argument} \cite{griffiths} (see also
\cite[Sect.~1]{DLS}).  Because $[H,\aoo]\neq 0$, $W_{\mu,\lambda}$ is
not of this product form.  However, the weight for
$\Xi''(\mu,\lambda)$, which is $W''_{\mu,\lambda}(z)\equiv
\Xi''(\mu,\lambda)^{-1} \Tr_{\hip}\exp\{-\beta
H^{\prime\prime}_{\mu,\lambda}(z)\}$, does have the right form. In the
following we shall show that the two weights are equal apart from
negligible errors.

Equality (\ref{equality}) holds also for all $\lambda$, i.e.,
$p(\mu,\lambda) = p''( \mu,\lambda) = p^{\max}(\mu,\lambda)$ in the
TL.  In fact, since the upper and lower symbols agree for $\aoo$ and
$\aos$, the error estimates above remain unchanged.  (Note that since
$\sqrt{V}|\aoo+\aos|\leq \delta (N+\half) + V/\delta$ for any $\delta>0$,
$p(\mu,\lambda)$ is finite for all $\lambda$ if it is finite for
$\lambda=0$ in a small interval around $\mu$.) At any point of
differentiability with respect to $\lambda$, Griffiths' theorem
\cite{griffiths} (see \cite[Cor.~1.1]{DLS}), applied to the partition
function $\Xi''(\lambda,\mu)$, implies that $W''_{\mu,\lambda}(\zeta \sqrt
V)$ converges to a $\delta$-function at some point $\widehat \zeta$ on the
real axis as $V\to\infty$. (The original Griffiths argument can easily
be extended to two variables, as we have here. Because of radial
symmetry, the derivative of the pressure with respect to ${\rm Im\,
}\lambda$ is zero at any non-zero real $\lambda$.)  Moreover, by
comparing the derivatives of $p''$ and $p^{\max}$ we see that
$\widehat \zeta= \lim_{V\to\infty} z_{\rm max}/\sqrt{V}$, since $z_{\rm
  max}/\sqrt{V}$ is contained in the interval between the left and
right derivatives of $p^{\max}(\mu,\lambda)$ with respect to
$\lambda$.

We shall now show that the same is true for $W_{\mu,\lambda}$. To this
end, we add another term to the Hamiltonian, namely $\eps F\equiv \eps V
\int
d^2\!z\, \Pi(z) f(zV^{-1/2})$, with $\eps$ and $f$ real. If $f(\zeta)$ is
a
nice function of two real variables with bounded second
derivatives, it is then easy to see that the upper and lower symbols of
$F$ differ only by a term of order $1$. Namely, for some $C>0$
independent of $z_0$ and $V$,
\begin{equation}\nonumber
\left| V \int d^2\!z\, |\langle z|z_0\rangle |^2
\left( f(zV^{-1/2})- f(z_0 V^{-1/2})\right) \right| \leq C .
\end{equation}
Hence, in particular,
$p(\mu,\lambda,\eps)=p''(\mu,\lambda,\eps)$ in the TL. Moreover, if
$f(\zeta)=0$ for $|\zeta-\widehat \zeta|\leq \delta$, then the pressure is
independent of $\eps$ for $|\eps|$ small enough (depending only on
$\delta$). This can be seen as follows. We have
\begin{equation}\label{c21}
p''(\mu,\lambda,\eps) - p''(\mu,\lambda,0) = \frac 1{\beta V} \ln
\left\langle
e^{-\beta\eps V f(zV^{-1/2})} \right\rangle,
\end{equation}
where the last expectation is in the $H_\mu''$ ensemble at $\eps =0$.  The
corresponding distribution is exponentially localized at
$z/\sqrt{V}=\widehat
\zeta$ \cite{griffiths,DLS}, and therefore the right side of (\ref{c21})
goes to zero in the TL for small enough $\eps$.  In particular, the
$\eps$-derivative of the TL pressure at $\eps=0$ is zero.  By
convexity in $\eps$, this implies that the derivative of $p$ at finite
volume, given by $V^{-1}\langle F\rangle_{\mu,\lambda} = \int
d^2\!z\,f(zV^{-1/2}) W_{\mu,\lambda}(z)$, goes to zero in the TL.
Since $f$ was arbitrary, $V\int_{|\zeta-\widehat \zeta|\geq \delta}
d^2\!\zeta\,
W_{\mu,\lambda}(\zeta\sqrt V)\to 0$ as $V\to \infty$. This holds for all
$\delta>0$, and therefore proves the statement.

Our method also applies to the case when the pressure is not
differentiable in $\lambda$ (which is the case at $\lambda=0$ in the
presence of BEC). In this case, the resulting weights
$W_{\mu,\lambda}$ and $W''_{\mu,\lambda}$ need not be
$\delta$-functions, but Griffiths' method \cite{griffiths,DLS} implies
that they are, for $\lambda\neq 0$, supported on the real axis between
the right and left derivative of $p$ and, for $\lambda=0$, on a disc
(due to the gauge symmetry)\index{gauge symmetry} with radius determined by the right
derivative at $\lambda=0$.  Convexity of the pressure as a function of
$\lambda$ thus  implies that in the TL the supports of the weights $W_{\mu,\lambda}$
and $W''_{\mu,\lambda}$ for $\lambda\neq 0$ lie outside of this disc.
Hence $\langle \no\rangle_\lambda$ is monotone increasing in $\lambda$
in the TL. In combination with (\ref{19}) this implies in particular that
\begin{equation}\label{roep}
\lim_{V\to\infty} \frac 1V \langle
\no\rangle_{\mu,\lambda=0}\leq \lim_{\lambda\to 0} \lim_{V\to \infty}
\frac 1V |\langle \aoo\rangle_{\mu,\lambda}|^2 \,,
\end{equation}
a fact which is
intuitively clear but has, to the best of our knowledge, not been
proved so far \cite{roepstorff} in this generality.  In fact, the only
hypothesis entering our analysis, apart from the bound $\varphi$ on
the potential, is the existence of the TL of the pressure and the
density.

We note that by Eq. (\ref{roep}) spontaneous symmetry
breaking\index{spontaneous symmetry breaking} (in the sense that the
right side of (\ref{roep}) is not zero) always takes place whenever
there is BEC is the usual sense, i.e., without explicit gauge breaking
(meaning that the left side of (\ref{roep}) is non-zero). Conversely,
by Eq. (\ref{19}) spontaneous symmetry breaking is equivalent to BEC
in the sense of \lq quasi-averages\rq\ \cite{bogvol2}, i.e.,\index{quasi-averages}
\begin{equation}
\lim_{\lambda\to 0} \lim_{V\to\infty} \frac 1V \langle \no\rangle_{\mu,\lambda} > 0 \,.
\end{equation}
Note, however, that a non-vanishing of the right side of (\ref{roep})
does not {\it a priori} imply a non-vanishing of the left side. I.e.,
it is {\it a priori} possible that BEC only shows up after introducing
an explicit gauge-breaking term to the Hamiltonian. While it is expected on
physical grounds that positivity of the right side of  (\ref{roep}) implies positivity of the left side, a rigorous proof is lacking, so far. In the
example of the Heisenberg magnet quoted above, equality in
(\ref{roep}) does not generally hold, but still both sides are
non-vanishing in the same parameter regime.

To illustrate what could  arise mathematically, in principle, consider a weight function of the form
\begin{equation}\label{junk5}
W''_{\mu,\lambda=0} (\sqrt V \zeta) \equiv  w_V(\zeta) = \left\{ \begin{array}{cl}  V^2 - V + 1/V & {\rm for\ } |\zeta|\leq 1/V \\ 1/V & {\rm for\ } 1/V \leq |\zeta|\leq 1 \\ 0 & {\rm for\ } |z|>1\,. \end{array} \right.
\end{equation}
This distribution converges for $V\to \infty$ to a $\delta$-function at $\zeta=0$, and
consequently there is no BEC at $\lambda=0$. On the other hand, it is
easy to see that the weight function $w_V(\zeta)e^{-\beta \lambda V \zeta}$
(with an appropriate normalization factor) converges, for any
$\lambda>0$, to a $\delta$-function at $\zeta=-1$ as $V\to\infty$, and hence there is
spontaneous symmetry breaking. The open problem for the mathematician is to prove that examples like (\ref{junk5}) do not occur in realistic bosonic systems.
\newpage
\
\thispagestyle{empty}

\newpage
\
\thispagestyle{empty}

\newpage
\addcontentsline{toc}{chapter}{Index}
\printindex


\begin{thebibliography}{CCRCW}
\thispagestyle{empty}\markboth{Bibliography}{Bibliography}
\addcontentsline{toc}{chapter}{Bibliography}


\bibitem[ALSSY]{ALSSY}
M. Aizenman, E.H. Lieb, R. Seiringer, J.P. Solovej, and J. Yngvason,
{\it Bose-Einstein quantum phase transition in an optical lattice model},
Phys. Rev. A {\bf 70}, 023612-1--12 (2004).


\bibitem[ALSSY2]{ALSSY2}
M. Aizenman, E.H. Lieb, R. Seiringer, J.P. Solovej, and J. Yngvason,
{\it  Bose-Einstein Condensation as a Quantum Phase Transition in an Optical Lattice},
In:  Mathematical Physics of Quantum Mechanics, J. Asch
and A. Joye (Eds.), Lecture Notes in Physics {\bf
690}, 199--215, Springer (2006).


\bibitem[AN]{AN} M. Aizenman, B. Nachtergaele, {\it Geometric Aspects of
Quantum Spin States}, Commun. Math. Phys. {\bf 164}, 17--63 (1994).

\bibitem[AVZ]{AVZ} N.~Angelescu, A.~Verbeure, V.A.~Zagrebnov,
{\it On Bogoliubov's model of superfluidity}, J. Phys. A: Math. Gen. {\bf 25},
3473-3491 (1992); {\it Superfluidity III},  J. Phys. A: Math. Gen. {\bf 30},
4895-4913 (1997).


\bibitem[AA]{Auer}
E. Altman, A. Auerbach. {\it Oscillating Superfluidity of Bosons in
Optical Lattices}, Phys. Rev. Lett. {\bf 89}, 250404 (2002).


\bibitem[AG]{astra} G.E. Astrakharchik and S. Giorgini, {\it Quantum
Monte Carlo study of the three- to one-dimensional crossover for a
trapped Bose gas}, Phys. Rev. A {\bf 66}, 053614-1--6 (2002).


\bibitem[ABCG]{giorgini} G.E. Astrakharchik, J. Boronat, J.
Casulleras, and
S. Giorgini, {\it Superfluidity versus Bose-Einstein condensation in
a Bose gas with disorder},
Phys. Rev. A {\bf 66}, 023603 (2002).


\bibitem[Ba]{BA} B. Baumgartner, \textit{The Existence of
Many-particle Bound
States Despite a Pair Interaction with Positive Scattering
Length}, J. Phys. A \textbf{30}, L741--L747 (1997).


\bibitem[Bm]{baym} G. Baym, in: {\it Mathematical Methods in Solid State and
Superfluid Theory}, Scottish Univ. Summer School of Physics, Oliver
and Boyd, Edinburgh (1969).


\bibitem[Be]{Berezin72}
F.A. Berezin, Izv. Akad. Nauk, ser. mat., {\bf 36} (No. 5) (1972);
English translation: USSR Izv. {\bf 6} (No. 5) (1972). F.A. Berezin,
{\it General concept of quantization}, Commun. Math. Phys. {\bf 40},
153--174 (1975).



\bibitem[BL]{pule} M. van den Berg, J.T. Lewis,
{\it On generalized condensation in the free Bose gas},
Physica A {\bf 110}, 550--564 (1982).


\bibitem[BLP]{pule2} M. van den Berg, J.T. Lewis, J.V. Pul{\`e},
{\it A general theory of Bose-Einstein condensation},
Helv. Phys. Acta {\bf 59}, 1271--1288 (1986).


\bibitem[Bl]{blume} D. Blume, {\it Fermionization of a bosonic gas
    under highly elongated confinement: A diffusion quantum Monte
    Carlo study}, Phys. Rev. A {\bf 66}, 053613-1--8 (2002).

\bibitem[Bo]{BO} N.N.~Bogoliubov, {\it On the theory of superfluidity},
  Izv. Akad. Nauk USSR, {\bf 11}, 77 (1947). Eng. Trans.  J. Phys.
  (USSR), {\bf 11}, 23 (1947). See also {\it Lectures on quantum
    statistics}, vol. 1, Gordon and Breach (1967).


\bibitem[Bo2]{bogvol2} N.N.~Bogoliubov, {\it Lectures on Quantum Statistics: Quasi-Averages},
Gordon and Breach (1970).


\bibitem[Bo3]{Bog} N.N. Bogoliubov, {\it Energy levels of the imperfect Bose-Einstein gas},
Bull. Moscow State Univ. {\bf 7}, 43--56 (1947).


\bibitem[BZ]{BZ} N.N. Bogoliubov, D.N. Zubarev,
{\it Wave function of the ground state of interacting Bose-particles},
Sov. Phys.-JETP {\bf 1}, 83 (1955).


\bibitem[BZT]{b8b} N.N. Bogoliubov, D.N. Zubarev, Y.A. Tserkovnikov,
Soviet Phys. Doklady {\bf 2}, 535 (1957).


\bibitem[BBD]{bongs} K.\ Bongs, S.\ Burger, S.\ Dettmer, D.\ Hellweg,
  J.\ Artl, W. Ertmer, and K.\ Sengstok, {\it Waveguides for
    Bose-Einstein condensates}, Phys. Rev. A, {\bf 63}, 031602 (2001).


\bibitem[B]{Bose} S.N. Bose, {\it Plancks Gesetz und
    Lichtquantenhypothese}, Z. Phys. {\bf 26}, 178--181 (1924).

\bibitem[BSP]{BSP} E.~Buffet, Ph.~de Smedt, J.V.~Pul\`e, {\it The condensate equation for some Bose systems},
J.~Phys. A {\bf 16},
  4307 (1983).


\bibitem[C]{C} Y. Castin,
{\it Bose-Einstein condensates in atomic gases: simple theoretical
results},
in ``Coherent atomic matter waves'',
Lecture Notes of Les Houches Summer School, pp. 1--136, edited by R. Kaiser,
C. Westbrook, and F. David, EDP Sciences and Springer-Verlag (2001).

\bibitem[CS1]{CS01b} A.Y. Cherny, A.A. Shanenko, {\it Dilute Bose gas
    in two dimensions: Density expansions and the Gross-Pitaevskii
    equation}, Phys.\ Rev.\ E {\bf 64}, 027105 (2001).

\bibitem[CS2]{CS01a} A.Y. Cherny, A.A. Shanenko, {\it The kinetic and
    interaction energies of a trapped Bose gas: Beyond the mean
    field}, Phys.\ Lett.\ A {\bf 293}, 287 (2002).



\bibitem[CLY]{CLY} J. Conlon, E.H. Lieb, H.-T. Yau, \textit{The
    $N^{7/5}$ Law for Charged Bosons}, Commun.\ Math.\ Phys.\ \textbf{
    116}, 417--448 (1988).



\bibitem[CCRCW]{Cornish} S.L. Cornish, N.R. Claussen, J.L. Roberts,
  E.A. Cornell, C.E. Wieman, {\it Stable $^{85}Rb$ Bose-Einstein
    Condensates with Widely Tunable Interactions}, Phys. Rev. Lett.
  {\bf 85}, 1795--98 (2000).


\bibitem[DGPS]{DGPS}
F. Dalfovo, S.\ Giorgini, L.P.\ Pitaevskii, S.\ Stringari, {\it
Theory of Bose-Einstein condensation in trapped gases}, Rev. Mod.
Phys. \textbf{71}, 463--512 (1999).


\bibitem[Da]{Das1} K.K. Das, {\it  Highly anisotropic Bose-Einstein condensates:
  Crossover to lower dimensionality}, Phys. Rev. A {\bf{66}},
  053612-1--7 (2002).


\bibitem[DGW]{das2} K.K. Das, M.D. Girardeau, and E.M. Wright, {\it
Crossover from One to Three Dimensions for a Gas of Hard-Core
Bosons}, Phys. Rev. Lett. {\bf 89}, 110402-1--4 (2002).


\bibitem[DODS]{DODS}
D.B.M.\ Dickerscheid, D.\
van Oosten, P.J.H.\ Denteneer, H.T.C.\ Stoof,
{\it Ultracold atoms in optical lattices},
Phys.\  Rev.\  A {\bf 68}, 043623 (2003).



\bibitem[DLO]{dunjko} V. Dunjko, V. Lorent, and M. Olshanii, {\it
Bosons
in Cigar-Shaped Traps: Thomas-Fermi Regime, Tonks-Girardeau Regime,
and In Between}, Phys. Rev. Lett. {\bf 86}, 5413--5316 (2001).



\bibitem[D1]{dyson}
F.J. Dyson, \textit{Ground-State Energy of a Hard-Sphere Gas},
Phys. Rev. \textbf{106}, 20--26 (1957).


\bibitem[D2]{D2} F.J. Dyson, \textit{Ground State Energy of a Finite
System
of Charged Particles}, J. Math. Phys. \textbf{8}, 1538--1545
(1967).


\bibitem[DLS]{DLS} F.J. Dyson, E.H. Lieb, B. Simon, {\it Phase
Transitions
in Quantum Spin Systems with Isotropic and Nonisotropic
Interactions}, J. Stat. Phys. {\bf 18}, 335--383 (1978).



\bibitem[E]{Einstein} A. Einstein, {\it Quantentheorie des
einatomigen idealen
Gases}, Sitzber. Kgl. Preuss. Akad. Wiss.,
261--267 (1924), and 3--14 (1925).


\bibitem[FPV]{fannes} M. Fannes, J.V. Pul{\`e}, V. Verbeure, {\it On Bose condensation},
Helv. Phys. Acta {\bf 55}, 391 (1982).


\bibitem[FS]{fetter} A.L. Fetter and A.A. Svidzinsky, {\it Vortices
in a trapped dilute Bose-Einstein condensate},
J. Phys.: Condens.
Matter {\bf 13}, R135 (2001).


\bibitem[FH]{fishho}
D.S. Fisher, P.C.\ Hohenberg, {\it Dilute Bose gas in two dimensions},
Phys. Rev. B {\bf 37}, 4936--4943 (1988).


\bibitem[FWGF]{FWGF} M.P.A.\ Fisher, P.B.\ Weichman, G.\ Grinstein, D.S.\ Fisher,
{\it Boson localization and the superfluid-insulator transition},
Phys.\ Rev.\ B {\bf 40}, 546--570 (1989).


\bibitem[F]{FO} L.L. Foldy, \textit{Charged Boson Gas}, Phys. Rev.
\textbf{124}, 649--651 (1961); Errata \textit{ibid} \textbf{125},
2208 (1962).


\bibitem[FL]{FL} J. Fr\"ohlich, E.H. Lieb, {\it Phase Transitions in
        Anisotropic Lattice Spin Systems}, Commun. Math. Phys. {\bf 60},
      233--267 (1978).


\bibitem[FSS]{FSS}
J. Fr\"ohlich, B. Simon, T. Spencer, {\it Phase Transitions and Continuous
Symmetry Breaking},
Phys. Rev. Lett. {\bf 36}, 804 (1976); {\it Infrared bounds, phase transitions
and continuous symmetry breaking}, Commun. Math. Phys. {\bf  50}, 79 (1976).


\bibitem[Ga]{GCZKSD} J.J.\ Garcia-Ripoll, J.I.\ Cirac, P.\ Zoller,
C.\ Kollath, U.\ Schollwoeck, J.\ von Delft,
{\it Variational ansatz for the superfluid Mott-insulator transition in
optical lattices}, Opt. Express {\bf 12}, 42 (2004).


\bibitem[Ge]{G} G.M.\ Genkin,
{\it Manipulating the superfluid -- Mott insulator transition of a
Bose-Einstein condensate in an amplitude-modulated optical lattice},
arXiv:cond-mat/0311589 (2003).


\bibitem[Gin]{ginibre} J.~Ginibre, {\it On the asymptotic exactness of
    the Bogoliubov approximation for many boson systems}, Commun. Math.
  Phys. {\bf 8}, 26--51 (1968).


\bibitem[GA]{b3} M. Girardeau, R. Arnowitt,
{\it Theory of Many-Boson Systems: Pair Theory},
Phys. Rev. {\bf 113}, 755 (1959).


\bibitem[Gi1]{b4} M. Girardeau,
{\it Weak-Coupling Expansion for the Ground-State Energy of a Many-Boson System},
Phys. Rev. {\bf 115}, 1090 (1959).


\bibitem[Gi2]{gir} M.D.\ Girardeau, {\it Relationship between systems
of impenetrable bosons and fermions in one dimension}, J. Math. Phys.
{\bf 1}, 516 (1960).


\bibitem[Gi3]{ref2boulder} M.\ Girardeau,
{\it Simple and generalized condensation in many-boson systems},
Phys. of Fluids {\bf 5}, 1468 (1962).


\bibitem[Gi4]{b5} M. Girardeau,
{\it Ground State of the Charged Bose Gas},
Phys. Rev. {\bf 127}, 1809 (1962).


\bibitem[Gi5]{b6} M. Girardeau,
{\it Variational method for the quantum statistics of interacting particles},
J. Math. Phys. {\bf 3}, 131 (1962).


\bibitem[GW]{girardeau2} M.D.\ Girardeau and E.M. Wright, {\it
Bose-Fermi
variational Theory for the BEC-Tonks Crossover}, Phys. Rev. Lett. {\bf
87}, 210402-1--4 (2001).


\bibitem[GWT]{girardeau} M.D.\ Girardeau, E.M.\ Wright, and J.M.\
Triscari,
{\it Ground-state properties of a one-dimensional system of hard-core
bosons in a harmonic trap}, Phys. Rev. A {\bf 63}, 033601-1--6 (2001).



\bibitem[Go]{goerlitz} A. G{\"o}rlitz, {\it et al.}, {\it Realization
of
Bose-Einstein Condensates in Lower Dimension}, Phys. Rev. Lett. {\bf
87}, 130402-1--4 (2001).


\bibitem[GS]{GS} G.M.\ Graf and J.P.\ Solovej, \textit{A correlation
estimate
with applications to quantum systems with Coulomb interactions},
Rev. Math. Phys. {\bf 6}, 977--997 (1994).


\bibitem[G1]{greiner} M. Greiner, I. Bloch, O. Mandel, T.W. H\"ansch, T. Esslinger,
{\it Exploring Phase
Coherence in a 2D Lattice of Bose-Einstein Condensates},
 Phys.  Rev.  Lett.  {\bf 87}, 160405 (2001).


\bibitem[G2]{G1} M.\ Greiner, O.\ Mandel, T.\ Esslinger, T.E.\ H\"ansch,
I.\ Bloch,
{\it Quantum phase transition from a superfluid to a Mott
insulator in a gas of ultracold atoms},
Nature {\bf 415}, 39 (2002).


\bibitem[G3]{G2}   M.\ Greiner, O.\ Mandel, T.E.\ H\"ansch,
I.\ Bloch,
{\it Collapse and revival of the matter wave field of a
Bose-Einstein condensate},
Nature {\bf 419}, 51 (2002).


\bibitem[Gri]{griffiths} R.B. Griffiths,
{\it Spontaneous magnetization in idealized ferromagnets},
Phys. Rev. {\bf 152}, 240 (1966).


\bibitem[Gr1]{G1961}E.P.\ Gross, {\it Structure of a Quantized Vortex
in
    Boson Systems,} Nuovo Cimento {\bf 20}, 454--466 (1961).


\bibitem[Gr2]{G1963} E.P.\ Gross, {\it Hydrodynamics of a superfluid
condensate,} J. Math. Phys. {\bf 4}, 195--207 (1963).


\bibitem[HL]{HL} K.~Hepp, E.H.~Lieb,
{\it Equilibrium statistical mechanics of matter interacting with the quantized radiation field},
Phys. Rev. A {\bf 8}, 2517 (1973).


\bibitem[HFM]{hines} D.F.\ Hines, N.E.\ Frankel, D.J.\ Mitchell,
{\it Hard disc Bose gas}, Phys.\ Lett.\ {\bf 68A}, 12--14 (1978).


\bibitem[Ho]{Ho} P.C.\ Hohenberg,
{\it Existence of Long-range Order in One and Two Dimensions}, Phys.\
Rev.\ {\bf 158},
383--386 (1966).


\bibitem[HoM]{HM} P.C.~Hohenberg and P.C.~Martin, {\it Microscopic
theory of helium}, Ann.\ Phys.\ (NY) {\bf
34}, 291 (1965).


\bibitem[H]{huang} K. Huang, {\it Bose-Einstein condensation and
    superfluidity}, in: {\it Bose-Einstein Condensation}, A.  Griffin,
  D.W. Stroke, S. Stringari, eds., Cambridge University Press, 31--50
  (1995).


\bibitem[HY]{Lee-Huang-YangEtc}K.~Huang, C.N.~Yang, Phys. Rev. {\bf
105}, 767--775 (1957); T.D.~Lee, K.~Huang, C.N.~Yang,  Phys.
Rev. {\bf 106}, 1135--1145 (1957); K.A. Brueckner, K. Sawada, Phys.
Rev. {\bf 106}, 1117--1127, 1128--1135 (1957); S.T. Beliaev, Sov.
Phys.-JETP {\bf 7}, 299--307 (1958); T.T. Wu, Phys. Rev. {\bf 115},
1390 (1959); N. Hugenholtz, D. Pines, Phys. Rev. {\bf 116}, 489
(1959); M. Girardeau, R. Arnowitt, Phys. Rev. {\bf 113}, 755
(1959); T.D. Lee,  C.N. Yang, Phys. Rev. {\bf 117}, 12 (1960).


\bibitem[ISW]{BEC} M. Inguscio,
S. Stringari, C. Wieman (eds.),
{\it Bose-Einstein Condensation in Atomic Gases}, Italian Physical Society (1999).


\bibitem[JK]{jackson}
A.D. Jackson and G.M. Kavoulakis, {\it Lieb Mode in a
Quasi-One-Dimensional Bose-Einstein Condensate of Atoms}, Phys. Rev.
Lett. {\bf 89}, 070403 (2002).


\bibitem[JBCGZ]{JBCGZ}
D.\ Jaksch, C.\ Bruder,
J.I.\ Cirac, C.W.\ Gardiner, P.\ Zoller,
{\it Cold bosonic atoms in optical lattices},
Phys.\  Rev.\  Lett.\ {\bf  81}, 3108--3111 (1998).


\bibitem[KLS]{KLS} T. Kennedy, E.H. Lieb, S. Shastry, {\it The $XY$
Model
has
Long-Range Order for all Spins and all Dimensions Greater than
One}, Phys. Rev. Lett. \textbf{ 61}, 2582--2584 (1988).


\bibitem[KD]{TRAP} W. Ketterle, N.J. van Druten, {\it Evaporative
Cooling of Trapped Atoms}, in: B. Bederson, H. Walther, eds.,  Advances in
Atomic, Molecular and Optical Physics, {\bf 37}, 181--236,
Academic Press (1996).


\bibitem[KS]{klauder} J.~Klauder, B.-S.~Skagerstam, {\it Coherent states,
    applications in physics and mathematical physics}, World
  Scientific (1985).


\bibitem[KNSQ]{KoSt2000}
E.B. Kolomeisky, T.J. Newman, J.P. Straley, X. Qi,
{\it Low-dimensional Bose liquids:
beyond the Gross-Pitaevskii approximation}, Phys. Rev. Lett.
{\bf 85}, 1146--1149 (2000).


\bibitem[KT]{KT} M.~Kobayashi and M.~Tsubota, {\it Bose-Einstein
condensation and superfluidity of a
dilute Bose gas in a random potential}, Phys. Rev. B {\bf 66}, 174516
(2002).


\bibitem[KP]{komineas}
S. Komineas and N. Papanicolaou, {\it Vortex Rings and Lieb Modes in
a Cylindrical Bose-Einstein Condensate}, Phys. Rev. Lett. {\bf 89},
070402 (2002).


\bibitem[KB]{b9} A.J. Kromminga, M. Bolsterli,
{\it Perturbation Theory of Many-Boson Systems},
Phys. Rev. {\bf 128}, 2887 (1962).


\bibitem[Le]{Lenard} A.~Lenard, {\it Momentum distribution in the
ground state of the one-dimensional system of impenetrable bosons},
J. Math. Phys. {\bf 5}, 930--943 (1964).



\bibitem[Leg]{Legg}
A.J. Leggett,
{\it Bose-Einstein condensation in the alkali gases:
Some fundamental concepts},
Rev. Mod. Phys. {\bf 73}, 307 (2001).


\bibitem[Len]{lenz} W.~Lenz, {\it Die Wellenfunktion und
  Geschwindig\-keits\-verteilung des entarteten Gases}, Z. Phys. {\bf
  56}, 778--789 (1929).


\bibitem[L1]{LL2} E.H. Lieb,
\textit{Exact Analysis of an Interacting Bose Gas. II. The
Excitation Spectrum}, Phys. Rev.\ \textbf{130}, 1616--1624 (1963).


\bibitem[L2]{Lieb63} E.H. Lieb, \textit{Simplified Approach to the
Ground
State Energy of an Imperfect Bose Gas}, Phys. Rev. \textbf{130},
2518--2528
(1963). See also Phys. Rev. \textbf{133} (1964),
A899--A906  (with A.Y. Sakakura) and Phys. Rev. \textbf{134}
(1964), A312--A315 (with W. Liniger).


\bibitem[L3]{EL2} E.H.~Lieb, \textit{The Bose Fluid}, in:
W.E.~Brittin, ed.,
Lecture Notes in Theoretical Physics VIIC,  Univ. of Colorado
Press, pp.\ 175--224 (1964).


\bibitem[L4]{Lieb73}
E.H. Lieb, {\it The classical limit of quantum spin systems},
Commun.\ Math.\ Phys. {\bf 31}, 327--340 (1973).


\bibitem[L5]{li2} E.H.~Lieb,
 {\it Thomas-Fermi and related theories of atoms and molecules},
Rev. Mod. Phys. {\bf 53}, 603 (1981). Errata {\bf 54}, 311 (1982).


\bibitem[L6]{liebcoh} E.H.~Lieb,
{\it Coherent states as a tool for obtaining rigorous bounds},
in: {\it Coherent States}, D.H.~Feng,
  J.~Klauder, M.R.~Strayer, eds., p. 267, World Scientific (1994).


\bibitem[L7]{L3} E.H.\ Lieb, {\it The Bose Gas: A Subtle Many-Body
Problem},
in: {\it Proceedings
of the XIII International Congress on Mathematical Physics, London},
A.~Fokas, et al. eds., pp.~91--111, International Press (2001).


\bibitem[LL]{LL} E.H. Lieb, W. Liniger, \textit{Exact Analysis of an
Interacting Bose Gas. I.  The General Solution and the Ground
State}, Phys. Rev. \textbf{130}, 1605--1616 (1963).


\bibitem[LLo]{LL01}
E.H. Lieb, M. Loss, {\it Analysis}, 2nd ed., Amer. Math. Society,
Providence, R.I. (2001).


\bibitem[LN]{LN} E.H. Lieb, H. Narnhofer,  \textit{The Thermodynamic
Limit for Jellium}, J.  Stat. Phys. \textbf{12}, 291--310 (1975).
Errata J. Stat. Phys. \textbf{14}, 465 (1976).


\bibitem[LSe]{LS02} E.H.\ Lieb, R.\ Seiringer, {\it Proof of
Bose-Einstein Condensation for Dilute Trapped Gases},
Phys. Rev. Lett. {\bf 88}, 170409-1--4 (2002).


\bibitem[LSe2]{LS05} E.H.\ Lieb, R.\ Seiringer, {\it Derivation of the
    Gross-Pitaevskii Equation for Rotating Bose Gases},
  Commun. Math. Phys. {\bf 264}, 505--537 (2006).

\bibitem[LSS]{LSSfermi} E.H.\ Lieb, R.\ Seiringer, J.P.\ Solovej, {\it
    Ground-state energy of the low-density Fermi gas}, Phys. Rev. A
  {\bf 71}, 053605-1--13 (2005).

\bibitem[LSSY]{LSSY}
E.H.~Lieb, R.~Seiringer, J.P.~Solovej, and J.~Yngvason, {\it The
ground state of the Bose gas},  in: Current Developments in
Mathematics, 2001, 131--178,
International Press, Cambridge (2002).


\bibitem[LSSY2]{LSSY2} E.H.~Lieb, R.~Seiringer, J.P.~Solovej, and
  J.~Yngvason, {\it The Quantum-Mechanical Many-Body Problem: The Bose
    Gas}, in: Perspectives in Analysis, Essays in Honor of L.
  Carleson's 75th Birthday, Series: Mathematical Physics Studies, Vol.
  27, M. Benedicks, P. Jones, S. Smirnov (Eds.), Springer (2005).

\bibitem[LSeY1]{LSY1999} E.H.\ Lieb, R.\ Seiringer, J.\ Yngvason,
\textit{
Bosons in a Trap: A Rigorous Derivation of the Gross-Pitaevskii
Energy Functional},  Phys.  Rev A \textbf{ 61}, 043602 (2000).


\bibitem[LSeY2]{LSY2d} E.H. Lieb, R. Seiringer, J. Yngvason, \textit{
A Rigorous Derivation of the Gross-Pitaevskii Energy Functional
for a Two-dimensional Bose Gas}, Commun. Math. Phys. {\bf 224}, 17
(2001).


\bibitem[LSeY3]{LSYdoeb} E.H. Lieb, R. Seiringer, J. Yngvason,
\textit{The
Ground State Energy and Density of Interacting Bosons in a Trap},
in: \textit{ Quantum Theory and Symmetries}, Goslar, 1999,
H.-D.~Doebner, V.K.~Dobrev, J.-D.~Hennig and W. Luecke, eds., pp.
101--110, World Scientific (2000).


\bibitem[LSeY4]{LSYnn} E.H.\ Lieb, R.\ Seiringer, J.\ Yngvason,
\textit{Two-Dimensional Gross-Pita\-evskii Theory}, in: Progress in
Nonlinear Science, Proceedings of the International Conference
Dedicated to the 100th Anniversary of A.A. Andronov, Volume II, A.G.
Litvak, ed., pp. 582--590, Nizhny Novgorod, Institute of Applied Physics,
University of Nizhny Novgorod (2002).


\bibitem[LSeY5]{LSYsuper} E.H.~Lieb, R.~Seiringer, J.~Yngvason, {\it
Superfluidity in Dilute Trapped
Bose Gases}, Phys.\ Rev.\ B {\bf 66}, 134529 (2002).


\bibitem[LSeY6]{LSY} E.H.~Lieb, R.~Seiringer, J.~Yngvason, {\it
One-Dimensional Behavior of Dilute, Trapped Bose Gases},
Commun. Math. Phys. {\bf 244}, 347--393 (2004). See also: {\it
One-Dimensional Bosons in Three-Dimensional Traps}, Phys. Rev. Lett.
{\bf 91}, 150401-1--4 (2003).


\bibitem[LSeY7]{lsy02}
E.H.\ Lieb, R.\ Seiringer, and J.\ Yngvason, {\it Poincar\'e
Inequalities in
Punctured Domains}, Ann. Math. {\bf 158}, 1067--1080 (2003).


\bibitem[LSeY8]{lsyc}
E.H.\ Lieb, R.\ Seiringer, and J.\ Yngvason,
{\it Justification of $c$-Number Substitutions in Bosonic Hamiltonians},
Phys. Rev. Lett. {\bf 94}, 080401 (2005).


\bibitem[LSo]{LS} E.H.\ Lieb, J.P.\ Solovej, \textit{ Ground State
Energy
of the One-Component Charged Bose Gas}, Commun. Math. Phys.\
\textbf{217}, 127--163 (2001). Errata {\bf 225}, 219--221 (2002).


\bibitem[LSo2]{LSo02} E.H.\ Lieb, J.P.\ Solovej, {\it Ground State
Energy of the Two-Component Charged Bose Gas}, Commun. Math. Phys.
{\bf 252}, 485--534 (2004).


\bibitem[LT]{LT} E.H.~Lieb, L.E.~Thomas,
 {\it  Exact ground state energy of the strong-coupling polaron},
Commun.~Math.~Phys. {\bf 183}, 511 (1997).  Errata \textit{ibid} {\bf 188}, 499 (1997).


\bibitem[LY1]{LY1998}
E.H. Lieb, J.\ Yngvason, \textit{ Ground State Energy of the low
density
Bose Gas}, Phys. Rev. Lett. \textbf{80}, 2504--2507 (1998).


\bibitem[LY2]{LY2d} E.H.\ Lieb, J.\ Yngvason, \textit{The Ground
State Energy
of a Dilute Two-dimensional Bose Gas}, J. Stat. Phys. {\bf 103},
509 (2001).


\bibitem[LY3]{LYbham} E.H. Lieb, J. Yngvason, \textit{ The
Ground State Energy of a Dilute Bose Gas}, in: \textit{ Differential
Equations and Mathematical Physics, University of Alabama,
Birmingham, 1999}, R.~Weikard and G.~Weinstein, eds., pp. 271--282,
Amer. Math. Soc./Internat. Press (2000).


\bibitem[Lu]{b7} M. Luban,
{\it Statistical Mechanics of a Nonideal Boson Gas: Pair Hamiltonian Model},
Phys. Rev. {\bf 128}, 965 (1962).


\bibitem[Ma]{b10} P.C. Martin, {\it
A microscopic approach to superfluidity and superconductivity},
J. Math. Phys. {\bf 4}, 208 (1963).


\bibitem[MM]{MM} T.\ Matsubara, H.\ Matsuda, {\it A lattice model of
     liquid helium}, Progr.\ Theor.  Phys.\ {\bf 16}, 569--582 (1956).


\bibitem[MW]{MW} N.D. Mermin, H. Wagner,
{\it Absence of Ferromagnetism or Antiferromagnetism in One- or Two-Dimensional Isotropic Heisenberg Models},
Phys. Rev. Lett. {\bf 17}, 1133--1136 (1966).


\bibitem[MSKE]{esslinger} H. Moritz, T. St\"oferle, M. K\"ohl and
T. Esslinger, {\it Exciting Collective Oscillations in a Trapped 1D
Gas},
Phys. Rev. Lett. {\bf 91}, 250402 (2003).


\bibitem[MA]{MA} O.\ Morsch, E.\ Arimondo,
{\it Ultracold atoms and
Bose-Einstein condensates in optical lattices},
Lecture Notes in Physics Vol.\  602, Springer (2002).


\bibitem[M]{M} W.J.\ Mullin, {\it Bose-Einstein Condensation in a
Harmonic Potential},
J.\ Low Temp.\ Phys.\ {\bf 106}, 615--642 (1997).


\bibitem[NS]{NS}
Z.\ Nazario, D.I.\ Santiago,
{\it Quantum states of matter of simple bosonic systems: BECs, superfluids and quantum solids},
Phys. Lett. A {\bf 328}, 207--211 (2004).


\bibitem[Ol]{olshanii} M. Olshanii, {\it Atomic Scattering in the
Presence of an External Confinement and a Gas of Impenetrable
Bosons}, Phys. Rev. Lett. {\bf 81}, 938--941 (1998).


\bibitem[OS]{oster} K. Osterwalder, R. Schrader,
{\it Axioms for Euclidean Green's Functions},
Commun. Math. Phys. {\bf 31}, 83--112 (1973);
Commun. Math. Phys. {\bf 42}, 281--305 (1975).


\bibitem[O]{Ovch}  A.A. Ovchinnikov, {\it On the description of a
two-dimensional Bose gas at low densities}, J. Phys. Condens. Matter
{\bf 5}, 8665--8676 (1993). See also JETP Letters {\bf 57}, 477
(1993); Mod. Phys. Lett. {\bf 7}, 1029 (1993).


\bibitem[PO]{penrose} O. Penrose, L. Onsager, {\it Bose-Einstein
Condensation and Liquid Helium}, Phys. Rev. {\bf 104}, 576--84
(1956).


\bibitem[PS]{PS} C.~Pethick, H.\ Smith, \textit{Bose-Einstein
Condensation of
Dilute Gases}, Cambridge University Press (2001).


\bibitem[PHS]{Petrov} D.S. Petrov, M. Holzmann and G.V. Shlyapnikov,
{\it Bose-Einstein Condensates in Quasi-2D trapped Gases},
Phys. Rev. Lett. {\bf{84}},
2551 (2000).


\bibitem[PSW]{petrov} D.S. Petrov, G.V. Shlyapnikov, and J.T.M.
Walraven,
{\it Regimes of Quantum Degeneracy in Trapped 1D Gases}, Phys. Rev.
Lett. {\bf 85}, 3745--3749 (2000).


\bibitem[Pi]{P1961} L.P. Pitaevskii, {\it Vortex lines in an imperfect
    Bose gas}, Sov. Phys. JETP. {\bf 13}, 451--454 (1961).


\bibitem[PiSt]{PiSt} L. Pitaevskii, S. Stringari, {\it Uncertainty
Principle,
Quantum Fluctuations, and Broken
Symmetries}, J. Low Temp. Phys. {\bf 85}, 377 (1991).


\bibitem[PiSt2]{PiSt2003} L. Pitaevskii and S. Stringari, {\it Bose-Einstein
Condensation}, Oxford Science Publications, Oxford (2003).



\bibitem[Po]{popov} V.N. Popov, {\it On the theory of the
superfluidity of
two- and one-dimensional Bose systems}, Theor. and Math. Phys. {\bf
11},
565--573 (1977).


\bibitem[PrSv]{PrSv} N.V.~Prokof'ev and B.V.~Svistunov, {\it Two
definitions of superfluid density},
Phys.\ Rev.\ B {\bf 61}, 11282 (2000).



\bibitem[R]{RBREWC}
A.M.\ Rey, K.\ Burnett, R.\ Roth, M.\ Edwards, C.J.\ Williams, C.W.\ Clark,
{\it Bogoliubov approach to superfluidity of atoms in an
optical lattice},
J. Phys. B {\bf 36}, 825--841 (2003).


\bibitem[Ro]{roepstorff} G.~Roepstorff,
{\it Bounds for a Bose condensate in dimensions $\nu\geq 3$},
J. Stat. Phys. {\bf 18}, 191 (1978).


\bibitem[Ru]{ruelle} D.\ Ruelle, {\it Statistical Mechanics. Rigorous Results},
World Scientific (1999).


\bibitem[Sa]{Sa} S.\ Sachdev, {\it Quantum Phase Transitions}, Cambridge
University Press (1999).



\bibitem[S]{schick} M.\ Schick, \textit{ Two-Dimensional System of
Hard Core Bosons}, Phys. Rev.\ A \textbf{ 3}, 1067--1073 (1971).


\bibitem[SY]{SY} K. Schnee, J. Yngvason, {\it Bosons in disc-shaped
    traps: from 3D to 2D}, preprint, arXiv:math-ph/0510006,
  Commun. Math. Phys. (in press).


\bibitem[Sc]{schreck}  F.\ Schreck, {\it et al.}, {\it Quasipure
Bose-Einstein Condensate Immersed in a Fermi Sea},
Phys. Rev. Lett. {\bf 87}, 080403 (2001).


\bibitem[Se1]{S1999}
R.\ Seiringer, Diplom thesis, University of Vienna (1999).

\bibitem[Se2]{S4}
R.\ Seiringer, {\it Bosons in a Trap: Asymptotic Exactness of the
Gross-Pitaevskii Ground State Energy Formula}, in: {\it Partial
Differential Equations and Spectral Theory}, PDE2000 Conference in
Clausthal, Germany, M. Demuth and B.-W. Schulze, eds., 307--314,
Birkh\"auser (2001).


\bibitem[Se3]{rot1} R.\ Seiringer, {\it Gross-Pitaevskii Theory of the
    Rotating Bose Gas}, Commun. Math. Phys. {\bf 229}, 491--509
  (2002).

\bibitem[Se4]{rot2} R.\ Seiringer,
{\it Ground state asymptotics of a dilute, rotating gas},
J. Phys. A: Math. Gen. {\bf 36}, 9755--9778 (2003).

\bibitem[Se5]{SeirNorway} R.\ Seiringer, {\it Dilute, Trapped Bose Gases and Bose-Einstein Condensation},
  in: {\it Large Coulomb Systems}, Lect. Notes Phys. {\bf 695}, 251--276, J. Derezinski, H.
Siedentop, eds., Springer (2006).

\bibitem[Se6]{SfermiT} R.\ Seiringer, {\it The Thermodynamic Pressure
    of a Dilute Fermi Gas}, Commun. Math. Phys. {\bf 261}, 729--758 (2006).

\bibitem[Se7]{SboseT} R.\ Seiringer, {\it Free Energy of a Dilute Bose
    Gas: Lower Bound}, preprint, arXiv:math-ph/0608069.

\bibitem[Sh]{Shev}
S.I. Shevchenko, {\it On the theory of a Bose gas in a nonuniform
field}, Sov.\ J.\ Low Temp.\ Phys.\ {\bf 18}, 223--230 (1992).



\bibitem[Si1]{S79} B.\ Simon, {\it Trace ideals and their application},
Cambridge University Press (1979).


\bibitem[Si2]{simon}
B. Simon, {\it The classical limit of quantum partition functions},
Commun. Math. Phys. {\bf 71}, 247 (1980).


\bibitem[So]{So} J.P. Solovej, {\it Upper Bounds to the Ground
    State Energies of the One- and Two-Component Charged Bose Gases},
  Commun. Math. Phys. {\bf 266}, 797--818 (2006).


\bibitem[SR]{SR} L. Spruch, L. Rosenberg,
{\it Upper bounds on scattering lengths for static potentials},
Phys. Rev. {\bf 116}, 1034 (1959).


\bibitem[Su]{suto} A.~S\"ut\H o,
 {\it Bose-Einstein condensation and symmetry breaking}, Phys. Rev. A {\bf 71}, 023602 (2005).


\bibitem[Su2]{suto2} A.~S\"ut\H o,
{\it Equivalence of Bose-Einstein Condensation and Symmetry Breaking}, Phys. Rev. Lett. {\bf 94}, 080402 (2005).


\bibitem[T]{TE} G.\ Temple, \textit{The theory of Rayleigh's
Principle as
Applied to Continuous Systems}, Proc.\ Roy.\ Soc.\ London A
\textbf{119},  276--293 (1928).


\bibitem[Th]{thir}
W.~Thirring,
{\it A lower bound with the best possible constants for Coulomb Hamiltonians},
Commun. Math. Phys. {\bf 79}, 1 (1987).



\bibitem[TT]{TT} D.R.~Tilley and J.~Tilley,  {\it Superfluidity and
Superconductivity}, third edition,
Adam Hilger, Bristol and New York (1990).

\bibitem[W]{b8} G. Wentzel,
{\it Thermodynamically Equivalent Hamiltonian for Some Many-Body Problems},
Phys. Rev. {\bf 120}, 1572 (1960).


\bibitem[Yu]{Yu} V.I. Yukalov,
{\it Principal problems in Bose-Einstein condensation of dilute gases},
Laser Phys. Lett. {\bf 1}, 435--461 (2004).


\bibitem[ZB]{zagrebnov} V.A.~Zagrebnov, J.B.~Bru {\it The Bogoliubov model of
weakly interacting Bose gas}, Phys. Reports {\bf 350}, 291--434 (2001).


\bibitem[ZFG]{feng}  W.-M. Zhang, D.H. Feng, R. Gilmore,
 {\it Coherent states: Theory and some applications},
Rev. Mod. Phys. {\bf 62}, 867 (1990).


\bibitem[Z1]{Ziegler} K. Ziegler,
{\it Phase Transition of a Bose Gas in
an Optical Lattice},
Laser Physics {\bf 13}, 587--593 (2003).


\bibitem[Z2]{Zi}
K.\ Ziegler,
{\it Two-component Bose gas in an optical lattice at single-particle
filling},
Phys.\  Rev.\  A {\bf 68}, 053602 (2003).


\bibitem[Zw]{Z} W.\ Zwerger,
{\it Mott-Hubbard transition of cold atoms in
optical lattices},
Journal of Optics B {\bf 5}, 9--16 (2003).


\end{thebibliography}
\end{document}